%% file: YourName-thesis.tex
\documentclass[11pt,          
               phd,           
               onehalfspacing, 
               ]{ncsuthesis}

\usepackage{booktabs}  
\usepackage{amsmath}
\usepackage{textcomp}  
\usepackage{lipsum}    
\usepackage{braket}
\usepackage{aas_macros}
\usepackage{pdflscape}

\RequirePackage[
			style=alphabetic,
			sorting=nyt,
			hyperref=true, %
			firstinits=true,%
			backend=bibtex,
			natbib=true,
			url=false,
			isbn=false,
			maxnames=2, 
			maxalphanames=1, 
			doi=false]{biblatex}

    \renewcommand*{\intitlepunct}{}
    \DefineBibliographyStrings{german}{in={}}
    \DefineBibliographyStrings{english}{in={}}
    \DeclareNameAlias{sortname}{last-first}
    \DeclareNameAlias{default}{last-first}
	
\DeclareFieldFormat[article,periodical]{volume}{\mkbibbold{#1}}
\makeatletter

\newrobustcmd*{\parentexttrack}[1]{%
  \begingroup
  \blx@blxinit
  \blx@setsfcodes
  \blx@bibopenparen#1\blx@bibcloseparen
  \endgroup}

\AtEveryCite{%
  }

\makeatother
\renewcommand{\cite}[1]{\parencite{#1}}

\renewbibmacro{in:}{%
  \ifentrytype{article}{}{%
  \printtext{\bibstring{in}\intitlepunct}}}
  
\AtEveryBibitem{\clearfield{month}}

\AtEveryBibitem{\clearfield{language}}

 \defbibheading{myheading}[BIBLIOGRAPHY]{
 \chapter*{#1}
 \markboth{#1}{#1}}

\usepackage{amsmath,amssymb,amsfonts} 
\usepackage{dcolumn}
\usepackage{bm}
\usepackage{cancel}
\usepackage{verbatim}
\usepackage{ifthen}
\usepackage{url}
\usepackage{sectsty}
\usepackage{balance} 
\usepackage{graphicx} 
\usepackage{lastpage}
\usepackage[format=plain,justification=RaggedRight,singlelinecheck=false,font=small,labelfont=bf,labelsep=space]{caption} 
\usepackage{fancyhdr}
\usepackage{dirtytalk}
\usepackage{threeparttable}
\usepackage{siunitx}
\usepackage{booktabs}
\usepackage{etoolbox}
\usepackage{braket}
\usepackage{threeparttable}
\usepackage{subcaption}
\usepackage{multirow}
\usepackage{comment}

\usepackage[ruled,vlined]{algorithm2e}
\usepackage{abbrevs}
\usepackage{etoolbox}
\robustify{\DateMark} 

\usepackage{units} 

\usepackage[sharp]{easylist} 

\usepackage[table]{xcolor}
\usepackage{array,booktabs}
\usepackage{colortbl}
\newcolumntype{L}{@{}>{\kern\tabcolsep}l<{\kern\tabcolsep}}

\usepackage{titlesec}

\titleformat{\paragraph}
{\normalfont\normalsize\bfseries}{\theparagraph}{1em}{}
\titlespacing*{\paragraph}
{0pt}{3.25ex plus 1ex minus .2ex}{1.5ex plus .2ex}

\usepackage[section]{placeins}

\dispositionformat{\bfseries}            

\headingformat{\large\MakeUppercase}   

\include{optional}
\committeesize{4}

\chair{Richard Longland}
\memberI{Arthur Champagne}
\memberII{Paul Huffman}
\memberIII{Gail McLaughlin}   

\student{Caleb A.}{Marshall} 

\program{Physics}

\thesistitle{Investigating Globular Cluster Elemental Abundance Anomalies Using $(^3\textnormal{He},d)$ Proton Transfer Reactions}

\usepackage{color}

\graphicspath{{./Chapter-1/figs/}{./Chapter-2/figs/}{./Chapter-3/figs/}{./Chapter-4/figs/}{./Chapter-5/figs/}{./Chapter-6/figs/}}

\usepackage{calc}
\newlength{\chaptercapitalheight}
\newlength{\chapterfootskip}
\renewcommand{\bibname}{BIBLIOGRAPHY}

\newlength\graphht

\usepackage{pdflscape}
\usepackage{tikz}
\fancypagestyle{lscapedplain}{%
  \fancyhf{}
  \fancyfoot{%
    \tikz[remember picture,overlay]
      \node[outer sep=1cm,above,rotate=90] at (current page.east) {\thepage};}

}

\begin{document}
\pagestyle{plain}

\frontmatter

\include{front}

\mainmatter

\pagestyle{plain}
\include{Chapter-1/Chapter-1}

\include{Chapter-2/Chapter-2}
\include{Chapter-3/Chapter-3}
\include{Chapter-4/Chapter-4}
\include{Chapter-5/Chapter-5}
\include{Chapter-6/Chapter-6}
\include{Chapter-7/Chapter-7}


%
%



\begin{spacing}{1}
 \setlength\bibitemsep{11pt} 
 \phantomsection
 \addcontentsline{toc}{chapter}{{\uppercase{\bibname}}} 
\titleformat{\chapter}[display]{\bf\filcenter
}{\chaptertitlename\ \thechapter}{11pt}{\bf\filcenter}
\titlespacing*{\chapter}{0pt}{0.0in-9pt}{22pt}

\printbibliography[heading=myheading]
\end{spacing}

\restoregeometry
\appendix

\include{Appendix-A/Appendix-A}

\restoregeometry

\backmatter

\end{document}

%% file: optional.tex
\usepackage[Rejne]{fncychap}

\usepackage[
  pdfauthor={Caleb Marshall},
  pdftitle={Stars and Stuff},
  pdfcreator={pdftex},
  pdfsubject={NC State ETD Thesis},
  colorlinks=true,
  linkcolor=black,
  citecolor=black,
  filecolor=black,
  urlcolor=black,
]{hyperref}

\usepackage{microtype}

\usepackage[T1]{fontenc}
\usepackage{times}

\usepackage{longtable}

%% file: front.tex
\begin{abstract}

Globular clusters are dense aggregates of stars that evolve in relative isolation. For the better part of $40$ years these clusters have been known to possess unique chemical signatures called abundance anomalies. Recent observations have found these abundance anomalies to be the result of distinct stellar populations with the youngest population undergoing an unknown enrichment process. Understanding these chemical signatures requires a precise understanding of the thermonuclear reaction rates at relatively low temperatures. At these low temperatures reaction rates suffer from large uncertainties arising from poorly understood resonances in several key reactions. Transfer reactions provide key constrains on nuclear inputs for these resonances. This thesis presents an updated understanding of the sodium and potassium destroying reactions $^{23}$Na$(p, \gamma)$ and $^{39}$K$(p, \gamma)$, respectively. A newly reevaluated rate for $^{39}$K$(p, \gamma)$ indicates that it is less precisely known than previously thought, and future experimental study is needed to reduce its impact on globular cluster nucleosynthesis. Novel Bayesian techniques are discussed that help accurately assess the uncertainties arising from transfer measurements. These techniques are applied to the transfer reaction $^{23}$Na$(^3\textnormal{He}, d)$, which was carried out at Triangle Universities Nuclear Laboratory. Results of this experiment indicate that the energy of an important resonance in $^{23}$Na$(p, \gamma)$ is much lower than previously thought at $E_r =132(3)$ keV. The transfer measurement also indicates tension between previous direct studies of the resonance strength and the current transfer measurement. The impact of these uncertainties on the $^{23}$Na$(p, \gamma)$ reaction rate is investigated, and it is shown that this rate requires more intensive study to provide the precision needed to constrain nucleosynthesis in globular clusters.

\end{abstract}

\makecopyrightpage

\maketitlepage

\begin{dedication}
 \centering To Ma, Pa, and Rachel.
\end{dedication}

\begin{biography}
Caleb Marshall was born in Sanford, N.C in 1992 to parents Andrew and Bonnie Marshall. His mother, a middle and high school science teacher, fostered an interest in science from an early age. During the last years of high school, the connections between mathematics and physical laws transformed his lifelong interest into a firm career goal. He attended North Carolina State university to study physics, and graduated in 2014. Soon thereafter, he continued his graduate studies at the same institution, fully committed to the study of nuclear astrophysics.      
\end{biography}

\begin{acknowledgements}

First I would like to thank my advisor, Richard Longland. Richard, though he was perhaps unaware of it at the time, took a chance hiring me fresh off of a rather unspectacular undergraduate career. From day one, he always treated me with respect and understanding, which allowed me to develop the confidence necessary to carry out my research. His hands on approach during the long and seemingly never-ending recommissioning of the spectrograph was probably the best introduction to experimental nuclear physics anyone could hope for.  

Although she was present for an all too brief period of time during my graduate school experience, I would, without prompting, refer to Kiana Setoodehnia as my second adviser (of course this designation carries no official recognition). Her experience and work ethic are staggering, and for the life of me I cannot actually recall what the lab was like before her arrival in 2015. I am thankful to her for showing me the ends and outs of FRESCO, which really set the stage for most of the work in this thesis. It was a pleasure to work with her.       

Special thanks are in order for John Kelley, who always took time out of his day to help me with any nuclear data question I might have. John's help with the DAQ was invaluable, and he taught me a lot about how to set up electronics, think about signal processing, and trouble shoot silicon detectors. He was also kind enough to sit on shift on multiple occasions, and to retrieve my lab notebooks in the final days of writing when the pandemic prevented me from having access to the lab.  

I would also like to thank the UNC astrophysics group: Christian Iliadis, Art Champagne, Tom Clegg, and more recently Robert Janssens. Christian was instrumental in helping me develop an understanding of Bayesian techniques, and his careful approach to research has served as a template for me, allowing me to curb my natural tendencies towards disorder and chaos. Art's guidance on the focal plane detector was key, and his insights into the field are truly unique. It was not an uncommon occurrence during the writing of this document for me to think back to a comment that Art made and come to a sudden realization that he had expressed an idea in a single sentence that I had just written paragraphs about. Although I had fewer chances to work with Tom than I would have liked, his enthusiasm is one of a kind. This enthusiasm was responsible for showing me how to overcome my cynicism and press forward with an experiment in the face of numerous setbacks. Robert has offered invaluable science and career advice. While everyone I have mentioned and will mention has graciously offered me their time, Robert's willingness to give advice on any issue, at seemingly any time, is unprecedented.        

None of this experimental work would have been possible without the TUNL technical staff: Tom Calisto, Bret Carlin, John Dunham, Richard O'Quinn, and Chris Westerfeldt. During night shifts I always looked forward to the moment John would arrive at the lab in the morning. It signaled both the end of my shift, and the relief that the experts had finally arrived to deal with the equipment that, seemingly without exception, always broke at 3 a.m. Bret helped create an excellent $\Delta E$ preamp, fixed the TPS on multiple occasions, and tolerated the constant influx of burnt out corona control boards. Chris was instrumental in helping our group better understand the $90 \text{-} 90$ system, and holding the lab together no matter the technical challenges. Richard taught me nearly everything I know about operating the tandem, and Tom was essential in fabricating and designing many of the parts needed to successfully run these experiments. The technical staff are the heart of the lab, and their kindness and enthusiasm are a large part of why TUNL is such a special place.        

There is not enough space to list all of the graduate students and friends that helped me during this time. I am truly thankful to all of them.  

Finally, I would like to thank my family. In particular, my parents have always provided me their love and support. My older sister, Rachel, has always been an inspiration for me, and I have learned as much from her as anybody mentioned before.
\\
\\
\\
\\
\textit{THERE IS NOTHING INNATELY IMPRESSIVE ABOUT THE UNIVERSE OR ANYTHING IN IT.} - Thomas Ligotti
\end{acknowledgements}

\thesistableofcontents

\thesislistoftables

\thesislistoffigures

%% file: Chapter-1/Chapter-1.tex
\chapter{INTRODUCTION}
\label{chap:astro}


The pioneering work of Burbidge, Burbidge, Fowler, and Hoyle \cite{b2fh} and independently Cameron \cite{cameron} established the field of nuclear astrophysics. These works came to the realization that the chemical elements we find in the Solar System are the remnants of nuclear burning happening inside of stars. While understanding the origin of the elements is still one of the primary questions of the field, this thesis will focus on another implication of stellar nucleosynthesis. Specifically: observed elemental abundances in the cosmos are the unique signatures of nuclear burning process. In this way, nuclear physics becomes an additional quantitative tool to gain insight into astronomical observations. By studying nuclear reactions in the laboratory terrestrially, we can answer questions about stellar phenomena that would otherwise be inaccessible to us. However, at the energies found in stellar plasmas the coulomb barrier dominates, making direct study of these reactions exceptionally difficult. Thus, in order to fully understand stellar nucleosynthesis, a variety of experimental approaches must be used to construct a composite picture of these nuclear reactions.             
The work presented in this thesis examines the use of nuclear transfer reactions to constrain the nuclear reactions responsible for destroying sodium and potassium in globular clusters. The present chapter will give an overview of the observation evidence for elemental abundance anomalies in globular clusters and discuss their importance for our theories of stellar evolution. Chapter~\ref{chap:reactions} will provide the necessary details for how nuclear reactions occur in stars and how we can better understand them using transfer reactions, Chapter~\ref{chap:nuclear_unc} will show how the uncertainties that naturally arise from nuclear physics experiments impact the astrophysical predictions, Chapter~\ref{chap:tunl} details the experimental methods used in this work, Chapter~\ref{chap:bay_dwba} presents novel Bayesian methods that were developed to quantify uncertainties from transfer reactions, and Chapter~\ref{chap:sodium} summarizes the analysis and results of the $^{23}$Na$(^3 \textnormal{He}, d)$ transfer reaction.

\section{Globular Cluster Abundance Anomalies}
\label{sec:abund_anomalies}

Of all the observed astronomical objects, globular clusters are perhaps the closest we can get to a laboratory-like environment in the galaxy. They are some of the brightest and oldest objects in our sky, which consist of hundreds of thousands of stars that are gravitationally bound within $\sim 100$ parsecs. These facts mean that we have a relatively easy to observe object that has a large, isolated population of stars. If our theories of stellar evolution and intra-cluster dynamics were complete, then any observed property of a cluster could be explained by just a few parameters \cite{gratton_2010}. At this time we have no such theory, and many properties of globular clusters remain uncertain.

\begin{figure}
    \centering
    \includegraphics[width=.7\textwidth]{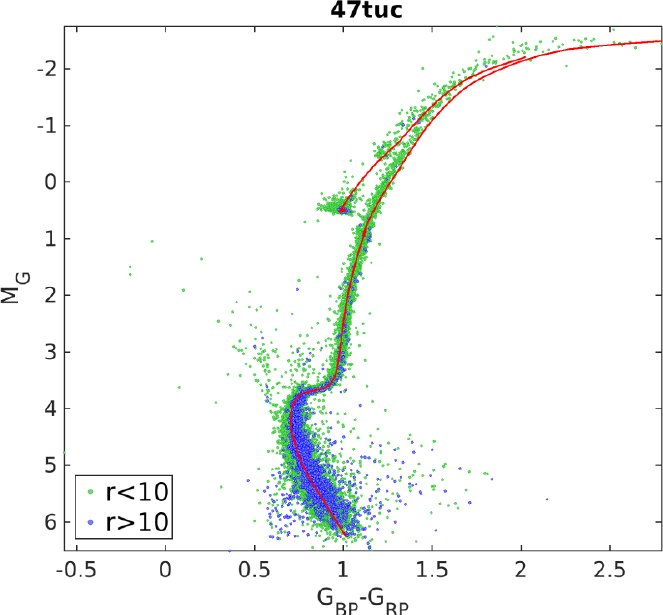}
    \caption{HR diagram for the globular cluster 47 Tuc. Plot is taken from Ref.~\cite{gaia_2018}. The red line is a calculated isochrone. The green and blue dots represent the inner and outer regions of the cluster.}
    \label{fig:hr_diagram}
\end{figure}

Since the 1980's, the simple perspective on globular clusters mentioned above has shifted significantly. Prior to detailed spectroscopy of the stars within the cluster, it was thought that these objects constituted a \textit{single stellar population}. Under this assumption, each star in the globular cluster would form at the same time from gas of similar chemical composition \cite{kraft_1979}. 
Operating under the assumption of a single stellar population, all of the observed properties of a globular cluster would arise from the starting composition and age of the cluster. If we look at the \textit{Hertzsprung-Russell diagram} (HR diagram) in Figure~\ref{fig:hr_diagram}, and assume a single stellar population, we could deduce the current evolutionary stage of any star in the cluster as a function of its initial mass. Specifically, more massive stars burn through their nuclear fuel more quickly, and are therefore more advanced in their evolution at the particular moment in time. Stars formed at the same time, but with different initial masses, can be compared to a calculated isochrone, i.e, the curve through the HR diagram that shows stars of the same age but different initial mass, as seen in Figure~\ref{fig:hr_diagram}. The relatively good agreement between the isochrone and observations demonstrate that the concept of a single stellar population does hold some merit for globular clusters. Further conclusions can be drawn using similar methods, most notably that the cluster can be used as a lower bound on the age of the universe \cite{Jimenez_1998}.    

\begin{figure}
    \centering
    \includegraphics[width=.7\textwidth]{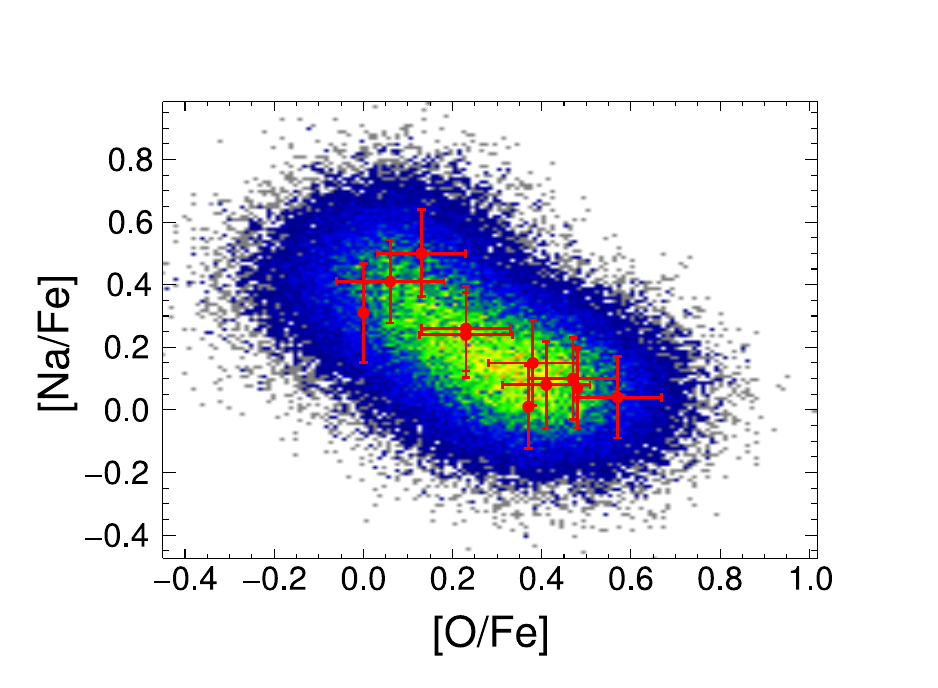}
    \caption{The observed Na-O anticorrelation for 13 stars on the RGB from 47 Tuc from Ref.~\cite{2014_Thygesen}. The red points are the observed abundance while the green blue contour shows the probability density inferred from these points using a bootstrap method.}
    \label{fig:na_o_tuc}
\end{figure}

Despite the success of the single population hypothesis, 
advances in high resolution spectroscopy have proven it to be an inadequate theory. Evidence against this hypothesis started to accumulate with the observation that a significant enhancement of sodium was present in some members of the red giant branch (RGB) in the cluster M13 \cite{perterson_1980}. Continued observational work eventually confirmed that globular clusters, to varying degrees, had specific star-to-star correlations and anticorrelations between light elements \cite{kraft_1994}. In particular, the anticorrelation between sodium and oxygen has been observed in every globular cluster that it has been looked for \cite{gratton_2004}. An example of this anticorrelation is shown in Fig.~\ref{fig:na_o_tuc}. These Na-O anticorrelations were originally dubbed \textit{abundance anomalies}, since they could not be easily explained in the framework of a single stellar population. This enriched material could only come from stellar burning happening \textit{in-situ} or from some initial inhomogeneity in the cluster material.   

Whatever the astrophysical source of the observed anticorrelations, they are clear signatures of nuclear burning processes. They have been uniquely identified as being produced by hydrogen burning at elevated temperatures \cite{d_and_d_1989, langer_1993, kudryashov_1988}. Since stars on the RGB can not produce these temperatures, the enriched material must have been produced at another site prior to their observation.    

\begin{figure}
    \centering
    \includegraphics[width=\textwidth]{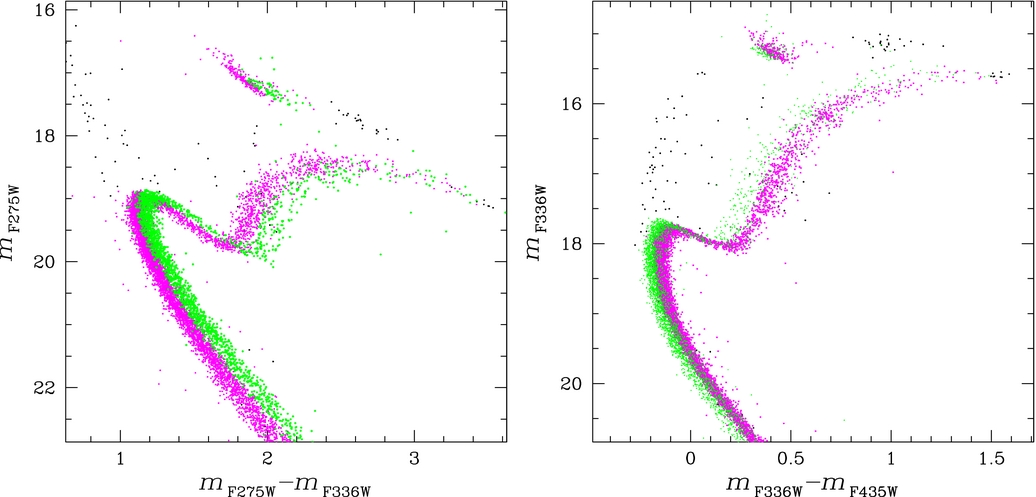}
    \caption{Figure is taken from Ref.~\cite{milone_2012}. HR diagrams of 47 Tuc. The x-axis shows the color index in terms of the filters used on the Hubble space telescope \cite{filters_2012}, while the y-axis shows the apparent magnitude. The two colors distinguish the multiple populations present within the cluster.}
    \label{fig:multi_ms}
\end{figure}

Continued advances in high resolution photometry led to the first unambiguous evidence of multiple stellar populations existing within globular clusters \cite{gratton_2012}. These measurements made it possible to resolve different tracks among the stars on the main sequence, with each track corresponding to a variation in helium content \cite{piotto_2007, villanova_2007}. Figure~\ref{fig:multi_ms} taken from Ref.~\cite{milone_2012} again shows 47 Tuc, but by using the high resolution photometric data, two distinct stellar populations can be identified. These observations unambiguously demonstrate that globular clusters are not simple stellar populations, but instead they possess multiple generations of stars, with the newest generation being enriched in light elements by some unknown polluter from the previous generation \cite{gratton_2012, gratton_2019}. 

These sets of observations have overturned the traditional view of globular clusters, and effectively open up new avenues of research. By trying to positively identify the source of the polluted material, determining how this material is ejected back into the cluster, and the dynamics of the second generation of stars, we have the opportunity to significantly advance our understanding of the formation of some of the oldest objects in our galaxy.   

{}

\section{Polluter Candidates and Nucleosynthesis}

As established in Section \ref{sec:abund_anomalies}, the enrichment of sodium in second generation stars is undoubtedly from hydrogen burning at elevated temperatures. Unfortunately, the astrophysical environment that can provide these temperatures and eject the processed material back into the cluster is unknown. Proposed environments include intermediate and massive Asymptotic Giant Branch (AGBs) Stars ($5 \textnormal{-} 9 \textnormal{M}_{\odot}$) \cite{Ventura_2001, dercole_2010}, fast rotating massive stars (FRMS, $ \geq 25 \textnormal{M}_{\odot}$) \cite{decressin_2007}, and very massive stars (VMS, $ \geq 10^4 \textnormal{M}_{\odot}$) \cite{denissenkov_2014, denissenkov_2015}.
While all of these environments can replicate the Na-O anti-correlation, they also share a common issue by overproducing He relative to Na \cite{problems_with_hbb, renzini_2015}.  

\begin{figure}
    \centering
    \includegraphics[width=.7\textwidth]{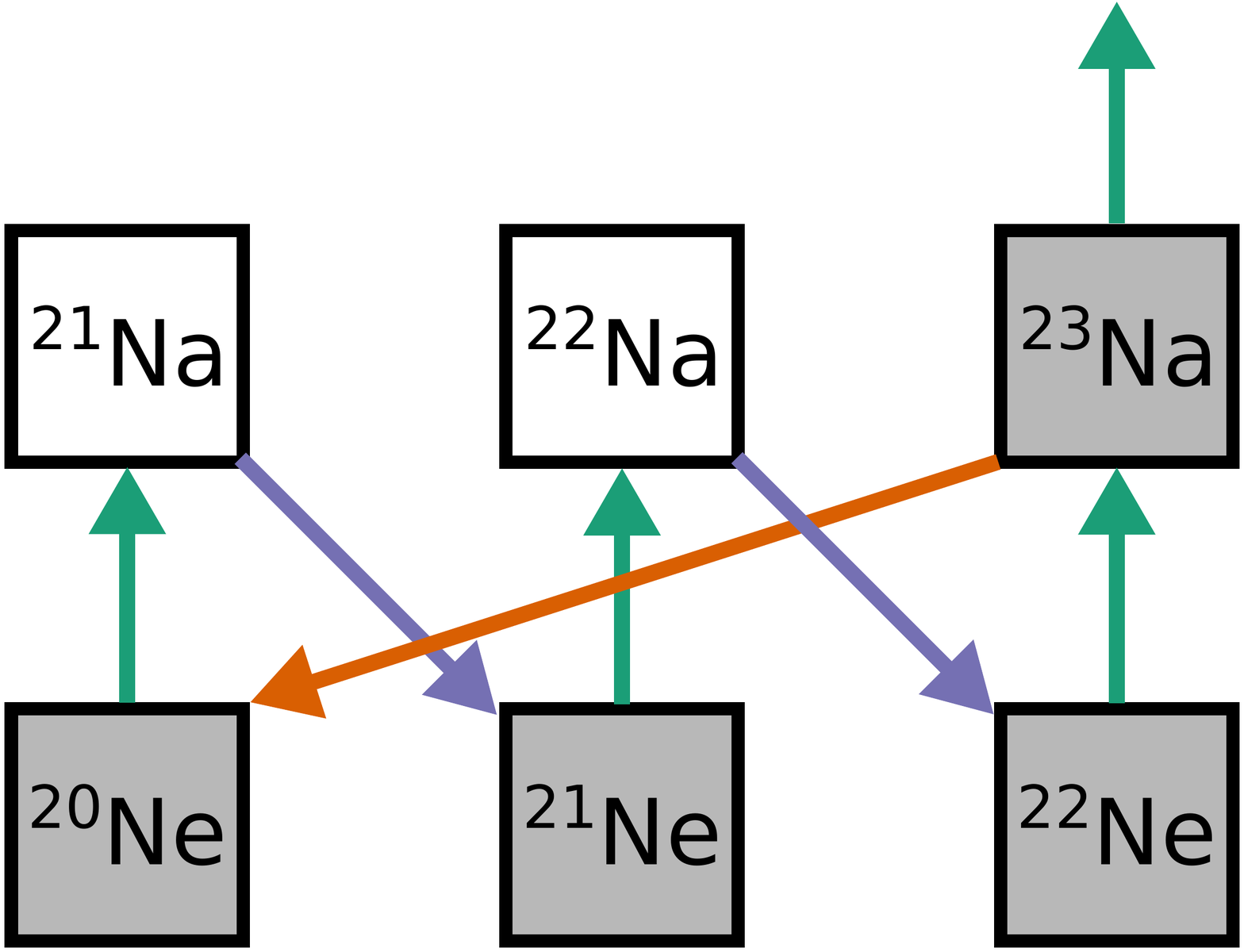}
    \caption{Illustration of the Ne-Na cycle. The $(p,\gamma)$, $\beta^{+}$, and $(p, \alpha)$ reactions and decays are shown in green, purple, and orange, respectively. Stable isotopes are shaded in grey.}
    \label{fig:ne_na_cycle}
\end{figure}

By examining different burning conditions, i.e., temperature and time, and holding density and metallicity constant, Ref.~\cite{prantzos_2017} found that all of the observed correlations and anti-correlations in NGC 2808, including the Na-O anti-correlation, were only reproducible at temperatures of $\textnormal{T} \sim 70-80$ MK. These results are notably independent from the complications present in more advanced models such as convection and mixing \cite{Ventura_2005}. As such, their results provide an excellent starting point for nuclear physicists looking to identify key reactions to study. In this narrow temperature regime, any oxygen destroyed via $^{17}$O$(p, \alpha)^{14}$N or $^{18}$O$(p, \alpha)^{15}$N will become trapped in the main CNO cycle, thereby causing the overall oxygen abundance to drop. Meanwhile, proton captures on stable isotopes of Ne lead to a series of reactions called the (Ne-Na cycle). These series of reactions and decays are give by: 
\begin{equation}
    ^{20}\textnormal{Ne}(p, \gamma)^{21}\textnormal{Na}(\beta^+)^{21}\textnormal{Ne}(p, \gamma)^{22}\textnormal{Na}(\beta^+)^{22}\textnormal{Ne}(p, \gamma)^{23}\textnormal{Na}(p, \alpha)^{20} \textnormal{Ne}, 
\end{equation}{}
and are presented in Figure~\ref{fig:ne_na_cycle}. The amount of material trapped in this cycle depends sensitively on the strength of $^{23}\textnormal{Na}(p, \alpha)^{20} \textnormal{Ne}$ versus that of $^{23}\textnormal{Na}(p, \gamma)^{24} \textnormal{Mg}$. H burning at these temperatures makes it impossible for material that has been turned into $^{24}$\textnormal{Mg} to reenter the Ne-Na cycle. These two processes, the destruction of oxygen and the conversion of Ne into Na happen concurrently causing the observed anticorrelation, but it should be noted that material does not flow between these two mass regions because of the $^{19}$F$(p, \alpha)^{16}$O reaction trapping material in the CNO region.         

{}

\section{NGC 2419 and K Enrichment}

\begin{figure}
    \centering
    \hspace*{-1.5cm}
    \includegraphics[width=.7\textwidth]{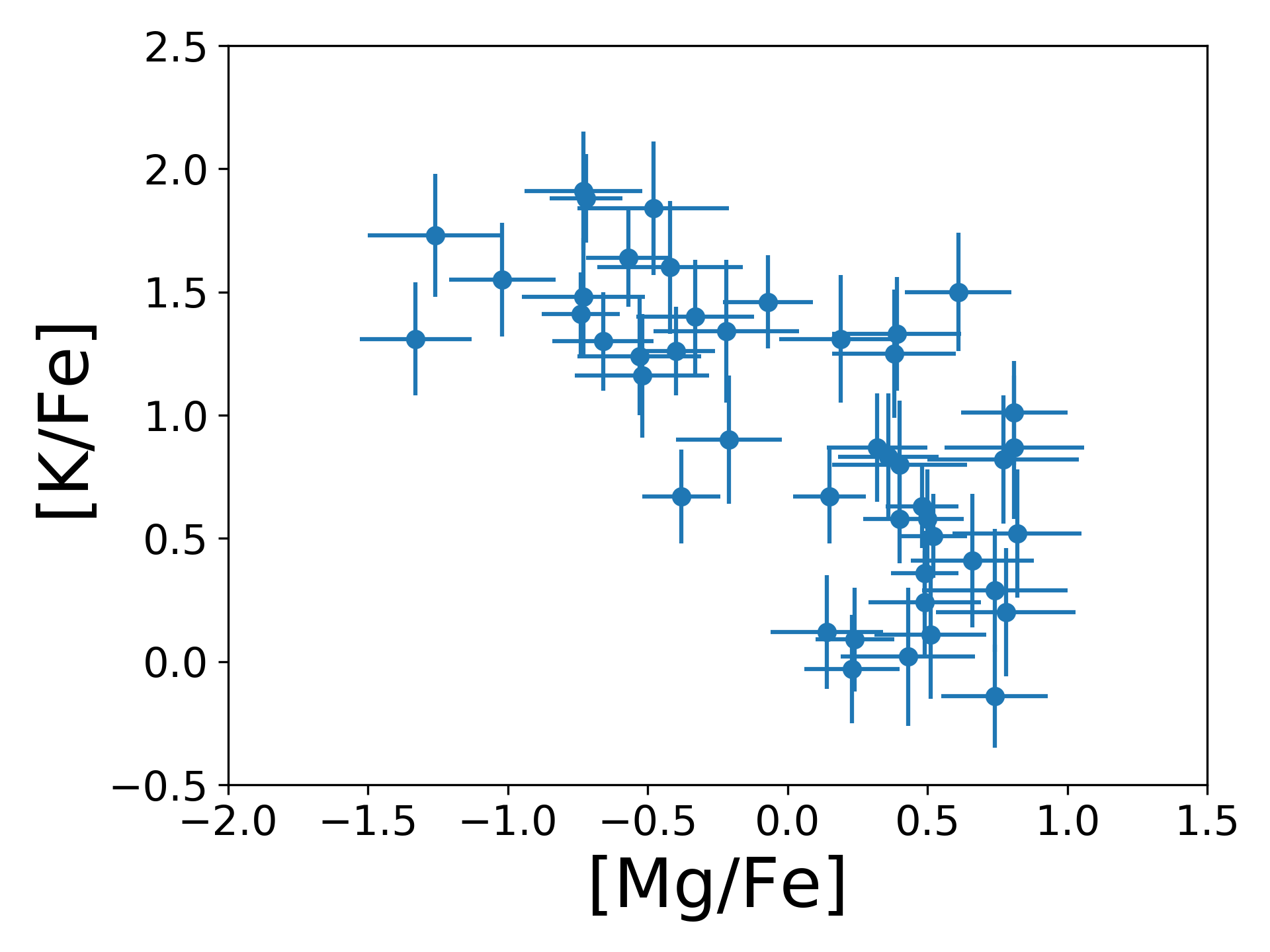}
    \caption{Plot of the potassium versus magnesium abundances observed in NGC 2419 \cite{mucciarelli_2012}. Approximately $40 \%$ of the stars within the cluster are shown to have a significant enrichment of potassium correlated with a depletion of magnesium.}
    \label{fig:mg_k_abund}
\end{figure}

The Na-O anti-correlation, as discussed above, is a common feature in every observed globular cluster, but there are rich varieties of phenomena that are specific to individual or a limited number of clusters. One of the most recently observed, and most puzzling, is the evidence of K enrichment in NGC 2419 \cite{cohen_2011, cohen_2012}. The full correlation was established in \cite{mucciarelli_2012}, and Figure \ref{fig:mg_k_abund} shows the potassium versus magnesium abundances from \cite{mucciarelli_2012}. 

{}

So far observations point towards NGC 2419 being a unique case. A small portion of the stars observed in NGC 2808 have shown a depletion of Mg correlated with an enhancement of K \cite{mucciarelli_2015}, but to date, there have been no other clear determinations of a K-Mg anti-correlation in any cluster \cite{mucciarelli_2017}. Interestingly, while the degree at which K and Mg are anti-correlated in NGC 2419 is unprecedented, similar trends have been reported in many field stars \cite{kemp_2018}. These observations raise important questions with regards to galactic chemical evolution, which our current stellar models are not able to match \cite{timmes_1995, romano_2010}. This could point to some common mechanism that is not isolated to clusters. However, even if the K-Mg anticorrelation is a cluster specific phenomenon, its study shares the same promises as the study of other abundance anomalies: a greater understanding of both nucleosynthesis in globular clusters and their multiple stellar populations. 

\begin{figure}
    \centering
    \includegraphics[width=.7\textwidth]{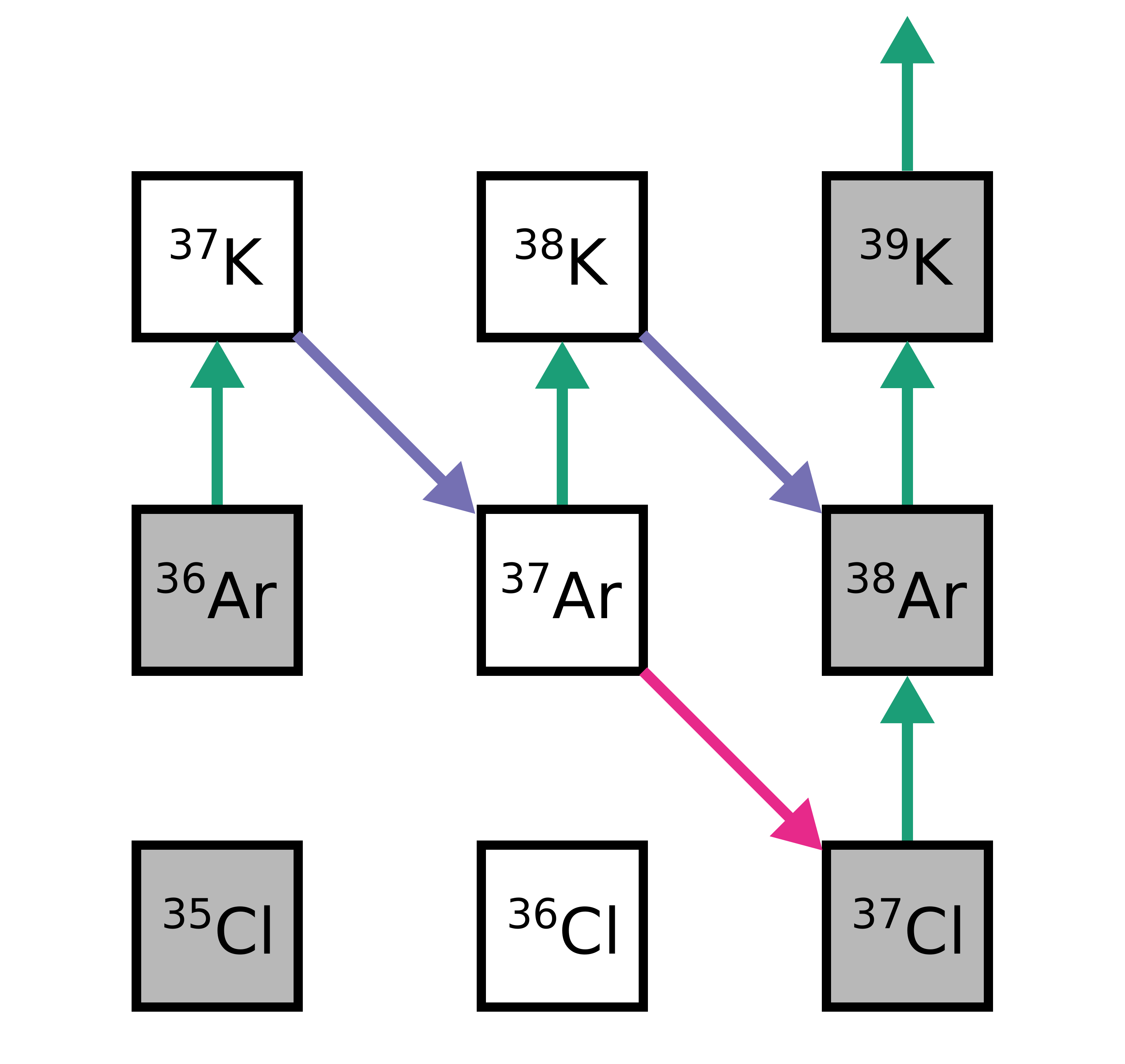}
    \caption{A pictorial representation of the reactions responsible for the synthesis of $^{39}$K due to elevated H-burning. $(p, \gamma)$, $\beta^+$, and electron capture $(e^-)$ are shown in green, purple, and pink, respectively. Although the electron capture on $^{37}$Ar is possible, its rate is sufficiently low which makes its contribution to the final abundance of $^{39}$K negligible.}
    \label{fig:39K_net}
\end{figure}

Regardless of how widespread this chemical signature is, its origin is unknown. Without appealing to any specific astrophysical environment, it is clear that in order for K to be synthesized,  H-burning at temperatures above those of the case for Na enrichment must be present. At these elevated temperatures a series of proton captures will lead to the following chain of reactions, shown in Fig.~\ref{fig:39K_net}:

\begin{equation}
\label{eq:k_reacs}
        ^{36}\textnormal{Ar}(p, \gamma)^{37}\textnormal{K}(\beta^+)^{37}\textnormal{Ar}(p, \gamma)^{38}\textnormal{K}(\beta^+)^{38}\textnormal{Ar}(p, \gamma)^{39}\textnormal{K}.
\end{equation}{}

Unlike the sodium production case discussed before, this chain of reactions does not form a cycle, due to the negligible contribution of $^{39} \textnormal{K} (p, \alpha) ^{36} \textnormal{Ar}$. It is also the case that $^{37} \textnormal{Ar}$ is unstable to electron capture, and has a half-life of $35$ days. This half-life compared with the relative strength of 
$^{37}\textnormal{Ar}(p, \gamma)^{38}\textnormal{K}$ means that this decay has little effect on the synthesis of $^{39} \textnormal{K}$.

{}

So far, the temperatures required for the sequence of reactions in Eq.~\ref{eq:k_reacs} have only produced a handful of potential polluter candidates. The scenario of hot bottom burning in AGB stars and super AGB stars ($5 \textnormal{-} 9 \textnormal{M}_{\odot}$) was proposed in Ref.~\cite{ventura_2012}. The authors noted that if the reaction rate of $^{38}\textnormal{Ar}(p, \gamma)^{39}\textnormal{K}$ was a factor of 100 times larger, then the observed levels of K could be explained by this scenario. An interesting idea was put forth in Ref.~\cite{carretta_2013}. The authors of this study hypothesize that because of the unique nature of NGC 2419, a similarly rare event could be a good candidate. They suggest  a Pair-instability Supernova ($140 \textnormal{-} 260 \textnormal{M}_{\odot}$) as a sufficiently rare phenomenon capable of producing K. However, this scenario is expected to have an odd-even effect \cite{woosley_2002}, i.e., that nuclei with even $Z$ are expected in much greater abundance than those with odd $Z$. This is not the case in NGC 2419, which presents major complications for this scenario.  

\begin{figure}
    \centering
    \includegraphics{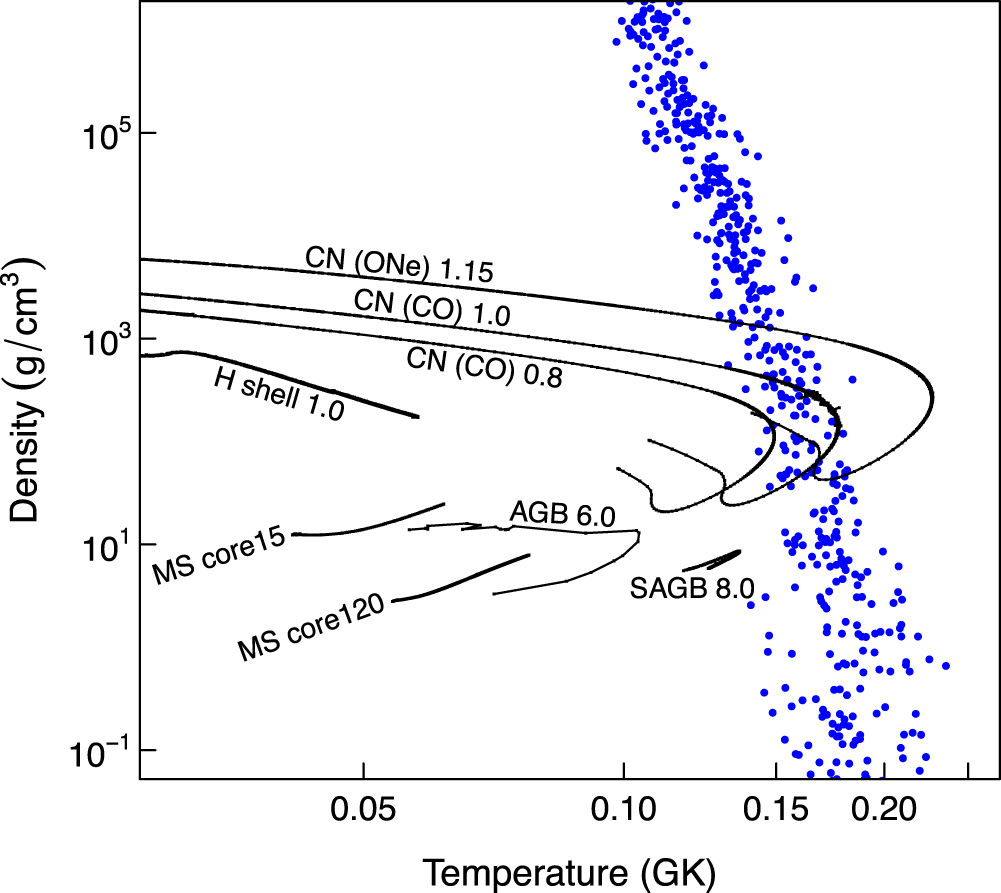}
    \caption{Monte-Carlo samples capable of reproducing the elemental abundances observed in NGC 2419 plotted as a function of their temperature and density. Temperature and density conditions of some of the proposed polluter environments are plotted and labelled. Plot is taken from \cite{iliadis_2016}. }
    \label{fig:trajectory}
\end{figure}

Instead of approaching this problem by selecting a potential polluter, developing and running a stellar model, and then comparing to observation, Iliadis et al. \cite{iliadis_2016} approached the problem by remaining agnostic towards the source of the pollution. Their study used a Monte-Carlo method to sample temperature, density, amount of hydrogen burned, and each of the thermonuclear reaction rates (taken from \cite{sallaska_2013}). For each of these samples, a nuclear reaction network calculation was carried out. The output from each network was then mixed by some fractional amount, $f$, of unprocessed material according to:
\begin{equation}
    X_{\textnormal{mix}} = \frac{X_{\textnormal{proc}} + f X_{\textnormal{pris}}}{1+f},
    \label{eq:mixing}
\end{equation}{}
where $X$ is the mass fraction of the material, "proc" refers to the processed material, and "pris" is the pristine cluster material. Finally, after the material was mixed using fixed values of $f$, it was compared to the observations from \cite{mucciarelli_2012, cohen_2012}. Samples that were able to reproduce all the observed correlations of NGC 2419 were kept. Figure~\ref{fig:trajectory} shows the accepted solutions plotted as a function of their temperature and density, while the lines show the conditions present in some of the proposed polluters. 

{}

A further analysis in \cite{dermigny_2017} more closely examined the effect individual reaction rates had on the range of acceptable temperatures and densities. This study clearly defined the role nuclear physicists could play in resolving the anomalous chemical signature of NGC 2419. The authors concluded that only a handful of rates played a significant role in increasing the uncertainties in the astrophysical parameters. They are $^{30} \textnormal{Si} (p, \gamma)^{31} \textnormal{P}$, $^{37} \textnormal{Ar} (p, \gamma)^{38} \textnormal{K}$, $^{38} \textnormal{Ar} (p, \gamma)^{39} \textnormal{K}$, and $^{39} \textnormal{K} (p, \gamma)^{40} \textnormal{Ca}$. 

\section{Summary}

The above discussion highlights the interesting questions posed by the observation of multiple, chemically distinct stellar populations in globular clusters. Without a precise knowledge of the nuclear reactions responsible for the enriched material, it is impossible to pin down the stellar environment that produced it. This work will be concerned with what appear to be two distinct processes. In the case of the Na-O correlation, an unknown pollution mechanism 
appears to be common to nearly every globular cluster. Thus, it is a fundamental element to our understanding of these objects. The K-Mg anti-correlation is associated with a higher temperature burning environment, but is still capable of informing our knowledge of nucleosynthesis within these unique objects. Critical to both of these scenarios is a better understanding of how quickly these elements are destroyed through the $^{23} \textnormal{Na} (p, \gamma)^{24} \textnormal{Mg}$ and $^{39} \textnormal{K} (p, \gamma)^{40} \textnormal{Ca}$ reactions, respectively.

%% file: Chapter-2/Chapter-2.tex
\chapter{Nuclear Reactions and Astrophysics}
\label{chap:reactions}

\section{Introduction}

This chapter draws the link between the thermonuclear reactions occurring in stellar interiors and the nuclear properties we can measure in the lab. Special emphasis is placed on the theory underpinning single-nucleon transfer reactions. It is shown that these reactions provide single-particle information essential to constraining the stellar reaction rate.        

\section{Thermonuclear Reaction Rates}

For the majority of a star's life, it will steadily burn its nuclear fuel in order to balance the force of gravitational attraction between the stellar gas. This state of hydrostatic equilibrium is reached as the infalling gas begins to increase in temperature and density to a point where nuclear fusion reactions begin to occur   \cite{cauldrons_in_the_cosmos}. The probability of fusion occurring is determined by the nuclear cross section, which itself is a function of energy. Thus, in a thermal environment where particles have a distribution of energies, the rate of fusion, called the \textit{thermonuclear reaction rate}, requires a convolution of two probabilities: the energy distribution of the reacting particles and the nuclear cross section. Assuming the stellar interior is in thermal equilibrium, then the particles will have kinetic energies dictated by the Maxwell-Boltzmann distribution, and the stellar reaction rate between nuclei $a$ and $A$, denoted $\langle \sigma v \rangle_{aA}$, is given by \cite{iliadis_book}:   
\begin{equation}
    \label{eq:reaction_rate}
     \langle \sigma v \rangle_{aA} = \bigg(\frac{8}{\pi \mu_{aA}}\bigg)^{1/2} \frac{1}{(kT)^{3/2}} \int_0^{\infty} E \sigma(E) e^{-E/kT} dE.
\end{equation}{}
This equation is expressed in the center of mass frame where $\mu_{aA}$ is the reduced mass of particles $a$ and $A$, $E$ is their center of mass energy, $k$ is Boltzmann's constant, $T$ is the temperature of the stellar plasma, and $\sigma(E)$ is the nuclear cross section.

\begin{figure}
    \centering
    \includegraphics[width=.8\textwidth]{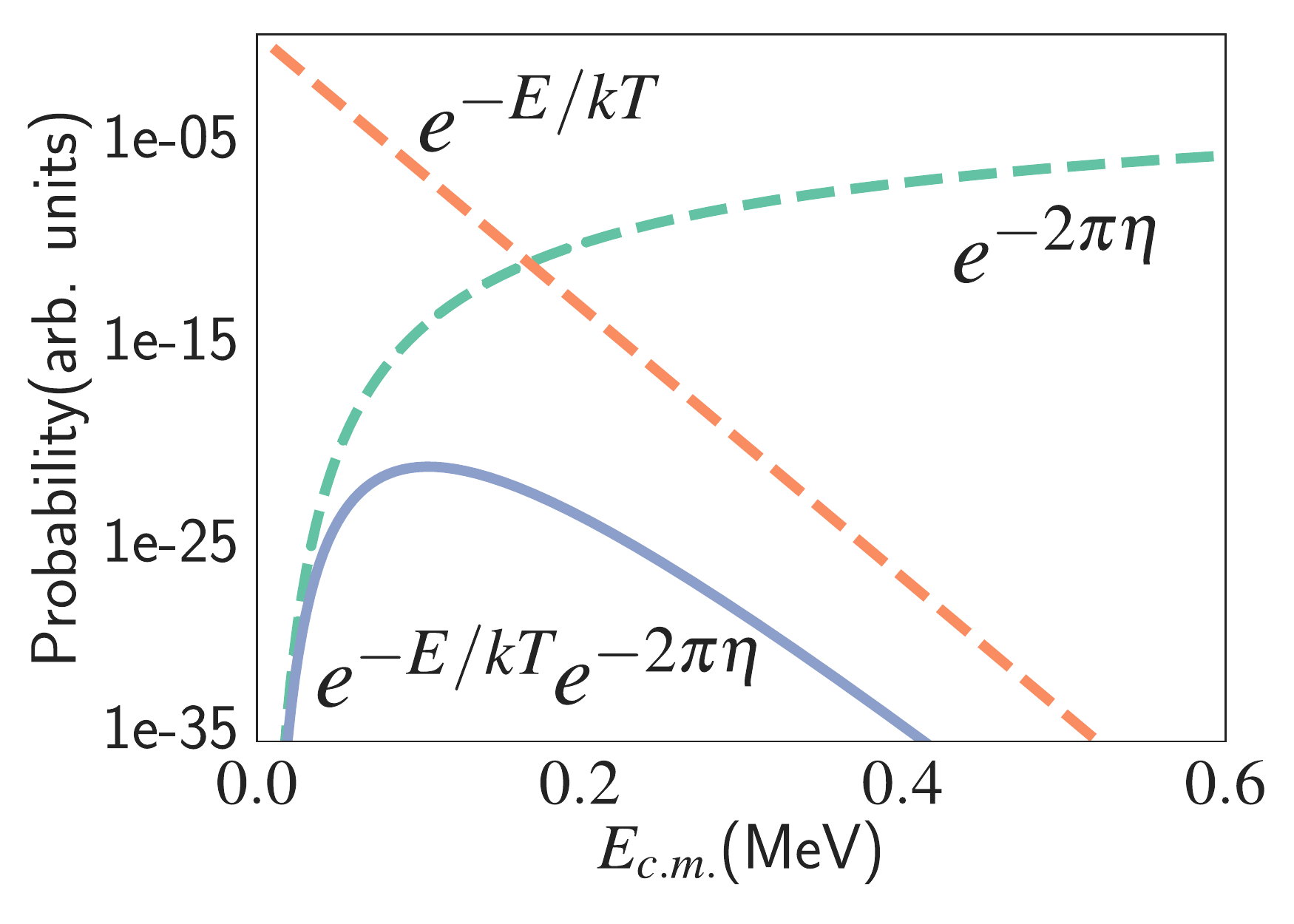}
    \caption{The Gamow peak for $^{23}$Na + $p$ reaction at $75$ MK.}
    \label{fig:gam_peak}
\end{figure}

Because the integral in Eq.~\ref{eq:reaction_rate} goes from $E = 0 \rightarrow \infty$, it is not immediately clear which energy range is relevant for a specific temperature. The peak of the Maxwell-Boltzmann distribution occurs at $E = kT$. For temperatures of $70 \text{-} 80$ MK relevant to globular cluster nucleosynthesis,  this peak occurs at $E \approx 6 \text{-} 7$ keV. However, at these low energies the coulomb repulsion between the positively charged ions will severely inhibit the reaction rate at the peak of the Maxwell-Boltzmann distribution. Classically, the reaction could only proceed once the particle has enough energy to overcome the coulomb barrier, but quantum mechanics gives a finite probability for the particle to tunnel through this barrier. This probability can be approximated for zero orbital angular momentum, i.e, $\ell =0 $, tunneling as:
\begin{equation}
    \label{eq:tunnel_prob}
    P = e^{-2 \pi \eta} , 
\end{equation}
where $P$ is called the Gamow factor and $\eta$ is the Sommerfeld parameter, which is defined as:
\begin{equation}
    \label{eq:sommerfeld}
    \eta =  \frac{1}{\hbar}\sqrt{\frac{\mu_{aA}}{2 E}} Z_a Z_A e^2.
\end{equation}
$Z_a$ and $Z_A$ are the atomic numbers of $a$ and $A$, respectively, while $e$ is the elementary charge. Multiplying the Gamow and the Boltzmann factors together gives a peaked probability distribution that approximates the energy range in which the majority of the nuclear reactions will occur. This peak is called the Gamow peak, and an example for $^{23}$Na + $p$ at $75$ MK is shown in Fig.~\ref{fig:gam_peak}. The center of this peak is around $\sim 100$ keV, showing that the effect of the Coulomb barrier is to sample from the high energy tail of the Maxwell-Boltzmann distribution.   

At the low stellar burning temperatures under consideration in this work, the behavior of $\sigma(E)$ within this narrow energy window is the dominant factor for calculating the thermonuclear reaction rate.   

\section{Resonant Reactions}
\label{sec:resonances}

At the relatively low center-of-mass energies where nuclear reactions in stellar interiors occur, the primary contributions to $\sigma(E)$ will come from direct radiative capture and resonant capture. The former plays a minor role in the $^{23}$Na$(p, \gamma)$ rate, but is not studied in this work. The latter is the primary mechanism by which both $^{23}$Na$(p, \gamma)$ and $^{39}$K$(p, \gamma)$ proceed.

Resonant reactions involve the formation of a compound nucleus. When the energy of the reacting particles is such that the center-of-mass energy matches an excited state of the compound system, there is an enhancement of the reaction cross section. This resonance energy is related to the excited state energy via:
\begin{equation}
    \label{eq:resonance_energy}
    E_r = E_x - Q,
\end{equation}
where $E_r$ is the resonance energy, $E_x$ is the excited state energy, and $Q$ is the reaction Q-value for the formation of the compound nucleus. Once the compound nucleus is formed, there is a sufficient period of time for the nucleons to rearrange themselves and effectively lose any information about how the compound nucleus was formed \cite{blatt_1991}. This means that the de-excitation of the compound system will be independent of the manner in which it was formed, a theory first proposed by Bohr \cite{bohr_1936}. Under these assumptions the nuclear cross section for a single isolated resonance, called the Breit-Wigner formula \cite{breit_1936}, can be written as: 
\begin{equation}
    \label{eq:resonance}
    \sigma(E)_{\textnormal{BW}} = \frac{\lambda^2}{4 \pi}\frac{(2J+1)}{(2j_a+1)(2j_A+1)}\frac{\Gamma_a \Gamma_b}{(E-E_r)^2 + \Gamma^2/4},
\end{equation}
where $\lambda$ is the de Broglie wavelength of the system at the center-of-mass energy $E$, $J$ is the spin of the resonance, $j_a$ is the spin of the incident particle, $j_a$ is the spin of the target, and $E_r$ is the energy of the resonance.
This formula, consistent with the assumptions that have just been laid out, only depends on the resonance energy ($E_r$), level width ($\Gamma$), and spins of the particles involved. Of particular importance are the partial widths, $\Gamma_a$ and $\Gamma_A$, and the total width $\Gamma$. These widths are expressed in terms of energy, and using the uncertainty principle can be related to lifetime, $\tau$, of the resonance:
\begin{equation}
    \label{eq:unc_principle}
    \Gamma = \frac{\hbar}{\tau}.
\end{equation}
Additionally, $\Gamma$ is equal to the sum of all the energetically allowed decay channels' partial widths:
\begin{equation}
    \Gamma = \sum_c \Gamma_c,
\end{equation}
where the index c runs over the channels. These widths are in general energy dependent quantities. This complication gives two distinct cases: \textit{narrow} resonances where the widths do not vary appreciably over the total width of the resonance and \text{broad} resonances where this does not hold.

It is now possible to calculate the astrophysical rate in the case of an isolated narrow resonance. Plugging the expression for the cross section, Eq.~\ref{eq:resonance}, into that of the reaction rate, Eq.~\ref{eq:reaction_rate}, yields:
\begin{equation}
    \label{eq:rate_and_bw}
    \langle \sigma v \rangle_{aA} = \bigg(\frac{8}{\pi \mu_{aA}}\bigg)^{1/2} \frac{1}{(kT)^{3/2}} \int_0^{\infty} \frac{\lambda^2}{4 \pi}\frac{(2J+1)}{(2j_a+1)(2j_A+1)}\frac{\Gamma_a \Gamma_b}{(E-E_r)^2 + \Gamma^2/4} E  e^{-E/kT} dE.
\end{equation}
The simplification takes the partial widths and Maxwell-Boltzmann distribution to be constant over the width of the resonance. Pulling these out of the integral leaves the expression:
\begin{equation}
    \label{eq:narrow_integral}
    \int_0^{\infty} \frac{dE}{(E-E_r)^2 + \Gamma^2/4}.
\end{equation}
 To explicitly show one more assumption, this integral can be be evaluated with trigonometric substitution and reduces to:
 \begin{equation}
     \label{eq:arctan_nonsense}
     \frac{2}{\Gamma} \arctan \bigg(\frac{E-E_r}{\Gamma/2}\bigg) \bigg|^{\infty}_{0} ,
 \end{equation}
 the upper limit can readily be calculated and comes out to $\frac{\pi}{2}$. The lower one invokes another property of narrow resonances, that being $E_r >> \Gamma$, which effectively makes the lower bound $-\infty$, giving $-\frac{\pi}{2}$. Thus, this integral yields $\frac{2}{\Gamma} \pi$. Combining these factors gives:
 \begin{equation}
    \label{eq:narrow_rate}
    \langle \sigma v\rangle_{aA} = \bigg(\frac{2 \pi}{\mu_{aA} kT} \bigg)^{3/2} \hbar^2 e^{-E_r/kT} \frac{(2J+1)}{(2j_a+1)(2j_A+1)}  \frac{\Gamma_a \Gamma_b}{\Gamma}.
\end{equation}
 The product of the spin multiplicities and widths is frequently denoted $\omega \gamma$ and is referred to as the resonance strength because it is proportional to the area under the resonance cross section \cite{iliadis_book}. Using this definition, the above becomes:  
 \begin{equation}
    \label{eq:narrow_rate_with_resonance_strength}
    \langle \sigma v \rangle_{aA} = \bigg(\frac{2 \pi}{\mu_{aA} kT} \bigg)^{3/2} \hbar^2 e^{-E_r/kT} \omega \gamma.
\end{equation}
 
 Although the experimental methods in this thesis will not focus on the direct determination of the resonance strength, the results presented can be compared to such measurements. It is therefore enlightening to discuss how such measurements are performed. From Eq.~\ref{eq:narrow_integral}, we can see that any process that integrates over a narrow resonance will be proportional to $\omega \gamma$. In the lab this energy variation comes not from the random thermal motion of particles, but from the energy loss of a particle-beam traveling through a target material with stopping power, $\epsilon(E)$. In such a case, the number of nuclear reactions that occur per beam particle, i.e. the yield, $Y$, of the reaction, will be:
 \begin{equation}
     \label{eq:resonance_yield}
     Y = \frac{\lambda_r^2}{2} \frac{\omega \gamma}{\epsilon_r},
 \end{equation}
 where the subscript $r$ means the quantities are evaluated at the resonance energy. This formula only holds if the target material can be considered infinitely thick, meaning that the target thickness in energy units is much greater than $\Gamma$, i.e. $\Delta E >> \Gamma$ \cite{iliadis_book}.

 \section{Connection to Nuclear Structure}

Resonances as discussed above make no assumptions about the detailed structure of the compound system. In cases where the resonance strength can be determined directly, this is a very satisfactory state of affairs, where almost no knowledge of the nuclear force beyond its short range and strong interactions has to be assumed. However, if it is the case that the resonance cannot be determined directly, as is often true for low energy resonances of interest to astrophysics, indirect determinations of the resonance parameters have to be made. Because of the relative simplicity of Eq.~\ref{eq:resonance}, it can be guessed that any information about the underlying nuclear structure must be contained within the partial widths. As will be shown, these particle partial widths can be expressed in terms of single-particle states within the shell model. Historically, this connection between the resonance partial widths and single-particle quantities related to the shell model is quite remarkable, since the proposal and success of the Breit-Winger formula \cite{breit_1936} predated the evidence of the nuclear shell model \cite{mayer_1948} by more than a decade.

As an example, lets turn towards the proton partial width, $\Gamma_p$, which, despite being expressed in energy units, can be thought of as being proportional to the probability of a proton to be emitted from the compound nucleus. In order for this process to happen, three things must occur: the proton must overcome the coulomb barrier to arrive at the nuclear surface, at the surface the proton must occupy a single-particle state that has the same quantum numbers of the corresponding resonance, and this occupancy must be weighted according to how much the mean field effect of the other nuclear interactions fragments the total single-particle strength of a proton shell among its constitute states. Thus, $\Gamma_p$ can be written:       
\begin{equation}
    \label{eq:proton_partial_width}
    \Gamma_{p} = 2\frac{\hbar^2}{\mu_{pA}R^2} C^2S_{\ell}P_{\ell} \theta_{\textnormal{sp}}^2, 
\end{equation}
where $C^2$ is the isospin Clebsch-Gordan coefficient for the $p + A$ system, 
$S$ is the spectroscopic factor of the single particle state, $P_{\ell}$ is the penetration factor, and $\theta_{\textnormal{sp}}$ is the single particle reduced width \cite{ILIADIS_1997}. Additionally, there is the parameter $R$, which is the channel radius. Matching these terms with the above probabilities: $P_{\ell}$ is the probability of the proton tunneling through the coulomb barrier, $\theta_{sp}^2$ is the probability of the proton being found at the nuclear surface, and $C^2S$ is the weight for the single-particle state.

\subsection{Calculation of Partial Widths}
\label{sec:calc_partial_widths}

Of the terms in Eq.~\ref{eq:proton_partial_width}, $P_{\ell}$ and $\theta_{sp}$ stand separately from $C^2S$ because they can be calculated accurately theoretically. $P_{\ell}$ is  calculated from: 
\begin{equation}
    \label{eq:penetrability}
    P_{\ell} = R\bigg( \frac{k}{F_{\ell}^2 + G_{\ell}^2} \bigg),
\end{equation}{}
where $F_{\ell}$ and $G_{\ell}$ are the regular and irregular coulomb functions, respectively \cite{Abramowitz_1974}. $k$ is the wave number of the particle, and $R$ is the channel radius \cite{iliadis_book}. $\theta_{sp}$ is   
the dimensionless single-particle reduced width and is defined as:
\begin{equation}
    \label{eq:dimensionless_reduced_width}
    \theta_{sp} = \frac{R}{2} \big| u_{\textnormal{sp}}(R) \big|^2.
\end{equation}
Again $R$ is the channel radius, while $u_{sp}$ is the radial wave function for a single particle in a nuclear potential. The channel radius is defined as:
\begin{equation}
    \label{eq:channel_radius}
    R = r_0 (A_t^{1/3} + A_p^{1/3}),
\end{equation}
where $A_t$ and $A_p$ are the atomic mass numbers for the target and projectile, respectively. For this work $r_0 = 1.25$ fm.

While it is possible to calculate the above factors individually, they, as well as the constant factors, can be absorbed into a single term, $\Gamma_{sp}$. \, Eq.~\ref{eq:proton_partial_width} is now written as:
\begin{equation}
    \Gamma_p = C^2S \Gamma_{\textnormal{sp}}. 
\end{equation}
The benefit of such an equation is that it can be calculated numerically by varying the parameters of a single-particle potential to produce a resonance at the observed energy. The width of this single-particle resonance corresponds to $\Gamma_{\textnormal{sp}}$. One such method is utilized by the code \texttt{BIND}, which is described in detail in Ref.~\cite{ILIADIS_1997}. \texttt{BIND} finds $\Gamma_{\textnormal{sp}}$ by solving:
\begin{equation}
    \label{eq:bind}
    \frac{2}{\Gamma_{\textnormal{sp}}} \approx \bigg( \frac{d \delta}{d E} \bigg)_{\pi/2} = \frac{2 \mu_{aA}}{\hbar^2 k} \bigg( \int_0^{R_{max}} |u(r)|^2 dr + \frac{G_{\ell}(R_{max})}{2k} \frac{d}{dk} \bigg[ \frac{d G_{\ell}}{dr} \bigg | _{r=R_{max}} / G_{\ell}(R_{max}) \bigg] \bigg),
\end{equation}
where $\delta$ is the scattering phase shift, $R_{max}$ is a cutoff radius, and the other symbols have been defined previously. One shortcoming of the above method is that as the resonance energy approaches the proton threshold, the numerical method becomes extremely unstable. This is expected since partial widths drop off exponentially as $E_{c.m.}$ approaches zero. These problems at lower energies can be worked around by using the methods presented in \cite{ILIADIS_1997} and \cite{BARKER_1998}, these studies use the high energy values of $\Gamma_{\textnormal{sp}}$ and Eq.~\ref{eq:penetrability} to give fits for the energy and mass dependence of $\theta_{\textnormal{sp}}^2$. These fits also allow the calculation of \textit{reduced particle width} for sub-threshold resonances:     

\begin{equation}
    \label{eq:formal_reduced_width}
    \theta^2 = C^2S \theta_{sp}^2. 
\end{equation}

The only missing unknown quantity at this point is $C^2S$, which scales the single-particle quantities to the physical resonance parameters. $C^2S$ can in principle be calculated using the shell model, but it is highly sensitive to the model chosen for the nucleon-nucleon interactions. The situation is further complicated for resonances since the single particle states are unbound, which requires treatment of states in the continuum. Due to the high uncertainty involved with their determination theoretically, it is advantageous to determine $C^2S$ experimentally. Single-nucleon transfer reactions have long been used to experimentally test the shell model due to their sensitivity to the single-particle structure of excited states \cite{satchler}.
In particular, if a transfer reaction is performed at a high enough energy, then the dominant contribution to the cross section will be a direct reaction process. Direct reactions, as opposed to compound reactions, do not form a compound nucleus; therefore, the particles in the exit channel carry information about the reaction mechanism itself, not just the decay of the excited nucleus \cite{satchler}. 
We can now look at how this information can be related to $C^2S$. 

\section{Transfer Reactions}

Transfer reactions are a broad class of nuclear reactions that refer to processes where either a single or a cluster of nucleons are moved between the target and projectile systems. In normal kinematics assuming a cluster of nucleons, $c$, we have either a pickup reaction:
\begin{equation}
    a + (A+c) \rightarrow (a+c) + A,
\end{equation}{}
or a stripping reaction:
\begin{equation}
    (a+c) + A \rightarrow a + (A+c).
\end{equation}{}
Describing this system theoretically requires a quantum mechanical treatment of each of these subsystems. For the case of a stripping reaction, this divides into the scattering of $(a+c)$ off of $A$ and $a$ off of $(A+c)$, while accounting for the interaction of $(a+c)$ and $(A+c)$.

Since the measurement presented in this thesis is a proton stripping reactions, i.e. $A(^3 \textnormal{He}, d)B$, I will tailor the following discussion around its description.

\subsection{Optical Model}

A full description of the scattering processes $A + ^{3} \textnormal{He}$ and $d + B$ would require specifying the Hamiltonian for the $N$-nucleon scattering problem. This would involve accounting for all excited states in both the projectile and target systems and coupling all available reaction channels. The nuclear optical model simplifies this multi-nucleon scattering problem by considering a single particle interacting with a complex potential, $\mathcal{U}(r)$. The complex potential removes flux from the elastic channel, which approximates the effect of the other open reaction channels on the cross section. This assumption is only valid when the number of open reaction channels is large enough such that their influence can be averaged over \cite{hodgson1971}.

The theoretical basis for the optical model was first established in Ref.~\cite{feshbach1958}, but fell short of actually prescribing the form of the complex potential. Through detailed analysis of elastic scattering from a range of targets and energies, Becchetti and Greenlees \cite{b_g_p} showed that a phenomenological optical model could successfully describe a wide variety of elastic scattering data. Further theoretical and historical details can be found in Ref.~\cite{hodgson1971}.
Any reference to or listing of optical model parameters in this thesis assumes the following form of the optical potential:
\begin{multline}
  \label{eq:optical_model}
  \mathcal{U}(r) = V_c(r; r_c)-Vf(r; r_0, a_0)
  -i(W-4a_iW_s\frac{d}{dr_i})f(r; r_i, a_i) \\
  + \bigg( \frac{\hbar}{m_{\pi}c} \bigg)^2V_{so} \frac{1}{r} \frac{d}{dr} f(r; r_{so}, a_{so}) \boldsymbol{\sigma}  \cdot \boldsymbol{\ell}, 
\end{multline}
where $f(r)$ is given by the Woods-Saxon form factor:
\begin{equation}
  \label{eq:ws_pot}
  f(r; r_0, a_0) = \frac{1}{1 + \textnormal{exp}\bigg({\frac{r-r_0A_t^{1/3}}{a_0}}\bigg)}.
\end{equation}
Each term, with the exception of the coulomb potential, in $\mathcal{U}$ is parameterized with a well depth, ${V, W, W_s}$, radius, ${r_0, r_i, r_{so}}$, and diffuseness, ${a_0, a_i, a_{so}}$. $A_t$ denotes the atomic mass number of the target nucleus.
The spin-orbit interaction is between the projectile orbital and spin angular momentum, $\boldsymbol{\ell}$ and $\bf{s}$, respectively. Additionally, the spin orbit term makes use of the common parameterization where $\boldsymbol{\sigma} = 2 \mathbf{s}$, and $(\frac{\hbar}{m_{\pi}c})^2$ is a constant with a value of approximately $2$ fm$^2$.
The Coulomb term, $V_c$, comes from the potential of a uniformly charged sphere with radius $R_c = r_c A_t^{1/3}$. 
These conventions follow those of the reaction code FRESCO \cite{fresco}, which is used for all calculations in this thesis, unless otherwise noted. A word of caution is in order due to many different conventions that exist for the parameter values in literature. Especially bothersome are the form of the imaginary surface and spin-orbit terms. For the imaginary surface term, the prefactor can either be $W_s$, $4W_s$, or as adopted here $4a_iW_s$. Both the subscript $D$ and $s$ are in use for the surface imaginary term, standing for either \textit{derivative} or \textit{surface}. For the spin-orbit term, the prefactor may omit the constant, $\mathbf{s}$ may be substituted for $\boldsymbol{\sigma}$, or the so-called Thomas form may be used which replaces $V_{so}$ and any constant by $\lambda$. These conventions change from author to author, study to study, and reaction code to reaction code. Beware!       

The selection of the parameter values for these potentials is critical to the successful theoretical description of transfer reactions. For the phenomenological optical model, they are determined from experimental data, typically differential elastic scattering cross sections and analyzing powers. These data can be local, like those listed in Ref.~\cite{perey_perey}, where only the elastic scattering from a single target nucleus at a single energy is considered. On the other hand, global studies, such as Refs.~\cite{varner, b_g_p, b_g_3he, pang_global, daehnick_global}, use a variety of targets and beam energies to derive relations between potential parameters and target mass, beam energy, and other nuclear properties. Often these two approaches are mixed, with global values used as a starting point for a local data set. These starting global values are then fit to best describe the local data set.

\subsection{Distorted Wave Born Approximation}

The optical model potentials described in the previous section allow a computationally tractable approach to the $N$-nucleon scattering problem. The Distorted Wave Born Approximation (DWBA) is a perturbative method that treats the transfer of a nucleon between nuclei as a perturbation of the elastic scattering process described using optical potentials \cite{satchler, thompson_nunes_2009, hodgson1971}. In order for this model to be valid, the transfer cross section must be small compared to the elastic scattering cross section and the transfer reaction must be consistent with a direct reaction process. A direct reaction is described to first order as a transition from an initial to final state via an interaction potential, $V$. The T-matrix element for such a process is:
\begin{equation}
    T = \braket{\mathbf{k}_f | V | \mathbf{k}_i},
\end{equation}{}
where $\mathbf{k}_i$ and $\mathbf{k}_f$ are the wave numbers for the incoming and outgoing wave functions, respectively. DWBA constructs these scattering wave functions using the entrance and exit optical model potentials according to:

\begin{equation}
    \label{eq:no_time_se}
    \bigg( -\frac{\hbar^2}{2 \mu} \nabla^2 + \mathcal{U} \bigg) \ket{\chi}  = E \ket{\chi}.
\end{equation}{}

\begin{figure}
    \centering
    \includegraphics[width=\textwidth]{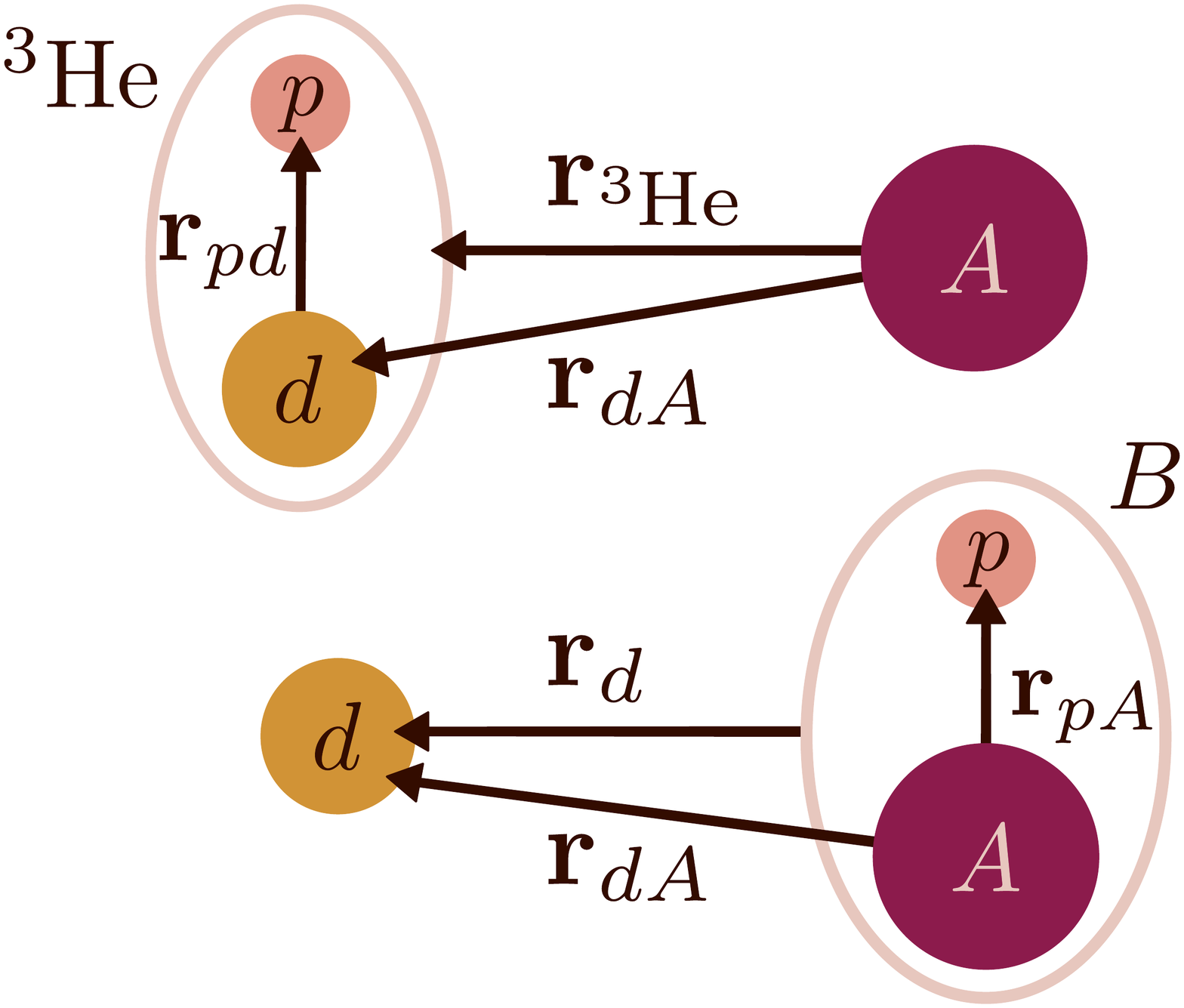}
    \caption{Diagram of the coordinates used in DWBA for a stripping reaction $A(^3\textnormal{He}, d)B$. $\mathbf{r}_{^3\textnormal{He}}$ and $\mathbf{r}_{d}$ are the coordinates for the optical model potentials shown explicitly in Eq.~\ref{eq:complex_t}, while the other coordinates are implicitly contained in the matrix element $\braket{B, d|V_{transfer}|A, ^3\!\textnormal{He}}$. }
    \label{fig:three_body}
\end{figure}

The T-matrix can then be written down for the transfer reaction using the incoming and outgoing wave functions for the final and initial states, $\chi_f^{*(-)}(\mathbf{k}_f)$ and $\chi_i^{(+)}(\mathbf{k}_i)$ :
\begin{equation}
    \label{eq:simple_t}
    T = \braket{\chi_f^{*(-)}(\mathbf{k}_f)|V_{transfer}|\chi_i^{(+)}(\mathbf{k}_i)}.
\end{equation}{}
This equation, while compact, masks much of the complexity of the problem. In particular, the transfer process needs to transform the incoming coordinates for the $A + ^{3}\!\textnormal{He}$ system to those of the outgoing $B + d$ system. A diagram of the coordinates considered in DWBA is shown in Fig.~\ref{fig:three_body}. Explicitly writing these out, we get:
\begin{equation}
    \label{eq:complex_t}
    T = J \int d \mathbf{r}_d \int d \mathbf{r}_{^3\!\textnormal{He}} \chi_f^{*(-)}(\mathbf{r}_d, \mathbf{k}_d) \braket{B, d|V_{transfer}|A, ^3\!\textnormal{He}} \chi_i^{(+)}(\mathbf{r}_{^3\textnormal{He}}, \mathbf{k}_{^3\textnormal{He}}).
\end{equation}{}
Eq.~\ref{eq:complex_t} reduces down to Eq.~\ref{eq:simple_t} once the integrals are evaluated. The perturbing potential $V_{transfer}$ is now acting on the states for the internal wave functions of the systems.

By construction, the distorted waves in Eq.~\ref{eq:complex_t} do not contain detailed nuclear structure information. Any angular momentum selection rules, nuclear structure information, or any other reaction information is thus contained in $\braket{B, d|V_{transfer}|A, ^3\!\textnormal{He}}$ \cite{Satchler_1964}. It is again the case that the Bra-ket notation hides the complex nature of this term, which requires evaluation of all the spins and internal coordinates of the particles involved in the transfer. Specifying the form of the perturbing potential is the last step before it becomes possible to fully describe the DWBA T-matrix.

Building the potential $V_{transfer}$ can be done in either the entrance or exit channels. This choice is referred to as either the \textit{prior} or \textit{post} representation, depending on if the interactions in the entrance or exit channel is chosen. First order DWBA is equivalent if either of these two representations is used to construct the perturbing potential \cite{thompson_nunes_2009}.
An example to help clarify this concept is the prior form of the $A(^3\textnormal{He}, d)B$ reaction. The valence particle is the proton, which in the entrance channel is bound to the deuteron. Since the transfer is dependent on the potential acting to remove the proton from the deuteron, it is necessary to find the energy difference between the individual and composite systems. The composite interaction is between $^3\textnormal{He}$ and $A$, while the individual interactions are for the $p+A$ and $d+A$ systems.     Solving these interactions exactly would require solving the $N$-nucleon systems, so it is again necessary to use simplifications to make the problem tractable. Optical potentials can be used for $d+A$ and $^{3}\textnormal{He} + A$ systems, and a single particle model can be used for $p+A$. This gives:    
\begin{equation}
    \label{eq:prior}
    \mathcal{V}_{prior} = V_{p+A} + \mathcal{U}_{d+A} - \mathcal{U}_{^3\textnormal{He}+A},   
\end{equation}{}
where $\mathcal{U}_{d+A}$ and $\mathcal{U}_{^3\textnormal{He}+A}$ are optical model potentials and $V_{p+A}$ is a real single particle potential. $\mathcal{U}_{d+A}$ is typically called the core-core potential because it describes the interaction between the two inert cores in the reaction. These terms can be readily written down for either stripping or pickup reactions in either the post or prior form if it is noted that the core nucleus in the selected channel must appear in every term. It is then a matter of writing down all of the interactions with that core, where the binding and core-core interactions are added together and the remaining term is subtracted off. As an example, the post form for the above reaction will depend on all of the interactions with the core in the exit channel, which is $d$. This gives:
\begin{equation}
    \label{eq:post}
    \mathcal{V}_{post} = V_{p+d} + \mathcal{U}_{A+d} - \mathcal{U}_{B+d},  
\end{equation}{}
where again it should be noted that $d$ appears in every term.
Plugging either of the above expressions into Eq.~\ref{eq:complex_t} will yield the DWBA cross section for a chosen excited state. This is known as a finite range transfer with full remnant \cite{thompson_nunes_2009}.

\subsection{Overlap Functions and Spectroscopic Factors}
\label{sec:overlaps}

It can be seen that any nuclear structure information that needs to be included in DWBA has to be contained in the matrix element $\braket{B, d|V_{transfer}|A, ^3\!\textnormal{He}}$. A full exploration of this term is beyond the scope of this thesis, but there are some details that merit discussion in order to draw the clear line from transfer reactions to nuclear astrophysics. If we return to the proton stripping example, it is easier to examine this matrix element in the post form, with the assumption that the remnant term is negligible. This gives:

\begin{equation}
    \label{eq:transfer_matrix_element}
    \braket{B, d|\mathcal{V}_{post}|A, ^3\!\textnormal{He}} = \braket{B|A} \braket{d|V_{d+p}|^3\!\textnormal{He}}.
\end{equation}{}

The term $\braket{B|A}$ is called the overlap function, and describes a state of the composite nucleus, $B$, in terms of the core nucleus and a proton, $A+p$. It is more accurate to express this quantity with its dependence on the relative coordinate between the proton and the core, $\mathbf{r}_{pA}$, as $\Phi_{I_A:I_B}(\mathbf{r}_{pA})$.  In general, this overlap can be viewed as the coefficient for a given state in the expansion:
\begin{equation}
    \label{eq:overlap_expansion}
    \psi^B_{I_B}(\zeta_A, \mathbf{r}_{pA}) = \sum_{I_A} \Phi_{I_A:I_B}(\mathbf{r}_{pA}) \psi^A_{I_A}(\zeta_A), 
\end{equation}
where the labels $I_A$ and $I_B$ can stand for any involved quantum number such as spin or isospin, while $\zeta_A$ is the symbol given to the internal coordinates of $A$. The spectroscopic factor, which when combined with the isospin Clebsch-Gordan coefficient is written $C^2S_{\ell s j}^{j I_A I_B}$, is defined as the integral of the square of the norm of the overlap function:

\begin{equation}
    \label{eq:overlap_integral}
    C^2S = \int_0^{\infty} \big|\Phi_{I_A:I_B}(\mathbf{r}_{pA})\big|^2 \, d \mathbf{r}_{pA}.
\end{equation}

The issue at this point, as has been the theme with this entire section, is that we are again confronted with a multi-nucleon problem when trying to determine $\Phi_{I_A:I_B}$. The problem is simplified by introducing a single-particle wave function, typically generated by a Woods-Saxon potential, that is normalized to unity. Thus, the overlap becomes:
\begin{equation}
    \label{eq:coeff_frac_parentage}
    \Phi_{I_A:I_B}(\mathbf{r}_{pA}) \approx A_{\ell s j}^{j I_A I_B} \phi_{\ell s j}^{j I_A I_B}(\mathbf{r}_{pA}).
\end{equation}
The coefficient $ A_{\ell s j}^{j I_A I_B}$ is called the \textit{coefficient of fractional parentage} (CFP). By Eq.~\ref{eq:overlap_integral}, we see that:
\begin{equation}
\label{eq:spec_factor_fraction_coeff}
    C^2S_{\ell s j} =  \big | A_{\ell s j}^{j I_A I_B} \big |^2.
\end{equation}

Stepping back from the math for a second, Eq.~\ref{eq:spec_factor_fraction_coeff} is the result of a fairly simple logic:
\begin{enumerate}
    \item System $B$ is made of an inert core $A$ plus a proton.
    \item System $B$ can thus be expanded into states of the system $A+p$.
    \item The coefficients of this expansion, called the overlap, can be modeled as single particle states with quantum numbers $n l s j$ coming from the shell model.
    \item The square of the norm of an overlap function is not equal to unity, but rather equal to the spectroscopic factor for that state.
    \item Since the single particle states from item $3$ are unit normalized, then they must be weighted by a factor $A_{\ell s j}^{j I_A I_B}$.
    
\end{enumerate}
The spectroscopic factor can be viewed as the fraction of the total strength of a single particle shell contained in a single state. 

Putting all of this information together, it can be seen that the matrix element in Eq.~\ref{eq:transfer_matrix_element} is proportional to the CFP for the $A+p$ system. The wave function for $^3$He can also be expanded via the overlap, yielding a CFP for the $d+p$ system as well. Once this is done, the DWBA formalism is complete and a cross section can be computed. The differential cross section will be proportional to the square of the T-matrix, and this cross section can be compared to the experimental cross section $\frac{d \sigma}{d \Omega}_{Exp}$. Pulling the spectroscopic factors out front gives:

\begin{equation}
    \label{eq:exp_to_theory}
    \frac{d \sigma}{d \Omega}_{Exp} = C^2S_{A+p}C^2S_{d+p} \frac{d \sigma}{d \Omega}_{\textnormal{DWBA}}. 
\end{equation}

The $d+p$ system spectroscopic factor is frequently omitted from this equation in the literature, which reflects the fact that it has been computed either by \textit{ab-initio} \cite{brida_ab_initio} methods, or simply from the relation for nuclei with mass number $ \leq 4$, which is $A/2$, e.g., $C^2S_{p+d} = 3 \times \frac{1}{2} $ \cite{satchler}.   

The upshot of DWBA is now clear, a transfer cross section measured in the lab can be used to extract $C^2S_{A+p}$. Using the spectroscopic factor, theoretical single-particle resonances are scaled to give $\Gamma_p$, which, in turn, can be plugged into Eq.~\ref{eq:narrow_rate_with_resonance_strength}. Transfer reaction experiments provide nuclear input essential to thermonuclear reaction rates. 

\section{Summary}

This chapter has laid out the foundation of how nuclear physics can be used to inform stellar burning. Resonant reactions were defined, and their contribution to thermonuclear reaction rates were made explicit in the case of narrow isolated resonances. The connection between the parameters in the Breit-Wigner cross section and nuclear structure led to an outline of how transfer reactions could constrain these reactions. Finally, the formalism for DWBA was presented, and the theory was examined in order to show how spectroscopic factors arise from the nuclear overlap functions and how they are determined from experiment.   

%% file: Chapter-3/Chapter-3.tex
\chapter{Nuclear Physics Uncertainties}
\label{chap:nuclear_unc}

\section{Introduction}

This chapter introduces the concepts of reaction rate uncertainties, which motivate much of the original research presented in this thesis. It will set the stage for the work on Bayesian DWBA that is presented in Chapter \ref{chap:bay_dwba}, and discuss the evaluation of the $^{39}$K$(p, \gamma)$ reaction rate which I was involved in. Within this framework nuclear physics uncertainties associated with quantities measured in the lab can be propagated all the way through a nuclear reaction rate network calculation. Thus, the results of nucleosynthesis calculations can be compared with astronomical observations taking into account the underlying nuclear physics uncertainties.  

\section{Reaction Rate Uncertainties}

As the nuclear inputs such as partial widths, resonance energies, and resonance strengths are measured in the lab, it is essential to quantify their impact on the resulting astrophysical reaction rate. At first glance, this calculation can appear to be a simple one, with the lab values plugged into the reaction rate formulas from Chapter \ref{chap:reactions}. This simple calculation, however, would ignore a fundamental part of these experimentally determined quantities, which is that they are subject to uncertainties. 
The danger in ignoring uncertainties is the same for nuclear astrophysics as it is for any other field of science, overconfidence in predictions and a loss of direction in how to better improve these predictions.
In other words, it is critical to estimate uncertainties both to
understand what is known and to develop plans on how to know it better.

The method described here for propagating nuclear uncertainties through reaction rate calculations was first described and utilized in a series of papers by Longland, Iliadis, and collaborators \cite{LONGLAND_2010_1, ILIADIS_2010_2, ILIADIS_2010_3, ILIADIS_2010_4}. While these methods will surely evolve in the future, they are of critical importance to the future of the field, and their foundation merits discussion.

Propagation of uncertainties through calculations can be done in several ways, with perhaps the most familiar way being:
\begin{equation}
    \label{eq:error_formula}
    \sigma_f = \sqrt{\sum_i \bigg( \frac{\partial f}{\partial x_i} \bigg)^2 \sigma_i^2},
\end{equation}
where for a function, $f$, dependent on random variables, $\mathbf{x}$. This equation relates the standard deviation of $f$, $\sigma_f$, to the standard deviations of the dependent variables, $x_i$. The issue with the above formula is that it is an approximation, which might not hold if there exist strong correlations between the variables or if the uncertainties are not normally distributed \cite{taylor_error_analysis}. Before moving on, an unfortunate situation in nuclear physics when discussing uncertainties should be mentioned. The context in this chapter should make it obvious whether $\sigma$ is referring to a cross section or a standard deviation. In order to be more explicit, cross sections will always be expressed as a function of another variable, i.e., $\sigma(E)$ or $\frac{d \sigma}{d \Omega}$.    

As will be shown in the next section, the uncertainties for many nuclear physics quantities, such as the cross section, are not normally distributed. Furthermore, the derivatives in Eq.~\ref{eq:error_formula} might require evaluation via numerical methods, eliminating much of the computational simplicity that makes this formula attractive in the first place. Due to these reasons, it is beneficial to use an alternative method when confronted with the specific issues that arise in reaction rate calculations.

A simple and flexible solution to uncertainty propagation is using a Monte Carlo approach. This method assigns explicit probability distributions for each $x_i$:
\begin{equation}
    \label{eq:prob_dist}
    x_i \sim X_i,
\end{equation}
where the symbol $\sim$ means \say{distributed according to.} If samples can be drawn from $X_i$, then these samples, in turn, can be used to calculate $f$. Thus, the Monte Carlo approach draws many samples from the distribution of all variables that are subject to random effects. These samples are fed one at a time through the formula for $f$, yielding samples for $f$. Finally, the samples of $f$ are collected and analyzed. This approach, while simple, has some definite drawbacks. If $f$ is computationally expensive, then it can be infeasible to perform the necessary number of evaluations, which can be on the order of $10^6$. Additionally, there is no guarantee that the randomly drawn samples will be representative of the underlying distribution. This issue becomes especially problematic when the tails of the distribution are important.

As a quick example, lets take $f$ to be $f = x + y$, where $x$ and $y$ are distributed according to:
\begin{align}
    \label{eq:example_dist}
    x \sim \mathcal{N}\big(4.0, 1.0 \big) \\
    y \sim \mathcal{N}\big(3.0, 1.0 \big),
\end{align}
where $\mathcal{N}(\mu, \sigma^2)$ stands for the normal distribution, which is parameterized by the mean, $\mu$, and the variance, $\sigma^2$. Thus, the above equations denote normal distributions with means $4$ and $3$, both with variance $1$.

\begin{figure}
    \centering
    \includegraphics[width=.6\textwidth]{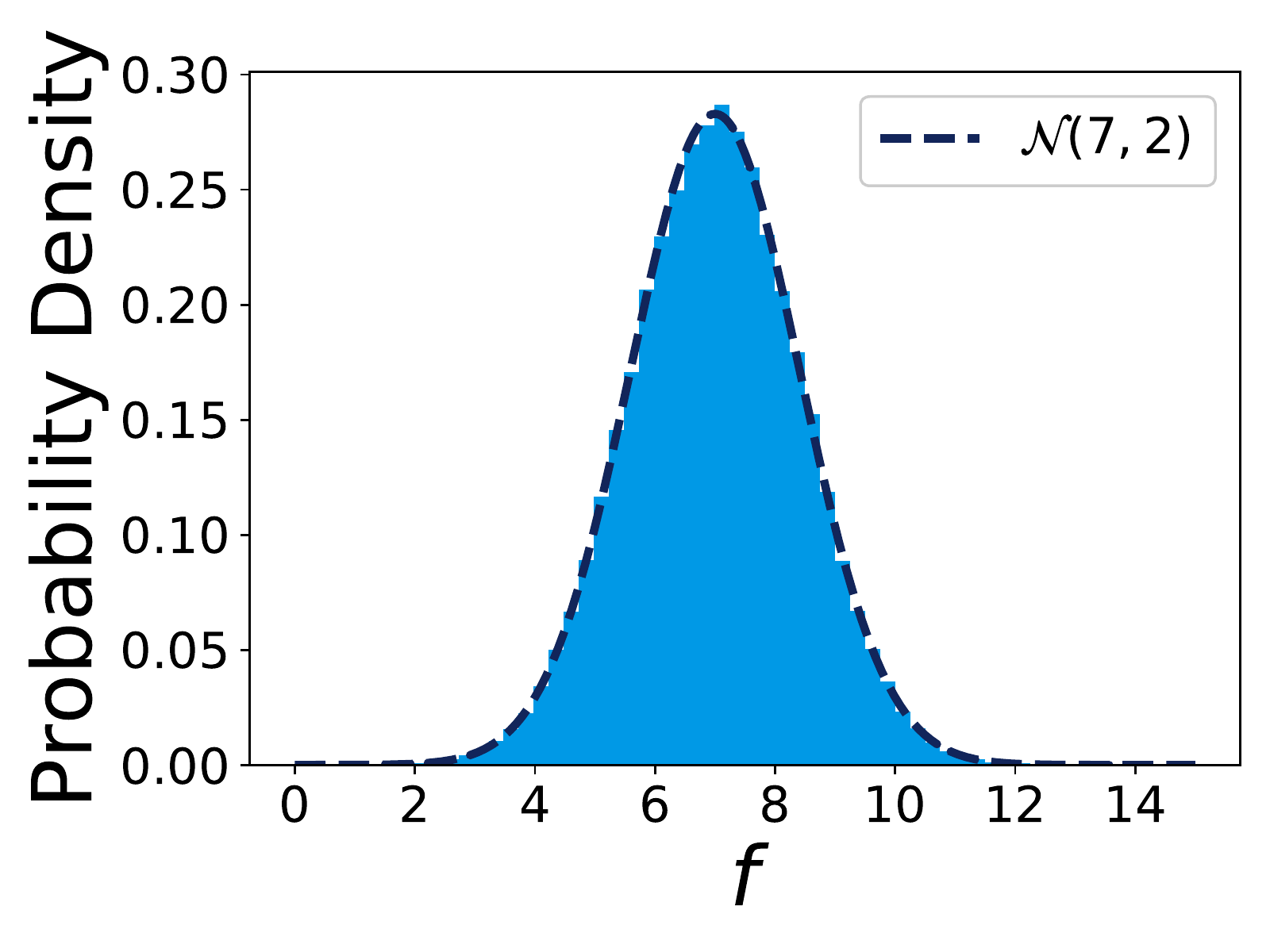}
    \caption{Histogram showing the comparison between the probability density of a normal distribution with parameters found from Eq.~\ref{eq:error_formula} and the samples obtained from the Monte Carlo method.}
    \label{fig:monte_carlo_example}
\end{figure}

Using Eq.~\ref{eq:error_formula} we can easily find that $f$ has $\mu = 7.0$ and $\sigma = \sqrt{1^2 + 1^2} = \sqrt{2}$. If instead we use a Monte Carlo method and draw $10^5$ samples from each $x$ and $y$, we get $10^5$ realizations of $f$. These samples are plotted in Fig.~\ref{fig:monte_carlo_example}. Once we have the samples of $f$ the other issue with Monte-Carlo procedures, or really any sampling based method, arises. How do we best summarize these samples? This is a rather trivial problem for this example since it is well described by a normal distribution, with the consequence that the mean, median, and mode are all equal, and that one standard deviation will contain $68 \%$ of the samples. In general this problem will arise again and again, and there is no general answer, so a deeper discussion is best left until we have a problem that requires more thought.  
Blissfully unconcerned with nuances of summary statistics, we take the mean and the standard deviation of the samples, which gives $\mu=7.00$ and $\sigma=1.41$.

It is clear from the above discussion that before a Monte Carlo procedure can be applied to the problem of calculating thermonuclear reaction rates, the inputs must be assigned probability distributions that can be sampled. Furthermore, once the samples are drawn for $\langle \sigma v \rangle$, a convenient way to summarize them will be necessary before they can be integrated into a reaction network.   

\section{Probability Distributions For Nuclear Inputs}

The form of the probability distributions for $E_r$, $\Gamma$, $\Gamma_x$ (partial width for particle $x$), and $\omega \gamma$ requires examination of how these quantities are derived from experiments. The central limit theorem states that the sum of random variable, no matter how these variable are distributed, will approach a normal distribution as more terms are added to the sum. This theorem, and some of its corollaries, provides a reasonable form for all of the above experimentally determined terms.

\subsection{Resonance Energies}

Energies in the lab are typically subject to uncertainties arising from the sum of many terms. Take for example proton elastic scattering off of a thin target of a heavy material. The protons are detected by a surface barrier silicon detector positioned close to $0^{\circ}$. In this case, the measured proton energy will be impacted by statistical uncertainties that comes from a variety of sources. Some examples are the incoming beam, which will have energy fluctuations due to the instability of the accelerator, $\delta E_{beam}$. Once this beam impinges on the thin target, it will lose some amount of energy, $\delta E_{target}$. The energy deposited within the detector by the scattered proton will be converted to an electric charge, which will be subject to thermal effects of the electrons moving through the semi-conductor, $\delta E_{thermal}$. Finally, this charge will be integrated by a charge sensitive preamplifier, which itself is subject to electronic noise, $\delta E_{electronic}$. We can think of all of these terms as additive effects of the energy of the particle $E_p$:
\begin{equation}
    \label{eq:energy_example}
    E_{detected} = E_p + \delta E_{beam} + \delta E_{target} + \delta E_{thermal} + \delta E_{electronic}, 
\end{equation}
where it should be understood that $E_p$ is a number, but all of the $\delta$ terms are randomly distributed quantities drawn from unknown distributions.

The above example excludes many potential factors, but the point is that a measured energy is most often the sum of many factors, all with uncertainty. Because of this fact, the central limit theorem can be invoked, which tells us that as these independent random effects are added up, the resulting quantity, i.e, energy, will tend towards a normal distribution. Thus, resonance energies can reasonably be assumed to be distributed normally.

\subsection{Quantities Derived From Cross Sections}

Although cross sections were discussed theoretically in Chapter \ref{chap:reactions}, we need to look how this quantity is measured to understand what the expected distribution is for a nuclear cross section. In its simplest form:
\begin{equation}
    \label{eq:cross_section_example}
    \sigma(E) = \frac{N_R}{(N_B/A) N_T},
\end{equation}
where $N_R$ is the number of reactions, $N_B$ is the number of particles in the beam, $N_T$ is the number of particles in the target, and $A$ is the area of the incident beam. Since all of these quantities are positive, we can look at logarithm of $\sigma(E)$:
\begin{equation}
    \label{eq:cross_section_log}
    \ln{\sigma(E)} = \ln{N_R} +\ln{A} - \ln{N_B} - \ln{N_T}.
\end{equation}
Random fluctuations in any of these quantities will therefore lead to a measured cross section of $\sigma_{measured}(E)$:
\begin{equation}
    \label{eq:cross_section_log_error}
    \ln{\sigma_{measured}(E)} = \ln{\sigma(E)} + \ln{\delta N_R} + \ln{\delta A} - \ln{\delta N_B} - \ln{\delta N_T}. 
\end{equation}

It can be seen that Eq.~\ref{eq:cross_section_log_error} is analogous to Eq.~\ref{eq:energy_example}. Thus, the \textit{logarithm} of $\sigma_{measured}(E)$ tends to be normally distributed according to the central limit theorem. The probability distribution for $\sigma_{measured}(E)$ is then the logarithm of a normal distribution. This is called the \textit{log-normal} distribution, and is given by:
\begin{equation}
    \label{eq:lognormal_dist}
    f(x) = \frac{1}{\sigma \sqrt{2 \pi}} \frac{1}{x} e^{(\ln{x} - \mu)^2/2 \sigma^2}, 
\end{equation}
where the parameters $\mu$ and $\sigma$ are the mean and standard deviation of the corresponding normal distribution, respectively. Unlike the normal distribution, the log-normal distribution's mean does not correspond to $\mu$, nor is its variance given by $\sigma^2$. Instead, the mean, $E[x]$, is given by:
\begin{equation}
    \label{eq:lognormal_mean}
    E[x] = e^{(2 \mu + \sigma^2)/2},
\end{equation}
and the variance, $V[x]$, by:
\begin{equation}
    \label{eq:lognormal_variance}
    V[x] = e^{(2 \mu + \sigma^2)} \big[ e^{\sigma^2} - 1 \big].
\end{equation}
It is also useful to find an easy way to express $68 \%$ coverage. One possible way is to give the median ($med.$):
\begin{equation}
    \label{eq:lognormal_median}
    med.~[x] = e^{\mu},
\end{equation}
which for the log-normal distribution has the property that it is also the \textit{geometric} mean, where Eq.~\ref{eq:lognormal_mean}
is the \textit{arithmetic} mean. The geometric standard deviation is given by what is called the factor uncertainty ($f.u.$): 
\begin{equation}
    \label{eq:lognormal_median}
    f.u.~[x] = e^{\sigma}.
\end{equation}

These definitions are given mostly for later reference, but some intuitive meaning does exist. Specifically, cross sections tend to come with a percentage uncertainty. This percentage uncertainty, let us say $20 \%$ for example, could be cast as a factor uncertainty of $f.u. = 1.20$. In this case, $68 \%$ coverage would be given by the $(med. / f.u.)$ to $(med. \times f.u.)$ . All of these quantities provide useful ways to summarize the shape and spread of a log-normal distribution.

It is also interesting to note what happens to a log-normally distributed variable as $\sigma$ becomes small. Take $Y$ to be log-normal, then there is a corresponding normal variable $X$ related by:
\begin{equation}
    \label{eq:log_normal_to_normal}
    Y = e^{X} = e^{\mu + \sigma Z},
\end{equation}
where $Z \sim \mathcal{N}(0, 1)$. If $\sigma$ is small, then $e^{\sigma Z} \approx (1+ \sigma Z)$, which yields $Y \approx e^{\mu}(1+\sigma Z)$. Defining new parameters $\mu' = e^{\mu}$ and $\sigma' = \mu' \sigma$, it can be seen that $Y \sim \mathcal{N}(\mu', \sigma'^2)$. Thus, in the limit of small $\sigma$, a log-normal distribution begins to resemble a normal distribution with mean $e^{\mu}$ and standard deviation $e^{\mu} \sigma$.    

The arguments presented in this section were made for a cross section measurement, but any nuclear quantity that is the factor of many randomly distributed, positive quantities will be well described by a log-normal distribution due to the power of the central limit theorem. From the last chapter, it can indeed be seen that resonance strengths, Eq.~\ref{eq:resonance_yield}, and partial widths, Eq.~\ref{eq:proton_partial_width}, do meet this requirement.

\subsection{Monte Carlo Reaction Rates}
\label{sec:mc_rates}

Once appropriate probability distributions have been assigned to all the relevant quantities, samples can be drawn for each of them to yield samples for $N_A \langle \sigma v \rangle$ (where $N_A$ is Avogadro's number) as a function of temperature. The program \texttt{RatesMC} was used to perform this Monte-Carlo sampling. Details of this program can be found in Ref.~\cite{LONGLAND_2010_1}.

An instructive example is to examine a simplified reaction rate that is dominated by narrow resonances. One of the closest available examples to a narrow resonance dominated rate is the $^{21}$Ne$(p, \gamma)^{22}$Na rate. If, for now, the upper limits of the two lowest energy resonances are ignored, then this rate is determined completely by a direct capture component and $44$ narrow resonances. The input data for \texttt{RatesMC} is taken from Ref.~\cite{ILIADIS_2010_3}. At higher temperatures ($T \approx 30$ MK), the direct capture component becomes negligible and the rate is found from the sum of the resonant contributions:  

\begin{equation}
    \label{eq:narrow_rate_sum}
    N_A \langle \sigma v \rangle_{p \gamma} = N_A \bigg(\frac{2 \pi}{m_{p} kT} \bigg)^{3/2} \hbar^2 \sum_{n} e^{-E_n/kT} (\omega \gamma)_n.
\end{equation}

\begin{figure}
    \centering
    \includegraphics[width=.6\textwidth]{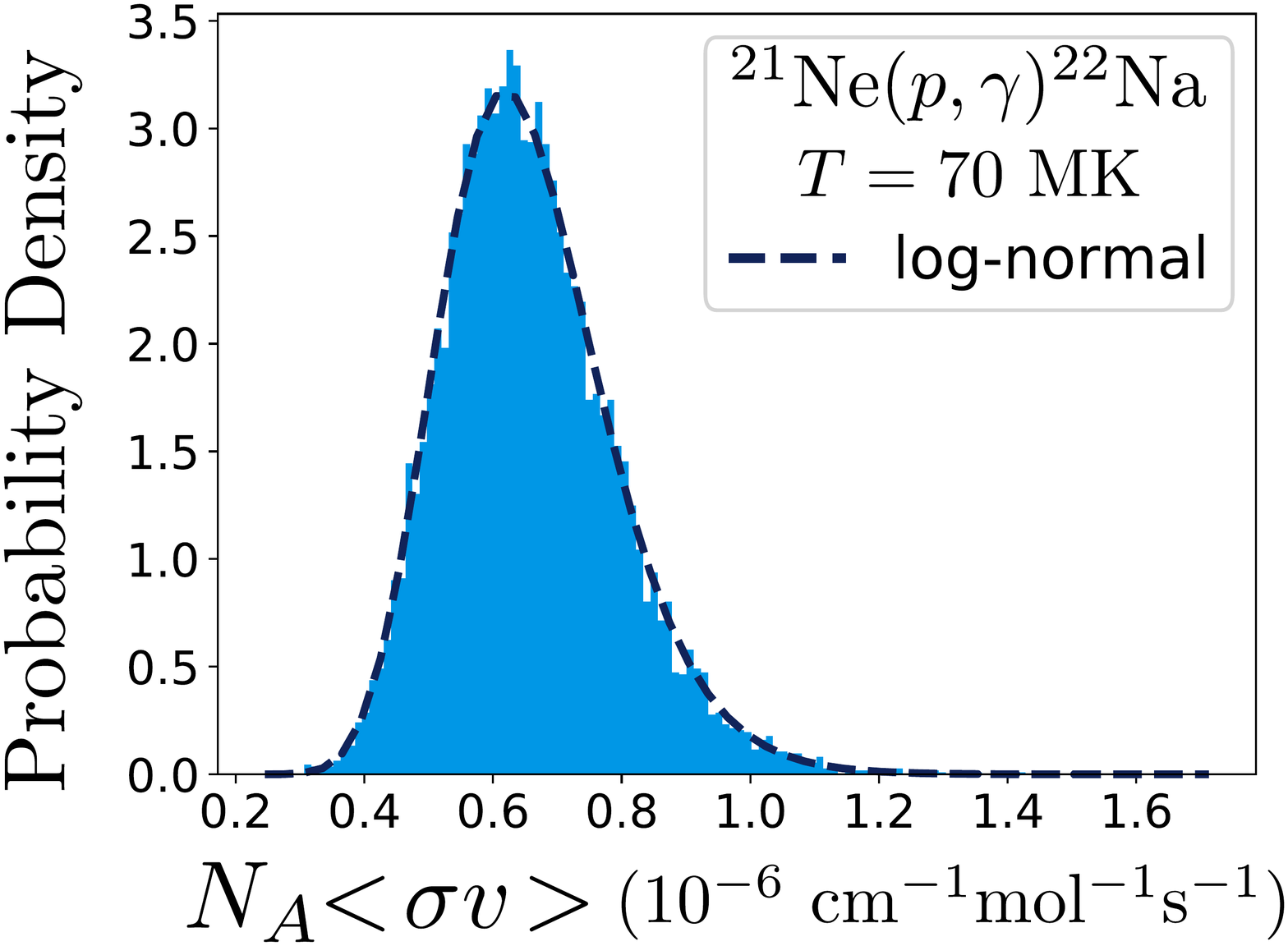}
    \caption{10000 samples for the simplified $^{21}$Ne$(p, \gamma)^{22}$Na reaction rate at $70$ MK. The log-normal approximation is shown as the dark blue, dashed line.}
    \label{fig:lognormal_na}
\end{figure}

It might be expected from this equation that the sum will cause the central limit theorem to take effect, resulting in a normally distributed reaction rate. However, at lower temperatures the Gamow peak effectively truncates this sum, meaning that the rate is dominated by individual resonances. In this case the rate is best viewed as a product of several variables, i.e, it will be log-normally distributed. This is indeed the case as seen in Fig.~\ref{fig:lognormal_na}, which shows the $10000$ Monte-Carlo samples drawn for $T = 70$ MK.      

\begin{figure}
    \centering
    \includegraphics[width=.6\textwidth]{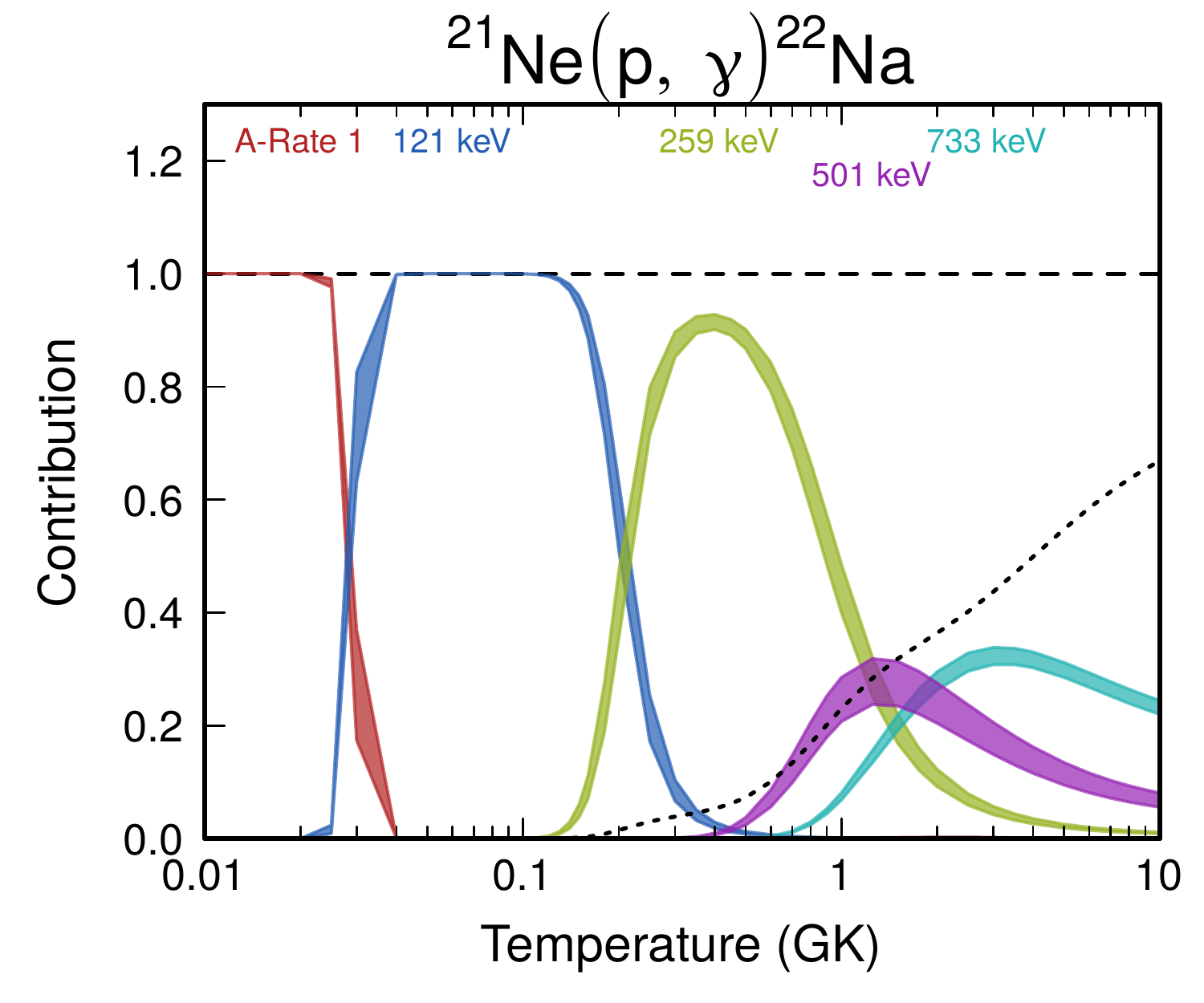}
    \caption{The individual contribution to the $^{21}$Ne$(p, \gamma)^{22}$Na reaction rate without the inclusion of upper limits. \textit{A-Rate $1$} refers to the non-resonant direct capture contribution, while the short-dashed line is the sum of all other resonances. The width of the lines shows the uncertainty in the contribution calculation. Only resonances that contribute more than $15 \%$ to the total rate are plotted. Furthermore, this rate ignores upper limits on two resonances for simplicity.}
    \label{fig:contribution_plot_ne}
\end{figure}

\begin{figure}
    \centering
    \includegraphics[width=.8\textwidth]{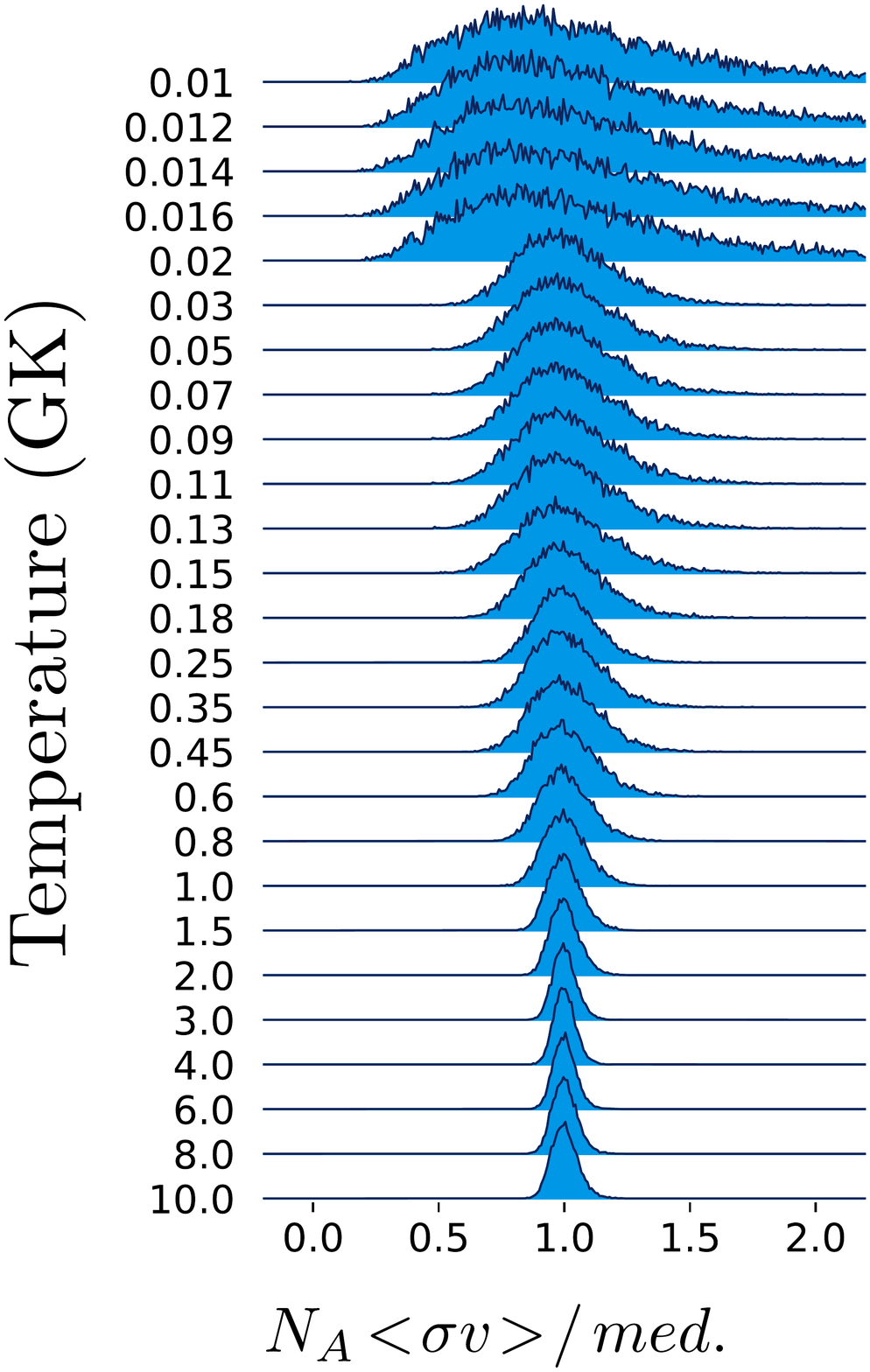}
    \caption{Plot of the Monte Carlo samples for the $^{21}$Ne$(p, \gamma)^{22}$Na reaction rate without upper limits normalized by the median of their corresponding log-normal distribution at various temperatures. As the temperature increases, no single resonance dominates, and the rate becomes more normally distributed.}
    \label{fig:ridge_ne}
\end{figure}

\begin{figure}
    \centering
    \includegraphics[width=.6\textwidth]{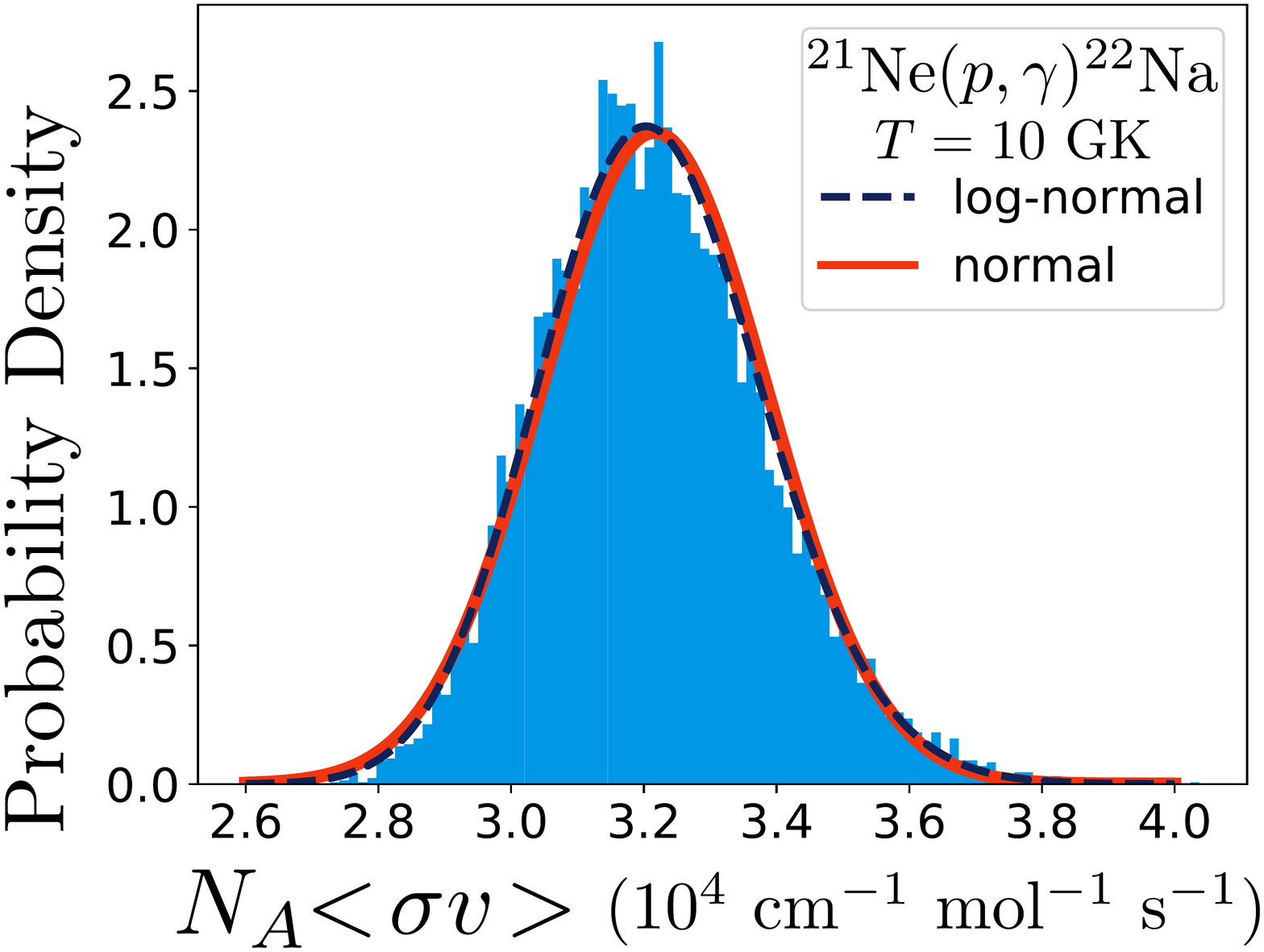}
    \caption{Monte-Carlo samples for the $^{21}$Ne$(p, \gamma)^{22}$Na rate at $10$ GK. The orange and dark blue, dashed lines show the normal and the log-normal approximations for the samples, respectively. }
    \label{fig:lognorm_norm_rate}
\end{figure}

To further examine this effect, the contribution of individual resonances can be plotted as a function of temperature using:
\begin{equation}
    \label{eq:rate_contribution}
    C(T) = \frac{\langle \sigma v \rangle_{n}(T)}{\sum_n \langle \sigma v \rangle_{n}(T)},
\end{equation}
where $C(T)$ denotes the relative contribution of the $n^{\textnormal{th}}$ resonance at temperature $T$. 
The results of this calculation for $^{21}$Ne$(p, \gamma)^{22}$Na are shown in Fig.~\ref{fig:contribution_plot_ne}. This plot emphasizes the impact of the Gamow peak, which causes the rate to be dominated only by resonances within this narrow energy window. Furthermore, as the temperature increases, the Gamow peak widens, the level density of the compound nucleus grows, and the sum of Eq.~\ref{eq:narrow_rate_sum} starts to become important. This effect can be seen in detail in Fig.~\ref{fig:ridge_ne}. This figure shows the samples from \texttt{RATESMC} at various temperatures divided by their median value. The distribution of the rate begins to become more normally distributed as the effects of individual resonances become less important. However, as Fig.~\ref{fig:lognorm_norm_rate} shows both the log-normal and normal distributions describe the reaction rate well at $10$ GK. This might seem odd, but it was shown that a normal distribution does arise from a log-normal when $\sigma$ becomes small. $\sigma$ becomes small in this case, even though the $f.u.$ values on the individual resonances might be quite large, because the sum does not preserve the percentage uncertainty. 
The log-normal distribution of reaction rates was first mentioned by Hix et. al \cite{HIX_2003}, but little justification was given beyond that the rates were manifestly positive. However, with development of the above Monte-Carlo method in Ref.~\cite{LONGLAND_2010_1}, it became possible to see why these distributions arise naturally in reaction rates, and, more importantly, connect them with the uncertainties measured in the lab. The log-normal distribution also has the benefit of being defined with only two parameters, making it easy to tabulate reaction rates, which, as will be shown, is necessary for reaction rate libraries.     

\subsection{Upper Limits}

Before this discussion is finished, it is important to note the impact of upper limits on reaction rates. An upper limit is reported from a study when the counts in the region of interest do not exceed background. As mentioned in the last section, I ignored the two low energy resonances in $^{21}$Ne$(p, \gamma)^{22}$Na, which only have upper limits reported, in order to simplify the discussion. However, it is not a trivial issue to define a probability distribution for these upper limits. Ref.~\cite{LONGLAND_2010_1} made the first arguments for how this should be approached. Without proof, the Porter-Thomas distribution (i.e. a chi-squared distribution with one degree of freedom \cite{porter_1956}) describes the distribution for reduced widths:
\begin{equation}
    \label{eq:porter_thomas}
    f(\theta) = \frac{c}{\sqrt{\theta^2}} e^{-\theta^2/ \langle \theta^2 \rangle},
\end{equation}
with $c$ being a normalization constant and $\langle \theta^2 \rangle$ denoting the \textit{local mean value} of the reduced width. This value can be estimated from experiments, with the most comprehensive study being Ref.~\cite{Pogrebnyak_2013}. 

\begin{figure}
    \centering
    \includegraphics[width=.8\textwidth]{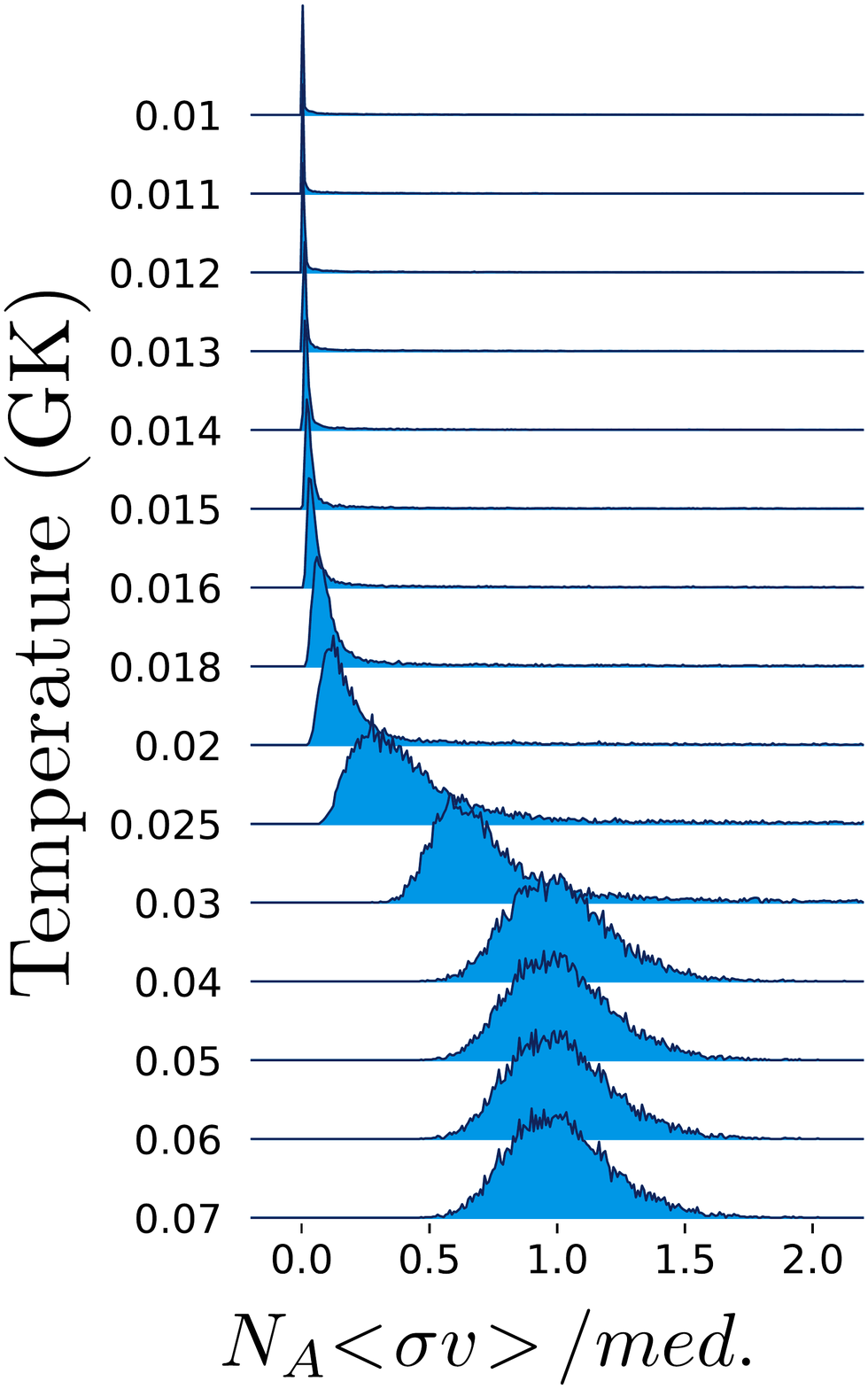}
    \caption{Plot of the Monte Carlo samples for the $^{21}$Ne$(p, \gamma)^{22}$Na reaction rate normalized by the median of their corresponding log-normal distribution at various temperatures. With the upper limits included, the rate is no longer well described by a log-normal distribution at low temperatures, which can be seen from the normalized rate deviating strongly from one.}
    \label{fig:ridge_ne_upper_limits}
\end{figure}

Including the two previously neglected low energy resonances, which only have upper limits, back into the Monte-Carlo calculation gives dramatically different behaviour at lower temperatures. Importantly, the log-normal distribution does not accurately represent the reaction rate in the scenario where a rate is dominated by upper limits. This means that while the log-normal distribution is convenient and appropriate in several situations, the reaction rate cannot be said in general to be log-normally distributed. This can be seen in Fig.~\ref{fig:ridge_ne_upper_limits}. Specifically, the rate normalized by $med.$ of the samples is no longer on the order of one. As the rate becomes dominated by measured resonances at $T = 40$ MK, the distribution returns to an approximately log-normal distribution.      

\section{Reaction Networks}

Accurate estimates of reaction rate uncertainties are, by themselves, of little importance. The ultimate goal of nuclear astrophysics is to relate theoretical calculations of stellar burning to observations. This requires that the rate of change of each element be tracked through the entirety of the stellar evolution. Considering just reactions between two particles, this can be done by solving a set of coupled differential equations of the form:
\begin{equation}
    \label{eq:network_rate}
    \frac{d N_{i}}{dt} = \sum_{j k} N_j N_k \langle \sigma v \rangle_{jk} - \sum_m N_i N_m \langle \sigma v \rangle_{im},
\end{equation}
where $N_i$ is the number of particles of nuclei $i$.
This equation can be made more general by including any additive or destructive rate, whether it is beta decay, electron capture, or photodisintegration \cite{iliadis_book}.

Solving this set of coupled differential equation requires each $\langle \sigma v \rangle$ be specified at every temperature considered. This is typically done by compiling a tabulated reaction rate library. This library specifies a rate for each reaction at specific temperatures. Examples are REACLIB \cite{Cyburt_2010}, BRUSLIB \cite{Xu_2013}, and STARLIB \cite{sallaska_2013}. Of these libraries, STARLIB is the only one, at this time, to incorporate statistical uncertainties. This is done by assuming the rate at each temperature is log-normally distributed, thus a rate is summarized by its recommended rate, $med.$, and factor uncertainty, $f.u.$, parameters. 

We now arrive at the final link in the chain, propagating the reaction rate uncertainties, i.e., those tabulated in STARLIB, to the predicted outputs of the network calculation. Again Monte-Carlo methods are the only effective way to do this due to the thousands of rates involved and the complexity of the network. However, the computational cost of Monte-Carlo procedures becomes most pronounced at this step in the process, since it is currently computationally infeasible to couple a full reaction network to a stellar evolution code and run it several thousand times. One recourse is to use \textit{post-processing}, where a temperature and density profile extracted from a more complex stellar evolution code is fed into a reaction network that separately evolves the nuclear material. This network calculation is referred to as a single-zone model because it only tracks nucleosynthesis at single location within the star and assumes that this is representative of the star as a whole, which necessarily excludes effects like convection and rotation.         

The code used for post-processing was originally developed by Nikos Prantzos and modified extensively by Christian Iliadis. It implements Gear's method to solve the coupled differential equations \cite{Longland_2014}, and is able to sample reaction rate uncertainties in the manner detailed in Ref.~\cite{Longland_2012}. This formalism draws a sample, $x(T)$, for the rates using a random variable $p \sim \mathcal{N}(0,1)$ to alter the median rate according to:
\begin{equation}
    \label{eq:rate_sample}
    x(T) = med. \times f.u.^p,  
\end{equation}
where $med.$ and $f.u.$ are the log-normal parameters tabulated in STARLIB. Thus, before a network is solved, a random sample is drawn for $p$ for each reaction in the network. This varies the rates by the same relative amount in light of the fact that $f.u.$ is temperature dependent. For example, take a calculation that involves a varying temperature profile. Looking at a single rate, $\langle \sigma v \rangle_i$, with a rate variation factor, $p_i = 1$. At each temperature, the median rate will be varied by $e^{\mu_i}e^{\sigma_i(T)}$. Thus, the median is being multiplied by $f.u.$ giving values of the rate at the upper end of the $68 \%$ coverage interval for every temperature. If instead $p_i = -1$, then $e^{\mu_i}/e^{\sigma_i(T)}$ giving the lower end of the $68 \%$ coverage interval. It should be noted that in general $p$ is dependent on temperature; however, Ref.~\cite{Longland_2012} showed that this simple scheme performs nearly as well as more complicated parameterizations of $p$.         

Once the network is run, the problem becomes identifying which reactions impact the abundance for a single element. Since the samples are drawn from meaningful statistical distributions, the relationship between a rate's variation and the final abundance can be understood in terms of correlations. To be more specific, given that we know the variations of each rate, which of these variations causes the largest corresponding variation in a selected nuclear species' final abundance. Unfortunately, there is no one way to measure statistical correlation. Two methods have been found to be most effective in the case of Monte-Carlo network calculations: the Spearman rank-order  correlation  coefficient, $r_s$, and mutual information, $MI$. The Spearman rank-order correlation coefficient converts the $n$ samples between random variables $X$ and $Y$ to ranks, $rg$. These ranks are integers where the smallest valued sample is assigned $1$ and the largest is given $n$. These ranks are assigned to each sample for both its $x$ and $y$ value. If none of the ranks are tied, i.e., all of the samples, $n$, are distinct, $r_s$ takes on the simple form:

\begin{equation}
    \label{eq:spearman}
    r_s = 1-\frac{6 \sum_i^n (rg(x_i) - rg(y_i))^2}{n(n^2-1)}.
\end{equation}

This quantity can vary from $-1$ to $1$ and measures any monotonic relationship between the variables $X$ and $Y$, or for this specific problem, a rate variation factor, $p_i$, and the final abundance of a selected nuclei. The use of this measure was first proposed in Ref.~\cite{Iliadis_2015}, and it is the method used for the results presented below.

The mutual information is a recently proposed method for identifying the correlations in Monte-Carlo network calculations \cite{Iliadis_mutual_information_2020}. While $r_s$ is a simple measure for quantifying correlation, its lack of sensitivity to non-monotonic functions is a serious deficiency for identifying network correlations where a simple relationship between $p_i$ and a final abundance might not exist (for examples see Ref.~\cite{Iliadis_2015}). The mutual information between two variables $X$ and $Y$ is given by \cite{Cover_2006}:
\begin{equation}
    \label{eq:mutual_information}
    MI = \sum_{i} \sum_{j} P(x_i, y_j) \log \bigg[ \frac{P(x_i, y_j)}{P(x_i)P(y_j)} \bigg], 
\end{equation}
where $P(x_i)$ and $P(y_j)$ are the marginalized probabilities for the samples and $P(x_i, y_j)$ is the joint probability for the two variables. If $X$ and $Y$ are independent, then $P(X,Y) = P(X)P(Y)$ and $MI = 0$. In this way, the mutual information is sensitive to any dependence of $Y$ on $X$ and vice versa. However, it does come with its own issues. The most pressing is: given a finite set of samples, how do we know $P(X)$, $P(Y)$, and the even more troublesome $P(X,Y)$? The difference between $r_s$ and $MI$ in this regard is that $r_s$ is \textit{nonparametric}, i.e., it does not assume what distribution the observed samples were pulled from, while $MI$ is \textit{parametric}, i.e., it requires the assumption that the observed samples were pulled from a specific probability distribution. The estimation of these distributions in order to calculate $MI$ reliably is well outside the context of this thesis, but more information can be found in Ref.~\cite{verdu_2019}.

Regardless of the issues with either $r_s$ or $MI$, they are only a means to an end. They are merely used to sift through the large amount of variables and data produced by the network calculations, in order to signal which reactions impact the final abundance of a given element the most strongly.

Now with a sampling scheme and correlation measure in place, the true advantage of Monte-Carlo reaction rates can be seen. The samples of $p_i$ for each rate can be examined for correlations between the final abundances of different isotopes, and, in turn, these correlations will signify that a given reaction rate has a strong impact on the final abundances of interest. Since the reaction rate uncertainties come directly from experimental uncertainties, these correlations in the network signal that the currently known experimental information for these important reactions is insufficient. Additional experiments can then be performed.

Demonstration of this method is shown in the next section, which details work originally presented in Ref.~\cite{Longland_2018} that uses the Monte-Carlo reaction rate formalism to reexamine $^{39}$K$(p, \gamma)$. For this study I performed all of the nucleosynthesis calculations, which show how the new rate dramatically impacts the spread in the amount of predicted $K$ in globular clusters.   

\section{The $^{39}$K$(p, \gamma)^{40}$Ca Rate}

The reevaluation of the $^{39}$K$(p, \gamma)^{40}$Ca rate was deemed necessary after the finding that it played an important role in the case of K enrichment in NGC 2419 \cite{dermigny_2017}. Prior to this reevaluation, the rate in STARLIB was based on an unpublished evaluation from 2014. However, it was found that this evaluation overlooked several measurements. By incorporating all available data, Longland et al. \cite{Longland_2018} found that the uncertainties were actually larger than previously thought.

The Monte-Carlo rate was incorporated into the STARLIB library to determine the astrophysical impact of the updated rate. A single-zone model was used for the Monte-Carlo reaction network. This model took initial mass fractions from Ref.~\cite{iliadis_2016}. For reference, the mass fraction is defined by:
\begin{equation}
    \label{eq:mass_fraction}
    X_i = \frac{N_i M_i}{N_A \rho},
\end{equation}
where $N_i$ is the number of nuclei of species $i$ per unit volume, $M_i$ is the atomic mass, $N_A$ is Avogadro's number, and $\rho$ is the density. Looking at the findings of Refs.~\cite{dermigny_2017, iliadis_2016} (see Fig.~\ref{fig:trajectory}), the observed abundances of all elements up to vanadium can be reproduced for hydrogen burning between $T = 100$ MK, $\rho = 10^8$ g/cm$^3$ and $T = 200$ MK, $\rho = 10^{-4}$ g/cm$^3$. A single representative burning environment with $T = 170$ MK and $\rho = 100$ g/cm$^3$ was selected. As a proxy for the timescale of the burning, the network was run until enough hydrogen was consumed such that it fell from its initial mass fraction, $X(^1 \textnormal{H}) = .75$, to $X(^1 \textnormal{H}) = .50$.         

\begin{figure}
    \centering
    \includegraphics[width=1.0\textwidth]{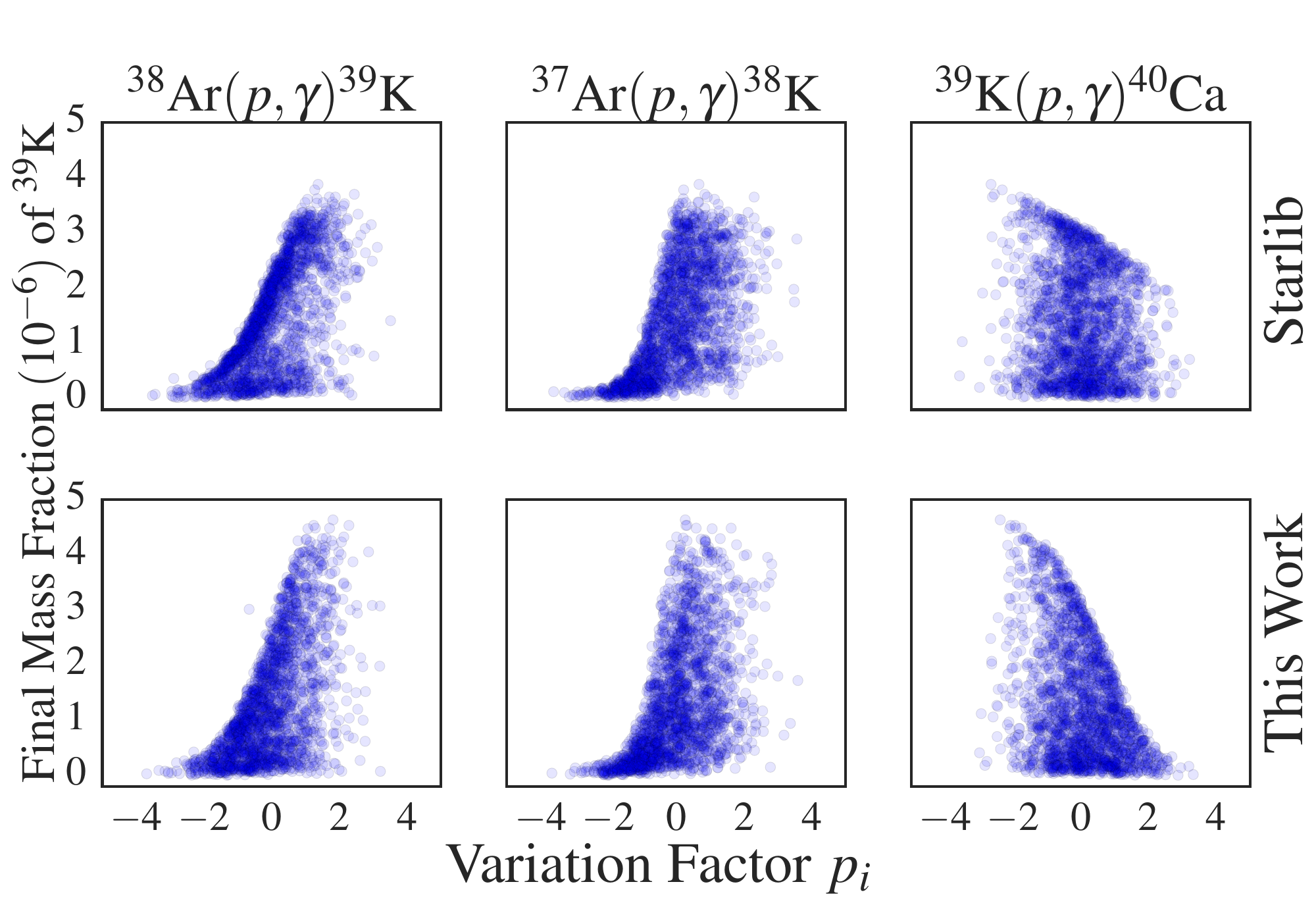}
    \caption{These scatter plots show the samples from $2000$ network runs plotted as a function of the rate variation factor $p_i$ and the final mass fraction of $^{39}$K. The three most strongly correlated rates from the previous evaluation and the updated one are shown. It can be seen that the correlation is much stronger for the $^{39}$K$(p, \gamma)$ rate once all of the experimental information has been included in the rate calculation. }
    \label{fig:dot_plot_39K}
\end{figure}

This network calculation was run $2000$ times with each run sampling every rate in the network, a total of $2373$ reactions up to $^{55}$Cr. All other parameters were held constant except for the rate variations. Two separate Monte-Carlo runs were performed with the first using the STARLIB rates and the second using the reevaluated rate. This was the only change between the two runs. After the $2000$ iterations for both sets of rates, $r_s$ was used to identify correlations between the $p_i$ and final mass fraction of $^{39}$K. The most strongly correlated reactions from both the STARLIB and reevaluated rates are shown in Fig.~\ref{fig:dot_plot_39K}. Each dot represents one run of the network. To highlight the importance of including all of the experimental information into the reaction rate calculation, the updated rate shows a markedly stronger correlation between $^{39}$K$(p, \gamma)$ and the final mass fraction of $^{39}$K. This correlation indicates that the nuclear uncertainties for the $^{39}$K$(p, \gamma)$ rate are large enough to hamper the predictions for a specific astrophysical environment.

\begin{figure}
    \centering
    \includegraphics[width=.6\textwidth]{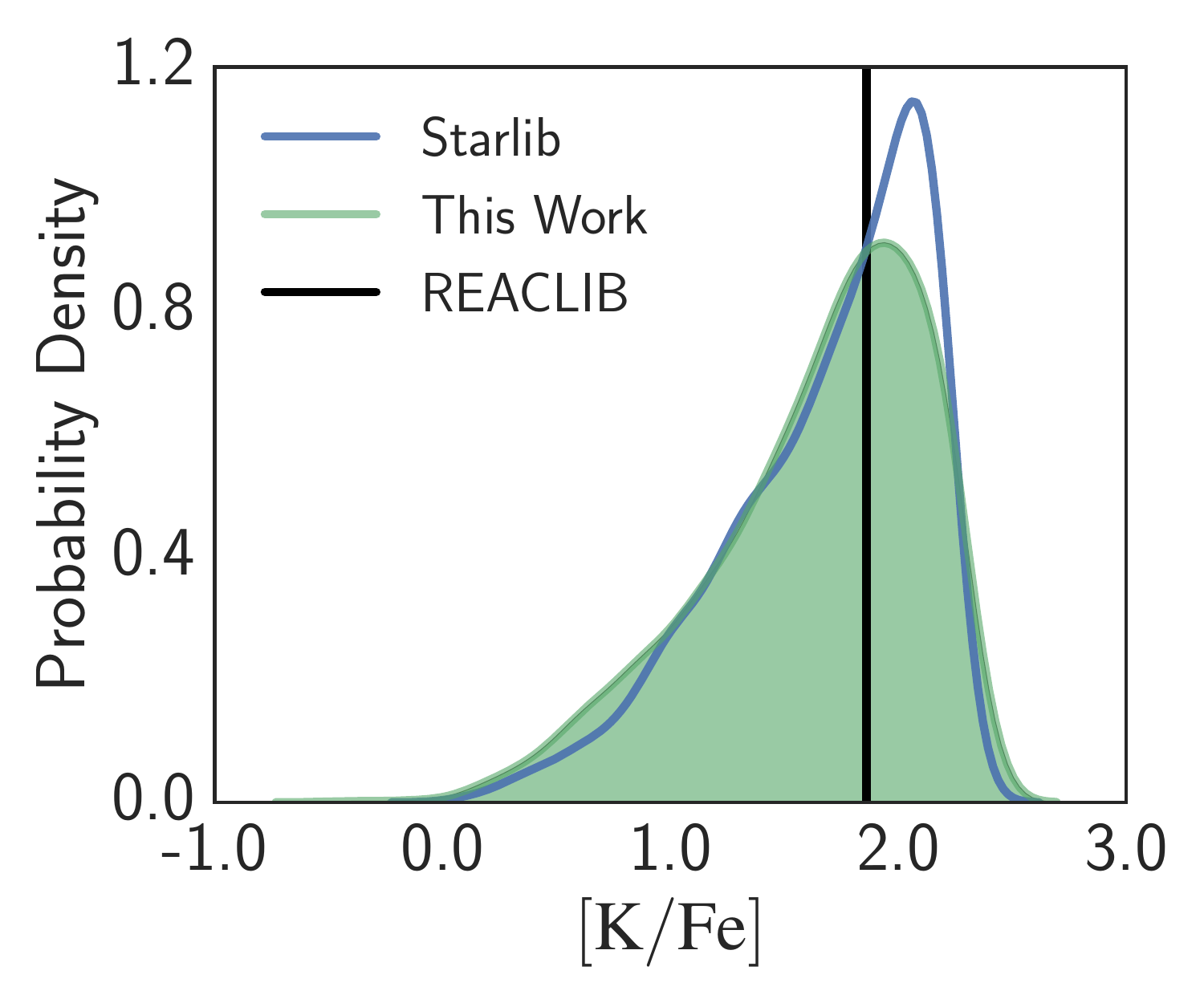}
    \caption{Predicted elemental abundances from the Monte-Carlo reaction network calculation. It can be seen that the updated rate is less peaked than the previous rate. The value from the commonly used REACLIB library is also shown. Note that REACLIB does not include uncertainties, and cannot properly account for the spread caused by the nuclear physics uncertainties.}
    \label{fig:K_final_abundances}
\end{figure}

Spectroscopic observations of globular clusters are only sensitive to elemental potassium, so transforming the results from these Monte-Carlo network calculations into elemental potassium is a key constraint on any future theoretical work.
In order to do this, all stable and long-lived isotopes of potassium must be considered. This includes $^{39}$K, $^{41}$K, and $^{40}$K. Additionally, any unstable nuclei that decay into these potassium isotopes must be accounted for. These radioactive nuclei are $^{39}$Cl, $^{39}$Ar, $^{41}$Ar, $^{39}$Ca,
$^{41}$Ca, and $^{41}$Sc. All of these elements contribute to the observed ratio $[\textnormal{K}/ \textnormal{Fe}]$, which is short hand for $[\textnormal{K}/ \textnormal{Fe}] = \log_{10} \big( \textnormal{K}/ \textnormal{Fe} \big)_{cluster} - \, \, \log_{10} \big( \textnormal{K}/ \textnormal{Fe} \big)_{sun} $. Thus, $[\textnormal{K}/ \textnormal{Fe}] = 1.0$  means that the amount of potassium relative to iron is $10$ times higher than the solar amount. The results of this calculation are shown in Fig.~\ref{fig:K_final_abundances}, where the results of this study are compared with the previous STARLIB rate and the rate found in REACLIB. The current rate and the previous STARLIB rate both have statistical uncertainties, so the samples from each network run are summarized by a Kernel Density Estimate (KDE) \cite{kde}. It can be seen that the updated rate has a larger spread in the predicted amount of $[\textnormal{K}/ \textnormal{Fe}]$, further demonstrating the need for a more precise reaction rate.

Finally, looking at the reaction rate contribution plot for this rate, the dominant resonances, and thus the uncertainty at the relevant temperatures, can be seen. In order for our predictions to improve, constraints must be put on the $337$-keV resonance.  

\begin{figure}
    \centering
    \includegraphics[width=.6\textwidth]{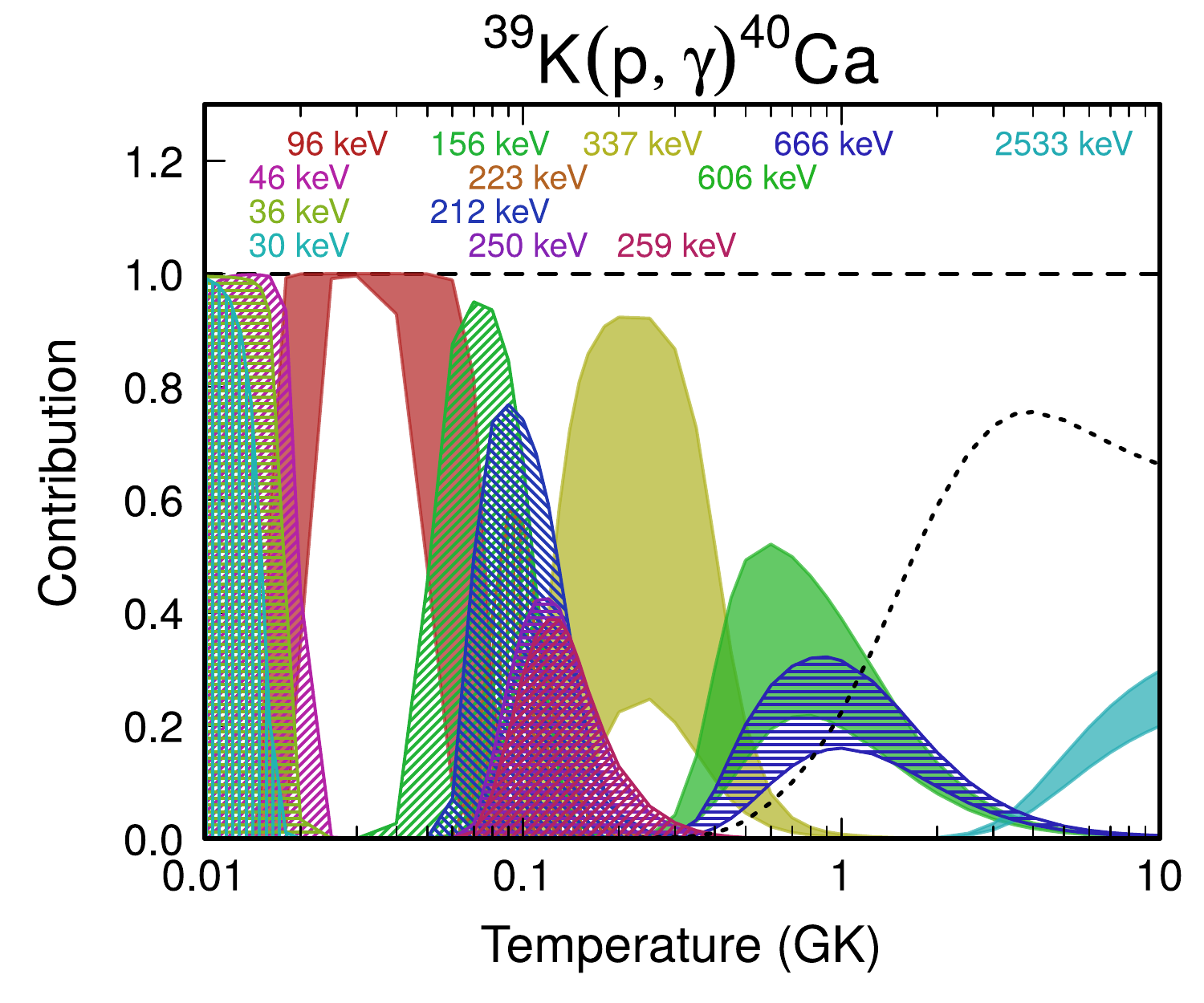}
    \caption{Contribution plot for the $^{39}$K$(p, \gamma)$ rate. The rate at $170$ MK is strongly determined by the properties of the $337$-keV resonance.}
    \label{fig:K_contribution}
\end{figure}

\section{Summary}

This chapter has outlined the modern approach to reaction rate uncertainties. Using Monte-Carlo methods, it was shown that nuclear uncertainties arising from particle partial widths, resonance energies, and resonance strengths can be propagated through the calculation of the reaction rate. Furthermore, the reaction rate library STARLIB allows these reaction rate uncertainties to be propagated through reaction network calculations. This means that at every step of the process, from the measurements in the lab to the predictions of stellar yields, uncertainties can be calculated and their impact properly assessed. The goals stated at the beginning of the chapter have been met, we can understand what we know and plan experiments to improve our predictions.    

%% file: Chapter-4/Chapter-4.tex
\chapter{Magnetic Spectroscopy at Triangle Universities Nuclear Laboratory}
\label{chap:tunl}

\section{Introduction}

This chapter details the facilities and equipment located at Triangle Universities Nuclear Laboratory (TUNL). Emphasis will be placed on the Split-pole spectrograph and its focal plane detector, which was used for the transfer measurements presented later. I was involved with the recommissioning of the spectrograph, which began in 2014, and led the work on the construction of the focal plane detector.  

\section{TUNL Tandem Lab}

TUNL has three primary experimental facilities: the Tandem lab, the High Intensity Gamma-ray Source (HIGS), and the Laboratory for Experimental Nuclear Astrophysics (LENA). The experimental work detailed in this document was performed at the Tandem lab. 

\subsection{Tandem Accelerator}

The tandem accelerator is so-called because it uses the "tandem" principle to create higher energy beams than a single ended accelerator would at the same voltage. This is done by holding the central terminal at a high positive voltage, $V_{term}$, and injecting the accelerator with negatively charged ions. An acceleration tube provides a smooth field gradient toward the central terminal, ensuring the ions in a negative charge state, $q$, are accelerated in a controlled way up to an energy of $qV$. At the center of the high voltage terminal is a thin carbon foil, with those used at TUNL typically having a value of $\sim 2 \frac{\mu \textnormal{g}}{\textnormal{cm}^2}$. This carbon foil strips the negative ions of their electrons, and its thickness impacts the amount of energy lost by the beam that is passing through it. Too thick of a target will degrade the energy resolution of the beam and lessen the lifetime of the foils, while too thin of a target will not ensure the ions reach a charge-state equilibrium, resulting in a smaller intensity of fully stripped ions \cite{SHIMA_2001}. After passing through the stripper foil, the now positively charged ions are repulsed by the positive terminal voltage and experience a second phase of acceleration, and finally leave the high energy end of the tandem. Thus, for these two phases of acceleration, a tandem accelerator with the terminal voltage, $V_{term}$, injected with an ion in a $-1$ charge state will produce a beam with energy (in units of eV):

\begin{equation}
    \label{eq:tandem_energy}
    E_{beam} = (Z + 1)V_{term}.
\end{equation}{}

The TUNL tandem is a High Voltage Engineering Corporation FN tandem. The maximum terminal voltage is 10 MV. The charging system consists of two Pelletron chains \cite{herb_1974}. These chains are made of metal pellets linked together with non-conductive nylon. The terminal is electrically isolated in a large cylindrical tank that is filled with a combination of CO, N$_2$, and SF$_6$.

\subsection{Ion Sources}

As emphasized above, a tandem accelerator must be injected with negatively charged ions. At TUNL two negative ion sources exist for hydrogen and helium, respectively. Using these two ion sources, beams of $^1$H, $d$, $^{3}$He, and $^{4}$He can be produced.

Hydrogen beams are created with the off-axis Direct-Extraction-Negative-Ion Source (DENIS). This ion source uses a duoplasmatron to extract negative ions from molecular gases \cite{duo}. The duoplasmatron uses a high voltage cathode to produce electrons. These electrons, in turn, break the molecular bonds of the H$_2$ gas, producing both positive and negative ions. A positively charged extraction electrode draws out the negative ions from the plasma, and accelerates them down the beam line.  

Unlike hydrogen, helium gas exists in an atomic form. Unique from other noble gases, He$^{-1}$ forms a metastable state with a binding energy of $.077$ eV \cite{he_binding_energy}. This means a negative ion beam of He$^{-1}$ can be produced. The TUNL helium-exchange source does this by first using a negatively charged extraction electrode to draw out positive ions from a plasma created by a duoplasmatron. These positive ions then pass through a sodium charge exchange canal. An oven is used to evaporate metallic sodium into a gas and pass it through the canal. The positive helium beam passed through this gas. Due to the low electron affinity of the gas, it is possible for the helium beam to pick up two additional electrons.  Approximately $1 \%$ of the beam passing through the sodium canal will form He$^{-1}$ \cite{charge_exchange_fraction}. The negatively charged beam is then accelerated using a positive potential towards the low energy side of the tandem. Beam currents of around $\sim 2$ $\mu$A are possible using this method.

\subsection{90-90 Beam Line}

\begin{figure}
    \centering
    \includegraphics[width=.85\textwidth]{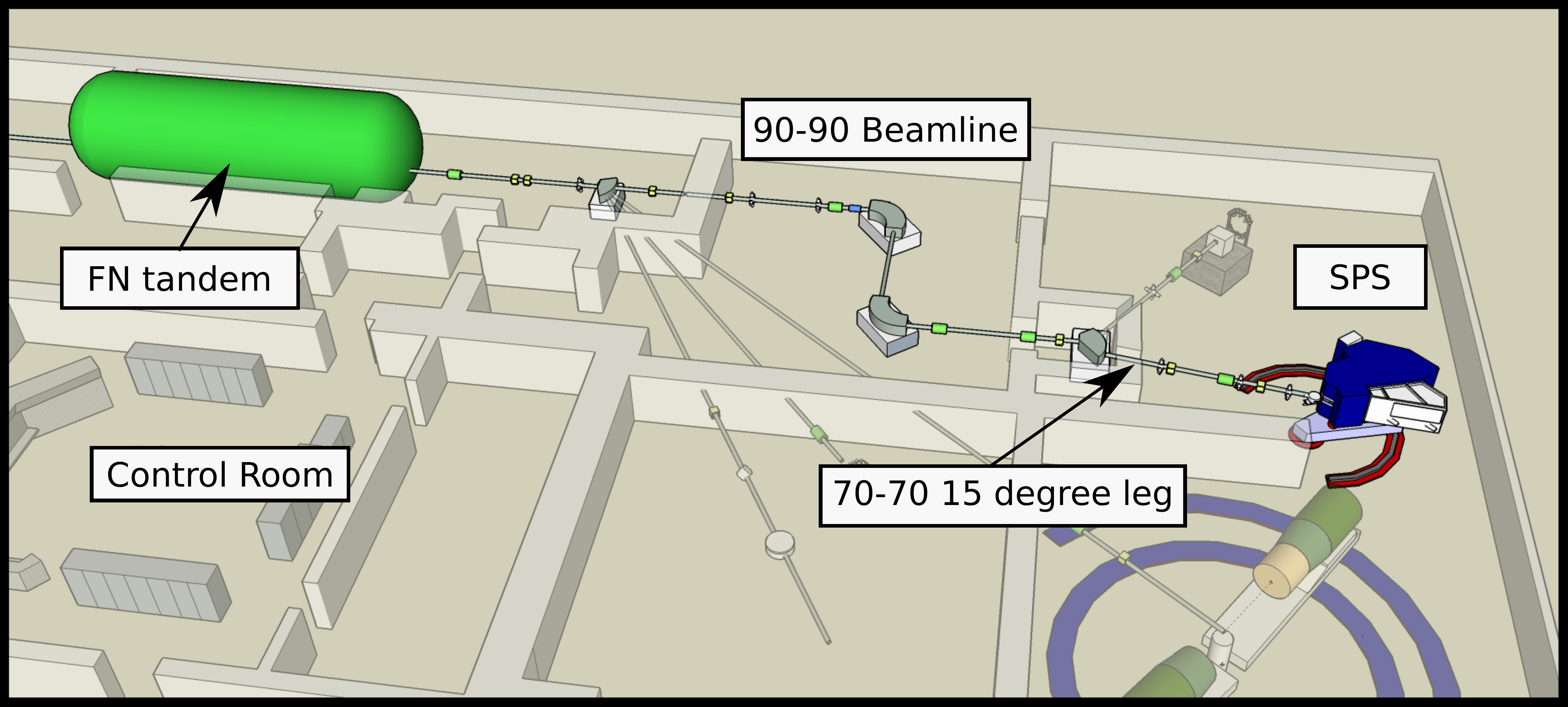}
    \caption{A $3$D model of the TUNL Tandem lab. The beamline used to deliver beam to the SPS is highlighted with major components labeled.}
    \label{fig:the_lab}
\end{figure}

\begin{figure}
    \centering
    \includegraphics[width=.85\textwidth]{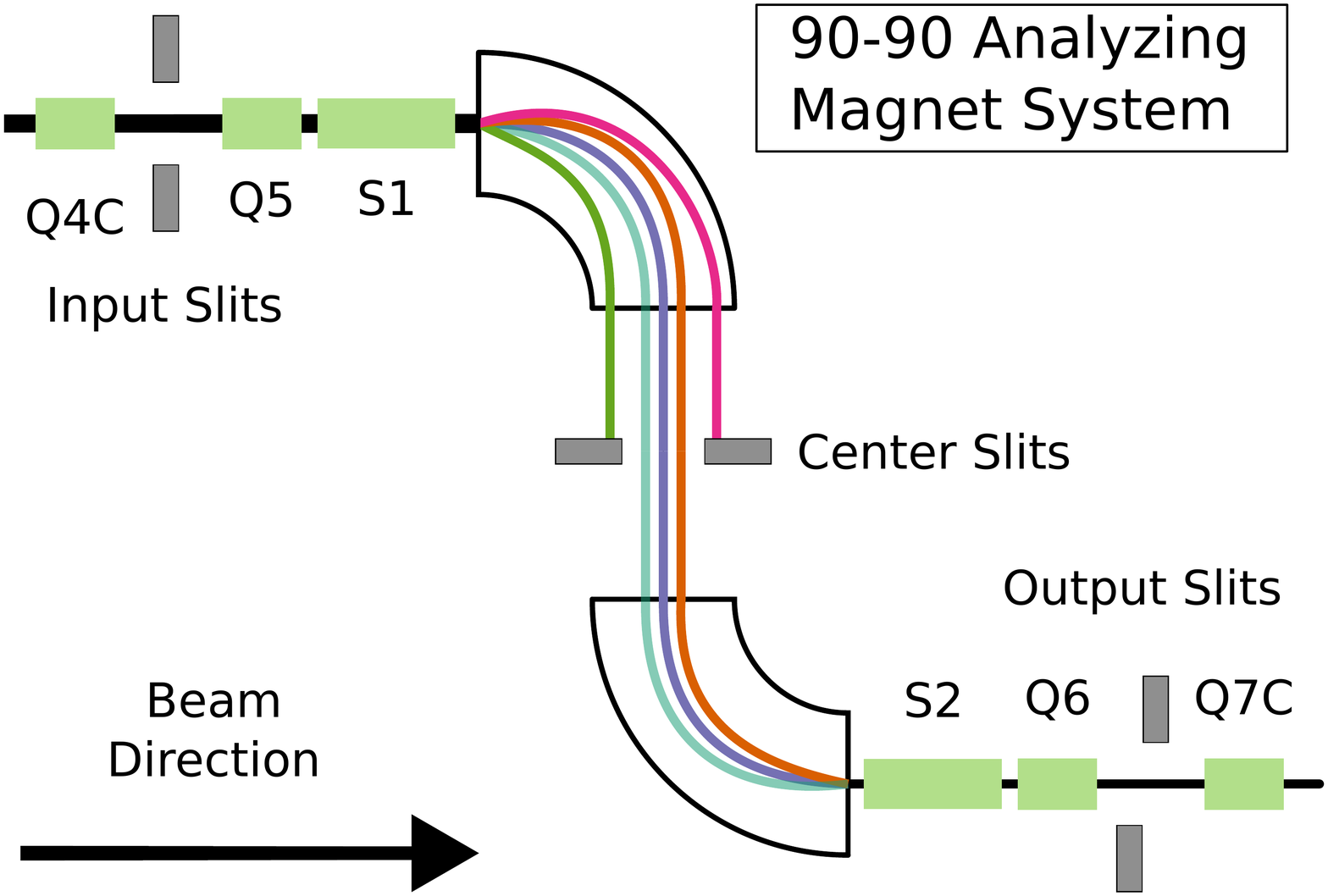}
    \caption{A top-down cartoon of the focusing and energy analyzing elements of the $90 \textnormal{-} 90$ system. Trajectories of particles with different energies passing through the two dipoles have been shown in order to demonstrate the operational principle of the center slits. Note that these trajectories are for illustration only, they do not represent ion-optic calculations.}
    \label{fig:90_90_system}
\end{figure}

The beam emerging from the tandem post-acceleration will have an energy that is inadequate for accurate and precise measurements. This problem can be remedied by passing the beam through an analyzing magnet. TUNL has two sets of analyzing magnets, a general purpose dipole that can bend the beam between $20^{\circ} \textnormal{-} 70^{\circ}$ and a set of $90^{\circ}$ dipole magnets for use with experiments requiring high energy resolution. A schematic of the high resolution beam line is shown in Fig.~\ref{fig:the_lab}. To deliver the beam to the Split-pole Spectrograph (SPS), it is passed through the $0^{\circ}$ port of the $20 \textnormal{-} 70$ magnet. The magnet is degaussed in order to prevent any deflection of the beam. A series of magnetic steerers and quadrupoles direct and focus the beam into the $90 \textnormal{-} 90$ system. The beam image is focused by the Q4C quadrupole through the entrance slits, which defines the spot size of the beam before energy analysis. Before the first dipole another quadruple, Q5, is used to increase the horizontal size of the beam to increase the resolving power of the system, while a sextupole, S1, provides higher order aberration corrections \cite{WILKERSON_1983}. After these elements the beam passes through the first $90^{\circ}$ dipole. The magnetic field of this dipole is controlled by a precision NMR probe. The energy dispersed beam is then defined further by the center slits. Upon entering the second dipole, the beam is bent again by $90^{\circ}$ and then passed through another sextupole and quadrupole. The presence of small differences between the field strengths of the two dipoles can be corrected by manually adjusting the current on the second dipole via a trim control in order to maximize beam current on the exit slits. This system and its operating principle are shown in Fig.~\ref{fig:90_90_system}. After all of these steps, the beam energy can be accurately determined using the NMR field reading and the known bending radius of the first dipole, which is $40$ inches \cite{9090_disc}. Further stability is provided by using either the center or exit slits to regulate the tandem terminal voltage. Beam current is measured on both of the horizontal slits. If the current on one slit starts to increase, the terminal voltage is increased or decreased automatically by a slit-current feedback system in order to balance the current reading between the slits \cite{WESTERFELDT_1984}. The spread in the beam energy is determined by how wide the entrance, center, and exit slits are. For a setting with the slit widths set to $\Delta x_{entrance} = 1$ mm, $\Delta x_{center} = 0.5$ mm, and $\Delta x_{exit} = 1$ mm, the beam spread, $\Delta E$, is given by $\Delta E \approx \frac{E}{5800}$. This relation scales linearly for the width of the slits, such that a setting with $\Delta x_{entrance} = 2$ mm, $\Delta x_{center} = 1.0$ mm, and $\Delta x_{exit} = 2$ mm gives $\Delta E \approx \frac{E}{2900}$. For the $^{23}$Na$(^3 \textnormal{He}, d)^{24}$Mg experiment the slits were set to $2$ mm, $1$ mm, and $2$ mm, which for a $21$ MeV beam, gives $\Delta E < 10 $ keV.               

\subsection{SPS Beamline and Target Chamber}

Once the beam is energy analyzed, it is directed to the $15^{\circ}$ SPS beam line via a $70 \textnormal{-} 70$ switching magnet. This beam line uses two sets of magnetic steerers that are controlled via feedback slits to stabilize the beam going through both the $70 \textnormal{-} 70$ magnet and another quadrupole magnet. A steerer is used to fine tune the beam through a $1$ mm collimator on the target ladder. With the $90 \textnormal{-} 90$ slits set as mentioned above, and with the focusing from the beamline quadrupole, nearly $100 \% $ of the energy analyzed beam can be passed through the collimator.

\section{Split-pole Spectrograph}
\label{sec:split_pole}

At the energies used to perform transfer reactions, there are many open reaction channels. As a consequence, any detection system will have to be able to identify the large variety of reaction products that are produced. Magnetic spectrographs separate reaction products based on:
\begin{equation}
    \label{eq:magnetic_rigidity}
    \rho B = \frac{p}{q},
\end{equation}{}
where $B$ is the magnetic field of the spectrograph, $\rho$ is the 
bending radius, $q$ is the charge of the particle, and $p$ is the particle's momentum. The product $B \rho$ is called the magnetic rigidity. It can be seen from Eq.~\ref{eq:magnetic_rigidity}
that, in the case of a constant charge, a magnetic spectrograph transforms differences of momentum into a spatial separation.

Several quantities need to be defined in order to characterize the capabilities of a magnetic spectrograph. Considering that the separation of particles based on their magnetic rigidity is analogous to an optical lens, ion-optics will provide the necessary tools to analyze a given spectrograph's performance. These optics can be expressed in a phase space that consists of the entrance angle $\theta_i$, the exit angle $\theta_f$, the entrance azimutal angle $\phi_i$, the exit azimutal angle $\phi_f$, the beam image in the horizontal direction $x_i$, the beam image in the vertical direction $y_i$, the image after the magnetic field $x_f$, the final vertical image $y_f$, and the momentum spread $\delta$, which is defined as:
\begin{equation}
  \label{eq:7}
  \delta = \frac{\Delta p}{p}.
\end{equation}
This quantity relates small changes in the momentum, $\Delta p$, to a reference momentum, $p$, through the small angle approximation in momentum space. It should also be noted that $\delta_i = \delta_f$ due to the conservation of momentum. Furthermore, a coordinate system that has the beam direction as $+z$, beam left as $+x$, and up as $+y$ is assumed. For all of these variables the initial and final coordinates can be related by an optical transfer matrix. Of particular interest is the coordinate $x_f$, which will dictate the final resolution of the spectrograph. Explicitly writing out all of the matrix elements to second order, we have:
\begin{equation}
\begin{split}
    \label{eq:opt_matrix}
    x_f = & (x_f|x_i)x_i + (x_f|\theta_i)\theta_i + (x_f|\delta)\delta \\
    & +(x_f|x_i,x_i)x_i^2 + (x_f|\theta_i, \theta_i) \theta_i^2 + (x_f| \delta, \delta) \delta^2 + (x_f|y_i, y_i)y_i^2 + (x_f|\phi_i, \phi_i) \phi_i^2 \\ 
    &+ (x_f|x_i, \theta_i)x_i \theta_i + (x_f|x_i, \delta) x_i \delta + (x_f | y_i, \phi_i) y_i \phi_i + (x_f|\theta_i, \delta) \theta_i \delta,     
\end{split}
\end{equation}
where the matrix coefficients are equivalent to partial derivatives \cite{enge, opt_matrix}. Azimutal symmetry eliminates all terms with odd powers of $\phi_i$ and $y_i$. First order focusing is achieved when $(x_f|\theta_i) \approx 0$, while second order double focusing requires $(x_f|\theta_i, \theta_i) \approx 0$.

The first order terms in this expansion have intuitive physical meaning. $(x_f|x_i)$ is the magnification, $M$, and it relates the initial and final beam spot sizes. $(x_f|\delta)$ is the dispersion, $D$, and it describes the shift in position corresponding to a unit shift in momentum. These two terms determine the first order resolving power, $\mathcal{R}_1 = \delta^{-1}$. A spectrograph is able to resolve particles with different momenta if the dispersion is greater than the peak width, $\Delta x_f$. If there is a finite size to beam spot, then the magnification of the spectrograph produces a final peak width according to $\Delta x_f = M \Delta x_i$, and spreads out particles according to $\Delta x_f = D \delta$. 
Setting these two terms equal, we have for the first order resolving powers:
\begin{equation}
    \label{eq:resolving_power}
    \mathcal{R}_1 = \frac{D}{M \Delta x_1}.
\end{equation}{}
Higher order aberrations will change the right hand side of this expression, but the definition of $\mathcal{R}$ does not change.

The SPS was originally designed by Harold Enge \cite{splitpole}. As implied by the name, the design encloses two dipole magnets into a single field coil. Configuring the magnets in this way creates three distinct magnetic field regions: the field of the first dipole, an intermediate field between the two dipoles, and the field of the second dipole \cite{enge_optics}. These fields create the conditions for second-order double focusing in the horizontal direction and first-order focusing in the vertical direction \cite{enge}. The resolving power of the SPS is $\mathcal{R} = 4500$.

An SPS can accept particles up to $8$ msr, but resolution rapidly degrades for solid angles greater than $2$ msr because of the
$(x_f|\theta^3)$ term \cite{enge}. In terms of the vertical and horizontal acceptance, this translates to $\Delta \theta = \pm 70$ mrad and $\Delta \phi = \pm 40$ mrad. The solid angle on the TUNL SPS is defined using brass apertures located $d = 22.58$ cm away from the target ladder. These apertures can be switched via motor control and they have a range of solid angles from, $0.122$ msr to $5.53$ msr. In order to minimize the resolution loss, we limited our solid angle to $< 1$ msr.    

Beam integration is a difficult problem for any spectrograph. A Faraday cup is necessary to properly integrate beam current, but the physical size of these devices can be prohibitive. In particular, the reaction products have to enter the spectrograph, which means that the larger the Faraday cup is, the fewer angles can be measured with the spectrograph. The dimensions of the target chamber are also limited in order to decrease the distance between the target ladder and the entrance aperture. For the TUNL Split-pole, this means that our beam integration is rather crude. The beam is stopped in a rectangular sheet of $1/16$-inch-thick tantalum, with dimensions of $3/8 \times 1/2$ inches. Even with these modest dimensions, the beam stop limits spectrograph angles $< 5^{\circ}$. There are two effects that will produce inaccurate current integration from a beam stop such as this: 
\begin{enumerate}
    \item Electrons from the target showering the beam stop.
    \item Electrons in the beam stop being boiled off due to the beam.
\end{enumerate}{}
Item 2 can be remedied by applying a positive charge to the beam stop in order to reabsorb the emitted electrons. For the TUNL SPS a $300$ V battery is used for the positive potential. The positive terminal of the battery is attached to the beam stop and the negative terminal is attached to the beam charge integrator (BCI). Though not implemented during the experiments for this thesis, it has been shown that rare earth permanent magnets attached to the target ladder can minimize the effects of item 1 \cite{wrede_thesis}.   

\section{Focal Plane Detector and Electronics}
Full details of the focal plane detector and construction can be 
found in Ref.~\cite{marshall_2018}, and this section reproduces portions of that work.

\begin{figure}
    \centering
    \includegraphics[width=.75\textwidth]{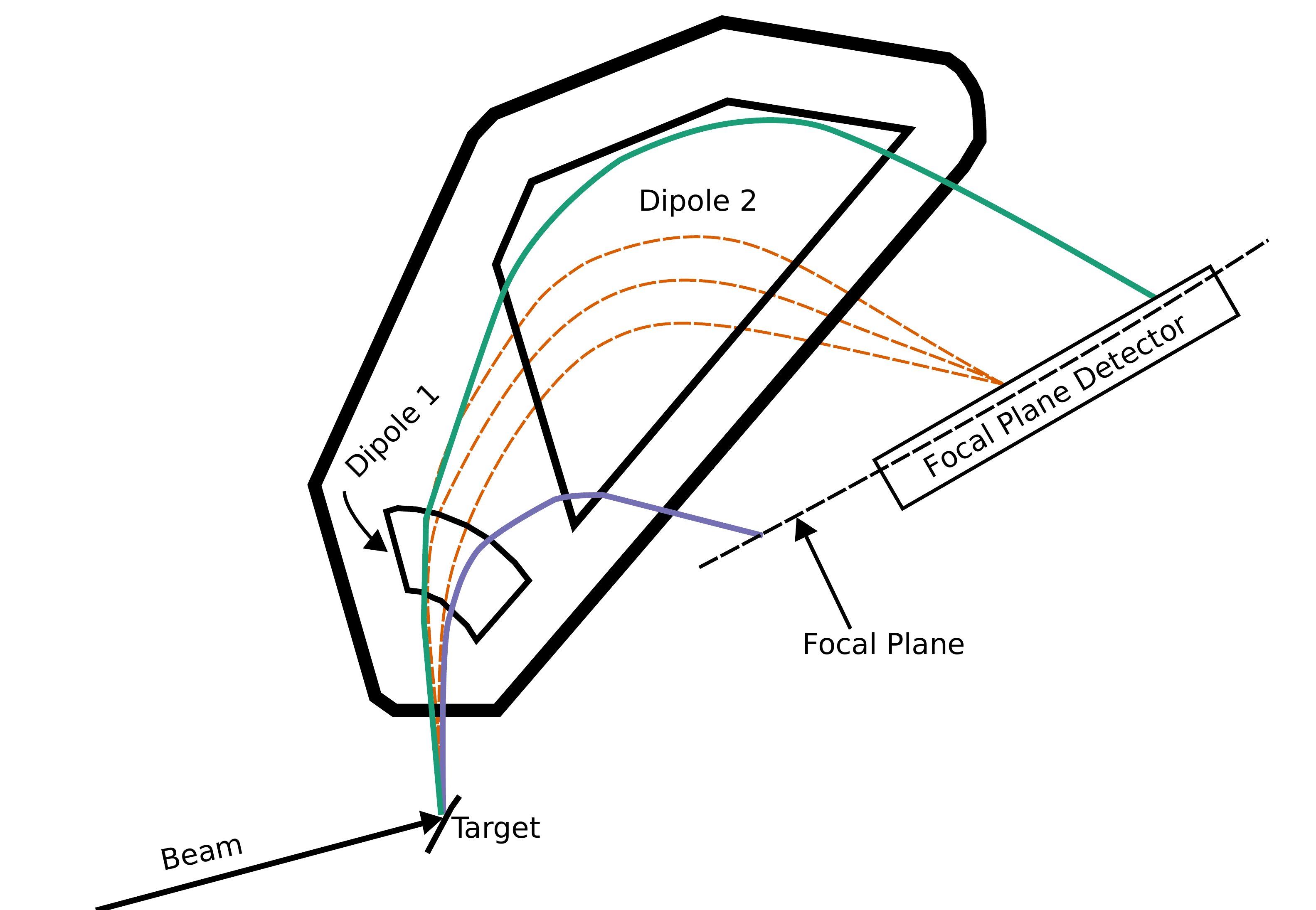}
    \caption{Top-down cross section of the SPS. The two dipoles enclosed in a single field coil produce a magnetic field that spatially separates particles based on their momentum to charge ratio. A position sensitive detector located at the curved focal plane of the spectrograph measures the bending radius of the reaction products.}
    \label{fig:enge_overview}
\end{figure}

As seen in Fig.~\ref{fig:enge_overview}, particles passing through the SPS are focused to a point. The focal points for particles with different magnetic rigidities form a dispersive image of the target on a plane. This focal plane is curved and lies at a $41.5^{\circ}$ angle to the magnetic exit. A high resolution, position sensitive detector located at this plane is needed in order to determine the magnetic rigidity, and thus energy, of the reaction products. This detector must also be able to distinguish between the various species of particles that have similar magnetic rigidities as those of the particles of interest.

\begin{figure}
    \centering
    \includegraphics[width=.75\textwidth]{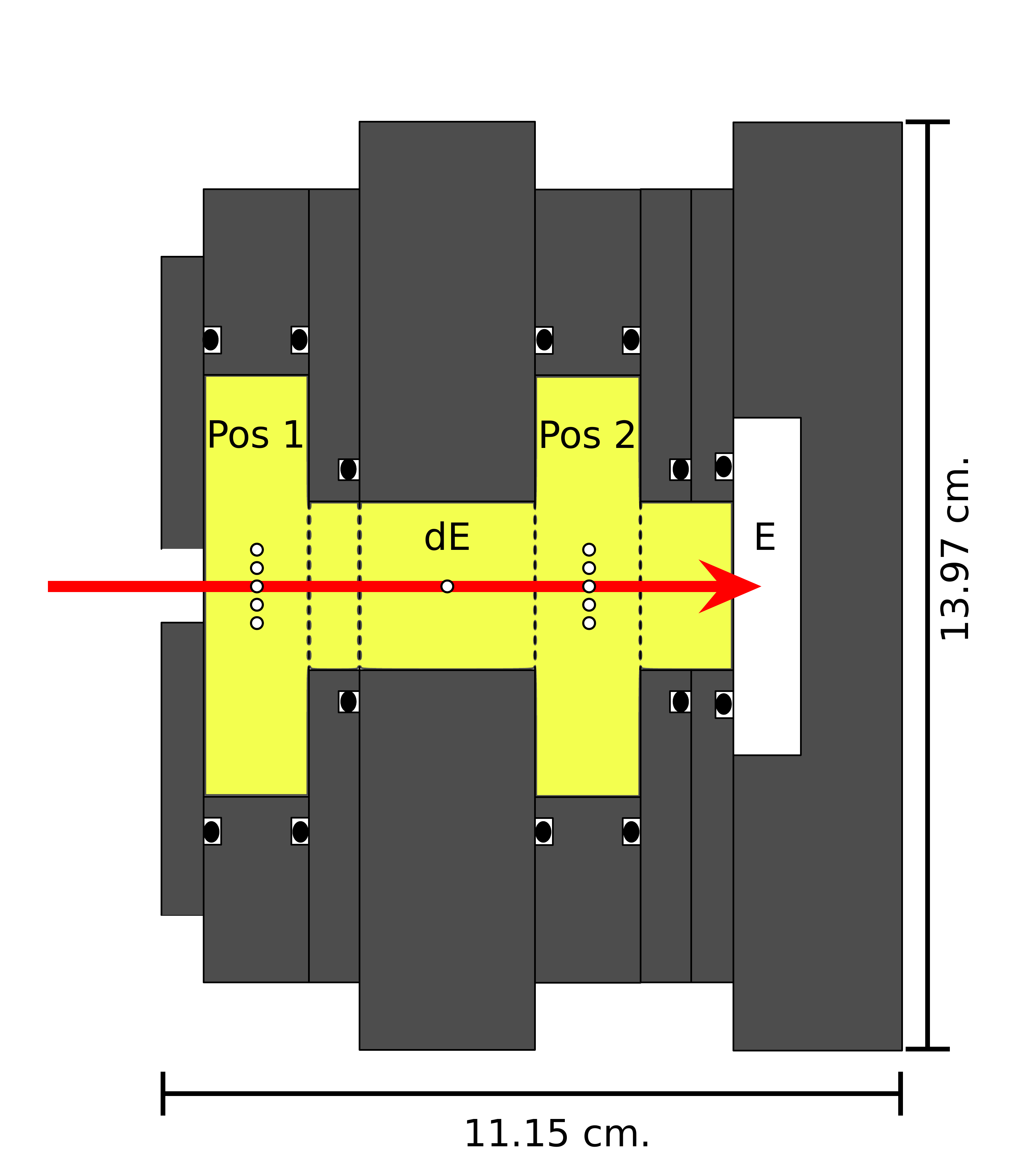}
    \caption{Side view of the cross section of the focal plane detector. The red arrow represents the direction of incident particles and
    the black ovals show the location of the o-ring seals. The approximate location of the anode wires throughout the detector are indicated by the white circles. The gas filled regions are indicated by light yellow shading. Though not indicated in the figure, the length of the detector is $71.12$ cm.}
    \label{fig:det_cross}
\end{figure}

Efforts to construct a focal plane detector capable of providing high spatial resolution and particle identification started in the summer of 2014. A non-functioning but completely fabricated detector was located at TUNL. The detector was an updated version of the one originally presented in Ref.~\cite{hale_thesis}. This design uses two position sensitive avalanche counters, a $\Delta E $ proportionality counter, and a plastic scintillator. All of these detectors are integrated into one assembly which is shown in Fig.~\ref{fig:det_cross}. Extensive work to return the detector to working order began in the summer of 2014, and the first experiments to characterize the detector began in 2015.           

{}

\subsection{Position Section}

Position measurements of particles leaving the high magnetic field region of the SPS are performed by two position sensitive avalanche counters. The positions are measured near the entrance of the detector and again before the particles are stopped in the total energy scintillator. The position-sensitive avalanche counters operate as follows, and are
represented pictorially in Fig.~\ref{fig:ion}. Five high voltage anode wires are located within each counter between two cathode foils made of aluminized Mylar. {These counters sit inside the detector chassis, which is pressurized to 200 Torr with circulating isobutane. The pressurized environment is isolated from the high vacuum of the spectrograph with a $12.7\textnormal{-} \mu m$ thick Kapton entrance window. Charged particles that enter the detector will ionize the isobutane as they pass through the detector volume. If an ionization event occurs within the electric field of the counters, electrons will be rapidly accelerated towards the positively charged anode wires setting off a series of secondary ionization events, thereby creating an electron avalanche \cite{knoll}. This avalanche is negatively charged and localized around the particle's position as it passes the anode. The cloud of negative charge induces a positive charge on both of the cathode foils. The foil closer to the entrance of the detector is electroetched \cite{etch}, and is described in detail below. Etching creates electrically isolated strips that are connected together via a delay line. Thus, as charge is carried out of the detector from each strip, it is exposed to a different amount of delay. Therefore, the timing difference between the two sides of the detector gives a relative measurement of position. If the charge was only distributed over one strip, then the position resolution would be limited to the strip width. However, distributing the charge over multiple strips allows an interpolation of the composite signal, thereby improving the spatial resolution to the sub-millimeter level. As the particle exits the position sections, it passes through the grounded cathode foil that helps shape and isolate the electric field from the anodes.  

Position sensitive avalanche counters are commonly designed to have
pickup strips parallel to the incident particle path \cite{msu,heavy,parikh, argonne_det}; however, the etched foils in the TUNL detector sit
perpendicular to the particle path. This type of setup has also been
used in the focal plane detector for the decommissioned Q3D spectrograph at the Maier-Leibnitz Laboratory
\cite{vert} and the decommissioned Q3D at Brookhaven National Laboratory \cite{BNLDetector}.
These designs have been shown to have excellent position resolution.
Additionally, if cathode foils are used, the number of wires required can be drastically reduced; thus, easing maintenance of the
detector. However, these designs are not suited towards heavy ion reactions, where
the cathode foils would provide additional scattering surfaces that degrade mass resolution.
The effects of these foils on the energy loss of light particles were examined via GEANT4 simulations in Ref.~\cite{marshall_2018} and found to have a negligible impact on the position resolution.

The methods used to fabricate the position section assemblies are discussed below with particular attention paid towards the etched cathode planes, delay line, and timing electronics.

\begin{figure}
  \centering
  \includegraphics[width=.6\textwidth,angle=90]{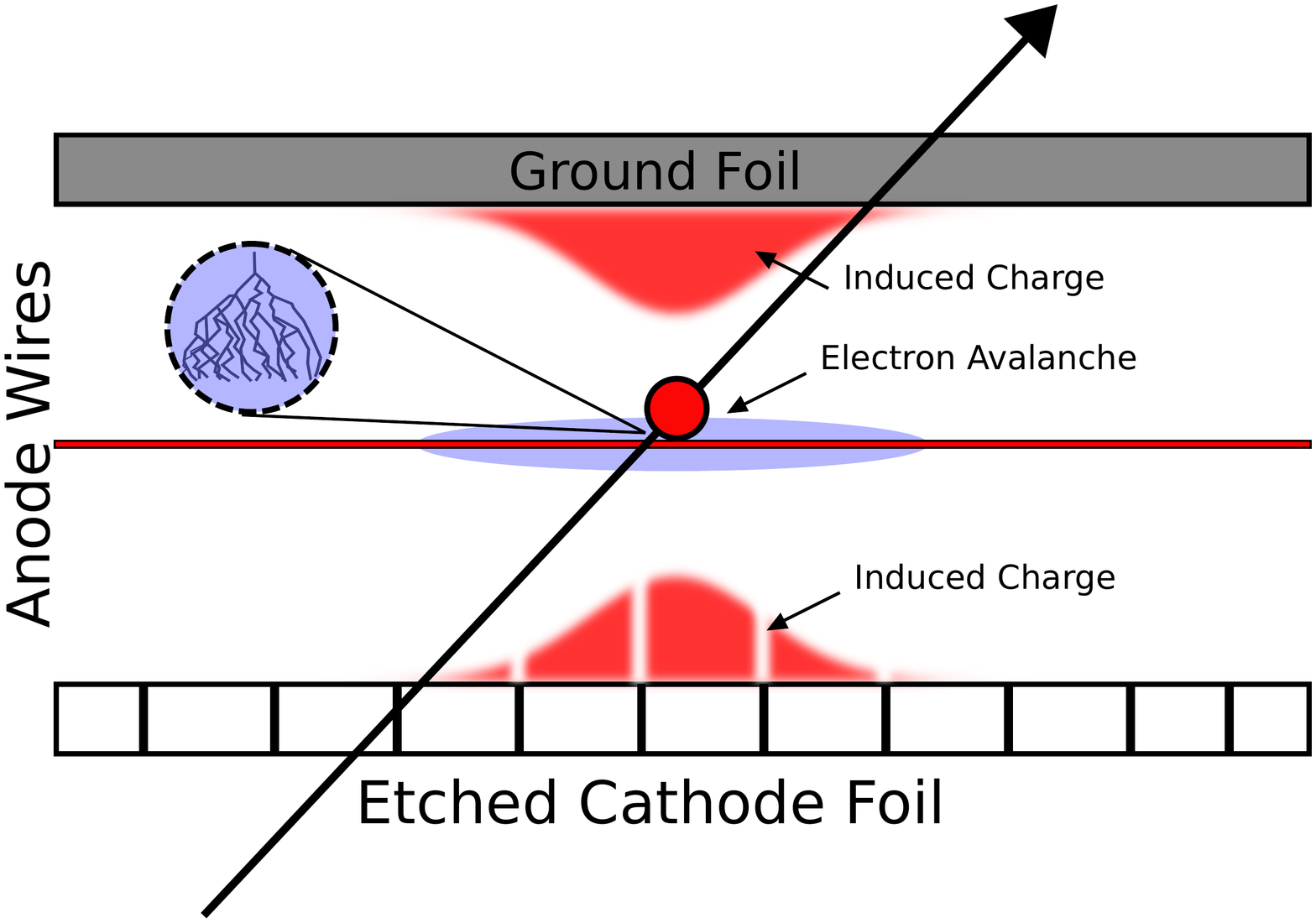}
  \caption{Cartoon of the principles of operation. When a charged particle enters the detector, ionization occurs on the fill gas, and an induced charge is created on the etched and grounded foils.}
  \label{fig:ion}
\end{figure}

\subsubsection{ElectroEtching Technique}

The design of the position sections is critically reliant on having precisely etched cathode foils. These foils should have evenly spaced, electrically isolated
strips, which necessitates a process to remove the aluminum coating from the Mylar foils. Ref.~\cite{etch} found that electric discharge etching produces clean, uniform lines. Electroetching uses a stylus held at a fixed voltage. By bringing the stylus into contact with the aluminum surface, an electric discharge is produced which removes the aluminum coating. It should be noted that Chemical etching with sodium hydroxide has also been shown to work, for example see Ref.~\cite{vert}, but difficulties arise with the precise application of sodium hydroxide and the cleaning of the reaction products. Considering the complications with the chemical etching, electroetching was chosen to create the cathode foils. Using the TUNL facilities, this technique was found to reliably produce etched foils in less than a day, which reduces the time and effort required for routine maintenance. Each strip is $2.54\textnormal{-}$ mm wide, with each etched line being $0.03\textnormal{-}$ mm wide. The strips are etched on $0.3\textnormal{-}\mu$m-thick single sided aluminized Mylar.

Our particular setup consists of a tungsten tipped stylus attached to a copper assembly pictured in Fig.~\ref{fig:etch}. {The etching is performed using a milling machine programmed with G-code. To isolate the copper rod from the spindle of the machine, a nylon covering rod was used.} During the machining process, it is of vital importance to keep
good electrical contact between the Mylar and stylus tip. To ensure this condition, several steps were taken. First, the stylus arm was attached to its base with a pivot. This design allows the tip to follow the natural curvature of the Mylar. Second, the Mylar is carefully clamped to the milling table with two $5.08\, \textnormal{cm} \times 5.08\, \textnormal{cm}$  grounded metallic plates. These plates were found to provide the proper grounding throughout the etching process. Finally, periodic sanding of the tungsten tip was found to be necessary to prevent aluminum buildup. The tip itself was held at $-15\ V$ during the process. This voltage was found to produce clean lines while reducing the possibility of damaging sparks. Ref.~\cite{etch} found that negative polarity produced cleaner lines when examined under an electron microscope.
       
\begin{figure}
  \centering
  \includegraphics[scale=.7]{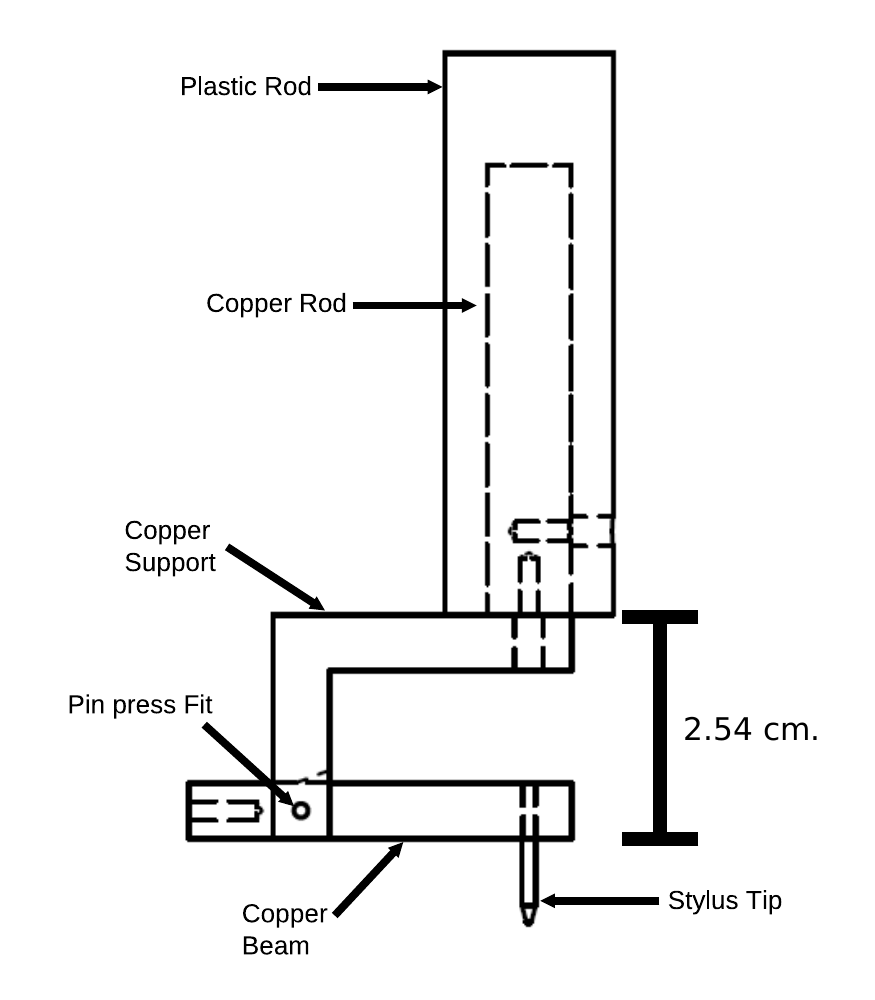}
  \caption{Drawing of the etching apparatus. The biased stylus tip is allowed to follow the curvature of the Mylar thanks to the pivoting copper arm. The plastic housing insulates the milling machine from the biasing potential. The dashed lines indicate threaded holes for screws.}
  \label{fig:etch}
\end{figure}

\subsubsection{Delay Lines}

The delay line consists of 20 delay chips with 10 taps per chip. Each tap provides $5$ ns of delay making the total delay across the line $1\, \mu$s.
Copper plated G-10 boards were machined to align the copper strips with the etched pickup strips, creating the necessary electrical contact between the etched pickup strips and the legs of the delay chips. The legs are attached via pin inserts on the back of the G-10 boards. The chips, Data Delay Devices 1507-50A \cite{chips}, have a 50 $\Omega$ impedance, which matches that of the signal cables.
Weldable BNC feedthroughs attached to NPT threads provide a vacuum tight method for connection to the delay line signal cables. It must also be noted that the error in delay per tap is quite high at $\pm 1.5\ ns$, which could lead to non-linearity in the delay to position conversion \cite{msu}.
Following the suggestions in Ref.~\cite{widths}, this effect is minimized by ensuring the ratio of the cathode strip width (2.54 mm) and the distance between the anode and the cathode (3.00 mm) is around $0.8$.    

\subsubsection{Position Section Assembly}

The delay line, cathode {foil}, anode wires, and grounded {foil} are all housed in the position section assembly shown in Fig. \ref{fig:pos}. Four metallic screws bring the copper plated top into electrical contact with the detector body, which is grounded. Plastic screws ensure proper contact between the cathode {foil} and the delay line, while maintaining electrical isolation with the rest of the board.
 
Five gold plated tungsten wires $25 \, \mu$m in diameter are used for the anodes. The wire spacing is $4$ mm, and they are surrounded by the cathode foils. These foils are made of $0.3\textnormal{-}\mu $m-thick aluminized Mylar, either single sided for the etched cathode foils, or double sided for the grounded cathode which were purchased from Goodfellow Cambridge Ltd. The Mylar sheets are secured to both the detector and position assemblies using double sided tape. Because the tape is an insulator, care was taken to ensure good electrical connection between the detector and the grounding foils . 

An accurate measurement of the charged particles' position requires a well localized electron avalanche. This requirement means that the position sections must be operated at a higher voltage than would be required of a proportional counter \cite{knoll}. In order to prevent sparking and allow voltages of ${\sim} 2000\ V$, insulating acrylic coating is applied to high risk areas and $1\ \textnormal{N}$ of tension is applied to the wires. The tension is necessary to keep the wires straight, which ensures the electric field is uniform and further reduces sparking. Isobutane was chosen as the fill gas, following the suggestions of Ref.~\cite{gas}.

\begin{figure}
  \centering
  \includegraphics[width=.6\textwidth]{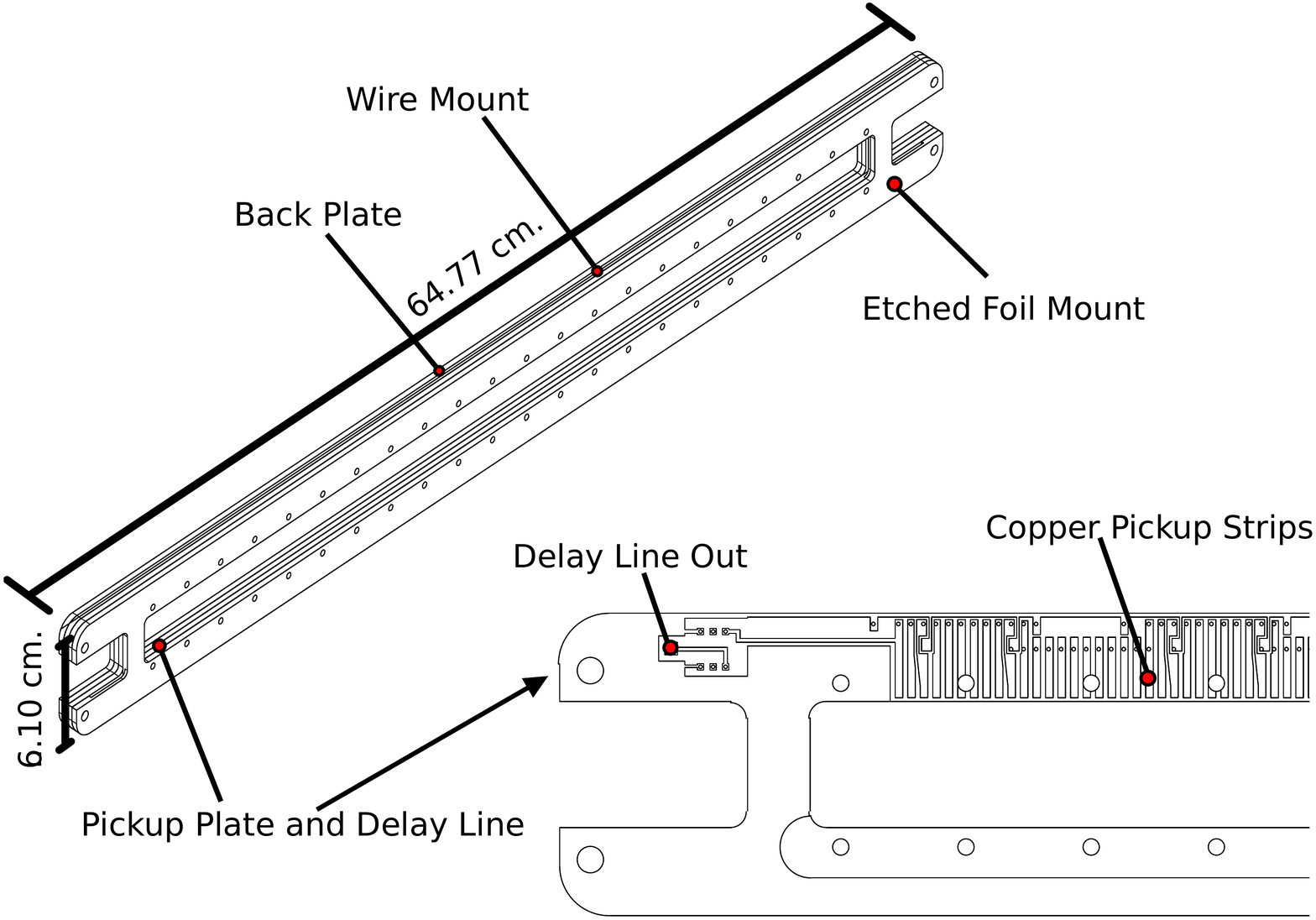}
  \caption{A model of the position section assemblies. From front to back we have: the copper plated front plate to which the etched cathode foil is taped, the copper
    plated G-10 board with pickup strips and mounts for the delay line chips, the anode wire plane board, and the back board to which the grounded plane is taped.
    {The expanded region shows the copper pickup strips that make contact with the cathode foil. The delay line is attached to the back via pin inserts that are
    at the top of each strip.}}
  \label{fig:pos}
\end{figure}

\subsubsection{Position Section Electronics}
Signal paths are visualized in Fig. \ref{fig:elec}. 
Hereafter signals will be referred to based on whether they exit the detector on the side corresponding to a high value of the magnetic rigidity (high energy) or a low one (low energy).
Inside the focal plane chamber, each of the four position signals are sent through fast timing preamplifiers. After preamplification, the signals are processed through an Ortec 863 quad Timing Filter Amplifier (TFA). The final shaping and noise rejection before our timing analysis is provided by a Constant Fraction Discriminator (CFD). Thresholds on each of the channels are adjusted to match the output levels of the TFA, which are on the order of $~300 $mV. After the final signal shaping, the signals from the high energy end of the detector are used to start an Ortec 567 Time to Amplitude Converter (TAC), while the low energy signals are all subject to a $1\ \mu \textnormal{s}$ delay and used to stop the TAC. This delay ensures that the stop signal always occurs after the start signal for a real event. The output from the TAC is sent into a CAEN V785 peak sensing Analog-to-Digital Converter (ADC), so that it can be recorded and later analyzed.

 \begin{figure}
   \centering
    \includegraphics[width=.95\textwidth]{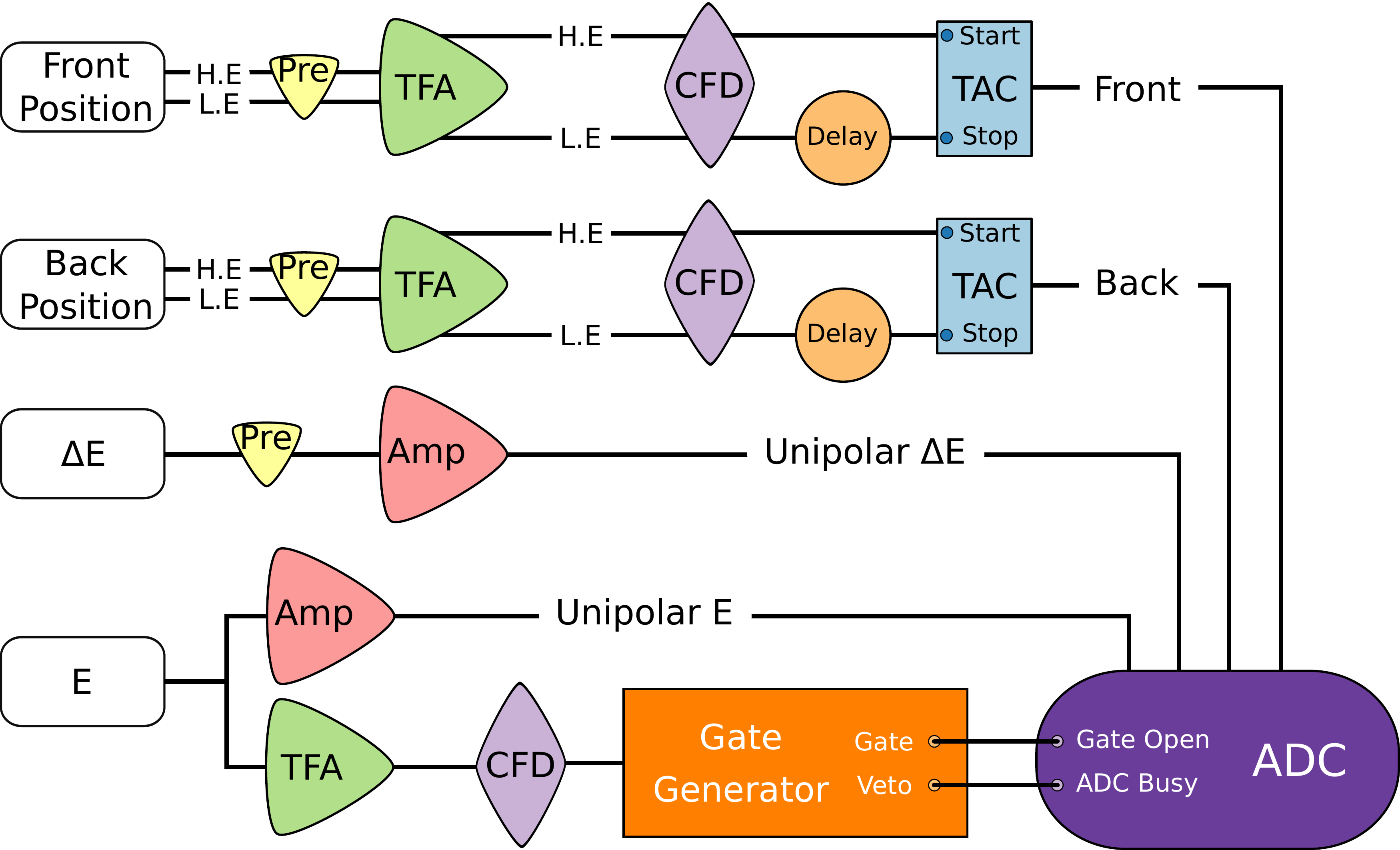}
   \caption{Diagram of the focal plane detector related electronics. Black lines show the flow of waveforms to a peak sensing ADC.}
   \label{fig:elec}
 \end{figure}

\subsection{$\Delta E $ Section}
\subsubsection{$\Delta E $ Assembly}

The $\Delta E$ section of the detector is a gas proportional counter, which consists of a single $12.7\textnormal{-} \mu$m-diameter anode wire and two grounded cathode planes.
The front cathode plane is the other side of the front position section's grounded cathode plane. The back plane is another strip of aluminized Mylar.
This Mylar plane is taped directly onto the detector body and checked for proper electrical contact. Due to low breaking tension of the anode wire, it is held taut by hand, and soldered onto NPT threaded feedthroughs. The wire is biased to $1000\ V$ to ensure that the charge collection is proportional to the energy loss of the particle.     

\subsubsection{$\Delta E$ Electronics}

The $\Delta E$ section of the detector's signal is processed with an in-house charge sensitive preamplifier based on the Cremat CR-110 operational amplifier, which provides a $1\ \mu s$ shaping time \cite{cremat}. After the preamplifier, an Ortec $572$A amplifier is used to shape the signal before it is sent to the ADC.  

\subsection{Residual Energy Section}

\subsubsection{Paddle Scintillator}
\label{sec:scint_paddle}

Particles are stopped, and residual energy deposited, in a Saint-Gobain BC-404 organic plastic scintillator. The BC-404 is sensitive to $\alpha$ and $\beta$ radiation, and is recommended for fast timing \cite{gobain}. The timing response makes it an ideal trigger for the current data acquisition system and planned $\gamma$-ray coincidence measurements. The dimensions are $71.76$ cm long by $5.08$ cm wide by $0.64$ cm thick. These dimensions are customized to cover the length of the detector and ensure all light particles will stop within the active volume. In order to maximize the amount of light collected along the entire length of the scintillator, it is wrapped in thin, reflective aluminum foil and Tyvek. Reference \cite{wrapping} demonstrated that Tyvek has an increased light output compared to the aluminum wrapping; however, it was unable to create a seal on the high pressure section of the detector, so a single sheet of aluminum foil on the sealing surface was used.

\subsubsection{Optical Fibers}

Early iterations of the $E$ section used a light guide to couple the paddle scintillator
to the Photomultiplier Tube (PMT); however, this design added significant weight and length to
the detector. To avoid the rigid constraints of light guides, optical fibers were chosen to gather and transmit light to the PMT.

The fibers are $1$-mm-diameter Bicron BCF-91A, which shift the wavelength of the violet/blue scintillated light ($380-495\ \textnormal{nm}$) of the BC-404 into the green spectrum ($495-570\ \textnormal{nm}$) \cite{fibers}.  
Following the suggestions of Ref. \cite{wrapping}, the optical fibers were spaced $5\ \textnormal{mm}$ apart
to maximize light collection. Eight $1$-mm-deep grooves were machined in the scintillator to hold the fibers.
The fibers were secured in place with BC-600 optical cement.
A light tight tube is used to bend the fibers to the PMT that sits on the top of the detector. 

\subsubsection{Photomultiplier Tube}
Matching the emitted light of the light fibers while maintaining a compact package were the main requirements for the PMT. 
The Hamamatsu H6524 has a spectral response of $300-650\ \textnormal{nm}$, a peak sensitivity of $420\ \textnormal{nm}$, and
a quantum efficiency of $27\%$ \cite{hamamatsu}. These features provide the highest quantum efficiency available for the wavelengths of interest.
The 10-stage dynode structure provides a gain of $1.7 \times 10^6$ with an anode bias of $-1500$ V.
Although the detector is located outside the high magnetic field region of the SPS, a magnetic shield was incorporated into the tube assembly to prevent possible interference.

\subsection{{$E$ Electronics and Event Structure}}

The dynode signal from the PMT is split to provide both timing and energy information. {Energy signals are processed through
an Ortec $572$A amplifier and then recorded.} Timing signals go through a TFA and a CFD to generate an event count.
A count from the $E$ detector triggers the master gate for the data acquisition system, which is vetoed if the ADC buffer is full. 
If a trigger is not vetoed, a $10\textnormal{-} \mu$s gate is generated, and the ADC records all
coincident signals. Using the $E$ signal to generate the ADC gate, as opposed to the position sections, avoids introducing a position dependent gate timing due to the delay lines.
Count rates are recorded for all detector signals, gates generated, and gates vetoed by the ADC busy signal. This setup allows us to easily diagnose electronic problems and adjust beam current to keep the acquisition dead time low ($<\!10\%$ with typical rates being $<\!5\%$).

\subsection{Improvement of the Delay Line}

The Data Delay Devices 1507-50A delay chips used in the initial design of the detector were found to be unsatisfactory for accurate work due to high differential non-linearity. It was determined that the likely cause was the high tap to tap variation of $5 \pm 1.5$ ns. A testing procedure was devised in which generated signals would be sent through each tap one at a time. These signals would be processed through the same electronics as the real signals. Thus, this testing procedure is only sensitive to issues in the delay line or the electronics. Fig.~\ref{fig:delay_chip_comp} shows residuals for the linear fit of these $200$ strips as a fraction of the strip. The purple points are the 1507-50A chips, and it can be seen that there is a large, systematic non-linearity in the measured delay time, which can result in the signals location being erroneously displaced by half a strip. From these results, it was decided to   purchase a new set of chips. The Allen Avionics 50P chips were selected due to having a $5.0 \pm 0.8$ ns tolerance \cite{allen_chips}. After installation, the delay line was tested again with the same tap by tap procedure. The results are shown as orange points in Fig.~\ref{fig:delay_chip_comp}, and show a significant improvement over the previous chips.   

\begin{figure}
    \centering
    \includegraphics[width=.75\textwidth]{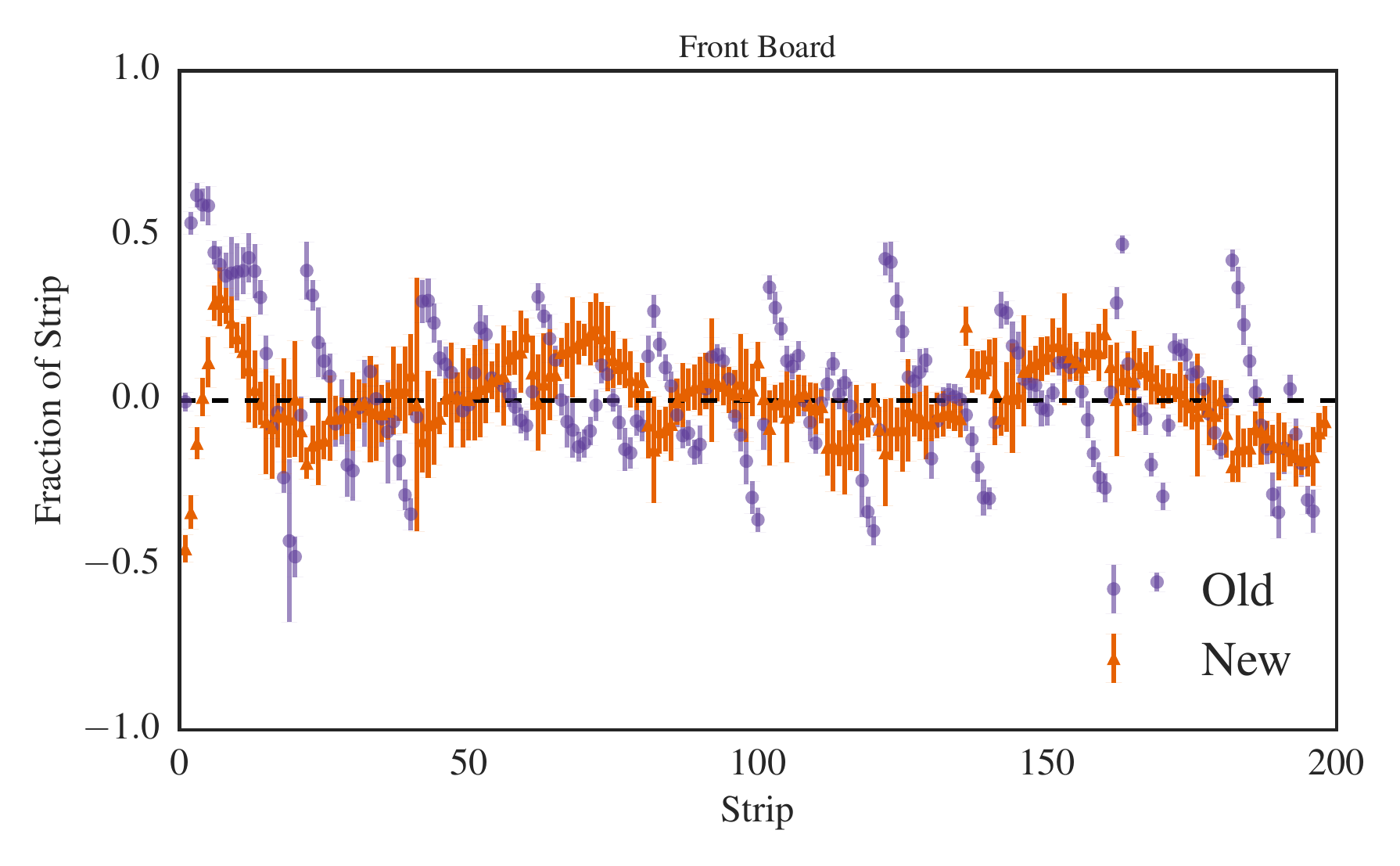}
    \caption{Residuals from the linear fit of the strip number to the measured strip position. The residuals are expressed in terms of a fraction of a strip, i.e., the residuals have been divided by the slope of the fit. The Data Delay Devices 1507-50A chips are shown in purple, while their replacements, the Allen Avionics 50P chips, are shown in orange.}
    \label{fig:delay_chip_comp}
\end{figure}{}

\subsection{Improvement of the Scintillator}

After a salvo of initial experiments, including the $^{23}$Na$(^{3} \textnormal{He}, d)^{24}$Mg experiment detailed in this document, an additional set of characterization runs were carried out. These runs used deuteron elastic scattering off of carbon to measure the relative efficiency of the detector as a function of focal plane position. This was done by changing the magnetic field of the SPS to move the elastic scattering peak from one side of the detector to the other. The magnetic field ranged from $B = 0.86 \textnormal{-} 1.02$ T in steps of $\Delta B = 0.02$ T, or roughly $400$ ADC channels. A severe linear dependence between position and efficiency was found, with the low energy side being the most efficient. This issue was traced back to the low light collection efficiency of the optical fibers.

In order to boost the efficiency, it was decided to return to the previous design. A scintillator paddle of the same type and dimensions described in Section~\ref{sec:scint_paddle} was coupled with optical cement to a light guide that adapts to a $2$ inch diameter PMT. Optical grease was used on the interface between the light guide and the photocathode. The PMT is a Thorn EMI Electron Tubes 9813B with $30 \%$ quantum efficiency, spectral response of $290 \textnormal{-} 630$ nm, and a peak sensitivity of $\approx 360$ nm \cite{emi_tube}.
The maximum bias is $(-) 3000$ V, but sufficient gain was found at an operational voltage of $(-) 2500$ V. The light guide and PMT were wrapped in vinyl tape to prevent light leaks.      
After installation, the efficiency experiments were repeated. The strength of the scintillator signal still exhibited a linear dependence, but there was no longer a loss in detection efficiency across the focal plane. 

\subsection{Kinematic Corrections}

The momenta of the reaction products entering a spectrograph have an angular dependence on $\theta_i$. This dependence will greatly degrade the resolution at the focal plane if no corrective action is taken \cite{enge, enge_optics, opt_matrix}.
This so-called kinematic broadening can be introduced into the optical matrix formalism from Section~\ref{sec:split_pole} by Taylor expanding $\delta$ around $\theta_i$ giving:
\begin{equation}
  \label{eq:kin_taylor_expansion}
  \delta(\theta_i) = \delta_0+\frac{\partial \delta}{\partial \theta_i}\theta_i = \delta_0 - K \theta_i ,
\end{equation}
where the kinematic factor $K$ is defined as the change in the momentum shift with a change in angle. The sign of $K$ is dependent upon both the direction the spectrograph is rotated and the reaction kinematics. For the TUNL SPS, the reaction angle is with respect to beam left; thus, $\Delta \theta_i$ is positive, and normal kinematics are used, meaning the outgoing particles have a lower momentum with increasing angle, leading to a negative sign on $K$.

In order to correct for this effect, the dependence of $x_f$ on $\theta_i$ must be removed. This can be done by displacing the detector in the $z$ direction. A change $\Delta z$ will introduce a dependence on $\theta_f$ into $x_f$. Two additional first order terms with $\theta_i$ will now be present in the optical matrix, which can be used to compensate for kinematic broadening since $z$ can be controlled by the experimenter. The expression for $x_f$ is now:
\begin{equation}
    \begin{split}
          x_f & =  (x_f|x_i)x_i + (x_f|\delta) \delta_0-K(x_f|\delta)\theta_i + \Delta z (\theta_f|\theta_i)\theta_i + \\
  & \Delta z (\theta_f|x_i) x_i + \Delta z (\theta_f|\delta) \delta_0 - K \Delta z (\theta_f|\delta) \theta_i.
    \end{split}{}
\end{equation}
Setting the $\theta_i$ terms equal to zero and solving for $\Delta z$ gives:
\begin{equation}
  \label{eq:10}
  \Delta z = \frac{KDM}{1-KM(\theta_f|\delta)} \approx KDM.
\end{equation}

The linear approximation is valid when $K$ is relatively small, so that the denominator is close to unity. While $M$ and $D$ can be calculated theoretically, it was decided to find an empirical fit between $K$ and $\Delta z$ in order to ensure maximum resolution.

\begin{figure}
  \centering
  \includegraphics[]{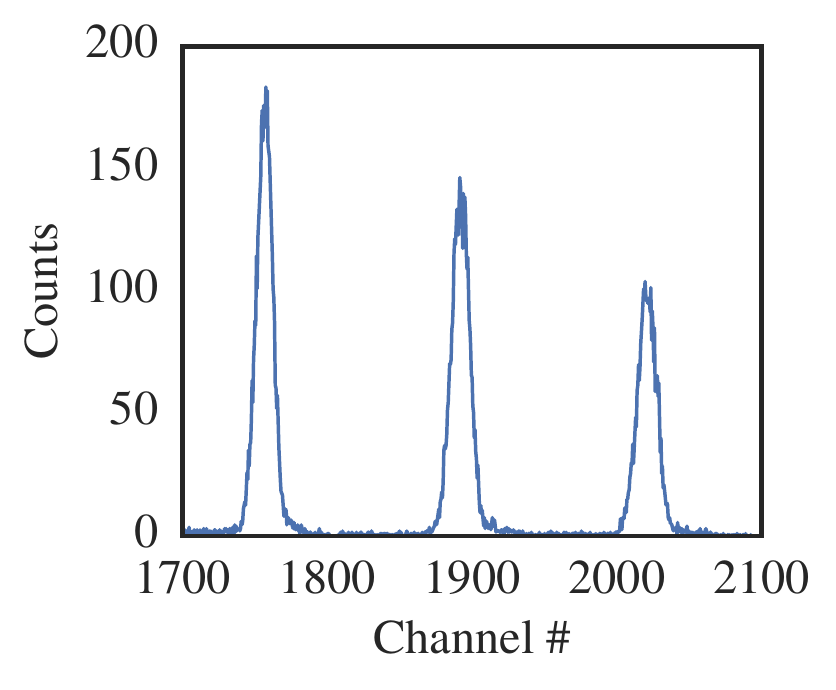}
  \caption{Example proton spectrum when detector is off focal plane with a 3-slit aperture. The data are from $^{12}$C$+$p elastic scattering at $\theta_{\textnormal{Lab}}=20^{\circ}$ and $E_{Lab} = 12$ MeV.
  The different peak intensities reflect the rapid variance of the cross section with the detection angle.}
  \label{fig:peaks}
\end{figure}

\begin{figure}
  \centering
  \includegraphics[]{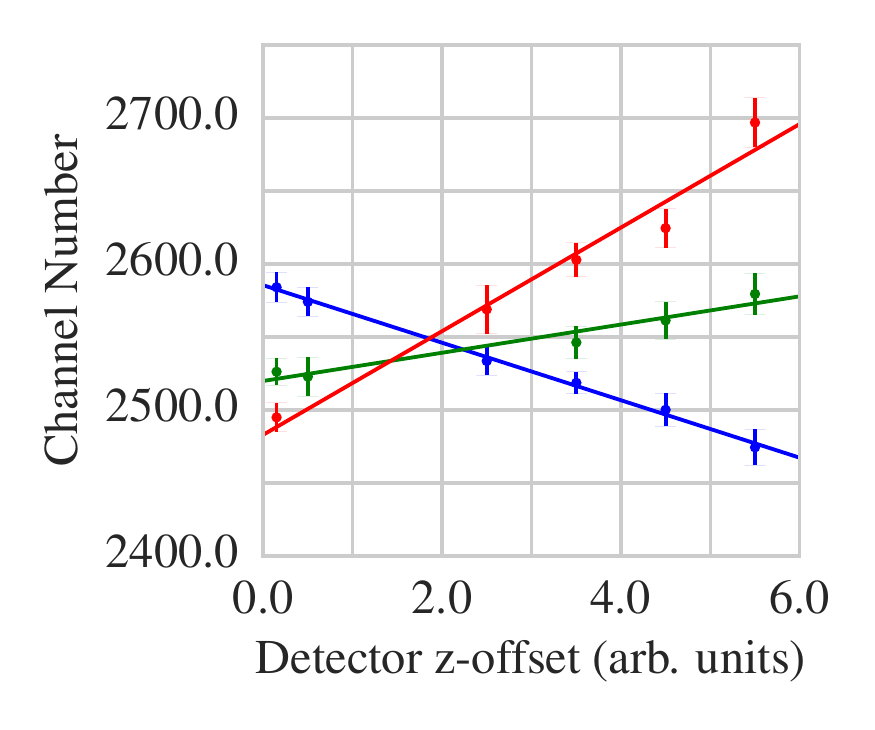}
  \caption{An example of measured peak centroids and the linear fit with respect to the $z$ position of the detector read in terms of a stepper motor voltage. Data are from the elastic scattering of protons off $^{197}\!$Au at $\theta_{\textnormal{Lab}} = 30^{\circ}$ and $E_{Lab} = 12$ MeV.}
  \label{fig:au}
\end{figure}

The empirical method fits the linear relationship between $\Delta z$ and $K$, both of which are known independently of the ion optics of the spectrometer. $\Delta z$ is just a relative change in the position of the focal plane detector. For this work, it was taken to be the distance from the exit of the magnet to the front face of the detector. $K$ can be determined for a given reaction using energy and momentum conservation, giving the formula \cite{enge}:

\begin{equation}
  \label{eq:2}
  K = \frac{(M_bM_eE_b/E_e)^{1/2} \sin \theta}{M_e+M_r-(M_bM_eE_b/E_e)^{1/2} \cos \theta} ,
\end{equation}

\noindent where e references the ejected particle, b is for beam, and r is for the residual particle. $M$ are the masses of the particles, while $E$ denote their kinetic energies. 

One possible method for finding an optimal $z$ for a given $K$ is described in Ref.~\cite{parikh}. Using this method, the optimal $z$ position is found by moving the detector through the focal chamber and minimizing the width of a chosen peak; however, this method does not give much feedback during the run, as peak width can be hard to determine without careful peak fitting. 

Instead, it was decided to carry out a series of experiments using C and Au targets to probe several $K$ values. A beam of $10$ MeV protons were impinged on the targets. Elastic scattering of these protons at $30^{\circ}$ was measured using a three-slit entrance aperture. This aperture serves to discretize the acceptance solid angle into three narrow ranges of $\theta$. When the detector is off the focal plane, three particle groups will be observed as shown in Fig.~\ref{fig:peaks}. When the detector is on the focal plane these groups should converge; thus, the detector is swept across the depth of the focal plane chamber and a linear fit of the accompanying peak positions is found, as shown in Fig.~\ref{fig:au}. The detector position is inferred based on a voltage reading from two motors, which displace it along the $z$ direction. There is considerable uncertainty on our fit between $K$ and $z$, but the relationship only needs to be approximately known to effectively compensate for kinematic broadening.

\section{Monitor Detector and Electronics}

Performing the $^{23}$Na$(^{3} \textnormal{He}, d)^{24}$Mg transfer reaction is complicated by the hygroscopic nature of the target material. Furthermore, as discussed in Section~\ref{sec:split_pole} electrons from the target could impact our integrated charge. For these reasons, it was decided to use a monitor detector to look for target degradation and to provide a relative normalization of the transfer cross sections.    

The monitor detector was a $\Delta E/E$ telescope, consisting of two surface barrier silicon detectors. The $\Delta E$ detector has a thickness of $150$ $\mu$m, and the $E$ has a thickness of $2000$ $\mu$m. A stand was designed to hold both of these detectors as a well as solid angle defining apertures. These apertures are made of brass and can be swapped out to change the solid angle. Two permanent magnets were attached at the front of the holder to deflect electrons away from the detectors.

The electronics for these detectors are shown in Fig.~\ref{fig:electronics_si}.
A Mesytec MSI-8 charge sensitive amplifier with built-in timing and shaping amplification was used. The timing signals from both detectors are sent into a CFD. In order to reduce background counts, a coincidence module is placed after the CFD signals, and requires a coincidence between the $\Delta E$ and $E$ detectors. If a coincidence is present, the ADC gate is opened and the output from the shaping amplifiers is recorded. A pulser is also fed through these electronics to verify the ADC dead time and to monitor for possible electronic noise.

\begin{figure}
    \centering
    \includegraphics[width=.95\textwidth]{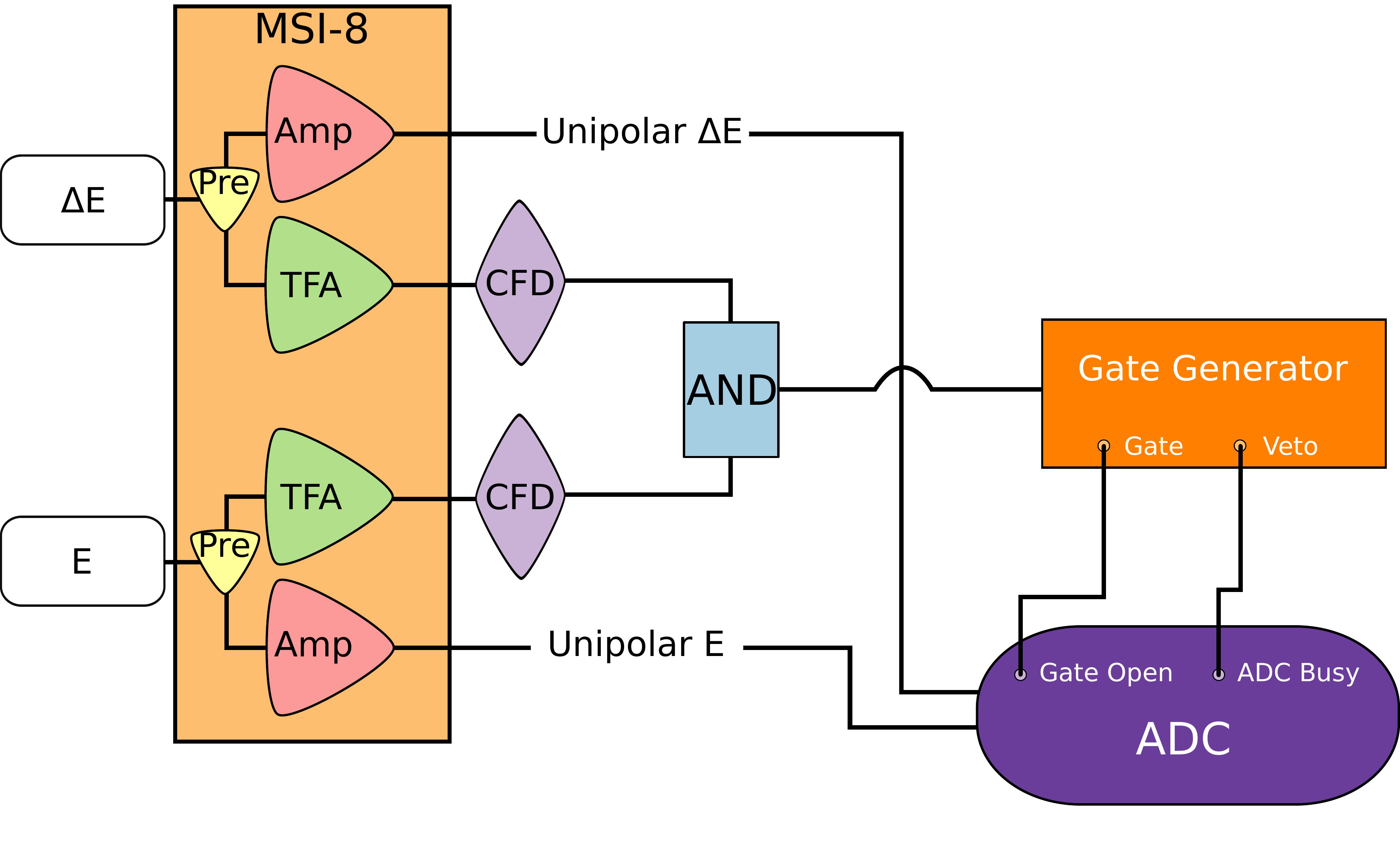}
    \caption{The electronics for the silicon telescope.}
    \label{fig:electronics_si}
\end{figure}

\section{Summary}

This chapter has laid out the technical details of the equipment used to carry out the experimental work presented in this thesis. As shown, measuring transfer reactions with high resolution requires a delicate interplay between the ion source, accelerator, beamline optical elements, magnetic spectrograph, and detector systems.       

%% file: Chapter-5/Chapter-5.tex
\chapter{Bayesian Methods For Transfer Reactions}
\label{chap:bay_dwba}

\section{Introduction}

This chapter will describe the novel Bayesian methods developed to analyze the experimental data from transfer reactions. As shown in Chapter~\ref{chap:nuclear_unc}, it is critical to properly assess  the nuclear uncertainties from experiments in order to know the limits of our astrophysical predictions. Transfer reactions, and in particular transfer reactions performed with magnetic spectrographs, come with their own unique uncertainties. The first part of this chapter introduces a Bayesian model for energy calibration that properly accounts for the statistical uncertainties present in both the dependent and independent variables. The second portion of this chapter is devoted to incorporating the uncertainties arising from optical model parameters into DWBA, thereby accounting for these effects when assigning $\ell$ values (the transferred angular momentum) and extracting spectroscopic factors. The energy calibration method was first presented in Ref.~\cite{marshall_2018}, but has since been updated. The Bayesian DWBA method was first presented in Ref.~\cite{Marshall_2020}. Both of these works represent my work to develop some of the first analysis methods of their kind for transfer reaction data.         

\section{Bayesian Statistics}

The problems that are addressed in this chapter are, in the most abstract sense, based around the concept of estimating parameters of a model from data that are subject to statistical uncertainties.
Bayesian statistics is a formulation of statistics that uses Bayes' theorem to update prior probability distributions of these parameters based on experimental data, $\mathbf{D}$, allowing experiments to improve our knowledge of these parameters. Despite the important distinctions between Bayesian methods and those of frequentist statistics, the topic is well outside the scope of this thesis.

While Bayesian statistics requires additional assumptions, it is centered around Bayes' theorem which itself is simply a logic expression to relate conditional probabilities, independent of statistical interpretations. A conditional probability between two variable $X$ and $Y$ is the probability of $X$ occurring given that $Y$ has occurred, and is symbolized by $P(X|Y)$. This is defined by the joint probability of $X$ and $Y$, i.e., the probability of both $X$ and $Y$ having occurred, over the probability of $Y$. This definition is symbolized as:
\begin{equation}
    \label{eq:cond_probs}
    P(X|Y) = \frac{P(X , Y)}{P(Y)}
\end{equation}
Bayes' theorem follows from the equality $P(X , Y) = P(Y , X)$. Thus, the conditional probability of $P(X|Y)$ can be written:
\begin{equation}
    \label{eq:simple_bayes}
    P(X|Y) = \frac{P(Y|X)P(X)}{P(Y)}.
\end{equation}

The goal for any quantitative experiment is to collect data, compare this data to a model, and finally make predictions using the parameters of the model. Bayesian statistics assumes that these model parameters, $\boldsymbol{\theta}$, can be represented by probability distributions. Thus, the data we observe in the lab, $\mathbf{D}$, are used to update these \textit{prior} probabilities for the model parameters through Eq.~\ref{eq:simple_bayes}. Reformulated into this framework, Bayes' theorem reads: 
\begin{equation}
  \label{eq:bayes_theorem}
  P(\boldsymbol{\theta}|\mathbf{D}) = \frac{P(\mathbf{D}|\boldsymbol{\theta}) P(\boldsymbol{\theta})}
  {P(\mathbf{D})},
\end{equation}
where $P(\boldsymbol{\theta})$ are the prior probability distributions of the model parameters, 
$P(\mathbf{D}|\boldsymbol{\theta})$ is the likelihood function, $P(\mathbf{D})$ is the evidence, and $P(\boldsymbol{\theta}|\mathbf{D})$ is the posterior \cite{bayes}. Expressing these terms in a more informal way: the priors are what we believe about the model parameters before the data are measured, the likelihood is the probability of the observed data given a set of model parameters, the evidence is the probability of the observed data, and the posterior is what we know about the model parameters after the data have been observed. In Bayesian statistics the general goal is to use the right hand side of the equation to calculate the posterior.
As will be shown, this is no simple task, and the problem has only become tractable because of the increased amount of computation power available in the last $40$ or so years. In particular, the evidence is only calculable if it is expanded into the integral:
\begin{equation}
    \label{eq:evidence_integral}
    P(\mathbf{D}) = \int_{\boldsymbol{\theta}} P(\boldsymbol{D}|\boldsymbol{\theta})P(\boldsymbol{\theta}) d \boldsymbol{\theta},
\end{equation}
which follows from Eq.~\ref{eq:cond_probs}. This integral will have the same dimension as the number of parameters that need to be integrated over. Once the integral is carried out, it can be seen that the evidence is just a number, and more specifically it represents a normalization constant that ensures the posterior is a properly normalized probability distribution. Further complications arise in Bayesian models with multiple parameters. In this case Bayes' theorem will yield the posterior joint probability distribution for all of the model parameters. Gaining posterior distributions for each parameter (i.e., knowing the uncertainties on each model parameter) independent of the other parameters requires that these other parameters be  \textit{marginalized} over. For example, take a two parameter model with parameters $a$ and $b$. Bayes' theorem for this expression is:
\begin{equation}
    \label{eq:bayes_2_parameters}
    P(a, b|\mathbf{D}) = \frac{P(\mathbf{D}|a, b)P(a, b)}{P(\mathbf{D})}.
\end{equation}
However, what is typically desired from a statistical analysis are the posteriors $P(a|\mathbf{D})$ and $P(b|\mathbf{D})$. The conditional probability $P(a, b| \mathbf{D})$ is equivalent to $P(a , b , \mathbf{D})/P(\mathbf{D})$, so marginalization over $a$ will yield $P(b|\mathbf{D}) = \int P(a, b|\mathbf{D}) da$. The mathematical difficulties the evidence and marginalization integrals present can frequently only be solved using numerical methods, as is the case for the two applications presented here.

\section{Markov Chain Monte Carlo}
\label{sec:bay_energy_cal}

Instead of numerically solving the above integrals directly, the goal will be to draw samples directly from the marginalized posterior distributions. This is related to the issues encountered when trying to propagate uncertainties through reaction rate calculations as  discussed throughout Chapter~\ref{chap:nuclear_unc}, and the solution is similar in nature. The goal is to draw representative samples from the posterior, but in this case randomly selecting samples from the priors and evaluating the likelihood will be insufficient for determining the posterior. This problem differs fundamentally from the Monte-Carlo uncertainty propagation because we are not trying to calculate a function of random variables, but, instead, we are trying to calculate the convolution of two probabilities. These probabilities need to be normalized by the evidence, which means these random draws need to sufficiently sample the integral as well.

In the most reductive terms, Markov chain Monte Carlo (MCMC) is a process to sample an arbitrary probability distribution. A Markov chain describes a system with a series of states $\dots x_{t-1}, x_{t}, x_{t+1} \dots$, where the transition from state $t$ to state $t+1$ only depends on the properties of the system at $t$ with no dependence on $x_{t-1}$ \cite{Mackay_2002, Cover_2006}. This abstract idea relates to the problem at hand, i.e., drawing samples from an arbitrary probability distribution, by trying to construct a Markov chain that produces a series of states that are distributed according to the desired probability distribution, $\pi$, i.e., each $x_t$ distributed according to $x_t \sim \pi$. The Monte Carlo portion of Markov Chain Monte Carlo refers to the fact that the Markov chain will be constructed using Monte-Carlo methods to generate the transitions from state $x_{t}$ to state $x_{t+1}$. The fact that this process will converge to the desired probability distribution is well beyond the scope of this thesis, but details can be found in Ref.~\cite{Christian_2005}.

Many variants of MCMC exist, but perhaps the most commonly encountered is the Metropolis-Hastings algorithm \cite{Metropolis_1953, Hastings_1970}. Say we want to estimate a target distribution, $\pi$. If the Markov chain is at a point $x_{t}$, then a new point is proposed according to a proposal distribution, $q(x^{\prime}|x_{t})$. This new point is accepted with a probability $\alpha$. The question now is what form must the product of these two probabilities, $T = \alpha q $, take in order for the samples to converge to $\pi$? A sufficient but not necessary condition for the Markov chain to converge to $\pi$ is that these transitions, $T$, to and from a state satisfy \textit{detailed balance} \cite{Mackay_2002}:

\begin{equation}
    \label{eq:detailed_balance}
    T(x^{\prime}|x_{t}) \pi(x_{t}) = T(x_{t}|x^{\prime}) \pi(x^{\prime}). 
\end{equation}

The total probability of a transition will be given by the product of the acceptance probability and the proposal distribution $T(x^{\prime}|x_{t}) = \alpha(x^{\prime}|x_{t}) q(x^{\prime}|x_{t})$. It follows that the proposal probability, whatever we choose, must satisfy:
\begin{equation}
    \label{eq:prop_form}
    \frac{\alpha(x^{\prime}|x_{t})}{\alpha(x_t|x^{\prime})} = \frac{\pi(x^{\prime})}{\pi(x_t)} \frac{q(x_t|x^{\prime})}{q(x^{\prime}|x_{t})}.
\end{equation}
The form chosen for $\alpha$ by Metropolis \textit{et al}. assumed that the proposal distributions were symmetric, $q(x_t|x^{\prime}) = q(x^{\prime}|x_{t})$, and enforced the condition that if a proposed state had a higher probability, then the proposed move should always be accepted \cite{Metropolis_1953}. Plugging these desired properties into Eq.~\ref{eq:prop_form} for the case that $\pi(x^{\prime}) > \pi(x_{t})$ gives: $\alpha(x^{\prime}|x_t)=1$, with detailed balance requiring $\alpha(x_t|x^{\prime}) = \frac{\pi(x_t)}{\pi(x^{\prime})}$. Therefore, the Metropolis acceptance probability is generally:
\begin{equation}
    \label{eq:metrop_acceptance}
    \alpha(x^{\prime}|x_t) = \textnormal{min} \bigg[ 1, \frac{\pi(x^{\prime})}{\pi(x_t)} \bigg], 
\end{equation}
where \say{min} means that $\alpha = 1$ if $\frac{\pi(x^{\prime})}{\pi(x_t)} > 1$ and otherwise will be equal to $\frac{\pi(x^{\prime})}{\pi(x_t)}$. The proposal distributions can be included in this definition as well if they are selected to be non-symmetric (this detail being Hastings' contribution \cite{Hastings_1970}). 

Once a proposal distribution has been selected, the algorithm becomes:

\begin{algorithm}[H]
\SetAlgoLined
\KwResult{$n$ samples drawn from probability distribution $P$ }
 set the total number of steps $n$\;
 set $t = 0$ \;
 select initial parameters for $x_t$ \;
 \While{$t < n$}{
  Propose a new set of parameter values for $x^{\prime}$\;
  \eIf{$P(x^{\prime})/P(x_t) \geq 1 $}{
   $t = t+1$\;
   accept new coordinates $x_t = x^{\prime}$\;
   }{
   draw a random number $r \in [0,1)$\;
   $t = t+1$\;
   \eIf{$P(x^{\prime})/P(x_t) \geq r$}{
   accept new coordinates $x_t = x^{\prime}$\;}{
   reject new coordinates $x_t = x_{t-1}$\;
  }
 }}
 \caption{Metropolis}
\end{algorithm}

\noindent This algorithm shows the power of MCMC, in a simple case all a computer has to do is evaluate the probability distribution function and draw a random sample from $[0,1)$. 

Of critical importance to the problems mentioned in Section 2 of this chapter is the fact that this method depends only on the \textit{ratio} of the probabilities. Since the Bayesian evidence is just a constant number, this ratio removes any dependence of the samples on knowing $P(\textbf{D})$. However, this is a double-edged sword because the samples will only be proportional to $P(\boldsymbol{\theta}|\textbf{D})$. So while we do not have to calculate $P(\textbf{D})$, we also learn very little about it. On a more positive note, the MCMC samples trivially yield marginal posteriors for each model parameter. If estimation of these parameters is the only concern, then MCMC is one of the most flexible, powerful, and widely used techniques \cite{mcmc_review}. 

There are some necessary complications with MCMC that merit discussion now. The first is that the initial position of the chain $x_{t=0}$ will not necessarily be a valid sample from $P(x)$. Furthermore, since each MCMC sample necessarily depends on the previous step, it is required that the first part of the chain is discarded in order to remove any effects of the initialization. This process is called \textit{burn-in}, and the length of a time it takes for MCMC to start drawing valid samples from $P(x)$ varies from problem to problem. A closely related issue is the concept of an \textit{effective sample size} ($ESS$). Again due to the correlated nature of the samples, if we draw $N$ samples, only a subset of these samples will be independent. The time it takes for a MCMC simulation to \say{forget} its history is the autocorrelation time, $\tau$, and the number of effective samples is $ESS = N/\tau$. These parameters all depend on $P(x)$ as well as $q(x^{\prime}|x_t)$. As an example, if the proposal distribution is a normal distribution centered around the current position of the chain, the variance of this distribution is a free parameter. If we have a $d$ dimensional problem, then there are $\approx d^2$ free parameters for each variance and covariance. Each of these parameters will affect $\tau$ and, therefore, $ESS$.

\section{Bayesian Energy Calibration}

\begin{figure}
    \centering
    \includegraphics[width=.6\textwidth]{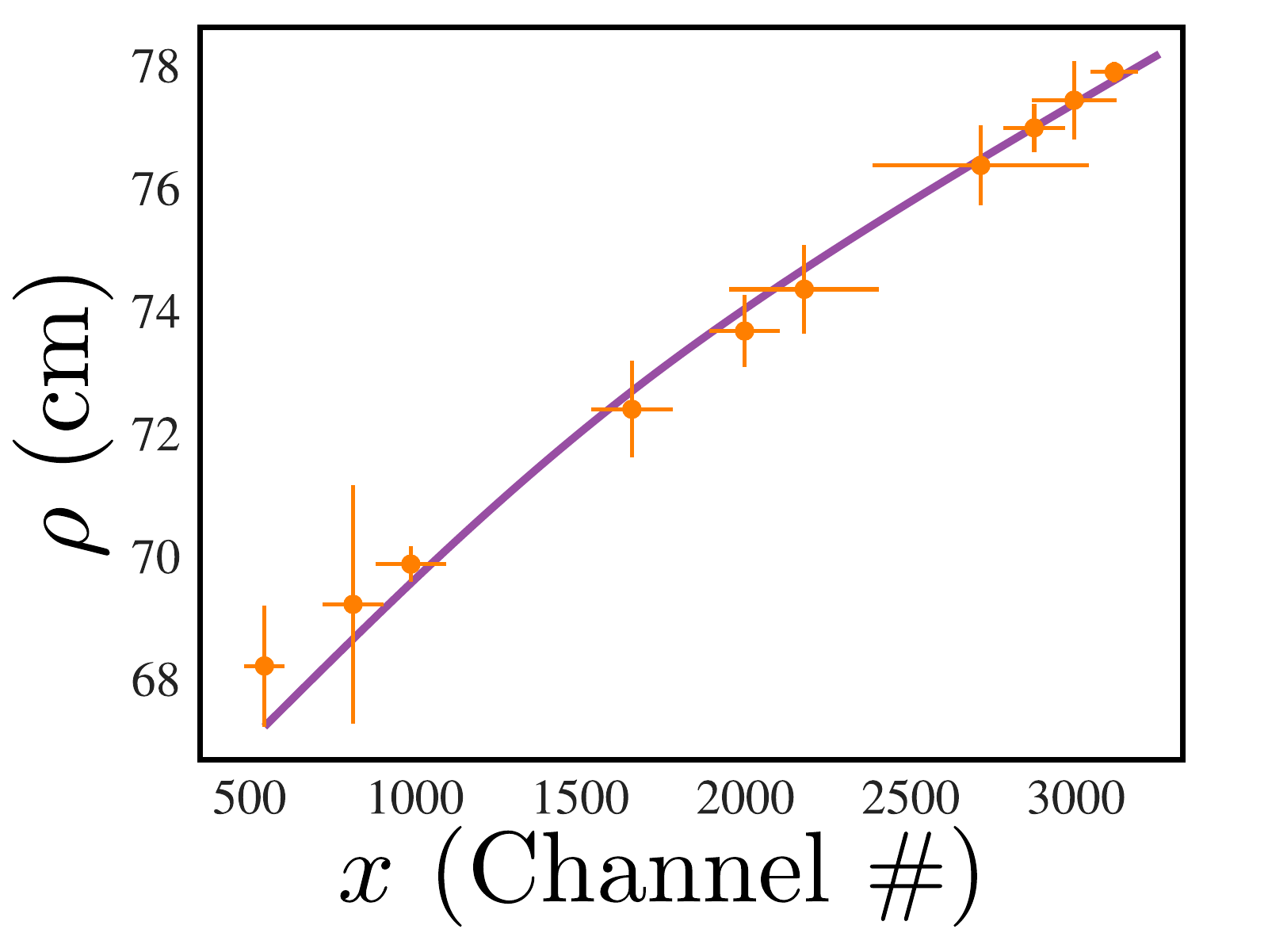}
    \caption{Cartoon that demonstrates the issue with energy calibrating the focal plane. The relationship between $x$ and $\rho$ is nonlinear, and uncertainties are present in both variables.}
    \label{fig:calibration_sketch}
\end{figure}

Focal plane detectors for magnetic spectrographs are typically energy calibrating using a simple, low-order polynomial fit. However, one of the most critical nuclear inputs for thermonuclear reaction rates are the resonance energies. This can be seen in Eq.~\ref{eq:narrow_rate_with_resonance_strength}, where $E_r$ enters the rate exponentially. Thus, despite the relative simplicity of the energy calibration process, it plays a critical role in determining reaction rates. If the uncertainties from the calibration process are underestimated, the associated impact on the reaction rates will be severely overconfident predictions. For a magnetic spectrograph, measured peaks on the focal plane represent energy levels in the residual nucleus. By using states of known energy, it is possible to calibrate these spectra, and thereby extract the energies of the other observed levels. Since the focal plane surface is curved, the relationship between $\rho$, the radius of curvature,
and ADC channel number, $x$, is not linear \cite{enge}. Using a polynomial calibration corrects for this curvature across
the focal plane, and takes the form:
\begin{equation}
    \label{eq:polynomial_fit}
    \rho = Ax^2 + Bx + C,
\end{equation}
for a second order fit. Energy values for calibration states are transformed to $\rho$ values using Eq.~\ref{eq:magnetic_rigidity}. Once a fit is found, $\rho$ can be predicted for any peak in the spectrum, and again with the use of Eq.~\ref{eq:magnetic_rigidity} an excitation energy can be predicted. 
Both the peak centroids and energy levels used for calibration will contribute to the statistical uncertainty of the calculated energy levels. Figure \ref{fig:calibration_sketch} shows a sketch of the problem, with the goal being to fit a polynomial in the presence of both $x$ and $\rho$ uncertainties.

The process outlined above offers two distinct problems: propagation of uncertainties through the relativistic collision kinematics and a statistically rigorous regression of the polynomial fit. For this work, uncertainty propagation for the kinematics was performed by using a Monte Carlo method, see Chapter \ref{chap:nuclear_unc} Section 2. To calculate $\rho$, this involves treating previous experimental values for energy levels as normal distributions. Random samples are drawn from these distributions and then used to solve the kinematic equations.
After enough samples have been drawn, a histogram of the solutions to the kinematic equations is made, and from this information, estimates of the probability distribution function
for $\rho$ can be made, which was found to be well described by a normal distribution. Each $\rho$ can then be associated with a peak in the spectrum, giving each calibration point a known $\rho$ and peak centroid, $x$. These points can then be used to find the coefficients of the polynomial through regression.    
 
The challenge associated with the polynomial regression in this case is that the uncertainties in $\boldsymbol{\rho}$ are comparable to those in the peak centroids $\mathbf{x}$. Fitting the polynomial needs to account for both of these uncertainties simultaneously. The uncertainties from both of these sources should be reflected in the posterior distributions for the polynomial coefficients, $(\theta_0,\ldots,\theta_N)$, where $N$
is the order of the polynomial.       

Following Eq.~\ref{eq:bayes_theorem}, a Bayesian analysis of this problem requires that prior distributions and a likelihood function are assigned. Uninformative priors are assigned to every polynomial coefficient. These distributions take the form of broad normal distributions centered around zero:
\begin{equation}
    \label{eq:coeff_priors}
    \theta_j \sim \mathcal{N}(0, 100^2),
\end{equation}
where $\theta_j$ is $j^\textnormal{th}$ order polynomial coefficient. This prior ensures that these parameters are allowed to vary over a wide range of values. In principle the intercept could be made to be more strict since $\rho \geq 0$, but the choice in priors, provided a wide enough coverage in values, was found to have no appreciable difference in the results.

For the peak centroids, a simplification is made that assumes these values are model parameters, and are therefore assigned informative priors. If $x_{i, obs}$ is the measured centroid in channel units for calibration peak $i$ with a measured standard deviation $\sigma_{i, obs}$, then this prior is given by:
\begin{equation}
    \label{eq:centroid_prior}
    x_i \sim \mathcal{N}(x_{i, obs}, \sigma^2_{i, obs}).
\end{equation}
An equivalent formulation of this statement could be made by assuming the measured values are subject to a likelihood function, but this model would be slightly more complicated in the present case. This formulation is the more general prescription but is mentioned here just for completeness.

The likelihood function makes the connection between the $\rho$ values that were selected for the calibration, and the polynomial function:
\begin{equation}
    \label{eq:poly_eq_for_model}
    f(\theta_j, x_i) = \sum_{j=0}^N \theta_j x_i.
\end{equation}
The residuals of the polynomial fit and the calibration $\rho$ values, $(f(\theta_j, x_i) - \rho_{i, cal})$, are assumed to be normally distributed with a standard deviation taken from the previously reported uncertainties, $\sigma_{i, cal}$. This gives:
\begin{equation}
    \label{eq:calibration_likelihood}
    \rho_{i, cal} \sim \mathcal{N}(f(\theta_j, x_i), \sigma^2_{i, cal}).
\end{equation}

Combining all of these terms, the full Bayesian model for the focal plane calibration becomes:
\begin{align}
  \label{eq:calibration_bayesian_model}
    & \textnormal{Priors:} \nonumber \\
    & x_i \sim \mathcal{N}(x_{i, obs}, \sigma^2_{i, obs}) \nonumber \\
    & \theta_j \sim \mathcal{N}(0, 100^2) \nonumber \\
    & \textnormal{Function:}  \\
    & f(\theta_j, x_i) = \sum_{j=0}^N \theta_j x_i \nonumber \\
    & \textnormal{Likelihood:} \nonumber \\
    & \rho_{i, cal} \sim \mathcal{N}(f(\theta_j, x_i), \sigma^2_{i, cal}). \nonumber
\end{align}
To summarize this model, each calibration peak in the spectrum has a measured channel mean $x_{i, obs}$ and variance, $\sigma_{i, obs}^2$. These values are used to assign an informative prior to the peak value used in the model, $x_i$.
The likelihood function is evaluated at the calibration points, $\rho_{i, cal}$, which are found by using excitation energies in the literature and are converted to $\rho$ using the kinematics and experimental parameters of the experiment being analyzed.  

Evaluation of the posterior distribution was performed using MCMC. The MCMC chain was initialized using values from a maximum likelihood estimate in order to decrease the burn in time.
The model was set up and evaluated using the \texttt{PyMC2} package \cite{pymc}.
Typical runs draw around $2 \times 10^5$ samples after $5 \times 10^4$ initial steps are discarded as burn in. Thinning is also employed as needed, but convergence times can vary greatly between
different nuclei depending on how well the calibration energy levels are known. Additionally, efficient sampling of the posterior was found to be greatly helped by scaling channel numbers
around their average value (i.e., for each of the N data points, $x_{i, obs}$: $x_{i, obs}^{\textnormal{scaled}} = x_{i, obs} - \frac{1}{N}\sum_i^N x_{i, obs}$).

To deduced excitation energies from the calibration fit, a Monte Carlo procedure is again used. In this case, the uncertainty contributions come from both the chosen peak centroid and the coefficients of the polynomial fit. The samples for the coefficients were drawn from a KDE constructed from the posterior samples to account for the non-normality of the sampled distributions. The resulting estimated energies, however, were found to be normally distributed. 

There is no guarantee that a chosen set of calibration points will produce a fit that accurately predicts energies.
Frequently this problem arises from misidentifying peaks in the spectrum.
Thus, a goodness-of-fit measure is necessary to help select a valid set of calibration points.
A reduced-$\chi^2$ statistic is available in a Bayesian framework, 
but comes out of a maximum likelihood approximation, with data that has normally distributed uncertainties, and priors that are uniformly distributed \cite{bayes}. 
However, the calibration above is also dependent on the uncertainties in the peak centroids. When these uncertainties are accounted for, a $\chi^2$ function that is only sensitive to the calibration energy uncertainties could lead to the rejection of an otherwise satisfactory
calibration set. In order to integrate these variations into a maximum likelihood estimate, a quantity, I will call $\delta^2$, is defined as: 
\begin{equation}
  \label{eq:delta_definition}
  \delta^2 = \frac{1}{2K}\sum_{\alpha=0}^{K}\bigg[\frac{1}{N-\nu} \sum_{i=0}^{N}
  \bigg(\frac{f(x_{\alpha i};\boldsymbol{\theta})-\mu_{\rho_i}}{\sigma_{\rho_i}}\bigg)^2 
  + \frac{1}{M} \sum_{j=0}^{M} \bigg(\frac{x_{\alpha j}-\mu_{x_j}}{\sigma_{x_j}}\bigg)^2 \bigg],
\end{equation}
where $N$ is the number of measured $\rho$ values, $\nu$ is the number of fitting parameters, $M$ is the number of centroids,
and $K$ is the number of centroid samples drawn. The factor of $1/2$ accounts for each term approaching unity
when the fitted parameters describe the data well. This quantity is again based on a maximum likelihood approximation applied to normally distributed
uncertainties with uniform priors, but it serves as a useful approximation for the goodness-of-fit of the $\rho$ versus channel calibration.
This method was found to clearly distinguish misidentified peaks without giving false negatives arising from channel uncertainties.
It was found that $\delta^2 < 5$ usually indicates a fit free from misidentified states and worthy of further examination.

The techniques outlined above define statistically sound procedures for uncertainty propagation and $\rho$ versus channel fitting for focal plane energy calibration.
These procedures have the advantage of not approximating the influence of the multiple sources of uncertainty, and creating a general framework which can be expanded as dictated by the experiment.

\subsection{$^{28}$Al Calibration}

\begin{figure}
  \centering
    \centering
    \includegraphics[width=.9\textwidth]{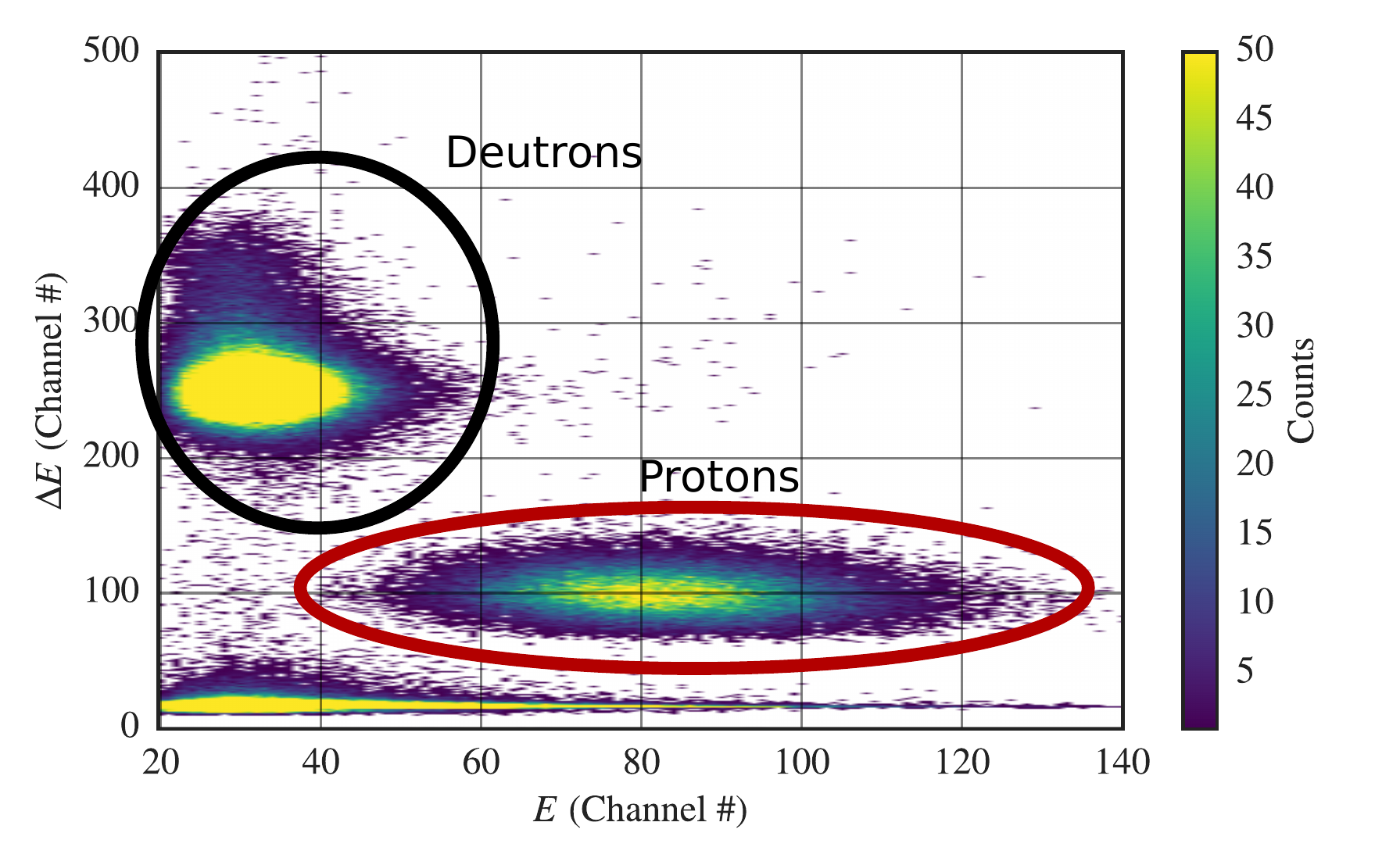}
    \caption{ $\Delta E$/E 2D spectrum from $^{27}\!$Al$(d,p)$ at $E_{Lab}=12$ MeV and $\theta_{Lab}=25^{\circ}$. The horizontal axis is the amount of energy deposited into the scintillator, while
      the vertical axis is the energy lost in the $\Delta E$ proportionality counter. {The two observed particle groups have been circled and labeled.}}
    \label{fig:2D}
  \end{figure}

\begin{figure}
    \centering
    \includegraphics[width=.9\textwidth]{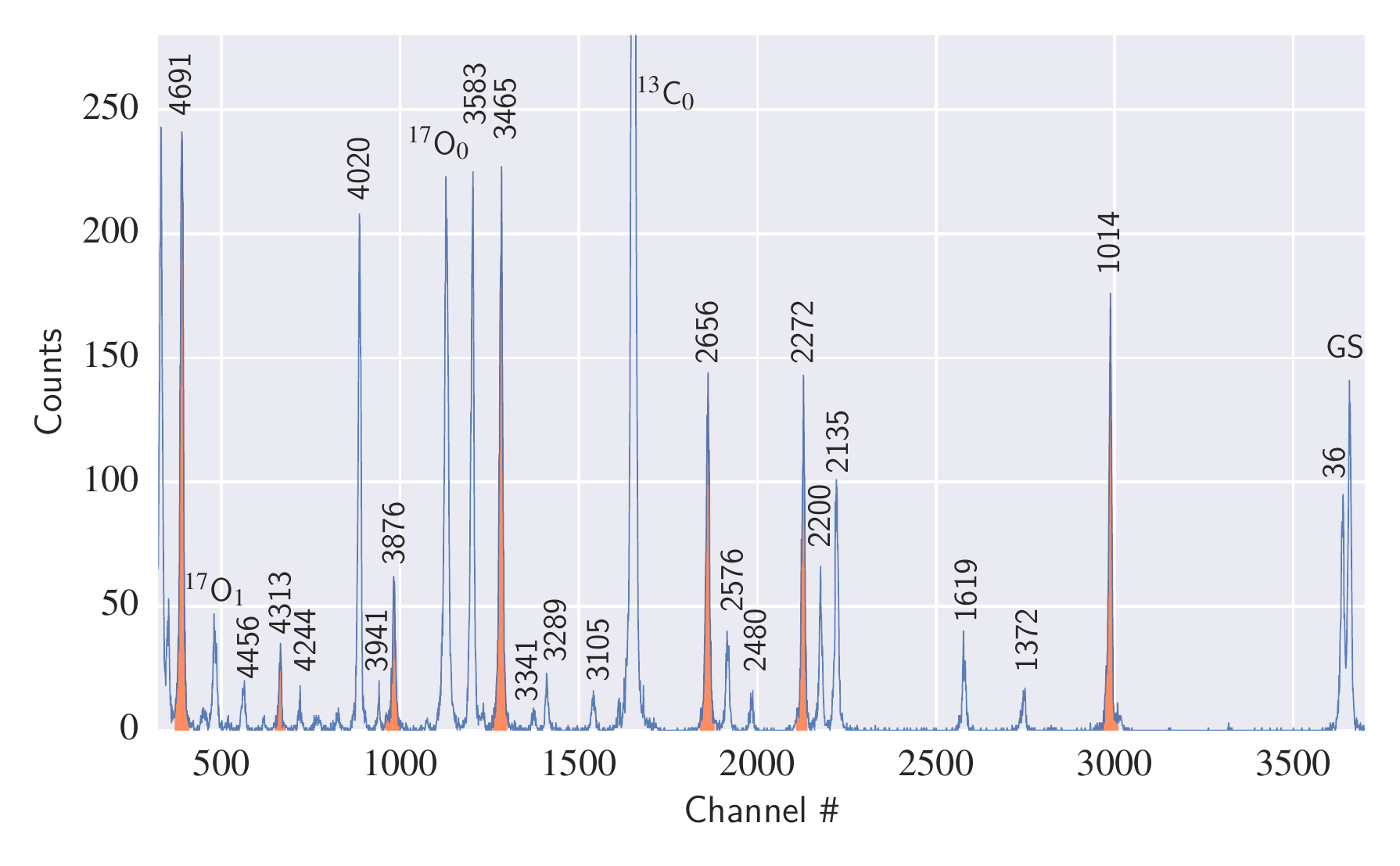}
    \caption{Front position section gated on the proton group at $E_{Lab}=12$ MeV and $\theta_{Lab}=25^{\circ}$. Peaks used for the energy calibration are highlighted and labeled by their energies in keV.
    All other labels are the deduced energies from this work, as reported in Section \ref{sec:al_results}. {Unlabeled peaks were unobserved at other angles due to lower statistics; thus, the reaction that produced them could not be identified with certainty and they are excluded from the reported energy values.}}
  \label{fig:spectra}
\end{figure}

The Bayesian energy calibration method presented above was originally conceived during the analysis of the $^{27}$Al$(d,p)$ experiment performed at TUNL to characterize the focal plane detector (see the work described in Chapter \ref{chap:tunl} Section 4). The experiment was done using DENIS to generate an approximately $ 1 \, \mu \textnormal{A}$ beam of $^2$H$^{-}$. This  
beam was accelerated through the tandem to generate a $12$-MeV $^2$H$^{+}$ beam. Energy analysis was done using the high transmission settings of the $90$-$90$ system. The SPS solid angle was fixed to $0.25$ msr to minimize the effects of kinematic broadening. The detector was filled to $225$ Torr and the position sections were biased to $1800-2000$ V.
The beam was impinged on a target of ${\sim} 80$ $\mu$g/cm$^2$  $^{27}\!$Al evaporated onto a 15.2 $\mu$g/cm$^2$ $^{\textnormal{nat}}$C foil.
Additionally, a $^{\textnormal{nat}}$C target similar to the Al target backing was used to identify contamination peaks arising from carbon and oxygen.
The spectrograph was positioned at three angles, $\theta_{Lab}=15^{\circ}$, $25^{\circ}$, and $35^{\circ}$; and its field was set to $0.75$ T. Example spectra from the $\Delta E$/$E$ and the front position gated on the proton group at $25^{\circ}$ are shown in Fig.~\ref{fig:2D} and Fig.~\ref{fig:spectra}, respectively.

States from $^{28}\!$Al were identified and matched to the peaks in each spectrum. Level energies from Ref.~\cite{levels} were used both as calibration values and as comparisons for predicted energies.
Initially, a set of seven calibration states were chosen for each angle. A second order polynomial was chosen due to the third order term being consistent with zero and a marginally better value of $\delta^2$.

\begin{figure*}
  \centering
  \includegraphics[width=\linewidth]{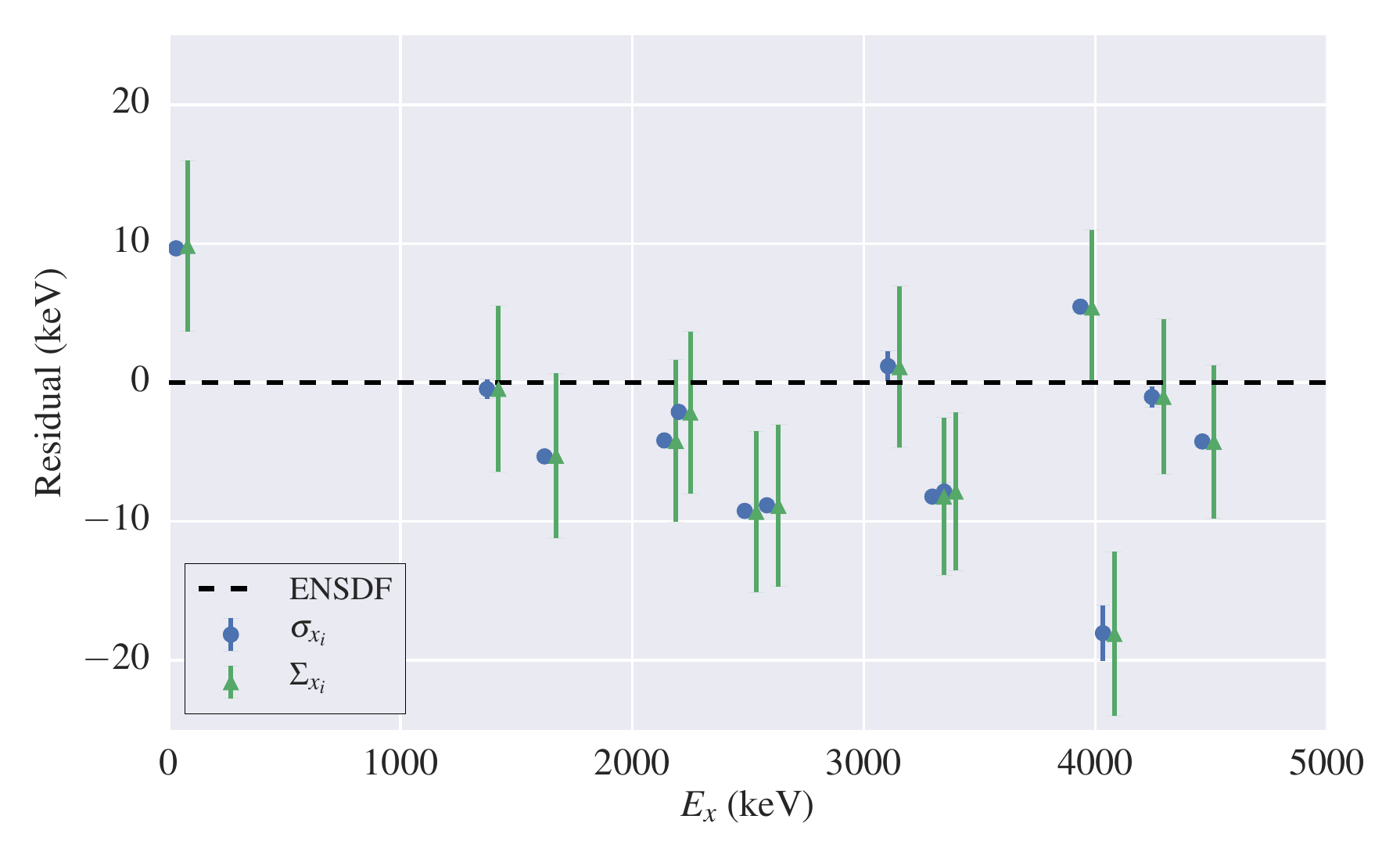}
  \caption{The residual plot of the Bayesian (blue) and {Bayesian with adjusted uncertainty (green)} calibrations at $\theta = 25^{\circ}$. {Calibration points are not included, so only predicted energies are shown.} The excitation energies of the points {with adjusted uncertainty} have been shifted to the right by 50 keV for visibility. The error bars represent the statistical uncertainty of the fit added in quadrature with the uncertainty reported in Ref. \protect\cite{levels}. It is clear that
    the statistical uncertainties from the fit alone are inconsistent with previously reported values. However, general agreement is found when an {additional uncertainty is fit during the calibration.}}
  \label{fig:nndc}
\end{figure*}

For the case of $^{28}$Al, most of the strongly populated states are known to sub-keV precision, which leads to small statistical uncertainties in the fit. These small uncertainties make the deduced values inconsistent with those reported in the Evaluated Nuclear Data Structure File (ENSDF) database \cite{levels}.
Thus, there is clearly an additional source of uncertainty present that is not accounted for in the Bayesian model of Eq.~\ref{eq:calibration_bayesian_model}. This could be due to systematic effects, but sources of systematic uncertainty from experimental parameters (i.e, reaction angle, beam energy, target effects)
have a minimal effect on the deduced excitation energies due to the calibration process.
For example, if the beam energy is off $\sim\! 5$ keV, then all of the calibration points, assuming they are from the
same reaction, will be shifted by roughly the same amount. Therefore, the calibration's intercept will change, and the effect is canceled out in the predicted energies.
The same arguments hold for any systematic effect that is equal for all points.  
Following these considerations, the main source of systematic uncertainty from these effects is the energy dependence of the straggling through the target.
This effect was estimated to be $0.2$ keV, which does not improve the agreement of our results with ENSDF.

The other possible sources of systematic uncertainty arise from the detector response.
In the polynomial model outlined above, the detector response is assumed to be linear; however, deviations from this assumption will cause the model to incorrectly predict
energies. In order to account for these possible effects, an extension to the Bayesian framework was used. In this model, the uncertainty
for the peak centroids {(which we refer to as the "adjusted uncertainty")} is considered to be of the form:
\begin{equation}
  \label{eq:1}
  \Sigma_{i}^2 = \sigma_{i, obs}^2+\sigma^{\prime}{}^{2} , 
\end{equation}
where $\sigma_{i, obs}$ is the observed statistical uncertainty in a given peak, $\sigma^{\prime}$ is an uncertainty that is not directly measured, and $\Sigma_i$ is the new {adjusted uncertainty} for a given peak.
The purpose of $\sigma^{\prime}$ is to broaden the
normal distribution associated with each peak to a degree dictated by the available data. This broadening accounts
for systematic effects in our position measurement, but does not assume a cause or a fixed value. Rather, it is merely another
model parameter to be estimated during calibration.  
In practice, this is done by extending the Bayesian model with a prior distribution for $\sigma^{\prime}$ and using the same MCMC method to infer its value during the polynomial regression.

The choice of prior for $\sigma^{\prime}$ is more nuanced than for the other parameters previously mentioned.
Recall that our data reflects the influence of an additional uncertainty that cannot be directly estimated.
Thus, the prior must encode a source of uncertainty that is larger than the observed statistical uncertainties,
but not large enough to affect the calibration process.
These considerations lead to the adoption of a Half-Cauchy distribution for the precision, $\tau^{\prime} \equiv 1/\sigma^{\prime}{}^2$,
which is a simple transformation of the standard deviation that was found to improve MCMC convergence.   
The Half-Cauchy distribution, written $\textnormal{HalfCauchy}(\alpha,\beta)$, 
is parameterized by $\alpha$, the location parameter, and $\beta$, the scale parameter. The Half-Cauchy distribution
has been found to give good behavior numerical behaviour close to 0, and also avoids the hard limits of the Uniform distribution \cite{gelman2006}.  
For this model:
\begin{equation}
    \label{eq:additional_prior_unc}
    \tau^{\prime} \sim \textnormal{HalfCauchy}(0,1).
\end{equation}
Additional analysis showed no noticeable dependence on $\beta$. With this addition, our model becomes:     
\begin{align}
  \label{eq:calibration_bayesian_model}
    & \textnormal{Priors:} \nonumber \\
    & \tau^{\prime} \sim \textnormal{HalfCauchy}(0,1) \nonumber \\
    & x_i \sim \mathcal{N}(x_{i, obs}, \Sigma^2_{i}) \nonumber \\
    & \theta_j \sim \mathcal{N}(0, 100^2) \nonumber \\
    & \textnormal{Function:}  \\
    & \sigma^{\prime}{}^2 = 1/\tau^{\prime} \nonumber \\
    & \Sigma_i^2 = \sigma_{i, obs}^2+\sigma^{\prime}{}^{2} \nonumber \\
    & f(\theta_j, x_i) = \sum_{j=0}^N \theta_j x_i \nonumber \\
    & \textnormal{Likelihood:} \nonumber \\
    & \rho_{i, cal} \sim \mathcal{N}(f(\theta_j, x_i), \sigma^2_{i, cal}). \nonumber
\end{align}

The comparison of the ENSDF values versus the original fit and the adjusted uncertainty fit
at $\theta_{Lab} = 25^{\circ}$ is shown in Figure \ref{fig:nndc}. Better agreement was found, indicating
the method produces reasonable estimates for the total uncertainty.

\subsection{Results}
\label{sec:al_results}

\begin{table}[]
\centering
\setlength{\tabcolsep}{12pt}
\caption{ Measured $^{28}$Al energy levels (keV)}
\begin{tabular}{c c l}

  \hline
  This Work & &ENSDF \cite{levels} \\
  \hline 
  \hline
 36(5)   & & 30.6382(7)   \\
 1372(5) & & 1372.917(20)  \\
 1619(5) & & 1621.60(4)*   \\
 2135(5) & & 2138.910(10)  \\
 2200(5) & & 2201.43(3)   \\
 2480(5) & & 2486.20(6)   \\
 2576(5) & & 2581.81(22)  \\
 3105(5) & & 3105(1)      \\
 3289(5) & & 3296.34(4)   \\
 3341(5) & & 3347.19(4)   \\
 3583(5) & & 3591.457(9)  \\
 3941(5) & & 3935.603(18)  \\
 4020(5) & & 4033(3)      \\
 4244(5) & & 4244.49(10)  \\
 4456(5) & & 4461.97(10)  \\
 4510(9) & & 4516.94(18)  \\
  \hline
  \multicolumn{3}{r}{* An average of two states at 1620 and 1622.} \\

\end{tabular}

\label{tab:energies}
\end{table}

Using the Bayesian framework with the addition of $\Sigma_i$, a calibration was produced for each angle.
The same calibration peaks were used for each angle, and they represent strongly populated states that are well-resolved in the spectra, which are not obscured by contamination peaks that shift with angle. The location of these states at $\theta_{Lab}=25^{\circ}$ are shown in Fig.~\ref{fig:spectra}

The reported values listed in Table \ref{tab:energies} are weighted averages of the energies over all angles, with
the requirement that any candidate state be observed at more than one angle. A total of 16 states (excluding the 7 calibration states) were measured in this way.

Finally, an estimation for the detector resolution can be found from the slope of the $\rho$ calibration.
A slope of $0.036 \frac{\textnormal{mm}}{\textnormal{channel}}$ was consistently found, with typical full width at half maximum (FWHM) values of $10$-$20$ channels. These values give resolutions between $0.36$-$0.72$ mm.
The separation on the ground and first excited state implies an energy resolution of $\approx 15$ keV.

\subsection{Additional Notes On Energy Calibration}

The method and results described were first presented in Ref.~\cite{marshall_2018}, since that time several improvements and insights have been made. The first is that the \texttt{PyMC2} package was abandoned in favor of the \texttt{emcee} package \cite{emcee}. The specifics of this sampler will be discussed in the next part of this chapter. \texttt{PyMC2} is a fairly simple implementation of the Metropolis Hastings algorithm, and therefore struggled severely even in this fairly simple setting. The chains suffered from high autocorrelation, which required the deduced energies be constructed from KDE representations of the samples. This KDE estimation was, by construction, insensitive to correlations between the polynomial parameters, which means that the deduced excitation energies did not account for these correlations. For the above results, this matters little because a majority of the states are inside the fitting region. These \textit{interpolated} values are less sensitive to the polynomial correlations; however, in the case of \textit{extrapolated} values, these correlations are critical. By implementing \texttt{emcee}, which easily sampled the model, the autocorrelation times decreased, allowing many more independent samples to be drawn. Independent samples can be used directly in the calculation of the excitation energies, and they naturally account for the correlation between model parameters.
 
The additional uncertainty, $\sigma^{\prime}$, was motivated above, but a few additional notes should be made. The description presented above focuses on the case of $^{28}$Al, but in general spectrograph measurements have often grasped for additional sources of uncertainty. For example, Wrede \cite{wrede_thesis} describes a \textit{reproducibility uncertainty} which adjusts the uncertainty based on the disagreement of the predicted values for calibration states versus their input values. Hale et al. \cite{hale_2004} describe a procedure of comparing the predicted energies for other states in the spectrum in order to estimate an additional uncertainty. These adjusted uncertainties and the additional parameter in the Bayesian model are both a response to the inadequacy of a simple polynomial fit for spectrograph data. Deviations from a polynomial fit are made worse because the uncertainties resulting from the fit are often smaller than the calibration uncertainties, which becomes a major issue when these precise values clearly conflict with the observations. This problem will likely grow worse with time, as the precision of the energies used for calibration improves, further decreasing the statistical uncertainties from the fit. Even if it was the case that the polynomial fit exactly described $\rho$ as a function of detector position, trouble can still arise due to systematic effects in the data used to calibrate. The Bayesian approach offers a clear advantage in this scenario because $\Sigma_i$ adds a small amount of robustness to the fitting procedure, which means that outliers will have a diminished impact on the fit results (for a larger discussion of robustness in the context of Bayesian analysis, see Chapter 15 of Ref.~\cite{Kruschke_2010}).

The final note is on the process of taking a weighted average between the measurements at different angles. This procedure can also produce unrealistically small uncertainties, which become more prevalent the more angles are measured. Again, if underlying systematic effects are thought to be present, a weighted average may not be a valid approach. A demonstration of this effect will be presented in Chapter \ref{chap:sodium}. 

\section{Bayesian DWBA}
\label{sec:bay_dwba}

This portion of the thesis details work originally publish in Ref.~\cite{Marshall_2020} and takes large portions from that paper. All of the calculations and methods were performed and developed by the author.

\subsection{Introduction to the Problem}

Chapter \ref{chap:reactions} discussed the need for a reaction theory in order to extract single-particle quantities from transfer reaction data. By using DWBA, it becomes possible to determine both the transferred angular momentum, $\ell$, and the
spectroscopic factor of the populated single particle or hole state \cite{satchler}.  
This structure information can in turn be used to answer questions in nuclear astrophysics \cite{transfers_in_astro},
and to test the shell model on isotopes located far from stability \cite{Wimmer_2018}. 

Despite the wide use of these methods, quantifying the uncertainties associated with
both the optical potentials and the reaction model has been a long standing issue. Previous studies have used statistical methods to
determine the uncertainty on the potential parameters \cite{varner}, but little work has been done to propagate
these uncertainties through DWBA calculations in a statistically meaningful way. To date, most spectroscopic factors are reported
with either no uncertainty, an assumed equivalence between the uncertainty in the data normalization and that of the spectroscopic factor, or a constant $25 \%$ determined from historical studies \cite{endt_cs}.

Over the last few years, these issues have led to a renewed focus on the impact of the optical model parameters on transfer reactions. A series of studies have focused on the nature and magnitude of these effects \cite{Lovell_2015, lovell_opt_unc, king_d_p}. The first steps have also been taken towards quantifying these uncertainties using Bayesian statistics \cite{lovell_mcmc, king_dwba}. These studies focus on the broad effects of optical potentials, but it is worthwhile to establish a Bayesian framework in which the results of a single experiment can be analyzed. This section details my work to establish such a framework, and to examine the possible implications on future experiments. This framework is critical to the analysis of the $^{23}$Na$ (^{3}\textnormal{He}, d)$ reaction in Chapter \ref{chap:sodium}.

The methods developed and presented here were first applied to the analysis of the proton pickup reaction $^{70} \textnormal{Zn} (d, ^{3} \textnormal{He}) ^{69} \textnormal{Cu}$,
which was originally reported in Ref. \cite{pierre_paper}.
This data set possesses many of the features typical of a transfer measurement study: the use of a high resolution magnetic spectrograph to resolve the excited states of interest, elastic data for the entrance channel collected with the same target and beam, experimental uncertainties coming from counting statistics,
and limited angular coverage in both the elastic scattering and transfer differential cross sections. The previous analysis assigned $\ell$ values and extracted spectroscopic factors for
the first eight excited states of $^{69} \textnormal{Cu}$. As such, this data set allows a proof of concept for the Bayesian analysis, since its methods closely parallel the experiments carried out at TUNL, but the data reduction has already been performed. This reanalysis aims to determine the uncertainties associated
with these quantities using Bayesian statistics, and in doing so provide a method that can be built upon in order to analyze transfer reactions of interest to astrophysics.

\subsection{Zero-range Approximation}

The Bayesian model that will be presented in the next part of this chapter will require many millions of likelihood evaluations in order to estimate the posterior distributions. Thus, it is critical to reduce the computational cost of the transfer calculations. In light of this requirement, the zero-range approximation has been adopted for these calculations \cite{satchler}. The approximation, in the specific case of the pick-up reaction $A(d, ^{3} \! \textnormal{He})B$, takes the prior form of the transfer potential, given by:
\begin{equation}
    \label{eq:prior_pickup}
    \mathcal{V}_{prior} = V_{p+d} + \mathcal{U}_{d+B} - \mathcal{U}_{d+A}
\end{equation}
and sets the remnant term, $\mathcal{U}_{d + B} - \mathcal{U}_{d+A}$, to zero. Experimental observation has justified considering the remnant term negligible \cite{first_dwba}. The projectile is then assumed to be absorbed and emitted from the same point giving. The light particle matrix element reduces to:
\begin{equation}
  \label{eq:1}
   \bra{d} V_{pd} \ket{^3\textnormal{He}} \sim D_o \delta(\mathbf{r_p}) , 
\end{equation}
where $\ket{^3\textnormal{He}}$ and $\ket{d}$ are the internal wave functions of the ejectile and projectile, respectively, $D_0$ is
the volume integral of the interaction strength, $V_{pd}$ is the binding potential of the proton to the deuteron,
and $\mathbf{r_p}$ is the coordinate of the proton relative to the deuteron. The use of this approximation gives the further benefit of a direct
comparison to the original analysis of $^{70}$Zn$(d, ^3 \textnormal{He}) ^{69}$Cu that used the zero-range code DWUCK4 for the extraction of $C^2S_{p+B}$ \cite{dwuck4}. It should be noted that Ref.~\cite{pierre_paper} also performed finite-range calculations, which solve the above matrix element by using the expansion from Eq.~\ref{eq:overlap_expansion} for $^3$He, but the computational costs are prohibitively expensive in the present analysis. The value of $D_0$ is calculated
theoretically, with the historical value for proton pick-up and stripping reactions being $D_0 = -172.8$ MeV fm$^{3/2}$ \cite{bassel}. Comparing the different models in Ref.~\cite{all_norms_3He}, an approximately $15 \%$ spread in the values of $D_0^2$ is observed. This is inline with the findings of Ref.~\cite{bertone}, which also noted an approximate $15 \%$ spread in the product $(C^2S_{p+d}) D_0^2$. The above value is adopted here with its associated uncertainty; however, \textit{Ab initio} methods, such as those in Ref.~\cite{brida_ab_initio}, now offer more precise determinations of the $\braket{d|^3 \textnormal{He}}$ overlap. If $D_0$ is deduced using these methods, then this additional source of uncertainty will effectively be eliminated. 

\subsection{Bayesian Considerations}

A Bayesian approach to this problem will, again, require prior probabilities are assigned for every optical model potential parameter and the spectroscopic factor itself. These prior probabilities will then be updated through the likelihood function using the experimentally measured cross sections for the elastic and transfer channels.   

One of the primary goal of this work is to give probabilities for each allowed transfer. The motivation for such a method is that states of astrophysical interest are frequently weakly populated and/or obscured by contaminants. In such a scenario, the angular distribution will often not be informative enough to give a unique $\ell$ assignment. As an example, Fig.~\ref{fig:hale_23na_8945} shows the data for the $8945$-keV state in $^{23}$Na populated via $^{22}$Ne$(^3\textnormal{He}, d)$ in Ref.~\cite{hale_2001}, as reproduced in Ref.~\cite{kelly_2017}. It is clear that a unique $\ell$ value cannot be assigned from such a distribution, but how much information about $\ell$ can we learn from such data? Ideally, these data could be used to give probabilities for each $\ell$ value, and these probabilities could be taken into account when calculating other quantities, such as the reaction rate.      

\begin{figure}
    \centering
    \includegraphics[width=.5\textwidth]{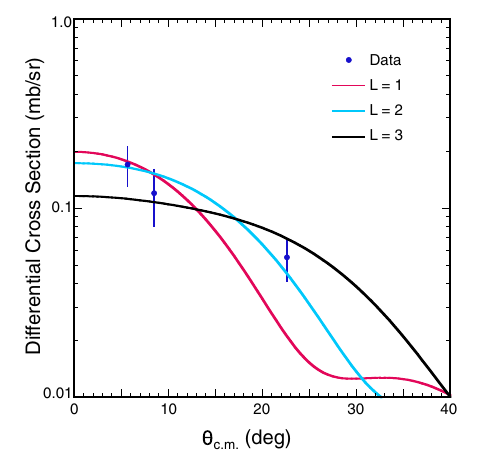}
    \caption{Angular distribution from the $8945$-keV state in $^{23}$Na populated via $^{22}$Ne$(^3\textnormal{He}, d)$ from Ref.~\cite{hale_2001}, as reproduced in Ref.~\cite{kelly_2017}. The state was weakly populated and obscured at several angles by a contamination peak. A unique $\ell$ value ($L$ in the figure) could not be determined.}
    \label{fig:hale_23na_8945}
\end{figure}

The problem of determining the $\ell$ value for a transfer requires a reformulation of Bayes' theorem. To be specific, this problem belongs to a subcategory of Bayesian inference called model selection. Computing the probability for a
model, $M_j$, can be done by restating Bayes' theorem:
\begin{equation}
  \label{eq:m_theorem}
  P(M_j|\mathbf{D}) = \frac{P(\mathbf{D}|M_j) P(M_j)}
  {\sum_i P(\mathbf{D}|M_i)P(M_i)}.
\end{equation}
The above expression is built on the same logical foundation as Eq.~\ref{eq:bayes_theorem}, but has been adapted to compute
posterior distributions for $M_j$, which means a comparison can now be made between different models. For each $M_j$, there is a set of model parameters $\boldsymbol{\theta}_j$, which have been marginalized over. Formally, we have:
\begin{equation}
  \label{eq:marg}
  P(\mathbf{D}|M_j) = \int P(\mathbf{D}|M_j, \boldsymbol{\theta}_j) P(\boldsymbol{\theta}_j|M_j) d\boldsymbol{\theta}_j.
\end{equation}
Examining Eq.~\ref{eq:marg}, it can be seen that $P(\mathbf{D}|M_j)$ is equivalent to the evidence integral from Eq.~\ref{eq:bayes_theorem}. Thus, in order to evaluate how probable different angular
momentum transfers are, the evidence integral must be calculated.

Once the evidence integral is calculated, there are several metrics to interpret model posterior probabilities. For simplicity, we will now refer to the evidence integral as $Z_j$, which corresponds to the model $M_j$, and additionally assume the model priors, $P(M_j)$, are equal for each model. This assumption means before the transfer data are analyzed, all $\ell$ transfers are considered equally likely. The most commonly used criterion for Bayesian model selection is called the Bayes Factor, which is defined by:
\begin{equation}
  \label{eq:bayes_factor}
  B_{ji} = \frac{Z_j}{Z_i},
\end{equation}
in the case of equal model priors.
If this ratio is greater than $1$, the data support the selection of model $j$, while values less than $1$
support model $i$. Judging the level of significance for a value of $B_{ji}$ is open to interpretation, but a useful
heuristic was given by Jefferys \cite{Jeffreys61}. For the cases where model $j$ is favored over $i$ we have the following
levels of evidence: $3 > B_{ji} > 1$ is anecdotal, $10 > B_{ji} > 3$ is substantial, $30 > B_{ji} > 10$ is strong, $100 > B_{ji} > 30$ is very strong, and $ B_{ji} > 100$ is decisive. 

It is also possible to calculate explicit probabilities for each model. Again assuming each of the models is equally likely, the probability of a given model can be expressed as:

\begin{equation}
  \label{eq:model_prob}
  P(M_j|\mathbf{D}) = \frac{Z_j}{\sum_i Z_i}.
\end{equation}

Through Eq.~\ref{eq:model_prob}, probabilities can be calculated for each physically allowed angular momentum transfer, $\ell_j$.
Using these definitions, Bayesian inference can be carried out after prior probabilities are assigned for each optical model parameter and a likelihood function for the data is chosen.

\subsection{Ambiguities in Potential Parameters}
\label{sec:amb_pots}

Any analysis involving potentials of the form in Eq.~\ref{eq:ws_pot} will suffer from so-called continuous and discrete ambiguities.
Both of these ambiguities arise because a single differential cross section at a single energy cannot uniquely determine the potential parameters.
The continuous ambiguity describes strong correlation between certain model parameters \cite{hodgson1971, vernotte_optical}. A well known example is the relation between the real volume depth, $V$, and the corresponding radius, $r_0$.
The relation has an approximate analytical form given by $Vr_0^n = const$, where the exponent $n \approx 1.14$ and the constant can vary depending on the reaction or chosen optical model \cite{vernotte_optical}. 
The continuous ambiguity can be remedied in part by a global analysis of the potential parameters
across a wide range of mass numbers and reaction energies, as noted in the comprehensive analysis of
proton and neutron scattering in Ref.~\cite{varner} and for $^3$He and $t$ scattering in Ref.~\cite{pang_global}.
Since the present analysis will be limited to a single elastic scattering data set, the model must be prepared to deal
with these parameter correlations.

The discrete ambiguity arises in optical model analysis due to the identical
phase shifts that are produced by different values of $V$ \cite{drisko_1963}.
This multi-modal behavior is perhaps the more problematic of the two ambiguities
since parameter correlation can be handled with standard statistical methods. However, interpretation
of uncertainties in a multi-modal problem requires care beyond standard credibility intervals.
The discrete families of parameters can be readily identified by the volume integral of the real potential:
\begin{equation}
  \label{eq:j_int}
  J = \frac{4 \pi}{A_{P}A_{T}} \int_0^{\infty} Vf(r; r_0, a_0)  r^2 dr ,
\end{equation}
where the mass numbers of the projectile and target, $A_P$ and $A_T$, respectively, ensure
that $J$ should be roughly constant for a family of potential parameters at a single energy.
Microscopic structure models such as the folding model can also be used to calculate $J$,
and the theoretical value can be used to identify the physical potential family \cite{daehnick_global}.
Trusting the efficacy of this method, the approach for this work is to adopt potential depths from global fits and to keep our prior values contained around these starting potential depths.

\subsection{Global Potential Selection}

The {initial} potentials used for the analysis of $^{70}$Zn$(d, ^3\textnormal{He}) ^{69}$Cu before inference can be found in Table~\ref{tab:opt_parms}.
In order to facilitate comparison with Ref.~\cite{pierre_paper}, the same global potentials have been used. These potentials are the Daehnick-F global $d$ optical model \cite{daehnick_global},
and the Becceheti and Greenless global $^3$He model of Ref.~\cite{b_g_3he}. It is also worth noting that elastic scattering with an
unpolarized beam does not provide a constraint on the parameters of a spin-orbit potential, so all spin orbit terms have been held fixed
in the current work \cite{hodgson1994, daehnick_global, thompson_nunes_2009}.

The bound state geometric parameters are assigned their most commonly used value of $r_0=1.25$ fm and $a_0=0.65$ fm, 
with the volume potential depth adjusted to reproduce the binding energy of the final state \cite{perey_params,hodgson1971,bjork_params}.
The bound state spin-orbit volume depth was fixed at a value of $V_{so} = 8.66$ MeV in order to
approximately correspond to the condition  $\lambda = 25$, where $\lambda \sim  \frac{180 V_{so}}{V}$
for the value of $V$ for the ground state. 

\begin{table*}[ht]
\centering
\begin{threeparttable}[e]
\caption{\label{tab:opt_parms}Optical potential parameters used in this work before inference.}
\setlength{\tabcolsep}{4pt} 
\begin{tabular}{ccccccccccccccc}
\toprule[1.0pt]\addlinespace[0.3mm] Interaction  & $V$ & $r_{0}$ & $a_{0}$ & $W$ & $W_{s}$ & $r_{i}$ & $a_{i}$ & $r_{c}$ & $V_{so}$ \\
                                                 & (MeV) & (fm) & (fm) & (MeV) & (MeV) & (fm) & (fm) & (fm) & (MeV)\\ \hline\hline\addlinespace[0.6mm]
$d$ $+$ $^{70}$Zn\tnote{a} & $86.76$ & $1.17$ & $0.75$ & $0.90$  & $11.93$ & $1.32$  &  $0.81$ & $1.30$ & $6.34$ &  \\
\hspace{0.15cm} $^{3}$He $+ ^{69}$Cu \tnote{b} & $156.5$ & $1.20$ & $0.72$ & $42.2$ & &$1.40$ & $0.86$ & $1.25$ & \\

\hspace{0.1cm}$p$ $+$ $^{69}$Cu & \tnote{c} & 1.25 & 0.65 & & & & & 1.25 & 8.66 & \\[0.2ex]
\bottomrule[1.0pt]
\end{tabular}
\begin{tablenotes}
\item[a] Global potential of Ref. \cite{daehnick_global}.
\item[b] Global potential of Ref. \cite{b_g_3he}.
\item[c] Adjusted to reproduce binding energy of the final state.
\end{tablenotes}
\end{threeparttable}
\end{table*}

\subsection{Bayesian Model}
\label{sec:model}

Following the above discussion and considerations, it is time to define the Bayesian model for this problem. The model will perform fits of each excited state simultaneously with the elastic scattering data. Again, using a Bayesian approach means each
parameter, whether from the optical model potentials or otherwise, has to be assigned a prior probability distribution.
Additionally, likelihood functions will need to be assigned for the data in both the elastic and transfer channels.
For this problem, only three distributions are used: normal, half-normal, and uniform. A half-normal distribution is equivalent to a normal distribution with $\mu=0$ and restricted to the interval $[0, \infty)$; thus, it only has one free parameter, which is the variance.
It is written as $\textnormal{HalfNorm}(\sigma^2)$. The uniform distribution will be given by its lower and upper limits, written
as $\textnormal{Uniform}(\textnormal{Lower}, \textnormal{Upper})$. This distribution gives equal probability to every value between the lower and upper limits.

The majority of parameters come from the optical model potentials. The elastic scattering data from $^{70} \textnormal{Zn} (d,d)$ should be able
to inform the posteriors for the entrance channel parameters, $\boldsymbol{\mathcal{U}}_{\textnormal{Entrance}}$. However, the ambiguities discussed in Section 5.3,  
combined with the lack of data at angles higher than $\theta_{c.\!m.}= 50^{\circ}$, means that the priors for the entrance channel must be weakly informative. In order to accomplish this, their radius and diffuseness parameters are focused around a reasonable range for both the real and imaginary potentials. If it is assumed that physical values for these parameters tend to lie within $r = 1.0-1.5$ fm and $a=0.52-0.78$ fm, then the priors can be constructed to favor these values. This is accomplished by assigning normal distributions with mean, $\mu_r = 1.25$ fm and $\mu_a=.65$ fm and variance  $\sigma^2_r = (0.20 \, \mu_r)^2$ and $\sigma^2_a = (0.20 \, \mu_a)^2$. These priors have $68 \% $ credibility intervals that are equivalent to $r = 1.0-1.5$ fm and $a=0.52-0.78$ fm, and importantly do not exclude values that lie outside of these ranges. This means that if the data are sufficiently informative, they can pull the values away from these ranges, but in the absence of strong evidence, the priors will bias the parameters toward their expected physical values. The depths of the potentials were also assigned standard deviations of $20 \% $ of their global depths, which favors the mode assigned by the global analysis, and was found to be sufficiently restrictive to eliminate the discrete ambiguity. These conditions are summarized in the prior:
\begin{equation}
  \label{eq:entrance_prior}
  \boldsymbol{\mathcal{U}}_{\textnormal{Entrance}} \sim \mathcal{N}(\mu_{\textnormal{central}, k}, \{0.20 \, \mu_{\textnormal{central}, k}\}^2), 
\end{equation}
where the symbol \say{central} refers to the global values for the depths and the central physical values of $r=1.25$ fm and $a=0.65$ fm defined above, and the index $k$ runs over the depth, radius and diffuseness parameters for the real and imaginary parts of the potential.

The exit channel, as opposed to the entrance channel, does not have elastic scattering data to constrain it directly. This means that informative priors based on a global analysis must be used, while also considering a reasonable range of values. Normal priors are used, again to avoid sharp boundaries on the values, with the global values of Table~\ref{tab:opt_parms} as the location parameters, and the scale parameter set to $\sigma^2 = (0.10 \, \mu)^2$. Values are thus focused around those of the global model, but are allowed a moderate amount of variation.
The prior choice can be stated: 
\begin{equation}
  \label{eq:entrance_prior}
  \boldsymbol{\mathcal{U}}_{\textnormal{Exit}} \sim \mathcal{N}(\mu_{\textnormal{global}, k}, \{0.10 \, \mu_{\textnormal{global}, k}\}^2), 
\end{equation}
with the \say{global} label referring to the values of Table~\ref{tab:opt_parms} and $k$ labeling each of the potential parameters for the exit channel.

At this point, it is worth emphasizing that the potential priors for both the entrance and exit potentials are essentially arbitrary. The $20 \%$ and $10 \%$ variation for the parameters are meant to make this computation tractable, since it is impossible with the limited amount of data to uniquely determine the parameters as discussed in Section 5.3. The influence of this choice on the entrance channel is limited since there are data to inform the parameters. However, the choice of $10 \%$ for the exit channel will influence our final calculated uncertainties. Lower or higher amounts of variation could be considered for these parameters, but a choice has to be made in order to account for their impact on DWBA calculations. I have also chosen to exclude variations in the spin-orbit and bound state potentials. However, the possible impact of the bound state potentials will be discussed later in this chapter.       
 
Since this model treats $C^2S$ as another parameter to be estimated, a prior must be specified. It has been assigned the mildly informative prior:
\begin{equation}
  \label{eq:cs_prior}
  C^2S \sim \textnormal{HalfNorm}(n_{nucleon}^2),  
\end{equation}
where $n_{nucleon}$ is the number of nuclei occupying the orbital that is involved in the transfer. The half-normal distribution ensures that $C^2S \geq 0$, while the scale parameter comes from the sum rules of Macfarlane and French \cite{sum_rules}. These rules have been found experimentally to be a robust constraint \cite{sum_rule_test}. However, it is likely that this prior is more conservative than necessary, since it is not expected that a single state will contain the entirety of the strength for a given shell. These considerations are just to provide a rough estimate to help construct the prior for $C^2S$.

The use of the zero-range approximation for the transfer channels also comes with an additional uncertainty from the strength parameter, $D_0$, as discussed in
Section 5.5.2. Our model explicitly accounts for this $15 \%$ uncertainty by using a parameter $\delta D_0^2$, which is assigned a normal and informative prior:

\begin{equation}
  \label{eq:d0}
  \delta D_0^2 \sim \mathcal{N}(1.0, 0.15^2).
\end{equation}

Two additional parameters are also introduced that are not a part of DWBA, but are instead meant to account for deficiencies in the reaction theory. The first is a normalization parameter, $\eta$, which allows for the adjustment of the theoretical predictions for both the elastic and transfer cross sections based on any observed normalization difference between the elastic channel data and optical model calculations. This can be in principle seen as treating the absolute scale of the data as arbitrary, which prevents biasing the potential parameters towards
unphysical values if a systematic difference is present. The posterior for this parameter will only be directly informed by the elastic data of the entrance channel, but will directly influence the posterior for $C^2S$. Since $\eta$ is multiplicative in nature, we do not want to bias it towards values less than or greater than $1$. This is done by introducing a
parameter, $g$, which is uniformly distributed according to:
\begin{equation}
  \label{eq:g_uni}
  g \sim \textnormal{Uniform}(-1, 1).
\end{equation}
$\eta$ is then defined as:
\begin{equation}
  \label{eq:eta}
  \eta = 10^{g}.
\end{equation}
Collecting all of these factors, the DWBA predictions can now be written at each angle $i$ as:
\begin{equation}
  \label{eq:cs_full}
  \frac{d \sigma}{d \Omega}^{\prime}_{\textnormal{DWBA}, i} = \eta \times \delta D_0^2 \times C^2S \times \frac{d \sigma}{d \Omega}_{\textnormal{DWBA}, i}.
\end{equation}

The second additional parameter comes from the consideration that the DWBA theory provides only an approximation to the true transfer cross section. If only the measured experimental uncertainties from the transfer channel are considered, then any deviation from DWBA will significantly influence the posteriors for the potential parameters. This is remedied by introducing an additional theoretical uncertainty, $\sigma_{\textnormal{theory}, i}$, where the index $i$ references the angle at which the differential cross section is evaluated. Because cross sections vary many orders of magnitude as a function of angle and are manifestly positive, uncertainties are best treated as percentages. Thus, this additional uncertainty is defined as a percentage uncertainty on the theoretical cross section, which is based on a single unknown parameter, $f$. The total uncertainty at an angle is:
\begin{equation}
  \label{eq:unc}
  \sigma_i^{\prime 2} = \sigma_{\textnormal{Transfer}, i}^2 +  \bigg(f\frac{d \sigma}{d \Omega}^{\prime}_{\textnormal{DWBA}, i}\bigg)^2.
\end{equation}
I use $\frac{d \sigma}{d \Omega}^{\prime}_{\textnormal{DWBA}, i}$ as defined in Eq.~\ref{eq:cs_full}, $\sigma_{\textnormal{Transfer}, i}^2$ is the experimental statistical uncertainty, and the adjusted uncertainty, $\sigma_i^{\prime 2}$,
assumes that the experimental and theoretical uncertainties are independent. Since $f$ is some fractional amount of the predicted cross section, it is assigned the weakly informative prior:
\begin{equation}
  \label{eq:f}
  f \sim \textnormal{HalfNorm}(1),
\end{equation}
so that it is biased towards values less than $1$.  

Finally, the likelihood functions for the experimental data must also be specified. The analysis of each excited state will
require two likelihood functions for both the elastic and transfer data. These likelihood functions use the normal distribution, and
take the form:
\begin{equation}
  \label{eq:likelihood}
  \frac{d \sigma}{d \Omega}_{\textnormal{Exp}, i} \sim \mathcal{N}\bigg(\frac{d \sigma}{d \Omega}_{\textnormal{Theory}, i}, \sigma_{\textnormal{Exp}, i}^2\bigg), 
\end{equation}
where $i$ again refers to a specific angle. The above expression assumes that the residuals between the experimental cross section and the ones calculated from theory are distributed normally. 

Taking into account all of the considerations and definitions listed above, the full Bayesian model can be written.
Experimental elastic scattering data are identified by the label \say{Elastic}, and the transfer data are labeled \say{Transfer}. The theoretical differential cross sections calculated with FRESCO are written $\frac{d \sigma}{d \Omega}_{\textnormal{Optical}, j}$ for elastic scattering and $\frac{d \sigma}{d \Omega}_{\textnormal{DWBA}, i}$ for the transfer reaction. The indices $i$ and $j$ refer to the transfer and elastic angles, respectively. The model is, thus:   
\begin{align}
  \label{eq:model}
 & \textnormal{Priors:} \nonumber \\
 & \boldsymbol{\mathcal{U}}_{\textnormal{Entrance}} \sim \mathcal{N}(\mu_{\textnormal{central}, k}, \{0.20 \, \mu_{\textnormal{central}, k}\}^2) \nonumber \\
 & \boldsymbol{\mathcal{U}}_{\textnormal{Exit}} \sim \mathcal{N}(\mu_{\textnormal{global}, k}, \{0.10 \, \mu_{\textnormal{global}, k}\}^2) \nonumber \\
 & f \sim \textnormal{HalfNorm}(1) \nonumber \\
 & \delta D_0^2 \sim \mathcal{N}(1.0, 0.15^2) \nonumber \\
 & C^2S \sim \textnormal{HalfNorm}(n_{nucleon}^2) \nonumber \\
 & g \sim \textnormal{Uniform}(-1, 1) \nonumber \\
 & \textnormal{Functions:} \\
 & \eta = 10^{g}  \nonumber \\
 & \frac{d \sigma}{d \Omega}^{\prime}_{\textnormal{Optical}, j} = \eta \times \frac{d \sigma}{d \Omega}_{\textnormal{Optical}, j} \nonumber \\
 & \frac{d \sigma}{d \Omega}^{\prime}_{\textnormal{DWBA}, i} = \eta \times \delta D_0^2 \times C^2S \times \frac{d \sigma}{d \Omega}_{\textnormal{DWBA}, i} \nonumber \\
 & \sigma_i^{\prime 2} = \sigma_{\textnormal{Transfer}, i}^2 +  \bigg(f\frac{d \sigma}{d \Omega}^{\prime}_{\textnormal{DWBA}, i}\bigg)^2 \nonumber \\
 & \textnormal{Likelihoods:} \nonumber \\
 & \frac{d \sigma}{d \Omega}_{\textnormal{Transfer}, i} \sim \mathcal{N}\bigg(\frac{d \sigma}{d \Omega}^{\prime}_{\textnormal{DWBA}, i}, \sigma_i^{\prime \, 2}\bigg) ,  \nonumber \\
 & \frac{d \sigma}{d \Omega}_{\textnormal{Elastic}, j} \sim \mathcal{N}\bigg(\frac{d \sigma}{d \Omega}^{\prime}_{\textnormal{Optical}, j}, \sigma_{\textnormal{Elastic}, j}^2\bigg) ,  \nonumber 
\end{align}
where the $k$ index runs over each of the potential parameters.

It should also be noted that the applicability of DWBA requires that the reaction is dominated
by a direct reaction mechanism occurring at the nuclear surface. Thus, transfer data must be collected at intermediate laboratory energies to suppress the contributions of isolated resonances and low angles to ensure a surface dominated reaction. Failure to adhere to these principles could introduce additional uncertainties into the extraction of $C^2S$. Practically, this work follows the suggestion of Ref.~\cite{thompson_nunes_2009} and only fits the transfer data up to the first observed minimum in the data.

\subsection{Posterior and Evidence Estimation}

It is clear from Eq.~\ref{eq:model} that our Bayesian model lives in a high dimensional space, which
presents a difficult challenge for all MCMC algorithms as discussed before. In particular, traditional Metropolis-Hastings samplers require tuning of the step proposals for each dimension. The problem of tuning the proposals is avoided with the Affine Invariant Ensemble sampler of Goodman and Weare \cite{ensemble_mcmc}.
This method uses an ensemble of random walkers to 
sample the posterior, and has been designed to perform well with linearly correlated
parameters. We use the Python package \texttt{emcee} to implement the algorithm \cite{emcee}.
Using \texttt{emcee} with a so-called stretch move requires only a single parameter, $a$, to be specified. \cite{emcee}.
The posteriors for each state are estimated using an ensemble of $400$ walkers which take $> 4000$ steps.
Burn in periods were found to take approximately $1000$ steps.
Final parameter estimates are taken from the final $2000$ steps, which are then thinned by $50$ in order to
give $1.6 \times 10^4$ samples. The autocorrelation in the samples before thinning was estimated to be roughly $400$ steps. $2000$ steps would then contain 5 autocorrelation lengths, with each length yielding one independent sample per walker. This means $\approx 2000$ independent samples are drawn from the posterior, ensuring that the statistical fluctuations of the sampling are negligible compared to the uncertainties in the posteriors. Thinning was only used to reduce the number of samples and thereby ease subsequent calculations such as the credibility intervals for the differential cross sections. 

MCMC methods draw samples directly from the posterior distribution which allows parameter estimation, but
they do not allow a straightforward estimation of the evidence integral. 
The model selection necessary to assign $\ell$ values requires the calculation of
Eq.~\ref{eq:model_prob}. Monte Carlo integration techniques solve the issue of calculating $Z$, but essentially reverse the previous issue by placing a diminished focus on the calculation of the posterior distributions. Thus, separate calculations have to be carried out for the two tasks of parameter estimation (spectroscopic factors)
and model selection ($\ell$ assignment). The evidence calculations presented here are carried out using the nested sampling
procedure introduced by Skilling \cite{skilling2006, skilling2004}, as
implemented in the \texttt{dynesty} Python package \cite{speagle2019dynesty}.

For this work, all nested sampling runs used $250$ live points bounded with multiple ellipsoids and updates performed through slice sampling.
The stopping criteria was set at $\Delta Z_i < .01$. Since nested sampling is subject to statistical uncertainties in $\ln Z$, it is necessary to propagate these uncertainties through to both $B_{ij}$ and the probabilities for each $\ell$ transfer defined by Eq.~\ref{eq:model_prob}. This was done by drawing $10^6$ random samples from the Gaussian distributions for each $\ln Z_i$, and then applying either Eq.~\ref{eq:model_prob} or Eq.~\ref{eq:bayes_factor} to each sample, yielding a set of samples for each quantity. From these samples, the $68 \%$ credibility intervals are reported, which are constructed from the $16$, $50$, and $84$ percentiles. 

\section{Analysis of $^{70} \textnormal{Zn}(d, ^{3} \textnormal{He})^{69} \textnormal{Cu}$}
\label{sec:results}

The Bayesian model detailed above allows the extraction of spectroscopic factors and the assignment of $\ell$ values
to transfer reaction data, while taking into account uncertainties associated with the optical potentials.
In order to test these methods, a reanalysis of the $^{70}$Zn$(d, ^{3}$He$)^{69}$Cu reaction data originally presented in Ref.~\cite{pierre_paper} was performed. For reference, data were collected by impinging a $27$-MeV deuteron beam onto
a thin target of enriched $^{70}$Zn. The reaction products were measured with a magnetic spectrograph. The original study should be referred to for complete experimental details. This reaction and the measured data set have two important conditions that simplify our study. First, since $^{70} \textnormal{Zn}$ has a $0^{+}$ ground state, only a unique $\ell$ transfer is allowed for a given final state. Second, only 8 low lying bound states were observed, meaning no additional
theoretical model is needed for treating transfers to the continuum. The results of the MCMC calculations
are summarized in Table~\ref{tab:cs}. Comparisons are made to the original values of the zero-range and finite-range calculations of the previous work.
Plots of the DWBA cross sections generated from the MCMC calculations are shown in Fig.~\ref{fig:states}. The purple and blue bands show the $68 \%$ and $95 \%$ credibility bands, respectively. Using samples directly from the Markov chain means that these credibility bands accurately account for all of the correlations present between the parameters. Each of these states will now be discussed in detail, with additional calculation details provided for the ground state in order to demonstrate
the use of our Bayesian method.

\begin{table}
 \centering
  \begin{threeparttable}
  \setlength{\tabcolsep}{4pt}
  \caption{\label{tab:cs} Summary of the spectroscopic factors derived in this work. {Comparisons to the zero-range (ZR) and finite-range (FR) calculations of} Ref.~\cite{pierre_paper} are made. {All calculations use the same bound state parameters}.}
  \begin{tabular}{cccccc}
    \addlinespace[0.5mm]
    \\ \hline \hline
    \\ [-1.0 ex]
    $E_x$(MeV) & $\ell$ & $J^{\pi}$ \tnote{a} & $C^2S(ZR)$ \cite{pierre_paper} &  $C^2S(FR)$ \cite{pierre_paper}   & $C^2S$(This work)           \\ [-.5ex] 
    \\ \hline 
    \\ [-1.5ex] 
$0.0$      & $1$    & $3/2^{-}$ & $1.40(15)$  & $1.50(17)$  & $2.06^{+0.87}_{-0.68}$            \\ [0.8ex]
$1.11$     & $1$    & $1/2^{-}$ &    -        & $0.35(11)$  & $0.48^{+0.52}_{-0.25}$            \\ [0.8ex]
$1.23$     & $3$    & $(5/2^{-})$ & $0.80(11)$& $0.70(10)$  & $1.10^{+0.81}_{-0.48}$   \\ [0.8ex]
$1.71$     & $3$    & $7/2^{-}$ & $2.00(11)$  & $2.50(14)$  & $2.37^{+1.36}_{-0.84}$            \\ [0.8ex]
$1.87$     & $3$    & $7/2^{-}$ & $0.40(10)$  & $0.50(10)$  & $1.07^{+0.93}_{-0.51}$            \\ [0.8ex]
$3.35$     & $3$    & $(7/2^{-})$ & $1.60(10)$ & $2.40(15)$  & $2.67^{+1.83}_{-1.06}$             \\ [0.8ex]
\multirow{2}{*}{$3.70$}     & $2$    & $(3/2^{+})$ &  $1.90(25)$ & $1.50(20)$  & $1.74^{+1.05}_{-0.62}$             \\ [0.8ex]
          & $3$    & $(7/2^{-})$ & -            & - & $2.90^{+2.75}_{-1.43}$             \\ [0.8ex]
    $3.94$     & $0$    & $1/2^{+}$ & $0.70(6)$    & $0.70(10)$  & $1.03^{+0.71}_{-0.44}$          \\ [-1.5ex]
    \\ \hline \hline
  \end{tabular}
  \begin{tablenotes}
\item[a] These assignments are discussed in depth in Section \ref{sec:gs_sec} through Section \ref{sec:394_sec}.
\end{tablenotes}
\end{threeparttable}
\end{table}

\afterpage{
\clearpage
\null
\hspace{0pt}
\vfill
\captionof{figure}{The DWBA calculations for the states of $^{69}$Cu. The $68 \%$ and $95 \%$ credibility intervals are shown in purple and blue, respectively. Only data points up to the first minimum were considered, and they are shown in orange. For the $3.70$ MeV state, the $68 \%$ bands are shown for the two most likely $\ell$
    transfers.}
\label{fig:states}
\vfill
\newpage
\clearpage
\begin{figure}
  \ContinuedFloat\centering
  \captionsetup[subfigure]{labelformat=empty}
    \vspace{-1\baselineskip}
    \begin{subfigure}[t]{0.45\textwidth}
        \includegraphics[width=\textwidth]{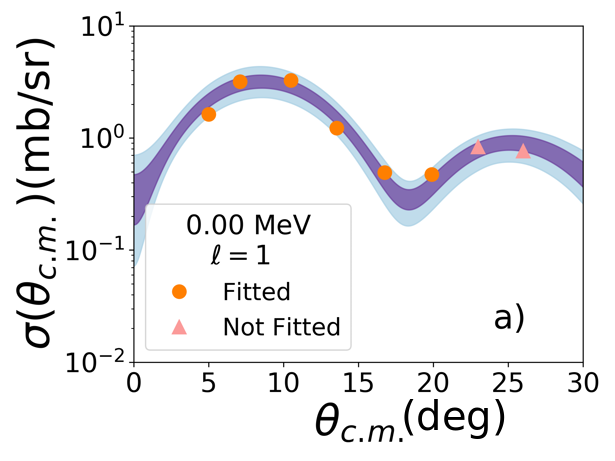}
        \caption{\label{fig:gs_fit}}
      \end{subfigure}
          \vspace{-1\baselineskip}
    \begin{subfigure}[t]{0.45\textwidth}
      \includegraphics[width=\textwidth]{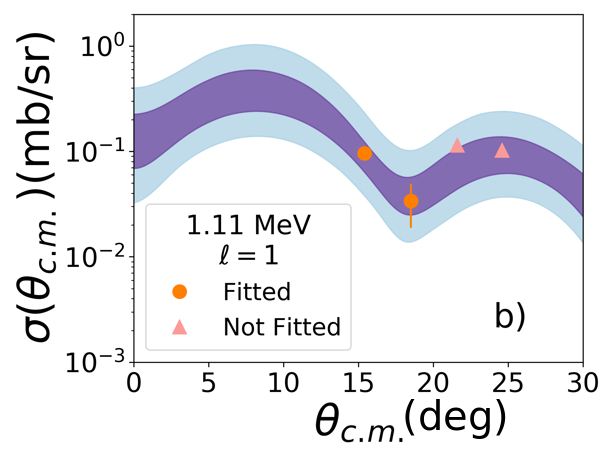}
          \vspace{-1\baselineskip}
      \caption{\label{fig:111_fit}}
    \end{subfigure}
    \vspace{-1\baselineskip}
    \begin{subfigure}[t]{0.45\textwidth}
      \includegraphics[width=\textwidth]{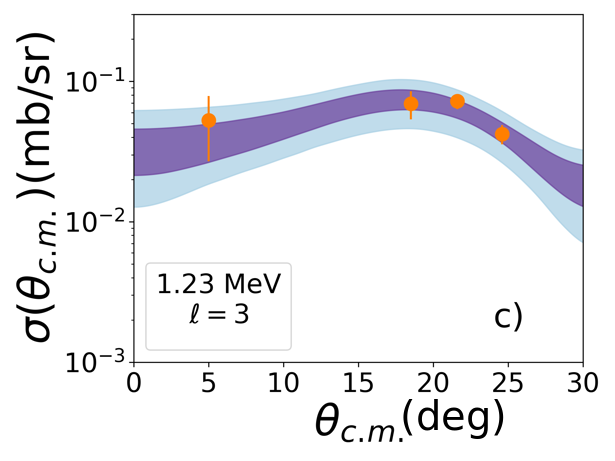}
          \vspace{-1\baselineskip}
      \caption{\label{fig:123_fit}}
    \end{subfigure}
    \begin{subfigure}[t]{0.45\textwidth}
      \includegraphics[width=\textwidth]{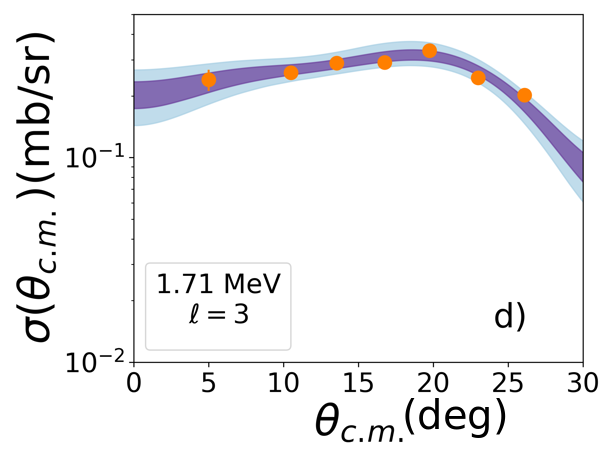}
          \vspace{-1\baselineskip}
      \caption{\label{fig:171_fit}}
    \end{subfigure}
    \vspace{-1\baselineskip}
    \begin{subfigure}[t]{0.45\textwidth}
      \includegraphics[width=\textwidth]{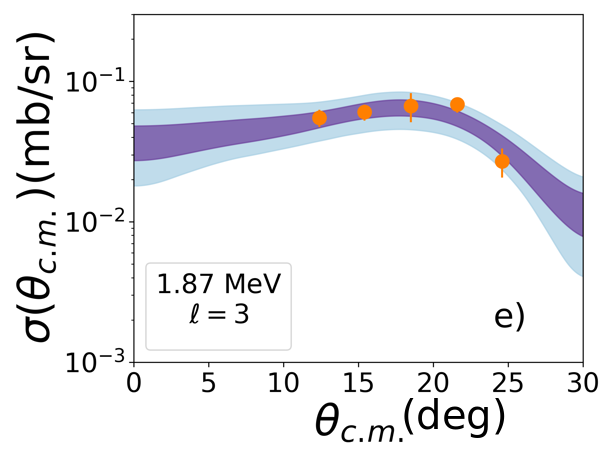}
          \vspace{-1\baselineskip}
      \caption{\label{fig:187_fit}}
    \end{subfigure}
    \begin{subfigure}[t]{0.45\textwidth}
      \includegraphics[width=\textwidth]{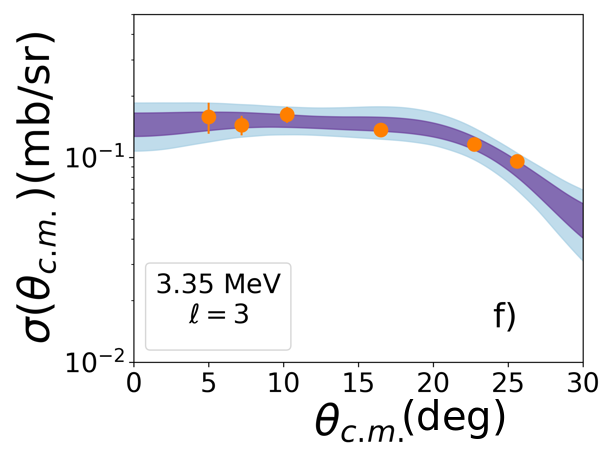}
          \vspace{-1\baselineskip}
      \caption{\label{fig:335_fit}}
    \end{subfigure}
\end{figure}
\begin{figure}[t]
 \ContinuedFloat\centering
  \captionsetup[subfigure]{labelformat=empty}
    \begin{subfigure}{0.45\textwidth}
      \includegraphics[width=\textwidth]{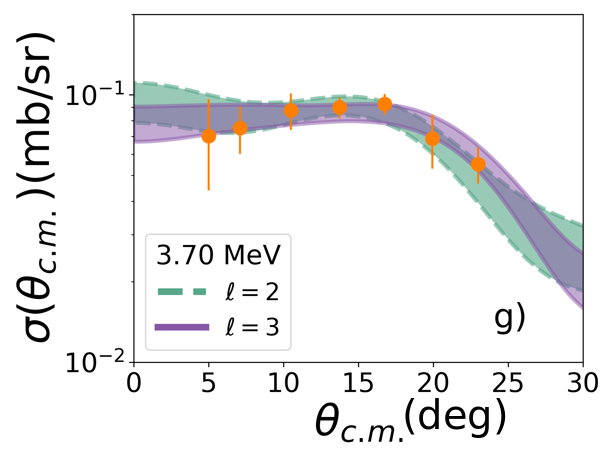}
          \vspace{-1\baselineskip}
      \caption{\label{fig:370_fit}}
    \end{subfigure}
    \begin{subfigure}{0.45\textwidth}
      \includegraphics[width=\textwidth]{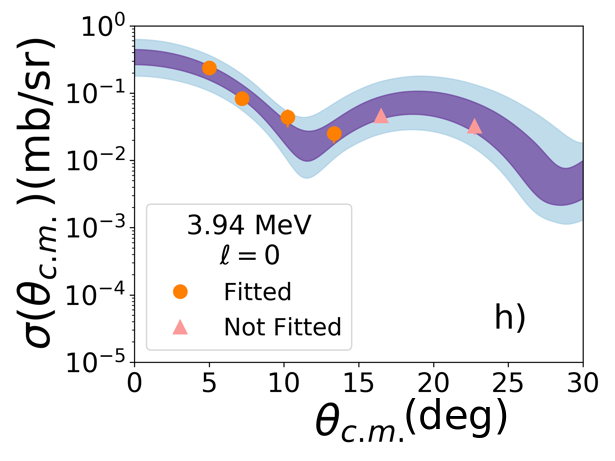}
          \vspace{-1\baselineskip}
      \caption{\label{fig:394_fit}}
    \end{subfigure}
    \vspace{-1.8\baselineskip}
\end{figure}
\clearpage
}
\newpage

\subsection{The Ground State}
\label{sec:gs_sec}

\begin{figure}
  \centering
  \includegraphics[width=.6\textwidth]{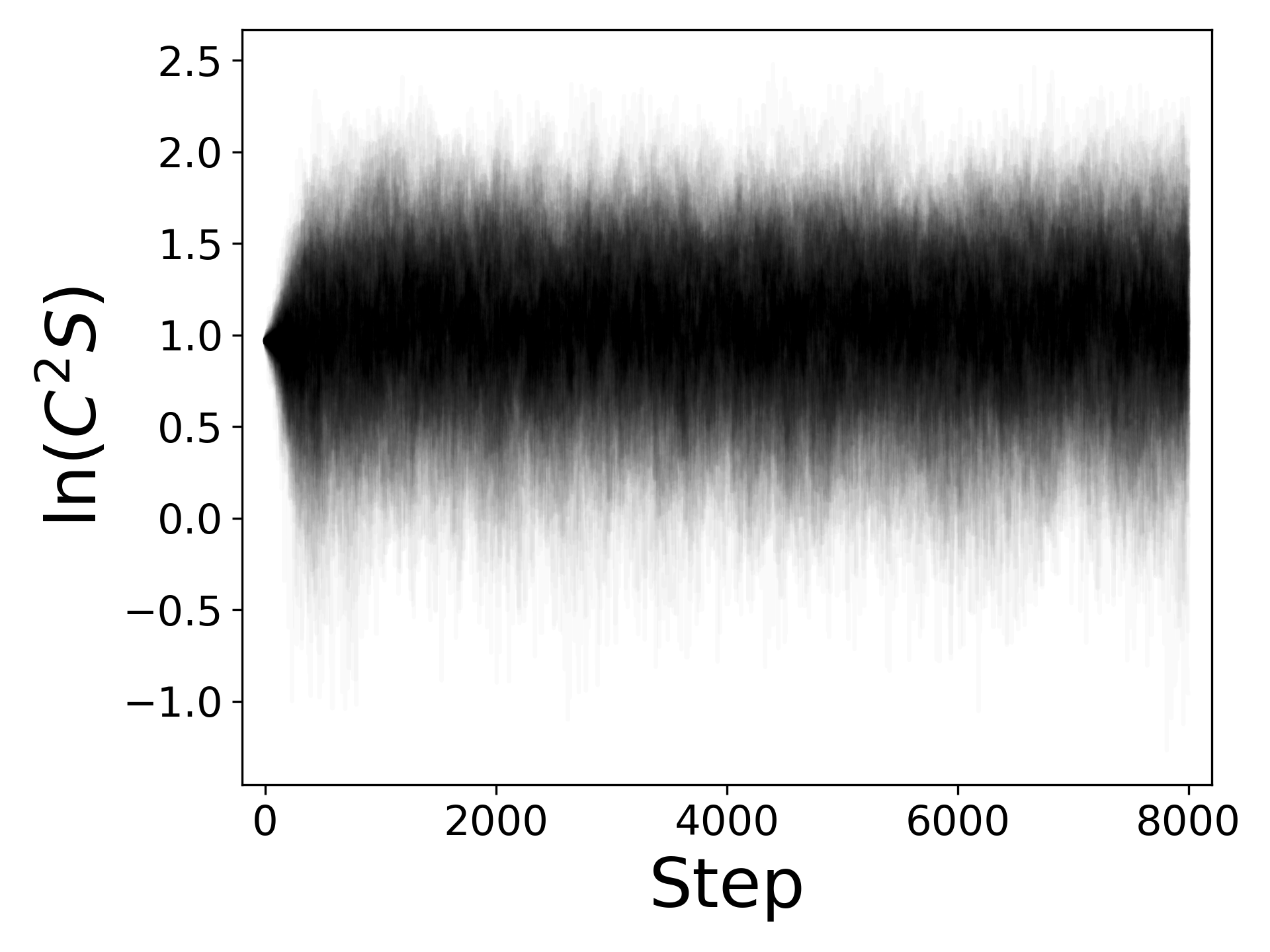}
  \caption{\label{fig:steps} Trace of the MCMC walkers as a function of step and $\ln(C^2S)$. Only the last 2000 steps were used for the posteriors.}
\end{figure}

The MCMC calculations for the ground state were carried out using $8000$
steps and $400$ walkers in the ensemble. As an example, the
trace plot for the value of $C^2S$ as a function of step is  provided in Fig.~\ref{fig:steps}.
Parameter values were estimated by using the last $2000$ steps and thinning by $50$.   

\begin{figure}
  \centering
  \includegraphics[width=.6\textwidth]{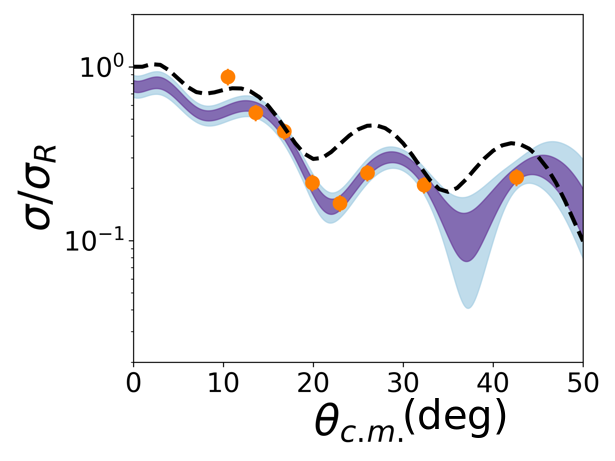}
  \caption{\label{fig:elastic_mcmc} Bayesian fit of the elastic data calculated simultaneously with the $0.00$ MeV state. The $68 \%$ and $95 \%$ credibility intervals are shown in purple and blue respectively, while the black dashed curve was calculated using the global values from Table~\ref{tab:opt_parms}.}
\end{figure}

\begin{figure*}
 \centering
 \includegraphics[width=\textwidth]{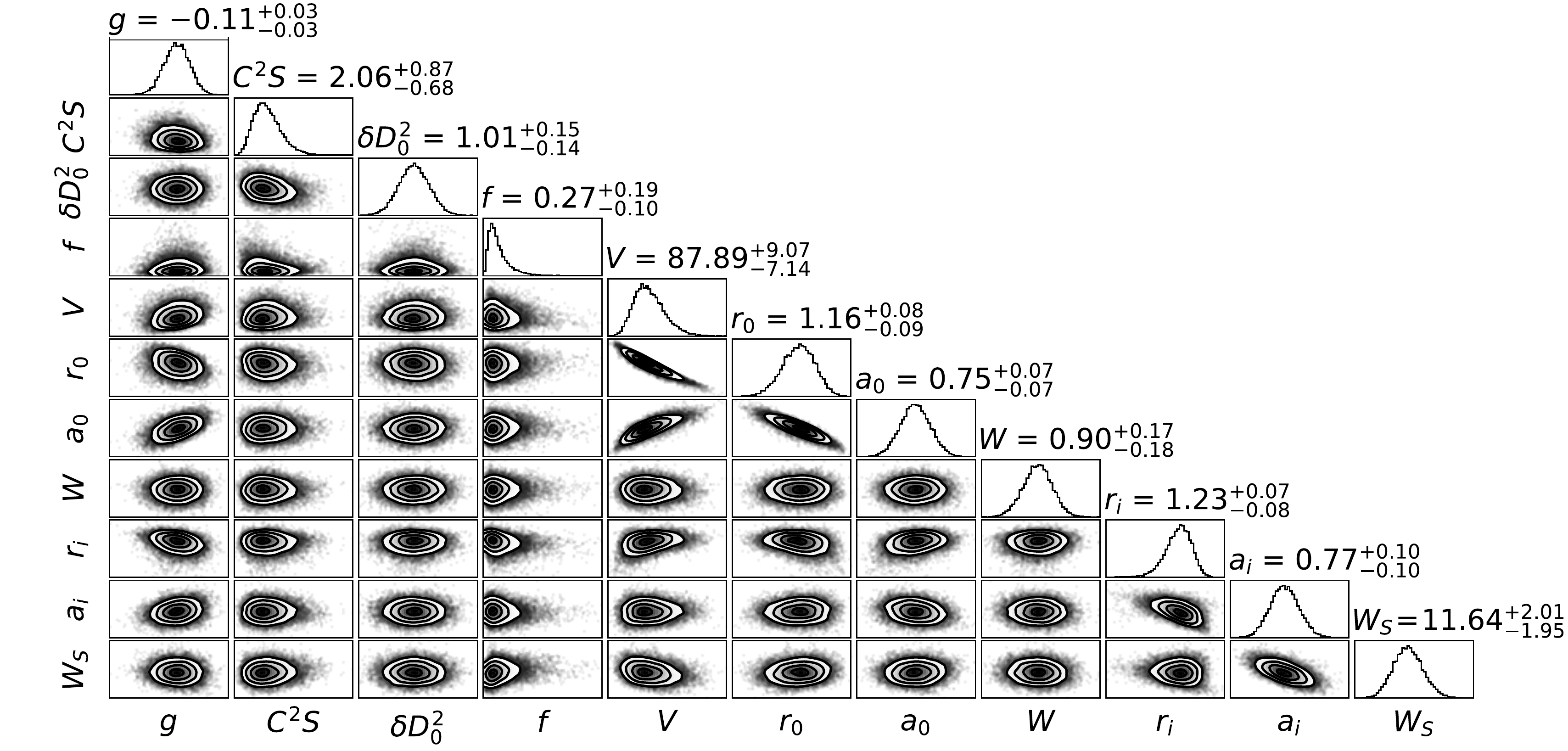}
 \includegraphics[width=\textwidth]{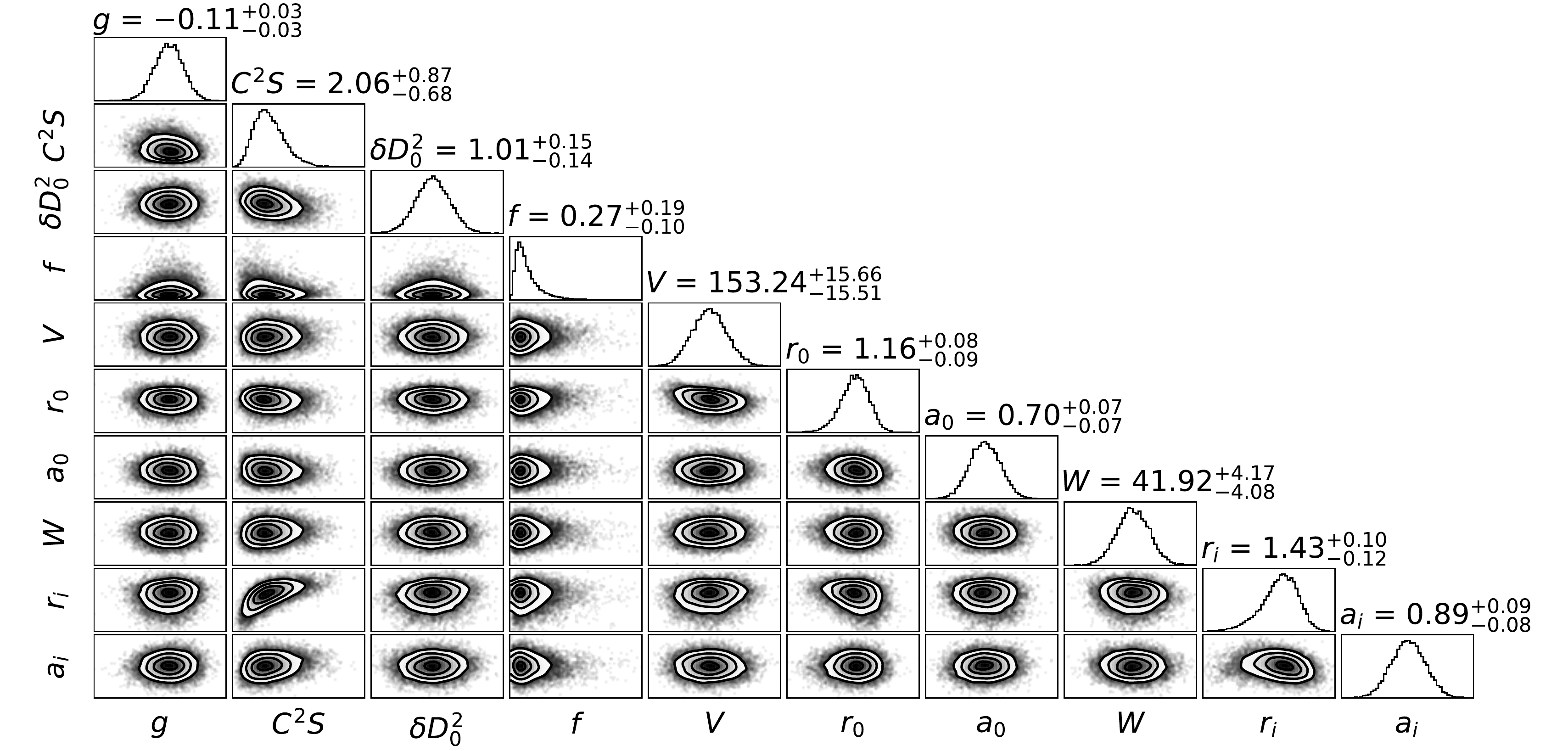}
 \caption{\label{fig:gs_corner} The pair-wise correlation plots for the ground state transfer. The top plot shows the entrance potential parameters, while the bottom shows the exit channel parameters. Both channels are compared to the model parameters as defined in Eq.~\ref{eq:model}. The $68 \%$ credibility intervals are listed at the top of each column with the dashed lines showing their relationship to the $1$-D parameter distributions. The plots were generated using the Python package \texttt{corner} \cite{Foreman-Mackey2016}.}
\end{figure*}

As noted before, all of the MCMC calculations simultaneously fit the elastic scattering and transfer data. This means that the posterior distributions shown in Fig.~\ref{fig:gs_corner} are functions of both the elastic and ground state transfer data. The impacts of the choice of potential parameters and the scale parameter $\eta$ on the elastic fit is quite dramatic. If the global values in Table~\ref{tab:opt_parms} were adopted without adjusting any parameters, the agreement between theory and experiment would be quite poor as shown by the dashed black line in Fig.~\ref{fig:elastic_mcmc}. It should also be noted that the experimental uncertainties for these points are roughly $10 \%$. On the other hand, the purple and blue bands in Fig.~\ref{fig:elastic_mcmc} show the fit obtained using the Bayesian model, which quite clearly provides a better description of the data. A significant difference is found between the normalization of the data and the optical model prediction, with $\eta \simeq 23 \% $.

By examining the correlations between the parameters, the model should display the continuous ambiguity discussed in Sec.~\ref{sec:amb_pots}. The pair-wise correlation plots in Fig.~\ref{fig:gs_corner} show the posterior samples from the entrance (top) and exit (bottom) channel potentials and how they relate to those of $g$, $C^2S$, $\delta D_0$, and $f$. The intra-potential correlations are quite striking for the entrance channel. All of the real potential parameters, $V, r_0,$ and $a_0$, show strong correlations with one another, and slightly weaker correlations existing between $V$, $r_0$, $r_i$, and $W_s$. Strong relationships also exist between $a_i$ and $W_s$, which is also another known continuous ambiguity \cite{perey_perey}. There is a much different situation for the exit potentials, where almost no intra-potential correlations
are seen. This result is expected since there are no elastic scattering data to constrain these parameters and because the Bayesian model parameter $f$ limits the amount of information that can be drawn from the transfer channel data. However, there is a surprisingly strong relationship between the exit channel imaginary radius and $C^2S$. A similar relationship can be seen with the entrance channel imaginary radius, but the effects on $C^2S$ are dramatically less.     

The results of the fit for the ground state are shown in Fig.~\ref{fig:gs_fit}.
The circular orange data points were the only data considered in the fit in order to
not bias our deduced spectroscopic factor as discussed in Section~\ref{sec:model}.  
The ground state of $^{69}$Cu is known to have a spin-parity of $\frac{3}{2}^-$, so the
transfer was calculated assuming a $2p_{3/2}$ state.

\subsection{The 1.11-MeV State}

The $1.11$-MeV state was only seen at four angles. Furthermore, only the first two data points lie within the
first minimum. The $J^{\pi} = \frac{1}{2}^-$ assignment is based on the observed angular distributions of Ref.~\cite{69cu_orig} and the analyzing power measurement of Ref.~\cite{fay_69cu}. In order to check that the data analyzed in the current work are consistent with these conclusions, the evidence integrals were calculated for $\ell = 0, 1, 2$, and $3$  transfers using all the data points.
The data support an $\ell = 1$ transfer, but do not rule out an $\ell = 3$ transfer. For this case, the median Bayes Factor defined in Eq.~\ref{eq:bayes_factor} is $B_{13} = 6.32$ (i.e, the fifty percentile of $Z_1/Z_3$), indicating that there is substantial evidence in favor of $\ell = 1$. Since the data are consistent with the $\ell$ assignments of Ref.~\cite{fay_69cu,fay_69cu}, the MCMC calculations were carried out assuming a $2p_{1/2}$ state. The results of this calculation are plotted in Fig.~\ref{fig:111_fit}.

\subsection{The 1.23-MeV State}

The state located at $1.23$ MeV is definitely associated with an $\ell = 3$
transfer. The previous analysis assumed a firm $J^{\pi} = \frac{5}{2}^-$; however,
the literature does not provide direct evidence for this. The analyzing power of Ref.~\cite{fay_69cu} was inconclusive, and the authors suggested the presence of a doublet based on the observed width of the peak in the spectrum. The $(d, ^3\textnormal{He})$ experiment of Ref.~\cite{69cu_orig} also suggested a doublet and noted the high spectroscopic factor obtained ($C^2S = 1.5$) if a $\frac{5}{2}^-$ assignment was assumed. Other studies have also assigned a firm $J^{\pi} = \frac{5}{2}^-$ \cite{franchoo_mono, coul_ex, beta_decay}, but it is unclear if these results are actually independent determinations, or if they follow Table II of Ref.~\cite{69cu_orig}. Therefore, the ENSDF evaluation is adopted \cite{a_69_ensdf}, which recommends $J^{\pi} = (\frac{5}{2}^-, \frac{7}{2}^-)$, but only present the $C^2S$ value for $1f_{5/2}$ with the fit shown in Fig.~\ref{fig:123_fit}.

\subsection{The 1.71 and 1.87-MeV States}

From the parity constraints of Ref.~\cite{69cu_orig, fay_69cu} and the $\gamma$-ray anisotropies observed in Ref.~\cite{deep_inelas}, a firm $J^{\pi} = \frac{7}{2}^-$ assignment has been made for the $1.71$-MeV state. The results from the DWBA fit for a $1f_{7/2}$ state are shown in Fig.~\ref{fig:171_fit}. The arguments from the $1.71$-MeV state also apply to the state at $1.87$-MeV. A firm 
$J^{\pi} = \frac{7}{2}^-$ was assumed and a fit for a $1f_{7/2}$ state is shown in Fig.~\ref{fig:187_fit}.

\subsection{The 3.35-MeV State}

The state at $3.35$ MeV was reportedly seen in Ref. \cite{69cu_orig}, but no information was presented other than its possible existence. The previous analysis found an $\ell = 3$ nature to the angular distribution, and made a tentative assignment of $J^{\pi} = (\frac{7}{2}^-)$. The current methods support this conclusion as shown in Table~\ref{tab:probs}. $B_{3 \ell} > 10$ for all other $\ell$ transitions, indicating strong evidence for the $\ell = 3$ transfer. The probability the final state was populated with an $\ell=3$ transfer is $P(\ell=3) = 91^{+3}_{-4} \%$. However, DWBA is still unable to discriminate between $J^{\pi} = (\frac{5}{2}^-, \frac{7}{2}^-)$. The fit assuming a $1f_{7/2}$ state is shown in Fig.~\ref{fig:335_fit}. 

\begin{table*}[]
\centering
  \caption{\label{tab:probs} Results of the model comparison calculations for the $3.35$ and $3.70$ MeV states. For each $\ell$ value, values are provided for the $\log{Z}$ value calculated
with nested sampling, the median Bayes factor when compared to the most likely transfer $\ell=3$, and the probability of each transfer .}
\begin{tabular}{ccccc}
  \\ [0.5ex] \hline \hline
  \\ [-2.0ex]

  & $\ell$ & $\log{Z}_{\ell}$  & $B_{3 \ell}$ & $P(\ell)$         \\ [0.5ex] \hline
  \\ [-2.0ex]
\multirow{4}{*}{$E_x = 3.35 $ MeV} & 0      & 3.856(330)                                          & $> 10^4$     & $< .01 \%$        \\ [1.0ex]
                                   & 1      & 10.662(359)                                         & $15.94$      & $6^{+3}_{-2} \%$  \\ [1.0ex]
                                   & 2      & 9.961(363)                                          & $32.14$      & $3^{+2}_{-1} \%$  \\ [1.0ex]
  & 3      & 13.431(349)                                         & 1.0          & $91^{+3}_{-4} \%$ \\ 
  \\  [-5.0ex]
      \\  \hline 
      \\  [-2.0ex]
\multirow{4}{*}{$E_x = 3.70 $ MeV} & 0      & 10.393(365)                                         & $> 10^3$     & $ < 0.02 \%$      \\ [1.0ex]
                                   & 1      & 14.947(351)                                         & $45.98$      & $2^{+1}_{-1} \%$  \\ [1.0ex]
                                   & 2      & 16.640(346)                                         & $8.47$       & $10^{+5}_{-4} \%$ \\ [1.0ex]
                                   & 3      & 18.776(336)                                         & 1.0          & $88^{+4}_{-6} \%$ \\ [0.5ex] \hline \hline
\end{tabular}
\end{table*}

\subsection{The 3.70-MeV State}

\begin{figure}
  \centering
  \includegraphics[width=.6\textwidth]{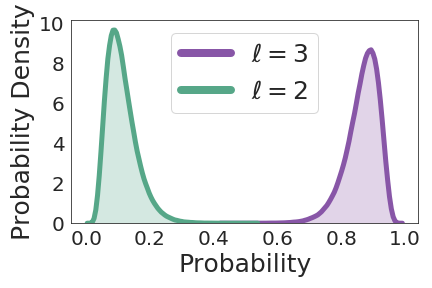}
  \caption{\label{fig:l_comp} The KDE representations of the probabilities of the $\ell=2$ and $3$ transfers for the $3.70$-MeV state.}
\end{figure}

The state at $3.70$ MeV was also seen for the first time in Ref.~\cite{pierre_paper}. However, the Bayesian method indicates an ambiguous $\ell$ assignment. As can
be seen in Fig.~\ref{fig:370_fit}, the measured angular distribution
is relatively flat, and does not appear to differ
from other states with $\ell = 3$. However, an assignment of $\ell = 2$ was
made in the previous analysis. Comparing the evidence integral for each case, it is found
that indeed the data effectively rule out $\ell = 0$ and $1$, while supporting an
$\ell = 2$ or $3$ assignment. Looking at Table~\ref{tab:probs}, a Bayes factor of $B_{32} = 8.47$ is found for
$\ell = 3$ over $\ell=2$, which suggests substantial evidence in favor of the $\ell = 3$ assignment. Using Eq.~\ref{eq:model_prob}, the $68 \%$ credibility intervals for the probabilities are $P(\ell=3) = 88^{+4}_{-6} \%$ and  $P(\ell=2) = 10^{+5}_{-4} \%$, with the uncertainties coming from the statistical uncertainties of the nested sampling evidence estimation. The KDE for
the two dominate transfers are shown in Fig.~\ref{fig:l_comp} \cite{kde}. The fits for both $\ell = 2$ and $3$ are shown in Fig.~\ref{fig:370_fit}.

\subsection{The 3.94-MeV State}
\label{sec:394_sec}

The $3.94$-MeV state was also observed for the first time in the previous study. The suggested $\ell = 0$ assignment was found to be supported by the data. The second most likely transfer was found to be $\ell = 1$. In this case, $B_{01} = 72.24$, indicating very strong evidence in favor of the $\ell = 0$ assignment. The transfer to a $2s_{1/2}$ state is shown in Fig.~\ref{fig:394_fit}.

\subsection{Spectroscopic Factors}
\label{sec:spec_factor_dis_cu}

The results of the previous sections merit closer examination, especially with regards to the spectroscopic factors. Comparing these results with
those previously obtained in Table~\ref{tab:cs}, two things are clear: our median values tend to be larger than those of Ref.~\cite{pierre_paper} and
the uncertainties are much larger. To the first point, a majority of the shift comes from the lower value of $W_s$ used in the previous analysis.
Though not stated in Ref.~\cite{pierre_paper}, the surface potential was given a value of $W_s \approx 7.5$ MeV, which has the effect of lowering the value of $C^2S$.
The values of this work are on average higher due to the Bayesian analysis favoring $W_s = 11.93$ MeV and the inclusion of $\eta$, but these are somewhat offset due to the posterior values of $r_i$ and $a_i$ being lower than their global values. To the second point, when all of the sources of uncertainty are included in the analysis, highly asymmetric and data-driven uncertainties on $C^2S$ ranging from $35-108 \%$ are found. This is a substantial increase with regards to the common assumption that the extraction of spectroscopic factors comes with an approximately $25 \%$ uncertainty \cite{endt_cs}. This may still be the case when the data are sufficiently informative, but the results of a single experiment should be viewed more conservatively. In particular, low angular coverage in the entrance channel elastic scattering data, the absence of any elastic scattering data in the exit channel, and transfer angular distributions with just a few points all play a role in final uncertainty that can be reported for $C^2S$.

To gain a clearer picture of the role each potential plays in the final uncertainty, the calculations for the ground state were repeated for the following cases:

\begin{enumerate}
\item Uncertainty in just the entrance channel potential parameters.
\item Uncertainty in both the entrance and exit channel potential parameters.
\item Uncertainty in the entrance, exit, and bound state potential parameters.
\end{enumerate}

\begin{figure}
  \centering
  \includegraphics[width=.6\textwidth]{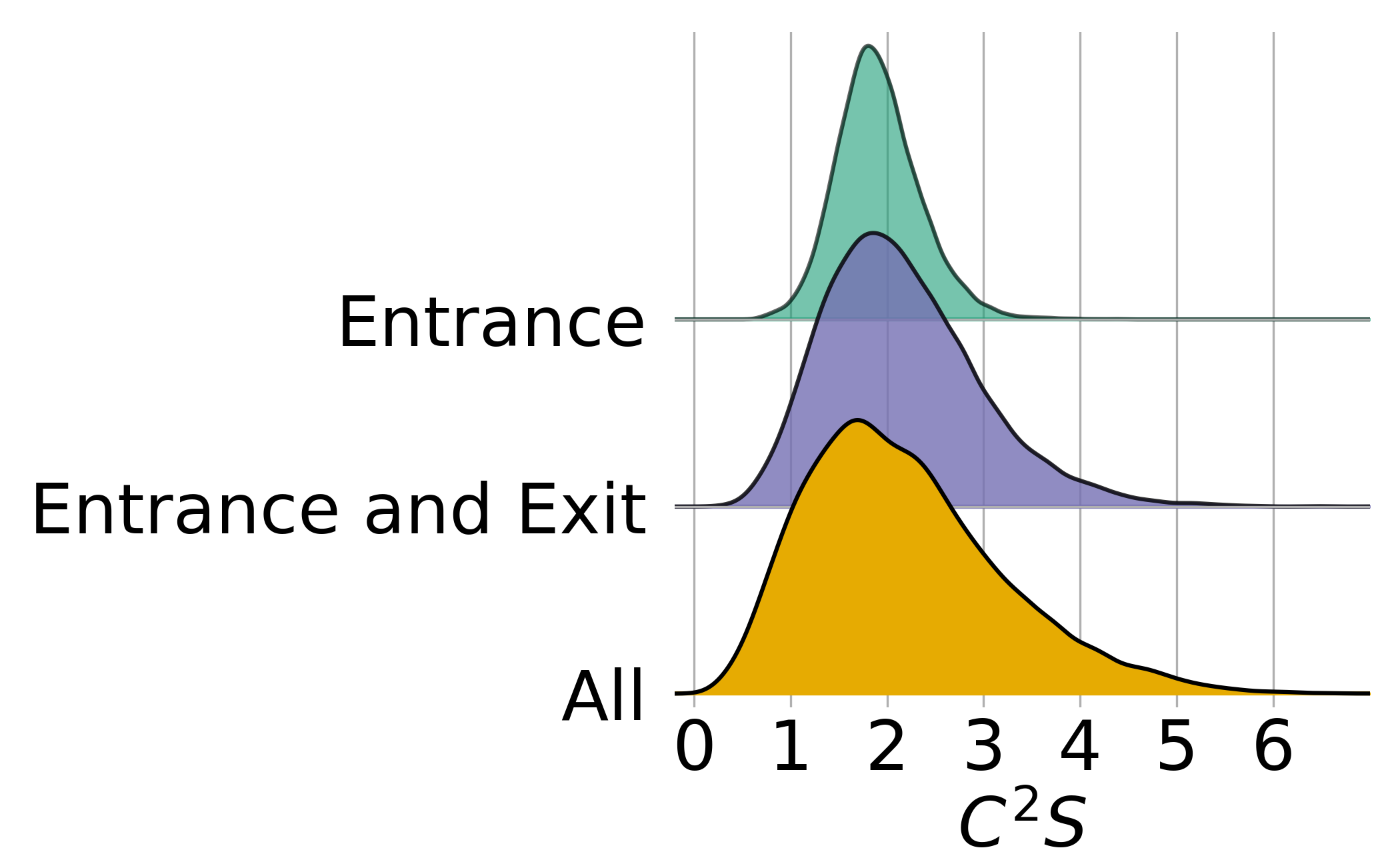}
  \caption{\label{fig:ridge} Ridge line plot that compares the KDE distributions for the ground state $C^2S$ when there is variation in the entrance potential; entrance and exit potentials; and in the entrance, exit, and bound state potentials. The percentage uncertainties go from $24 \%$, $42 \%$, and $56 \%$, respectively.}
\end{figure}

Case one has the lowest uncertainty with $C^2S = 1.88^{+0.44}_{-0.37}$ ($\approx \! 24 \%$).
Case two is the same model used for all of the states in Section.~\ref{sec:results}. This gives $C^2S = 2.06^{+0.87}_{-0.68}$ ($\approx \! 42 \%$). 
Case three first requires that priors for the radius and diffuseness parameters of the bound state potential are specified. Analogously to the exit channel, which also lacks data to directly constrain these parameters, they are assigned $V_{\textnormal{Bound}} \sim (\mu_{\textnormal{central}, k}, \{ 0.10 \mu_{\textnormal{central}, k}\}^2)$. Again, $k$ is an index that runs over the radius and diffuseness parameters, and \say{central} refers to $r = 1.25$ and $a=0.65$. This case has the largest final uncertainty with $C^2S = 2.04^{+1.15}_{-0.85}$ ($\approx \! 56 \%$).
The comparison between the final distribution for the spectroscopic factors obtained for just the entrance channel, entrance channel and exit channel, and all of the potentials including the bound state are shown in Fig. \ref{fig:ridge}. This demonstrates the strong dependence of $C^2S$ on each of these potentials.

These results point toward ways to improve the precision of $C^2S$. Examination of the correlations in the posterior samples in Fig.~\ref{fig:gs_corner} shows that the imaginary radius in the exit channel is the parameter responsible for much of the uncertainty in $C^2S$. The samples for the exit channel also show little intra-potential correlation between the parameters. This is expected since the only data that could inform these parameters are in the transfer channel. If elastic data for the exit channel were available, then the proper parameter correlations could be inferred, thereby, reducing the uncertainty in the extracted spectroscopic factors. This could bring the uncertainty closer to the roughly $24 \%$ seen in the case when just the entrance potential is considered.

Bound state parameter dependence could have significant impact on astrophysical applications as well. In these applications, the extraction of $C^2S$ is an intermediate step towards calculation of quantities relevant to astrophysics such as particle partial widths and direct capture
cross sections. It was noted in Ref.~\cite{bertone} that it is essential to use the same bound state parameters for both the extraction of $C^2S$ and calculation of the direct capture cross section or partial width. This procedure was found to significantly reduce the final uncertainties on these quantities. If the bound state parameters are included in a Bayesian model to extract $C^2S$, then it becomes possible to calculate these quantities not only using the same bound state parameters, but using fully correlated, statistically meaningful samples informed directly by the transfer reaction measurement. The preliminary steps towards investigating these claims are presented in Chapter \ref{chap:sodium}.        

\subsection{Nuclear Structure of $^{69} \textnormal{Cu}$}

Structure properties of $^{69} \textnormal{Cu}$ are also influenced by these results. The occupancy of orbitals tends to be higher than
expected for both the open $pf$ orbitals and for the closed $1f_{7/2}$ proton shell.
In order to propagate the uncertainties from each $C^2S$, the MCMC samples are used to
construct a KDE for each state. From these densities, $10^5$ samples are pulled to estimate the occupancy:
\begin{equation}
  \label{eq:occ}
  n = \sum_{i}^N C^2S_i,
\end{equation}
where $i$ refers to each of the $N$ states considered in the sum. Similarly,
the energy of the $1f_{7/2}$ shell can be determined from:

\begin{equation}
  \label{eq:energy}
  E(1f_{7/2}) = \frac{\sum_{i}^N C^2S_i(1f_{7/2}) E_i(1f_{7/2})}{n_{1f_{7/2}}}.
\end{equation}

The occupancy above the closed shell was found to be $n_{pf} = 3.90^{+1.03}_{-1.28}$,
which is consistent but systematically higher than the value of $2.55(23)$ from the finite range calculations of the previous analysis \cite{pierre_paper}.
For the $1f_{7/2}$ shell, there are two scenarios dependent on the identity of the state at $3.70$ MeV.
If the state does not belong to the $f$ shell, then $n_{1f_{7/2}} = 6.64^{+2.47}_{-1.79}$ and $E(1f_{7/2}) = 2.43^{+0.23}_{-0.25}$ MeV
, or if it does, $n_{1f_{7/2}} = 10.03^{+3.63}_{-2.66}$ and $E(1f_{7/2}) = 2.86^{+0.23}_{-0.26}$ MeV. 
Looking at the median value for $n_{pf}$, it would be expected that $n_{1f_{7/2}} = 6.10$. This may point to the $\ell=2$ assignment of the $3.70$-MeV state
being the correct one, but it must be recognized that there are still large uncertainties on all of these quantities.
Furthermore, since the optical model parameters are shared by these states, values derived through combinations of states are susceptible to significant systematic shifts.
In light of this fact, these credibility intervals should be viewed as approximations. Perhaps more importantly is that if the $3.70$-MeV state belongs to the $1f_{7/2}$, then the full strength of this shell has been observed. The shell model calculations in Ref.~\cite{pierre_paper} predict a much higher energy than $E(1f_{7/2}) = 2.86$ MeV due to the presence of more states at higher excitation energies. A future experiment with a higher incident
beam energy that would be capable of populating these predicted higher lying states could help clarify these discrepancies.

\subsection{Comparison to Other Bayesian Studies}

It is also worthwhile to compare these methods with those of several recent publications, which have also applied Bayesian methods to optical potentials \cite{lovell_mcmc, king_dwba}.
These papers differ from our approach in a few key ways: the data come from multiple experiments, exit and entrance channels are fitted separately, transfers are calculated using finite range effects, and the prior distributions are much wider than ours ($100 \%$ of the initial global values).
In a fully Bayesian framework, fitting the data in the entrance and exit channels separately or simultaneously is equivalent as long as the same model is used \cite{bayes}.
While the priors of this work are narrower, they could likely be made broader if there was elastic scattering data over a wider range of angles.
Full finite-range calculations could be important to include in future studies, but, as seen in Table~\ref{tab:cs}, for this reaction the average difference is roughly $16 \%$, well within the uncertainty arising from the optical potentials.
Including these effects will require a more efficient way to evaluate the likelihood function. Specifically, a finite-range calculation takes roughly $50$ times longer than a calculation using the zero-range approximation. For this work, $2 \times 10^6$ likelihood evaluations took approximately $22$ hours, meaning the finite-range calculation would take over $1000$ hours.
As well as those differences, these results differ from those of
Ref.~\cite{king_dwba} in one important aspect. Here, I confirmed the strong correlations between optical model parameters that are expected from historical studies \cite{hodgson1971}, and
treated them in a statistically meaningful way. It should be stressed that the Bayesian model does
not assume these correlations, but that they appear to be a consequence of the Woods-Saxon potential form factor.
On the other hand, in their comparison of frequentist and Bayesian
methods, Ref.~\cite{king_dwba} do not observe such
correlations, with the exception of the $V_0$ and $r_0$ anti-correlation, and ascribe their finding to the non-Gaussian
posterior distributions, which would be poorly described by the
frequentist model. The origin of this
disagreement is unclear, and further investigation is needed.

\section{Summary}

In this chapter, Bayesian analysis was introduced along with the basics of the numerically approximating the posterior distribution. This formalism was then applied to the two primary methods for extracting information from transfer reactions: energy calibration of the focal plane and DWBA. The Bayesian approach to the focal plane calibration allowed a fit that considered both calibration energy uncertainties as well as those from the experimentally determined peak centroids. An additional uncertainty could also be easily integrated into the model, which improves upon previous techniques that estimated this uncertainty after the calibration.

The Bayesian model for DWBA signifies a major advancement in analysis. Although the full implications of this method are still not known, it was shown in this chapter that Bayesian methods allow a more detailed analysis of transfer data. In particular, optical model uncertainties can be more fully understood with the Bayesian approach, and for the first time explicit probabilities can be assigned to each allowed $\ell$ value.    

%% file: Chapter-6/Chapter-6.tex
\chapter{$^{23}$N\lowercase{a}$(^{3}$H\lowercase{e}, $\lowercase{d})^{24}$M\lowercase{g}}
\label{chap:sodium}

\section{Previous Studies and Motivation for This Work}

The NeNa cycle (see Fig.~\ref{fig:ne_na_cycle}) has been the subject of experimental investigation for 40 years. Early direct measurements focused on constraining explosive nucleosynthesis happening in classical novae and were carried out for both the $(p, \gamma)$ and $(p, \alpha)$ reaction channels \cite{Zyskind_1981, goerres_1989}. The study of Goerres \textit{et al.} (Ref.~\cite{goerres_1989}) was one of the first to directly search for the resonance that corresponded to the $E_x \approx 11827$-keV state in $^{24}$Mg, which had been observed in the indirect measurements of Refs.~\cite{moss_1976, vermeer_1988}. However, they were only able to establish an upper limit of $\omega \gamma_{(p, \gamma)} \leq 5 \times 10^{-6}$ eV. A short time later, Ref.~\cite{1996_Eid} was the first to calculate a direct capture component of the reaction rate, and identify its importance in the $(p, \gamma)$ rate.

The current understanding of these rates is in large part due to the study of Ref.~\cite{hale_2004}. In that study, a $(^{3}$He$,d)$ transfer reaction was performed, and a state at $E_x = 11.831$ keV was observed, thereby placing constraints on the energy and strength of the $138$-keV resonance. However, the angular distribution for this state was inconclusive, leaving a large amount of uncertainty in the $(p, \gamma)$ and $(p, \alpha)$ rates. Additionally, the authors of that work presented a detailed evaluation of the literature to formulate the basis for the current $(p, \gamma)$ and $(p, \alpha)$ rates. Since that time, several direct searches have been performed, with the intent of measuring the $138$-keV resonance. Ref.~\cite{Rowland_2004} set a new upper limit at $\omega \gamma_{(p, \gamma)} \leq 1.5 \times 10^{-7}$ eV. Subsequently, Ref.~\cite{Cesaratto_2013} used a high intensity proton beam of $\approx 1$ mA to give a further reduced upper limit of $\omega \gamma_{(p, \gamma)} \leq 5.17 \times 10^{-9}$ eV, which ruled out this resonance's importance for the $(p, \alpha)$ channel. Finally, the first direct detection of the $138$-keV resonance with a statistical significance above $2 \sigma$ came in Ref.~\cite{BOELTZIG_2019}. That study reports $\omega \gamma_{(p, \gamma)} = 1.46^{+0.58}_{-0.53} \times 10^{-9}$ eV. 

At the present time, the $^{23}$Na$(p, \gamma)$ reaction rate has a greatly reduced uncertainty, now on the order of $30 \%$ at the temperatures of relevance to globular clusters, because of the intense study of the $138$-keV resonance using direct measurements. However, much of the rate is still dependent on the results and evaluation presented in Ref.~\cite{hale_2004}. The purpose of the current study is to use the $^{23}$Na$(^{3}\textnormal{He}, d)^{24}$Mg reaction to investigate the results and conclusions of the foundational work of Ref.~\cite{hale_2004}. Of particular interest is placing constraints on the spin and parity of the $138$-keV resonance using the model selection capabilities of the Bayesian DWBA method presented in Section \ref{sec:bay_dwba}.       
 
\begin{figure}
    \centering
    \includegraphics[width=.6\textwidth]{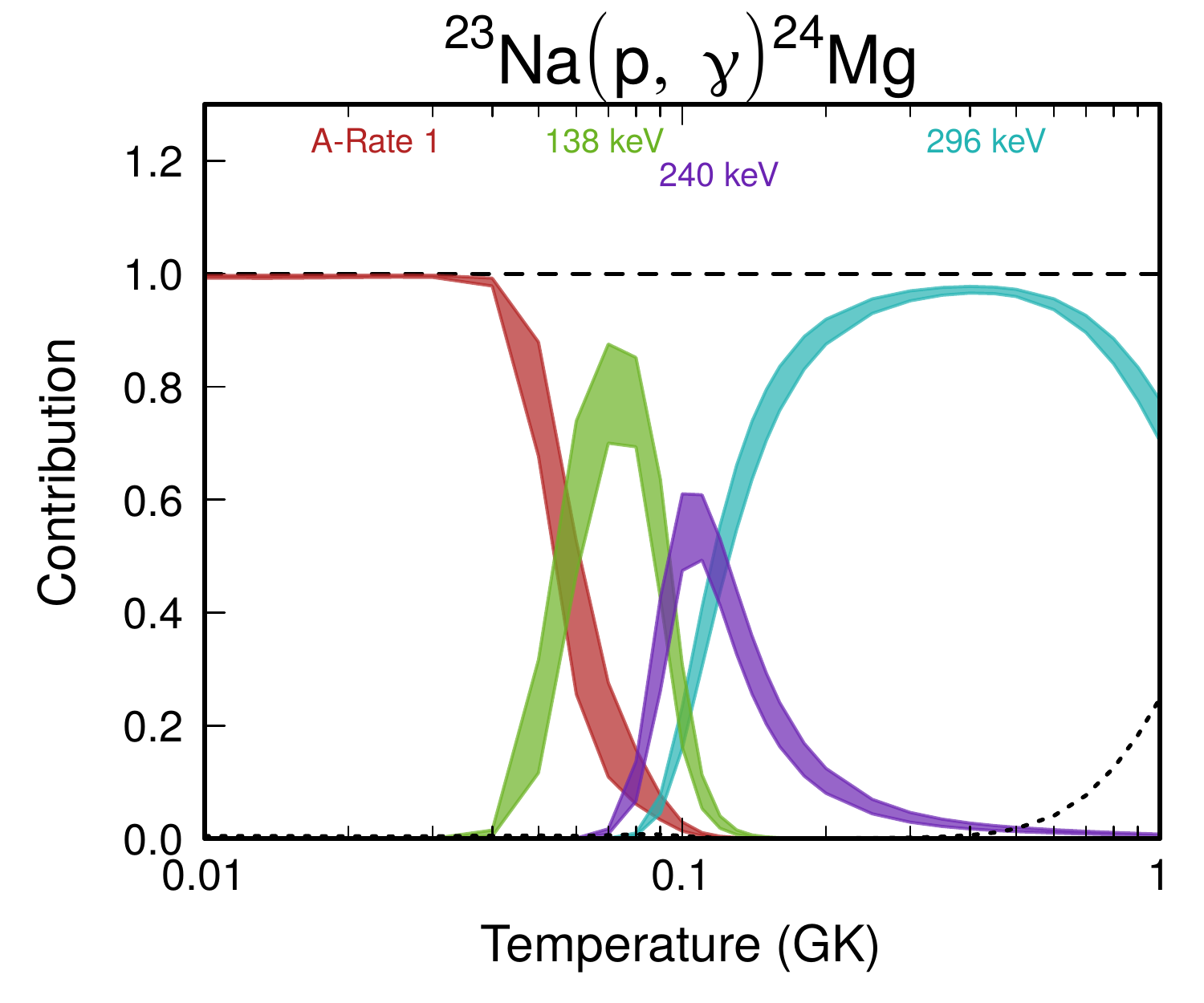}
    \caption{Rate contribution plot for $^{23}$Na$(p, \gamma)$ based on the current STARLIB rate updated with the resonance strengths of Ref.~\cite{BOELTZIG_2019}. \textit{A-Rate 1} is the direct capture rate, while the dashed curve is the summed contribution of all other resonances.}
    \label{fig:contribution_luna_p_g}
\end{figure}

\begin{figure}
    \centering
    \includegraphics[width=.6\textwidth]{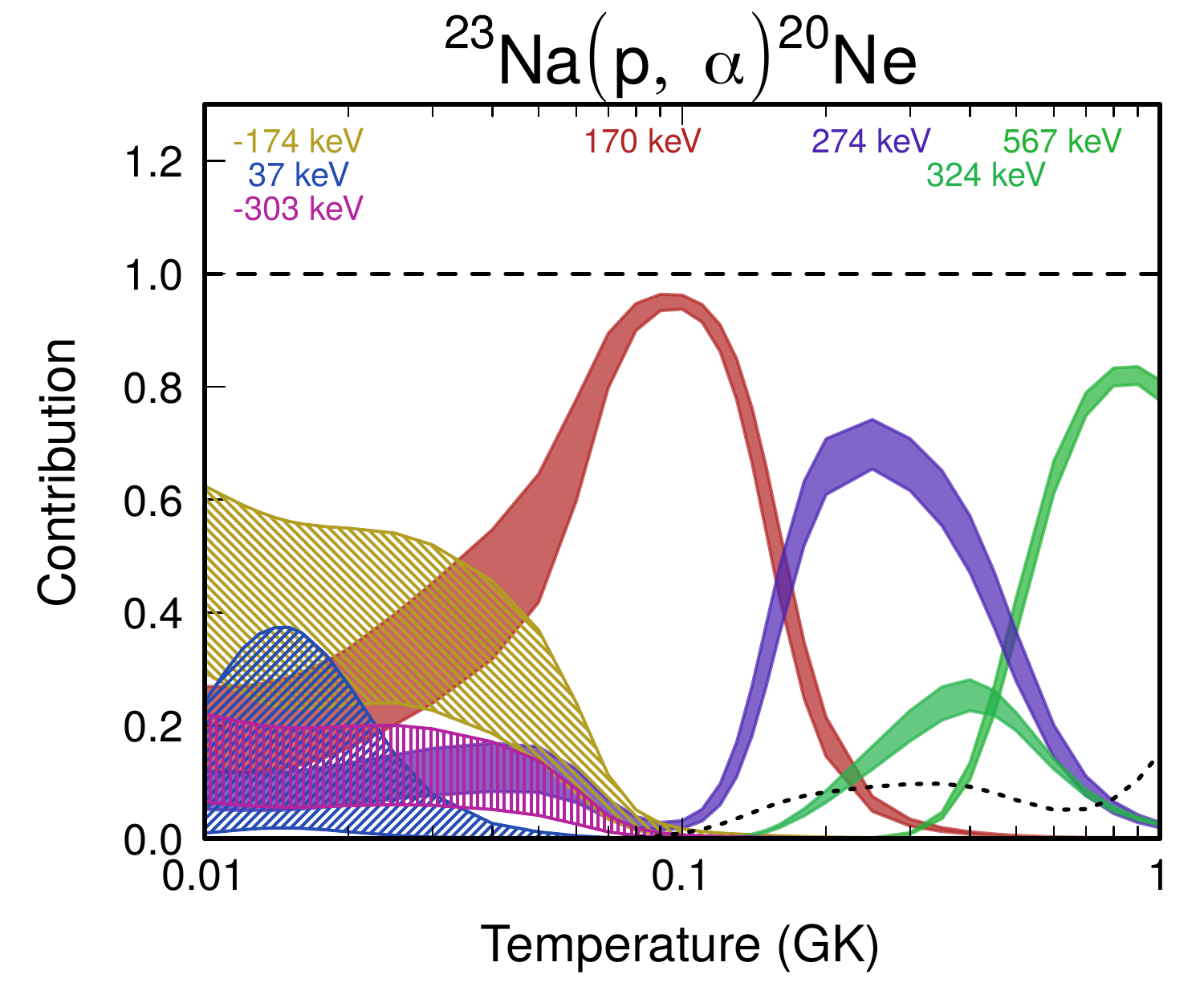}
    \caption{Rate contribution plot for $^{23}$Na$(p, \alpha)$ based on the current STARLIB rate updated with the resonance strengths of Ref.~\cite{BOELTZIG_2019}. Subthreshold resonances dominate at low temperatures in the absence of a direct capture rate. The dashed curve is the summed contribution of all other resonances.}
    \label{fig:contribution_luna_p_a}
\end{figure}

\begin{figure}
    \centering
    \includegraphics[width=.8\textwidth]{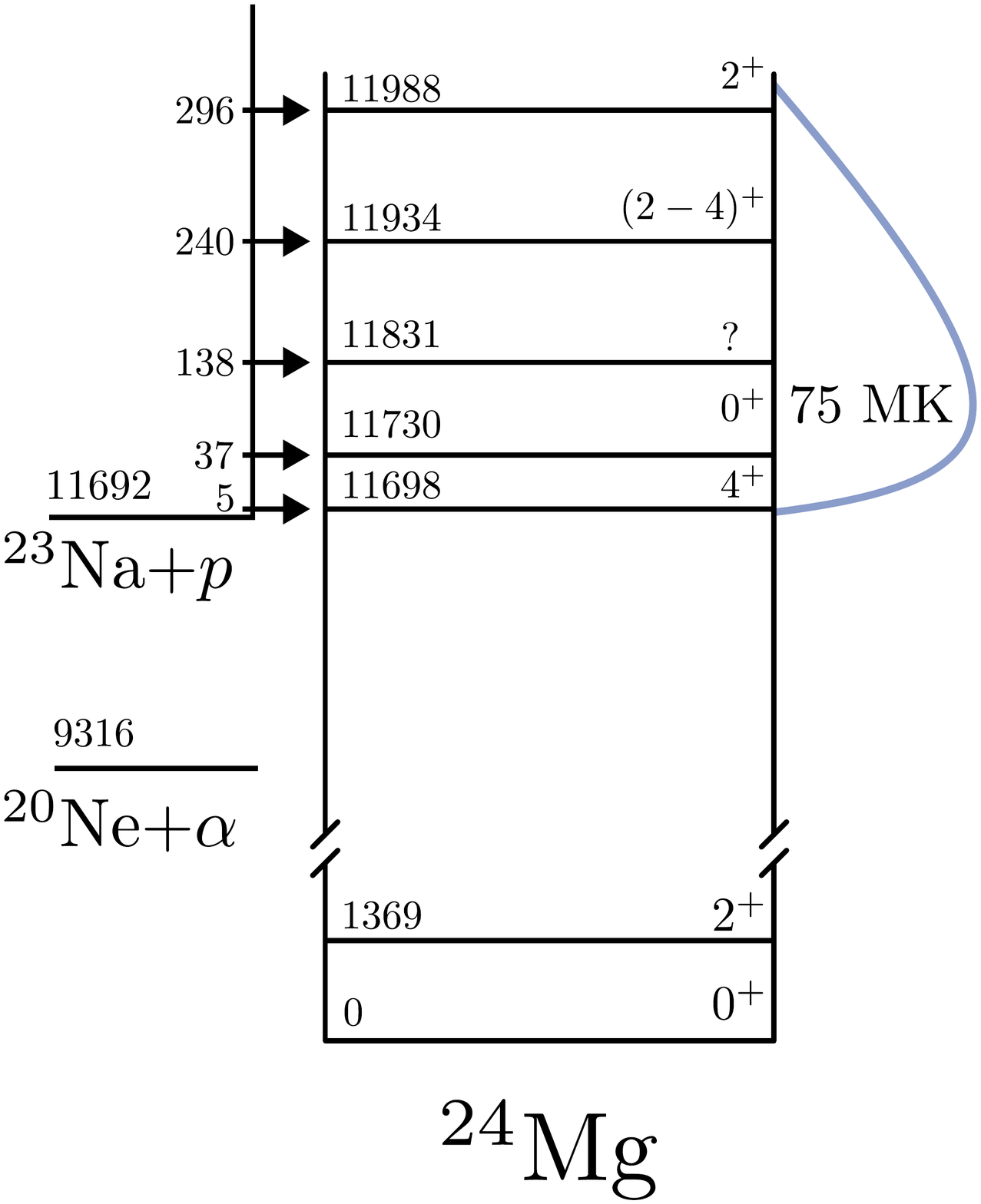}
    \caption{Level scheme of $^{24}$Mg with the relative locations of some of the astrophysically states shown.}
    \label{fig:level_scheme}
\end{figure}

The current rates in STARLIB has been updated for the measurement of Ref.~\cite{BOELTZIG_2019}. \texttt{RateMC} was used to generate the rate contribution plot for the $(p, \gamma)$ rate shown in Fig.~\ref{fig:contribution_luna_p_g} and the $(p, \alpha)$ rate shown Fig.~\ref{fig:contribution_luna_p_a}. As can be seen in the figure for the $(p, \gamma)$ rate, at the temperatures relevant to globular cluster nucleosynthesis, $70 \text{-} 80$ MK, the $138$-keV resonance dominates, with a lesser contribution coming from the direct capture rate and the $240$-keV resonance. For the $(p, \alpha)$ rate the dominant contributions are from subthreshold states and the $170$-keV resonance. Looking at Fig.~\ref{fig:level_scheme}, the relative locations of some of these resonances with respect to the excitation energy of $^{24}$Mg can be seen. Thus, the goal of the current experiment is to study the excited state of $^{24}$Mg in the region of $11 \lessapprox E_x \lessapprox 12$ MeV. Using the SPS at TUNL it is possible to extract excitation energies, angular distributions, and spectroscopic factors. From the spectroscopic factors, proton partial widths can be extracted of these states. It can often be assumed that $\omega \gamma_{(p, \gamma)} \approx \omega \Gamma_{p}$ for resonances below $500$ keV where $\Gamma \approx \Gamma_{\gamma}$. However, as can be seen in Fig.~\ref{fig:level_scheme}, $^{24}$Mg is $\alpha$ unbound by several MeV near the proton threshold. For the current experiment we cannot assume the proton partial width is directly proportional to the resonance strength, and as a consequence additional nuclear input will be needed to determine $\omega \gamma_{(p, \gamma)}$ with precision.       

This chapter will be roughly divided into five parts: the production of the transmission targets needed for this experiment, details of the transfer experiment, energy calibration of the focal plane, Bayesian DWBA analysis, and finally the calculation of the reaction rate and its impact on the globular cluster nucleosynthesis.

\section{Targets}
\label{sec:targets}

Several attempts were made to produce transmission targets for this experiment. These targets need to be thin enough for the outgoing deutrons to leave the target and be detected with high resolution. However, they must also be thick enough to give reasonable count rates, so that weakly populated states can be detected with decent statistics. This section details my attempts to produce such targets over a roughly two year period.

It was decided early on to focus efforts on producing NaBr targets based on the observations in Refs.~\cite{hale_2004, hale_thesis} that these targets were fairly stable to bombardment, reasonably resistant to oxygen contamination, and the astrophysical region of interest is free from contamination arising from $^{79, 81}$Br.   

\begin{figure}
    \centering
    \includegraphics[width=.6\textwidth]{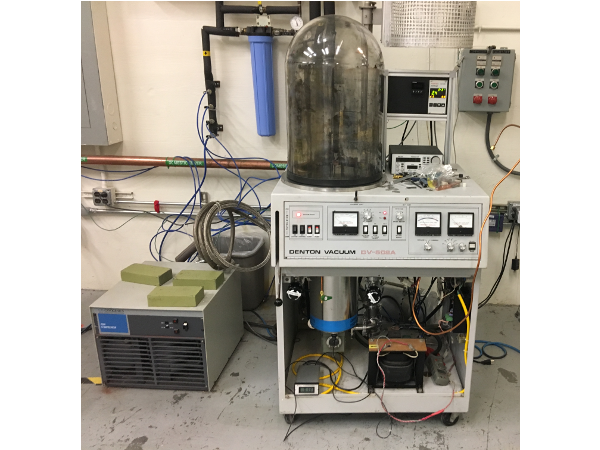}
    \caption{The LENA evaporator. Note that the bell jar is coated with a layer of tantalum from the target development of Ref.~\cite{HUNT_2019}.}
    \label{fig:evaporator}
\end{figure}

\begin{figure}
    \centering
    \includegraphics[width=.6\textwidth]{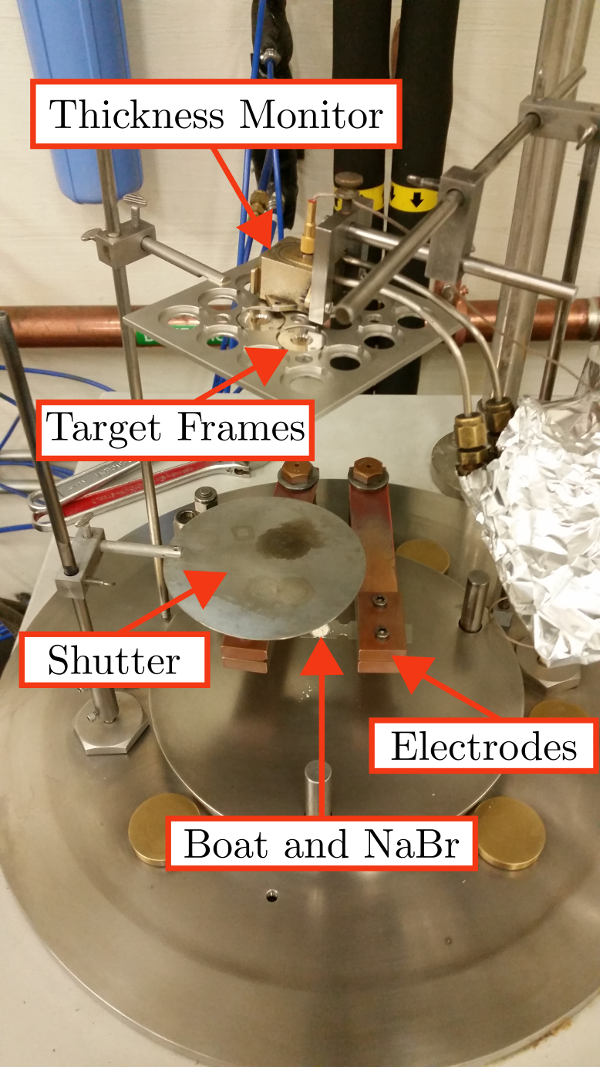}
    \caption{Evaporation setup for NaBr. The shutter covers the boat as the electrode current is increased. This prevents outgassing material from being deposited on the targets until it is removed.}
    \label{fig:nabr_setup}
\end{figure}

All targets were produced by thermal evaporation. The principle of this process is to place the material that we want to create a thin film of into a resistive boat that has a comparatively higher evaporation point. This boat is clamped between two copper electrodes that are water cooled. Several inches above the boat, a substrate holder is loaded with target backings mounted on target frames. An example of this setup is shown in Fig.~\ref{fig:nabr_setup}. A bell jar is placed over the substrates, electrodes, and boat. The bell jar is then brought down to high vacuum. Once under vacuum, current is gradually applied to the electrodes, thereby heating the boat. After the material reaches its evaporation point, the gaseous material leaves the boat as a result of thermal energy. In this way, the material from the boat comes to slowly condense on the cooler substrate, creating a thin layer of the target material. A quartz crystal monitor, which is also water cooled, tracks the rate and total deposition of the material. Once the desired thickness has been reached, the current is reduced to zero, and the bell jar is brought back up to atmospheric pressure and removed. For the NaBr evaporation, the NaBr was a reagent grade crystalline powder, the boat was made of tantalum, and the substrates were carbon foils of natural isotopic abundance floated onto a target frame. Although several batches of carbon foils were used during the course of the target making process, they were all purchased from The Arizona Carbon Foil Co., Inc., with thicknesses varying from $15 \text{-} 25$ $\mu$g$/$cm$^2$ \cite{acfmetals}. The LENA evaporator, pictured in Fig.~\ref{fig:evaporator}, was used for all target production. This evaporator is devoted towards target fabrication for low background experiments, making it an attractive alternative to the more general use TUNL evaporator. In particular, the bell jar is brought down to rough vacuum with a scroll pump, and to high vacuums of $1 \text{-} 5 \times 10^{-7}$ Torr by a cryopump, both of which reduce the potential of contamination coming from pump oils. A specific evaporation setup is shown in Fig.~\ref{fig:nabr_setup}.

\subsection{Rutherford Backscattering Spectroscopy}

The failure of several initial $^{23}$Na$(^3\textnormal{He}, d)$ experiments necessitated a better understanding of the NaBr targets. Several different methods were explored to characterize the targets, none of which yielded a satisfactory way to accurately measure the target thickness. However, enough information was available to reach the following conclusions about the targets:

\begin{enumerate}
    \item They contained $^{23}$Na.
    \item Exposure to atmosphere can be tolerated for short periods of time.
    \item After this time, the targets degrade heavily from oxidation.
\end{enumerate}

All of these conclusions are based on a series of Rutherford  backscattering spectroscopy (RBS) experiments that were carried out at TUNL. These experiments utilized the $52^{\circ}$ beamline, which is equipped with a general purpose scattering chamber. A $2$-MeV beam of $^4$He$^{2+}$ was accelerated down the beamline and impinged on the NaBr targets. Currents were typically on the order of $100$ nA (i.e., $50$ pnA). The backscattered $\alpha$ particles were detected by a $100$-$\mu$m-thick silicon surface barrier detector positioned at $165^{\circ}$ relative to the beam direction. Charge integration of the beam was carried out via a Faraday cup located downstream of target chamber. The silicon detector electronics were identical to those of the monitor detector setup shown in Fig.~\ref{fig:electronics_si}, with the exception of the second detector and coincidence circuit. Energy calibration for this detector was carried out using the peaks from a gold target of known thickness. 

\begin{figure}
    \centering
    \includegraphics[width=\textwidth]{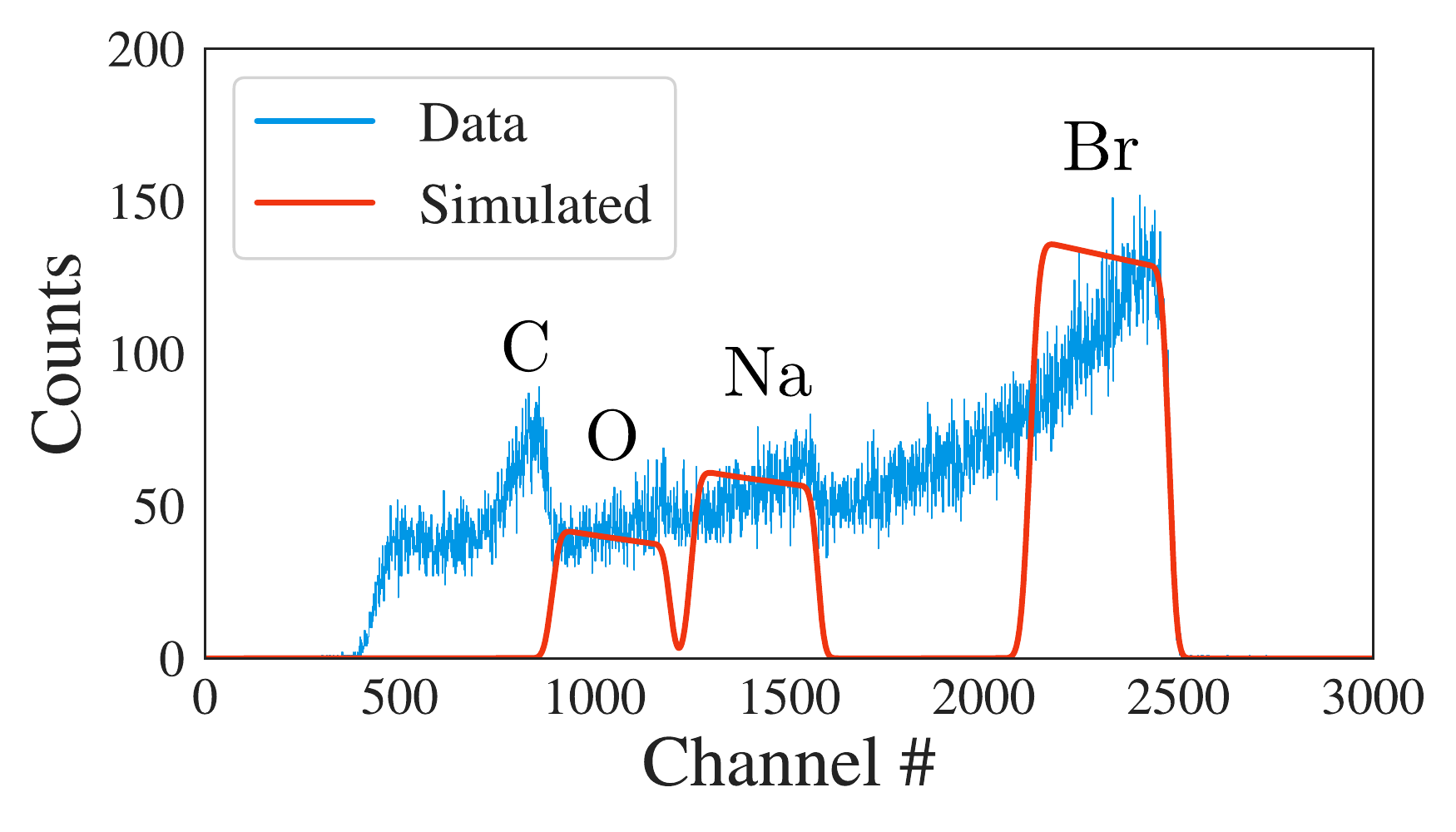}
    \caption{An example spectra from the first series of RBS measurements to attempt to characterize the NaBr targets. The simulated curves were produced using SIMNRA. The low energy tails on the data cannot be described by a single layer of NaBr. }
    \label{fig:nabr_rbs}
\end{figure}

A detailed analysis of this data set was attempted, but proved unsuccessful. Theoretical cross sections were calculated using SIMNRA \cite{Mayer_1999}. It was quickly found that the low energy tails on O, Na, and Br peaks could not be described by assuming a single layer of these materials. An example spectra is shown in Fig.~\ref{fig:nabr_rbs} along with the simulated cross section from SIMNRA. As can be seen in the figure, thick layers of material produce an increase in cross section at lower energies. The observed low energy tails can only be explained by introducing many layers with varying composition. This situation means that the observed RBS data for this target are nearly useless for predictive purposes due to the abundance of free parameters, i.e., the number of layers, the thickness of the layers, and the relative composition of the elements in the layers.

\begin{figure}
    \centering
    \includegraphics[width=\textwidth]{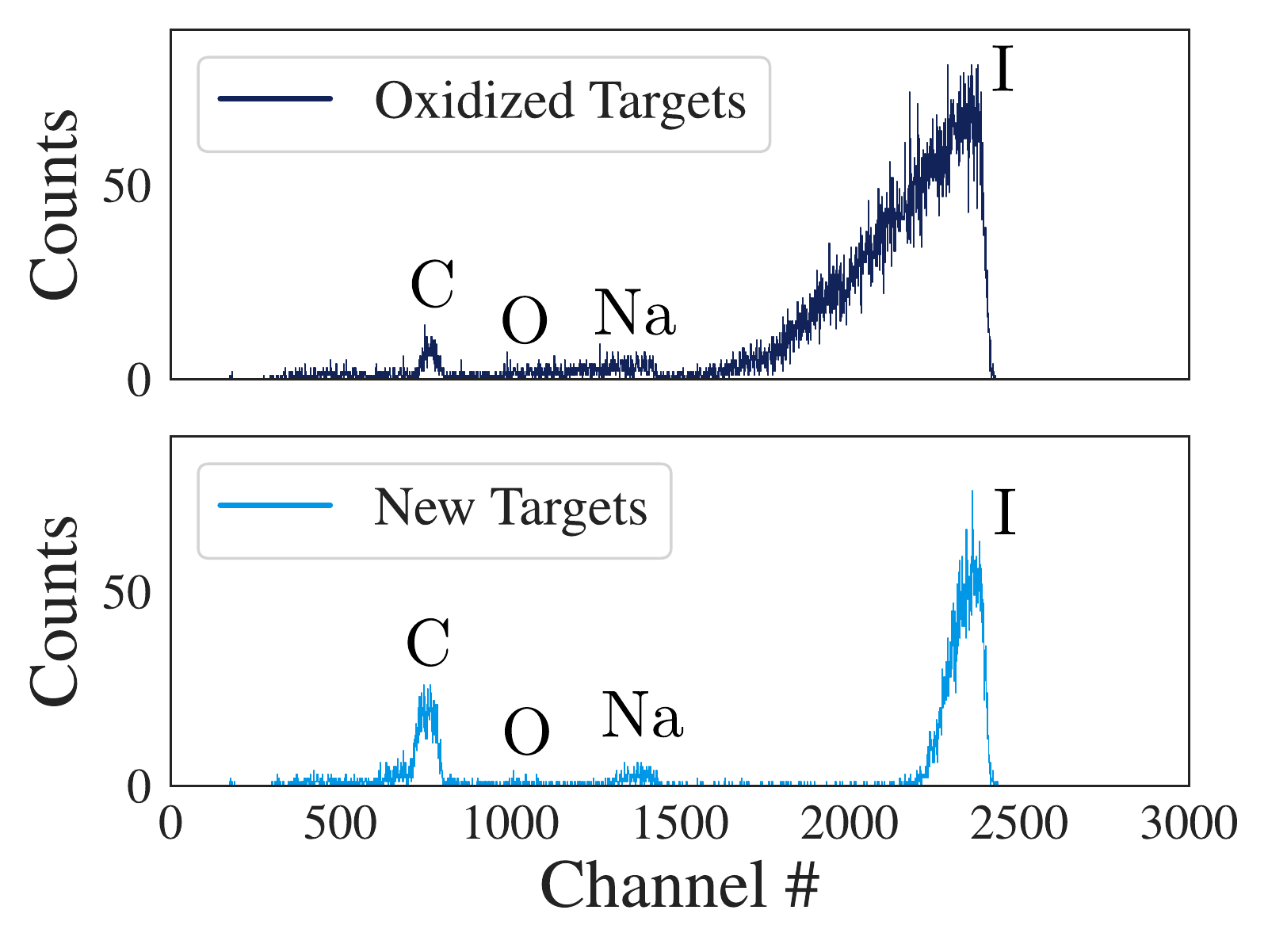}
    \caption{(top) The RBS spectra obtained for a NaI target exposed to atmosphere for a week, (bottom) and a NaI target evaporated the day of the RBS run and immediately transferred to a scattering chamber at high vacuum. Oxidation creates significant low energy tails in the spectrum, and prohibits accurate determination of the target composition and thickness.}
    \label{fig:na_I_rbs}
\end{figure}

Explaining the above results proved difficult. In time, it was gradually understood that the low energy tails were a result of the material becoming heavily oxidized. Definitive proof came during a later run, in which RBS was performed on a pair of NaI targets. One target was prepared the day of the run, and immediately transferred from the evaporator to the target chamber, while the other was evaporated over a week before and exposed to atmosphere. The change in target material aimed to test whether the low energy tails were a characteristic of the NaBr targets, or a more general phenomenon. As can be seen in Fig.~\ref{fig:na_I_rbs}, the exposure to atmosphere is linked to the low-energy tails seen in the RBS data.

Practically, this result points towards the issue with relying on RBS for an accurate determination of target thickness for the $^{23}$Na$(^3$He$,d)$ experiment. Even if the targets do not degrade during the course of the transfer experiment, the process of transferring them between the SPS target chamber and the chamber on the $52^{\circ}$ beamline can dramatically change their properties. If RBS was carried out before the transfer measurement, the resolution of the experiment will suffer because of the oxidation. If, instead, the RBS is carried out after the experiment, the resulting oxidation will render the analysis of the RBS data inconclusive.  

These results ultimately led to the conclusion that an absolute cross section scale would be hard to obtain both because of the poorly known target thickness and the previously mentioned issues with charge integration of the SPS beamstop (see Section~\ref{sec:split_pole}).

\section{Experiment Details}

Data taking for this experiment took place over a five-day period from October $24 \text{-} 28$, 2018. The run was limited to five days due to the sodium oven on the helium source. The oven can run for approximately $140$ hours before it has to be brought up to atmosphere and loaded with more sodium. Because of this constraint, the experiment focused on taking data at the maximum number of angles below $25^{\circ}$. These angles cover the region of the angular distribution that is best described by DWBA. While higher angles can provide useful information, they suffer from lower count rates, meaning more time would be needed per angle. The lower statistics combined with the expected inadequacies of DWBA offered little reason to sacrifice the limited run time towards these measurement. The initial plan was to measure from $3^{\circ} \text{-} 21^{\circ}$ in steps of $2^{\circ}$. Higher angles would only be attempted once these angles had accumulated sufficient statistics to constrain the $11831$-keV state.  

Taking into consideration the findings presented in Section \ref{sec:targets}, the NaBr targets used for the $^{23}$Na$(^3 \textnormal{He}, d)$ experiment were evaporated the morning of the run. $^{\textnormal{nat}}$C foils $22$ $\mu$g$/$cm$^2$ thick were used as the target backing. These were floated several days before evaporation to allow them time to dry.  The evaporation took place over a period of roughly $55$ minutes, with the rate of deposition fluctuating between $10-30$ ng/(cm$^2$ s). Evaporation was halted after the thickness monitor indicated a thickness of $\approx 70$ $\mu$g/cm$^2$. Six targets were produced in total. After the evaporation was complete, the bell jar was gradually brought up to atmosphere, and the targets were placed into a container to transfer them to the target chamber of the SPS. This container was brought down to rough vacuum to reduce exposure to air. At the SPS target chamber, three of the targets were mounted onto the SPS target ladder. The other targets consisting of a $1$ mm collimator for beam tuning, a $^{\textnormal{nat}}$C target identical to the backing of the NaBr targets for background, and thermally evaporated $^{27}$Al on a $^{\textnormal{nat}}$C backing to use for an initial energy calibration. After quickly mounting these targets onto the target ladder, the ladder itself was mounted in the target chamber and brought down to high vacuum. 

The nominal energy selected for the experiment was $21$ MeV, following Ref.~\cite{hale_2004}. The tandem was brought up to a voltage of approximately $6.7$ MV. The $90 \text{-} 90$ NMR was set to $\approx 568.28$ mT. The beam energy of $E_{^3\textnormal{He}} = 21.249(1)$ MeV was determined from the bending radius and the NMR setting. The uncertainty was estimated from the fluctuations in the NMR reading during the course of the experiment, with the average value being $\Bar{B} = 568.290(1)$ mT. The magnet settings were chosen based on a scaled set of values derived from an earlier beam development run of $d$, which had been used to maximize transmission through the collimator of the SPS using the high resolution setting of the $90 \text{-} 90$ system.  

A beam of $2$ $\mu$A $^3$He$^{-}$ was extracted out of the helium source at the start of the run. In order to minimize the amount of $^{3}$He needed, a gas recycling system was used. Details of this system can be found in Ref.~\cite{combs_2017}. New $^3$He has to be introduced into this system every $\approx 48$ hours in order to maintain beam current. The amount of beam that made it to the target varied, but was typically around $100-200$ nA of $^3$He$^{+2}$. The high resolution setting of the $90 \text{-} 90$ system meant that close to $100 \%$ of this beam could be passed through the $1$ mm collimator. It was noticed during the course of data taking that the beam would drift vertically on the target. In order to mitigate the potential impact of this drift, the beam was retuned through the collimator every few hours. The entirety of the run was plagued by source instability. The power supplies for the attachment and extractor electrodes would current limit every few seconds from a periodic discharge coming from the extractor. Each discharge was accompanied by a drop in the beam current, and as a result we were severely limited in the amount of quality data that could be collected.  

\begin{figure}
    \centering
    \includegraphics[width=.6\textwidth]{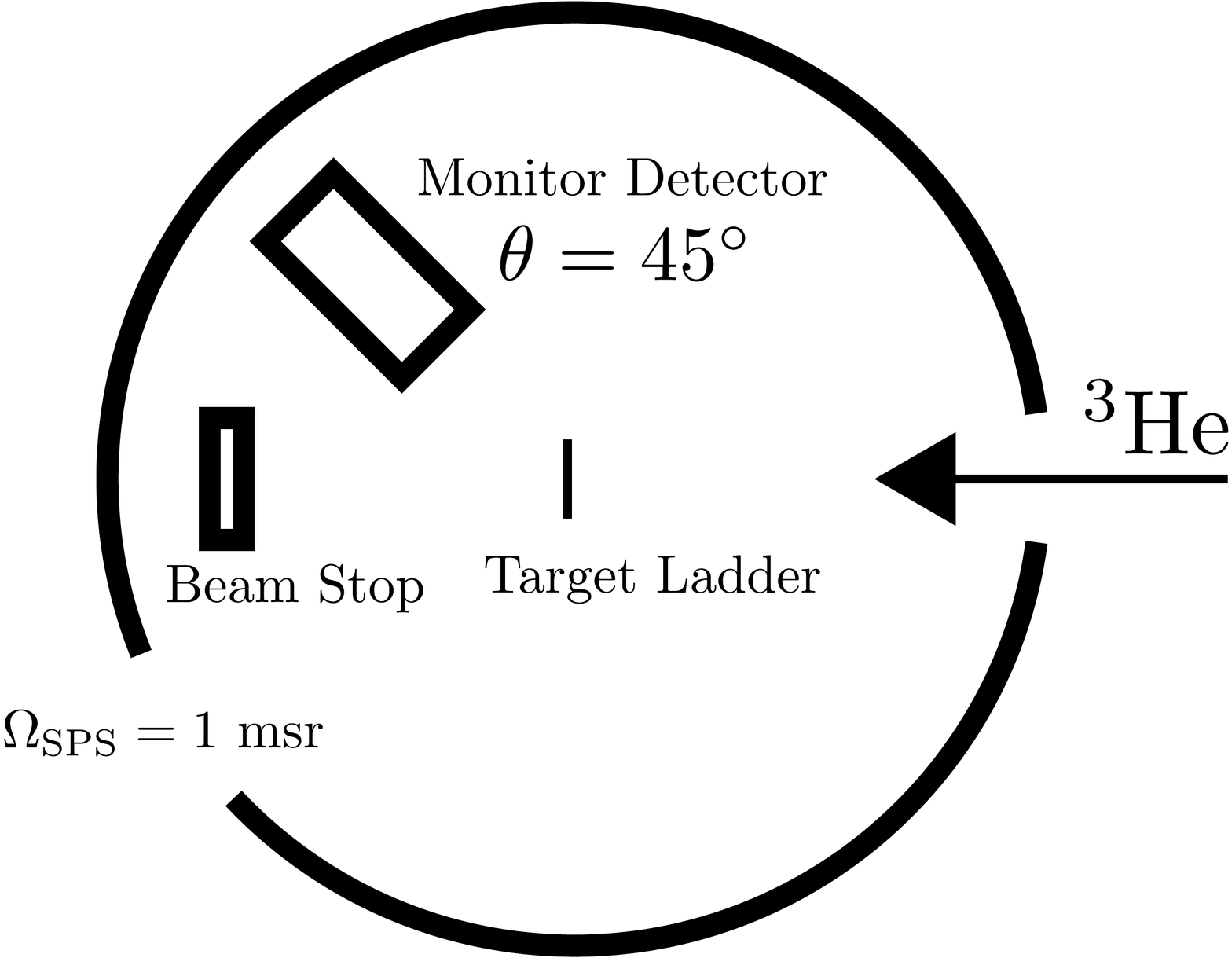}
    \caption{Top view drawing of the SPS target chamber.}
    \label{fig:target_chamber}
\end{figure}

A diagram of the SPS target chamber is shown in Fig.~\ref{fig:target_chamber}. At each angle the field of the spectrograph was set between $1.14 \text{-}  1.13$ T in order to keep the states of interest on the focal plane. The solid angle of the spectrograph was fixed throughout the experiment at $\Omega_{\textnormal{SPS}} = 1$ msr. The monitor telescope was positioned at $45^{\circ}$. The planned angles between $3^{\circ} \text{-} 21^{\circ}$ in steps of $2^{\circ}$ were measured in addition to $26^{\circ}$. The source instability severely limited the statistics collected at $3^{\circ}$ and $26^{\circ}$. The last $8$-hour shift was devoted towards elastic scattering measurements for the DWBA analysis. These measurements were performed by changing the field of the SPS to $0.75 \text{-} 0.80$ T and measuring angles between $15^{\circ} \text{-} 55^{\circ}$ in $5^{\circ}$ steps and a final angle at $59^{\circ}$. An attempt was made to measure at $60^{\circ}$, but the sliding seal on the target chamber started to lose vacuum around $59.5^{\circ}$.  

\section{Focal Plane Peak Fitting}
\label{sec:peak_fitting_fp}

\begin{figure}
    \centering
    \includegraphics[width=\textwidth]{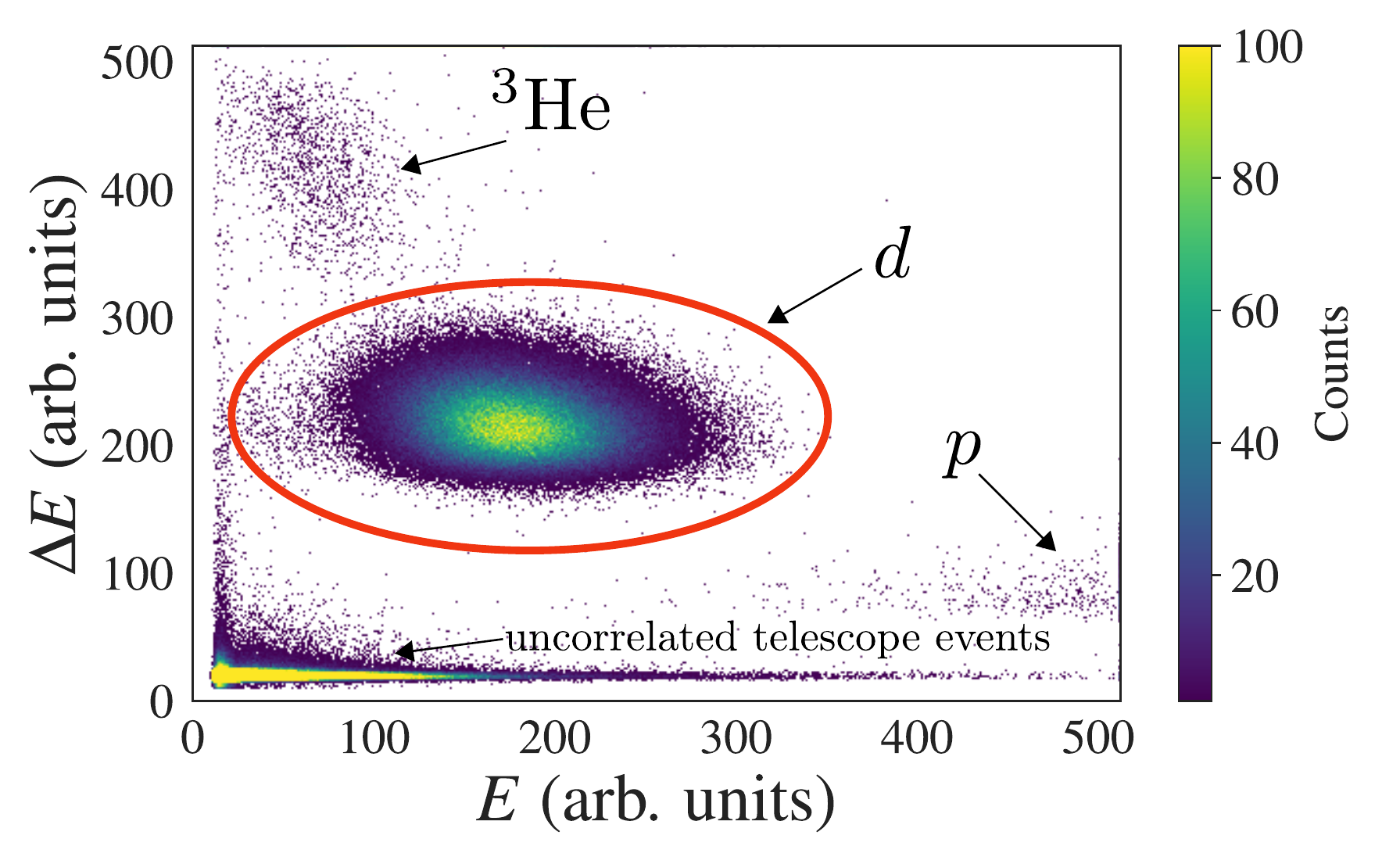}
    \caption{Histogram of the $\Delta E$ versus $E$ spectra. This example is from $\theta_{lab} = 11^{\circ}$}
    \label{fig:11_deg_e_de}
\end{figure}

\begin{figure}
    \centering
    \includegraphics[width=\textwidth]{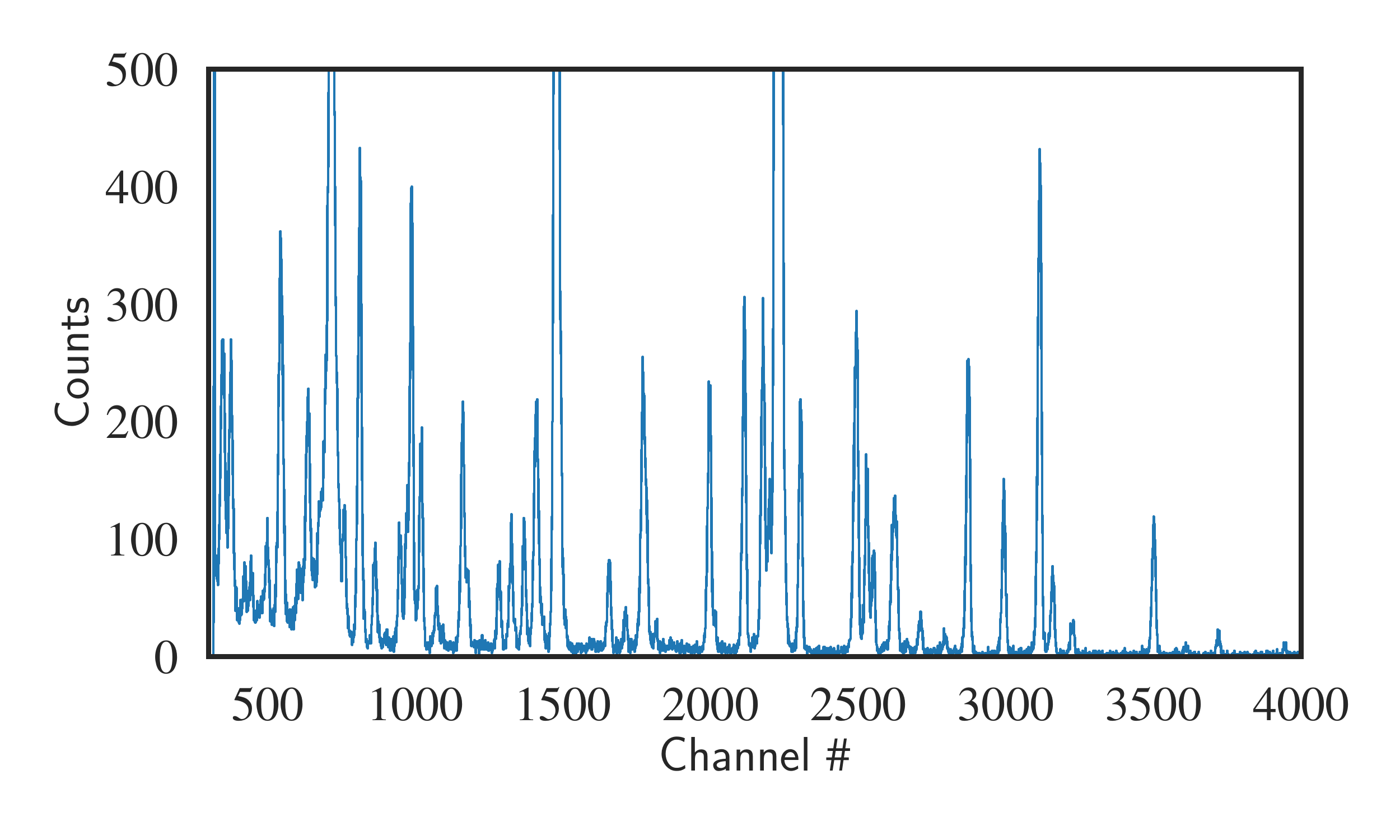}
    \caption{Focal plane position spectra gated on the deuteron group from Fig.~\ref{fig:11_deg_e_de}.}
    \label{fig:11_deg_fp_spectra_raw}
\end{figure}

The deuteron group was clearly resolved in the focal plane $\Delta E / E$ spectra, as can be seen in Fig.~\ref{fig:11_deg_e_de}. This group was gated on at each angle to produce the deuteron spectra in the focal plane. An example of the spectra produced with the above gate in shown in Fig.~\ref{fig:11_deg_fp_spectra_raw}.

\begin{figure}
    \centering
    \includegraphics[width=.8\textwidth]{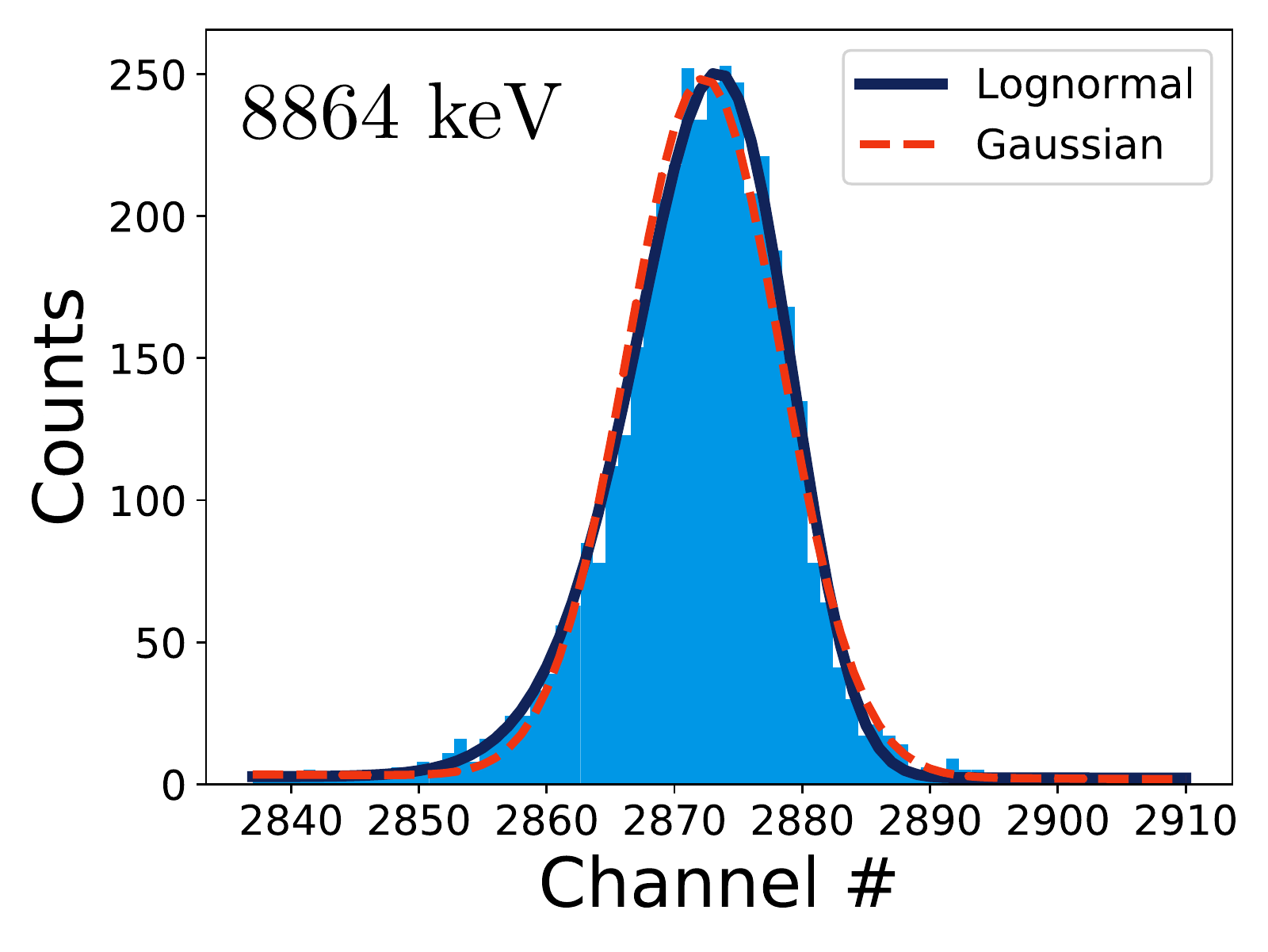}
    \caption{Example of the low energy tail seen in the high energy particles. This fit is from $\theta_{lab} = 11^{\circ}$. The log-normal fit is shown to provide a mildly improved description of this tail.}
    \label{fig:8864_fit}
\end{figure}

Peaks in the gated focal plane spectra were fit using Gaussian functions with a linear background. The Gaussian function takes the form:
\begin{equation}
    \label{eq:gaussian_peak_fit}
    f(x ; A, w, c) = \frac{A}{\sqrt{2 \pi w^2}} e^{\frac{(x-c)^2}{2 w^2}},
\end{equation}
where $A$ is the area, $w$ is the width, and $c$ is the centroid. Some of the peaks on the high energy side of the detector at higher angles showed a mild low energy tail. For computational ease, these peaks were fit with a log-normal shaped function. An example of this behaviour for the $8864$-keV peak at $\theta_{lab} = 11^{\circ}$ is shown in Fig.~\ref{fig:8864_fit}. Compared to a Gaussian fit, the log-normal fit resulted in an approximately $1$ channel change in the peak centroid and $2 \%$ change in area. Peak fitting with Bayesian statistics based on a Poisson likelihood was also explored. It was found that these results were consistent with the frequentist approach based on a $\chi^2$ function; however, the Bayesian method had larger uncertainties on average for the peak parameters. The additional computation overhead from the Bayesian method combined with the high number of states in the spectrum at each angle, led to a frequentist method being adopted. Parameter uncertainties adjusted by a factor $\sqrt{\chi^2/dof}$, where $dof$ is the degree of freedom of the fit, were adopted in the case that $\chi^2/dof > 1$. This correction was a marginally more conservative than the Bayesian method, but had the benefit of being computationally less expensive. All of the fits using the frequentist method were performed using the program \texttt{fityk} \cite{Wojdyr_2010}. Spectra were fit for the $10$ angles between $3^{\circ} \text{-} 21^{\circ}$. Centroids, areas, FWHM, and their associated uncertainties were tabulated for each peak in the spectrum. 

\section{Updates to Energy Levels Above $11$ MeV}
\label{sec:energy_level_update}

The current ENSDF evaluation, Ref.~\cite{firestone_2007}, is incomplete due to missing the measurements of Ref.~\cite{goerres_1989} and Ref.~\cite{hale_2004}, which are important in the astrophysical region of interest of $ 11 \lessapprox E_x \lessapprox 12$ MeV. This evaluation also has inaccuracies for many of the states above the proton threshold. It was found that \textit{deduced} $E_{\gamma}$ values from Ref.~\cite{schmalbrock_1983} were incorrectly used in the evaluation. Excitation energies can be derived by performing a least squares fits to all of the observed $E_{\gamma}$ values that decay from or feed into an excited state \cite{Firestone_1991}. However, deduced $E_{\gamma}$ values are inferred from excitation energies, meaning they cannot be considered as independent measurements. Additionally, many of the levels were calculated using calibration points from the spectrograph measurements of Ref.~\cite{moss_1976} and Ref.~\cite{zwieglinski_1978}. This error is long standing, and is present in every compilation and evaluation since 1978 \cite{ENDT_1978}. Both of these inaccuracies lead to overly precise recommended energies. Finally, the evaluation is outdated because many of the level energies have been determined through $^{20}\textnormal{Ne}(\alpha, \gamma)^{24}\textnormal{Mg}$ and $^{23}\textnormal{Na}(p, \gamma)^{24}\textnormal{Mg}$, and therefore
need to be updated based on the 2016 mass evaluation \cite{Wang_2017}. The measured energies of Vermeer \textit{et. al} \cite{vermeer_1988} also present challenges that are discussed in Appendix \ref{chap:rant_on_energies}, but have been adopted as is. Note that the measurements of Hale \textit{et al}. have been excluded from the compiled values. This decision will be discussed in depth in Section \ref{sec:hale_discussion}, but for now it is worth emphasizing that these compiled values are only for the purpose of accurately energy calibrating the current experiment and for calculating the astrophysical reaction rate.     

The compiled energies are presented in Table.~\ref{tab:energy_comp}. Note that in the case of Ref.~\cite{endt_1990}, resonant capture was used to excite $^{24}$Mg, but the excitation energies were deduced from gamma ray energies making these values independent of the reaction $Q$-value. For the measurements that report the lab frame resonance energies, the excitation energies are deduced from:
\begin{equation}
  \label{eq:lab_to_ex}
  E_x = Q + E_{P} \frac{M_T}{M_T + M_{P}}, 
\end{equation}
where $E_P$ is the projectile energy measured in the laboratory frame, and $M_{P}$ and $M_T$ are the \textit{nuclear} masses for the projectile and target nuclei, respectively. I have used the \textit{atomic} masses from Ref.~\cite{Wang_2017} assuming the difference is negligible compared to the statistical uncertainty in $E_{P}$. $Q$ is the $Q$-value for either the $(p,\gamma)$ or $(\alpha, \gamma)$ reaction. The column in Table \ref{tab:energy_comp} from Ref.~\cite{endt_eval_1990} shows energies deduced from a weighted average of several $(p, \gamma)$ measurements, and that paper should be referred to for additional details. For the present work, the suggested value of these weighted averages is treated as a single measurement that is updated according to Eq.~\ref{eq:lab_to_ex}. The weighted averages presented in the last column were calculated from:
\begin{equation}
    \bar{x} = \frac{\sum_i^N w_i x_i}{\sum_j^N w_j},
\end{equation}
with uncertainty given by:
\begin{equation}
    \bar{\sigma} = \frac{1}{\sqrt{\sum_j^N w_j}},
\end{equation}
where the weight is $w_i = 1/\sigma_i^2$, $\sigma_i$ is the uncertainty of measurement $i$, $x_i$ is the reported value of measurement $i$, and $N$ is the total number of measurements. In order to reduce the effects of potential outliers, the lowest measured uncertainty was used instead of $\bar{\sigma}$ in the case of $N \leq 3$.

\begin{lscapenum}
\begin{table}[]
  \centering
  \setlength\tabcolsep{3pt}
  \def\arraystretch{1}
  \caption{ \label{tab:energy_comp} Previously measured energies. An * indicates that the listed energy was used as a calibration point in the listed experiment. These values, therefore, have been excluded from the weighted average. Excitation energies derived from resonance energies have been updated based on the 2016 mass evaluation \cite{Wang_2017}. Note the $\dagger$ on the $12259.6$-keV state. This value was taken from Ref.~\cite{endt_eval_1990}, and is actually the unweighted average of a pair of states with updated energies of $12259.4(4)$ keV and $12259.8(4)$ keV, respectively.}

  \begin{tabular}{llllllllll}
    \toprule
    \toprule
         $(p, p^{\prime})$ & $(p, p^{\prime})$ & $(^{16} \textnormal{O}, \alpha)$ & $(p,\gamma)$ & $(p,\gamma)$ & $(\alpha,\gamma)$ & $(\alpha,\gamma)$ & $(\alpha,\gamma)$ & $(\alpha,\gamma)$ & Weighted Average \\
    \cite{moss_1976}      & \cite{zwieglinski_1978} & \cite{vermeer_1988}  & \cite{endt_1990}       & \cite{endt_eval_1990} & \cite{fiffield_1978} & \cite{schmalbrock_1983} & \cite{goerres_1989}  & \cite{smulders_1965}   \\  \hline
 11389(3)*                             & 11391(7)                                     & 11390(4)                                 &                                       &              &                                            & 11394(3)                                     &                                                               & 11393(5)                                  & 11392.6(21)      \\
11456(3)                              & 11452(7)                                     & 11455(4)                                 & 11452.8(4)                            &              &                                            & 11456(3)                                     &                                                               &                                           & 11452.9(4)       \\ 
11521(3)*                             & 11520(7)                                     & 11519(4)                                 &                                       &              &                                            & 11522(2)                                     &                                                               & 11523(5)                                  & 11521.5(16)      \\
11694(3)*                             & 11694(7)*                                    & 11694(4)                                 &                                       &              &                                            & 11699(2)                                     &                                                               & 11694(5)                                  & 11698(2)         \\
11727(3)*                             & 11727(7)                                     & 11727(4)                                 &                                       &              &                                            & 11731(2)                                     &                                                               & 11728(5)                                  & 11729.8(16)      \\
11828(3)                              &                                              & 11827(4)                                 &                                       &              &                                            &                                              &                                                               &                                           & 11827(3)         \\
11862(3)*                             & 11860(7)                                     & 11860(4)                                 &                                       &              & 11861(5)                                   & 11868(3)                                     & 11859.4(20)                                                   & 11862(5)                                  & 11861.6(15)      \\
11935(3)*                             &                                              & 11930(4)                                 &                                       & 11933.05(19) &                                            &                                              & 11933.2(10)                                                   &                                           & 11933.0(10)      \\
11967(3)*                             & 11965(7)                                     & 11963(4)                                 &                                       & 11966.6(5)   & 11967(5)                                   & 11974(3)                                     & 11966.7(10)                                                   & 11968(5)                                  & 11966.7(5)       \\
11989(3)*                             & 11990(7)                                     & 11985(4)                                 & 11988.0(3)                            & 11988.47(6)  &                                            &                                              & 11988.7(10)                                                   &                                           & 11988.45(6)      \\
12015(3)*                             & 12016(7)*                                    &                                          &                                       & 12017.1(6)   & 12016(5)                                   &                                              & 12016.5(10)                                                   &                                           & 12016.9(5)       \\
12050(3)*                             & 12050(7)                                     &                                          &                                       & 12051.3(4)   & 12050(5)                                   &                                              &                                                               &                                           & 12051.3(4)       \\
12121(3)*                             & 12124(7)                                     &                                          &                                       & 12119(1)     & 12121(5)                                   &                                              &                                                               &                                           & 12119(1)         \\
12181(3)*                             &                                              &                                          &                                       & 12183.3(1)   &                                            &                                              &                                                               &                                           & 12183.3(1)       \\
12258(3)*                             & 12261(7)                                     &                                          &                                       & 12259.6(4)$^{\dagger}$   & 12258(5)                                   &                                              &                                                               &                                           & 12259.6(4)       \\
12342(3)*                             &                                              &                                          &                                       & 12341.0(4)   &                                            &                                              &                                                               &                                           & 12341.0(4)       \\
12402(3)                              & 12402(7)                                     &                                          &                                       & 12405.3(3)   & 12405(5)                                   &                                              &                                                               &                                           & 12405.3(3)       \\
12528(3)*                             &                                              &                                          &                                       & 12528.4(6)   &                                            &                                              &                                                               &                                           & 12528.4(6)       \\
12577(3)*                             & 12578(7)                                     &                                          &                                       &              & 12578(5)                                   &                                              &                                                               &                                           & 12578(5)         \\
12669(3)                              &                                              &                                          & 12669.9(2)                            & 12670.0(4)   &                                            &                                              &                                                               &                                           & 12669.9(4)       \\
12736(3)                              & 12739(7)                                     &                                          &                                       & 12739.0(7)   & 12740(5)                                   &                                              &                                                               &                                           & 12738.9(7)       \\
                                      &                                              &                                          &                                       & 12817.77(19) &                                            &                                              &                                                               &                                           & 12817.77(19)     \\
12849(3)                              & 12850(7)                                     &                                          &                                       & 12852.2(5)   &                                            &                                              &                                                               &                                           & 12852.1(5)       \\
12921(3)*                             &                                              &                                          &                                       & 12921.6(4)   & 12923(5)                                   &                                              &                                                               &                                           & 12921.6(4)       \\
12963(3)                              &                                              &                                          &                                       & 12963.9(5)   &                                            &                                              &                                                               &                                           & 12963.9(5)      \\
   
    \bottomrule
    \bottomrule
\end{tabular}

\end{table}
\end{lscapenum}
\restoregeometry

\section{Energy Calibration}
\label{sec:energy_cal_na}

Using the position measurements of the focal plane detector, excitation energies were extracted using a third-order polynomial fit for the bending radius of the SPS in terms of the ADC channels, $x$:

\begin{equation}
  \label{eq:energy_fit}
  \rho = Ax^3 + Bx^2 + Cx + D.
\end{equation}

This was done using the updated version of the Bayesian method presented in Section \ref{sec:bay_energy_cal} that uses \texttt{emcee} to sample the posterior.

\begin{figure}
    \centering
    \includegraphics[width=\textwidth]{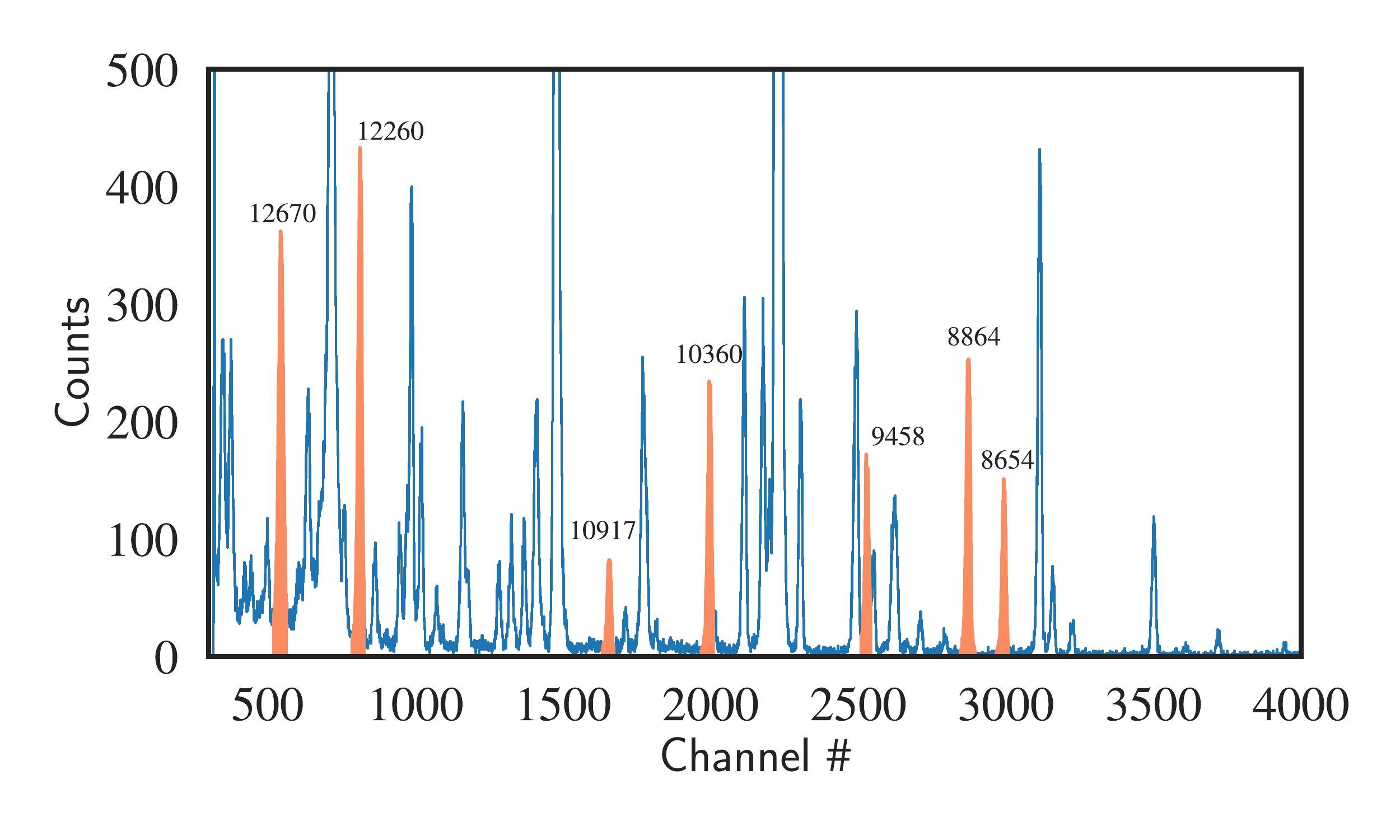}
    \caption{Calibration states (orange) for $\theta_{lab} = 11^{\circ}$. The energy values for the states below the proton threshold are taken from Ref.~\cite{firestone_2007}, while the rest are from Table \ref{tab:energy_comp}.}
    \label{fig:na_cal_spec}
\end{figure}

Calibration states were methodically selected by gradually extrapolating from bound states across the focal plane. This was done by first using a linear fit on the two states at $8654$, and $8864$ keV. By using these states as a starting point, it was possible to iteratively extend the fit by identifying new states to use as calibration points, incorporating these new points, increasing the order of the polynomial as needed, and repeating this process until the majority of the focal plane was calibrated. In a few regions, the most intensely populated peaks were excluded due to the possibility of closely spaced levels that differed in energy by more than a few keV, which could cause an additional source of error to be introduced into the calibration. The chosen calibration states at $\theta_{lab} = 11^{\circ}$ are shown in Fig.~\ref{fig:na_cal_spec}. The validity of this internal calibration in the astrophysical region of interest between $11$ and $12$ MeV was checked at $\theta_{lab} = 11^{\circ}$ against a separate external calibration using the $^{27}$Al$(^3$He$, d)^{28}$Si reaction. The aluminum states were selected based on the spectrum shown in Ref.~\cite{champagne_1986}. The two calibrations showed an energy offset of $\approx 7$ keV arising from the difference in the thicknesses of the Al and NaBr targets. Once the energy offset was corrected for, the two methods showed excellent agreement, and the internal calibration was subsequently adopted at each angle.

\begin{figure}
    \centering
    \includegraphics[width=.8\textwidth]{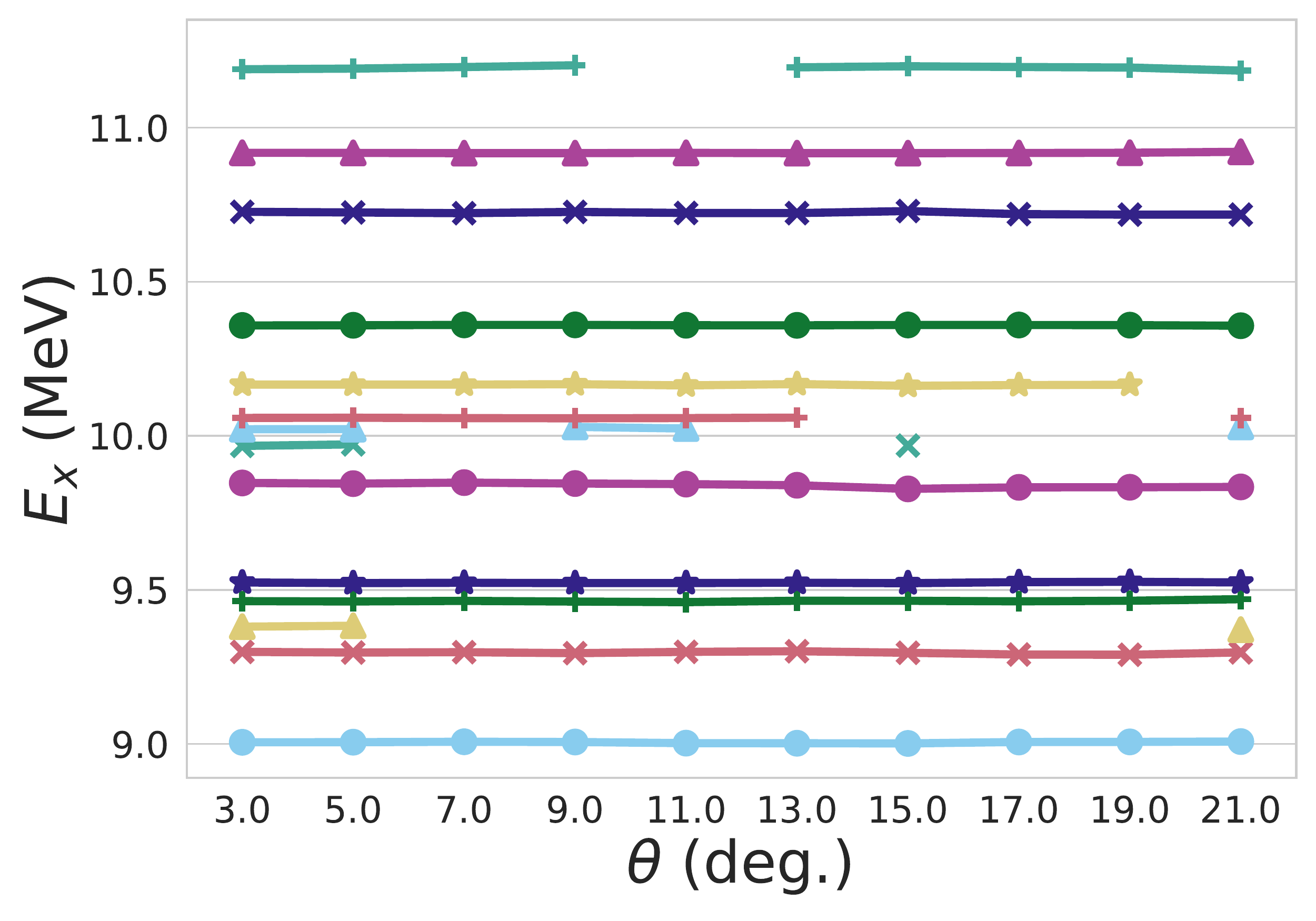}
    \caption{A subset of $14$ clusters found using \texttt{DBSCAN}. Colors and point shapes indicate which cluster the peak belongs to. Connecting lines are to guide the eye.}
    \label{fig:clustering_energy}
\end{figure}

Once initial calibrations for each angle were found, it was necessary to identify the states associated with excited states of $^{24}$Mg and those arising from contaminants. Known contamination peaks arise from $^{12}$C$(^3 \textnormal{He}, d)$, $^{13}$C$(^3 \textnormal{He}, d)$, $^{14}$N$(^3 \textnormal{He}, d)$, and $^{16}$O$(^3 \textnormal{He}, d)$. This process would require tracking $> 50$ peaks across $10$ angles. However, once an energy calibration is obtained, these peaks can readily be sorted. If a state belongs to $^{24}$Mg, its predicted energy will be consistent from angle to angle; however, if a state arises from another reaction besides $^{23}$Na$(^3 \textnormal{He}, d)$, its kinematic shift will cause a dramatically changing energy prediction from angle to angle. Thus, states that belong to $^{24}$Mg will tend to cluster around some energy. In order to ease the identification of states, a clustering algorithm was used to look for this behaviour in the peaks at each angle. For this problem the \texttt{DBSCAN} algorithm was chosen \cite{Ester_1996}. \texttt{DBSCAN} is well suited towards the problem because it does not require prior knowledge on the number of clusters and one of its free parameters sets the minimum number of points needed in a region to construct a cluster, i.e, it mimics the requirement for a state to be observed at a minimum number of angles before it can be considered to be coming from $^{24}$Mg. The minimum number was set to $min=3$ in this case. The statistical energy variation between angles also needs to be accounted for. Membership to a cluster is controlled by the free parameter $eps$, which was set to $10$ keV. The implementation of \texttt{DBSCAN} in \texttt{scikit-learn} was used \cite{scikit-learn}. An example of the clustering is shown in Fig.~\ref{fig:clustering_energy}. In total $54$ levels were found, with $49$ peaks classified as contaminants. It is still necessary to check the results of the method by hand, but it effectively transforms the problem of sorting the nearly $600$ peaks over all angles, to checking the consistency of the $54$ states and $49$ contaminates.                       

\begin{figure}
    \centering
    \includegraphics[width=\textwidth]{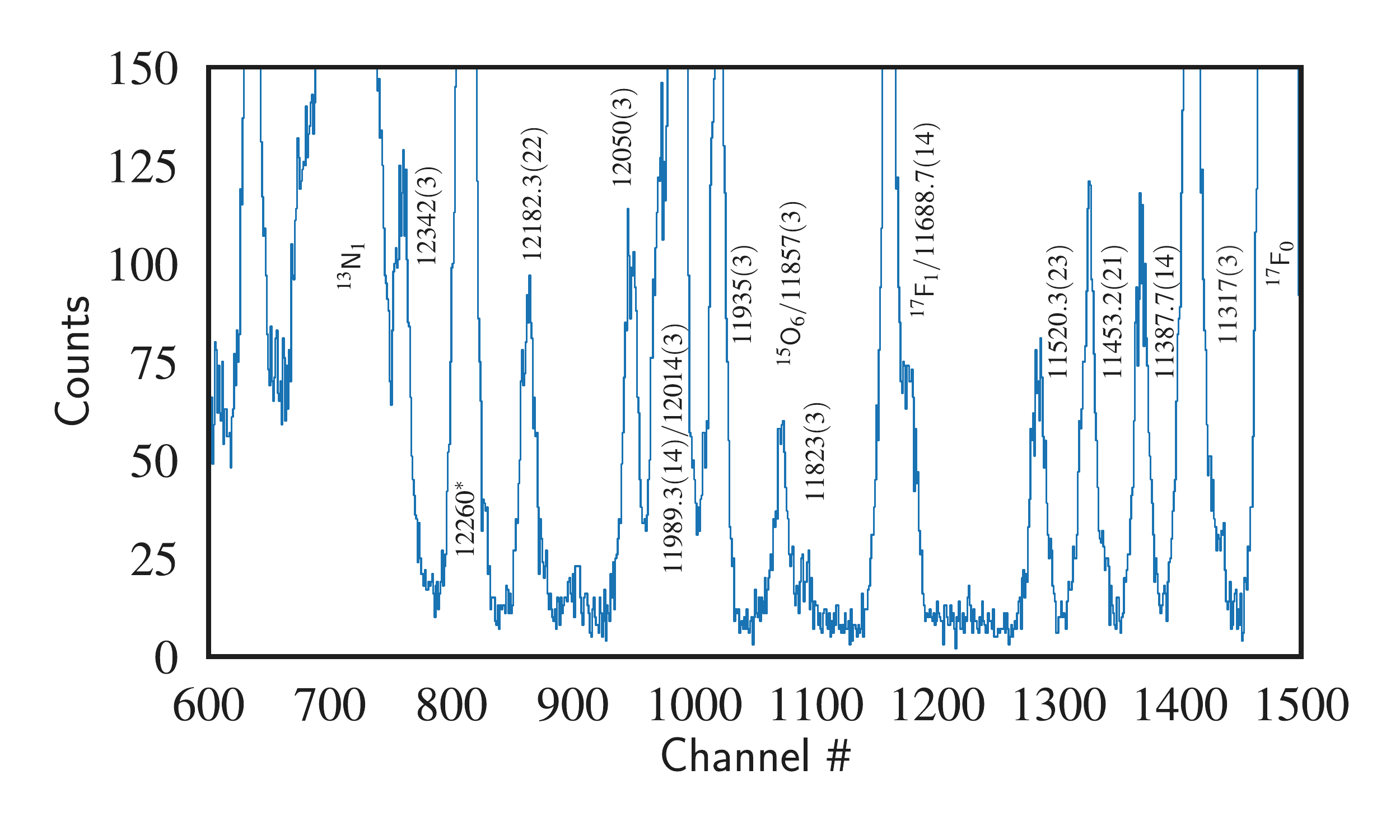}
    \caption{$\theta_{lab} = 11^{\circ}$ spectrum that has been zoomed in on the astrophysical region of interest. All of the peaks from $^{24}$Mg have been identified with the final weighted average energy value in keV.}
    \label{fig:calibrated_na_energies}
\end{figure}

The energies presented in Table \ref{tab:my_excitation_energies} are the weighted average of the energies deduced at each angle. Fig.~\ref{fig:calibrated_na_energies} shows the location of the peaks in the astrophysical region of interest at $11^{\circ}$.  Only states that were seen at three or more angles are reported. The additional uncertainty estimated by our Bayesian method, see Eq.~\ref{eq:calibration_bayesian_model}, also introduces a further complication into the weighted averaging between the angles. Since this uncertainty is estimated directly from the data, it will be influenced by systematic effects. These systematic effects introduce correlations between the deduced energies and uncertainties at each angle, which can become significant because of the high number of angles measured in this experiment. A clear indication of correlation was the observation that the deduced energies of our calibration points from the fit tend to agree with their input values at each angle, but a simple weighted average of these points yields a disagreement at a high level of significance. In order to account for possible correlations, the uncertainties on the weighted average were estimated using the methods of Ref.~\cite{Schmelling_1995}. This correction is done by calculating the $\chi^2$ value of the data with respect to the weighted average, $\bar{x}$, which is given by:
\begin{equation}
  \label{eq:chi_sq}
  \chi^2 = \sum_i^N \frac{(x_i - \bar{x})^2}{\sigma_i^2}.
\end{equation}
Since the expected value of $\chi^2$ is $N - 1$, the idea is to adjust the uncertainties from the weighted average, $\bar{\sigma}$, based on the deviation from $N-1$. For the case of positive correlations, $\chi^2 < 1$, and, therefore, $\bar{\sigma}$ will need to be adjusted by:
\begin{equation}
  \label{eq:positive_corr_chi}
  \sigma_{adj} = \sqrt{(N-\chi^2) \bar{\sigma}^2}.
\end{equation}
A separate estimate can also be made if the scatter in the data is not well described by the weighted average. In this case, $\chi^2 > 1$, which gives the adjustment: 
\begin{equation}
  \label{eq:negative_corr_chi}
  \sigma_{adj} = \sqrt{\frac{\chi^2}{N-1} \bar{\sigma}^2}.
\end{equation}
To be conservative, the larger of these two values is adopted.

\begin{lscapenum}
\begin{table}
  \caption{\label{tab:my_excitation_energies} $^{24}$Mg excitation energies from this work compared to those of Ref.~\cite{firestone_2007}. Because of the presence of a high number of states in certain regions, a unique identification of the observed state could not be made. States used for the energy calibration are reported in italics, marked with $*$, and listed without uncertainties, but they do represent the mean value obtained after calibration.}
  \begin{tabular}{lll|lll|lll}
    \toprule
    \toprule
This Work & ENSDF \cite{firestone_2007} & $J^{\pi}$ & This Work & ENSDF \cite{firestone_2007} &  $J^{\pi}$ & This Work & ENSDF \cite{firestone_2007} & $J^{\pi}$     \\ \hline \vspace{-2mm}
            &              &           &                 &              &            &             &              &  \\
7364(14)     & 7349.00(3)   & $2^+$     &     10660.1(21) & 10659.58(13)&  $1,2^+$    & 12050(3)    & 12049.1(20)  & $4^+$ \\ 
7555(13)     & 7555.04(15)  & $1^-$     &                 & 10660.03(4)&   $(4^+)$    & 12121.5(17) & 12119.5(18)  &  $(2,3,4)^+$  \\     
7752(10)     & 7747.51(9)   & $1^+$     &     10713.9(12) & 10711.74(17)&  $1^+$      &  &  12128.3(10)         & \\                      
8362(4)  & 8357.98(13)  & $3^-$     &     10732.5(16) & 10730.79(11)&  $2^+$          & 12182.3(22)  & 12181.23(23)  & $(1,2)^+$  \\      
8441(4)  & 8437.31(15)  & $1^-$     &     10824.3(13) & 10820.7(4)&    $3^+,4^+$      & \textit{12260}* & 12257.5(7)  & $(3^-)$   \\      
            & 8439.36(4)   & $4^+$     &     \textit{10918}* & 10916.96(17)&  $2^+$   &                 & 12257.69(21) & $2^+$    \\      
\textit{8654}*  & 8654.53(15)  & $2^+$     &     11011(3) & 11010.5(14)&   $3$        & 12342(3)        & 12339.00(24) & $2^+, 3, 4^+$ \\ 
\textit{8864}*  & 8864.29(9)   & $2^-$     &                 & 11015.8(7)&    $2^+$   &                 & 12344(3)     &          \\      
9002.9(24)  & 9003.34(9)   & $2^+$     &     11201(5) & 11186.82(24) &                & 12406.0(22)     & 12403.3(7)   &  $2^+$   \\      
9145.0(16)  & 9145.99(15)  & $1^-$     &                 & 11208.4(16)  &             & 12530.5(24)     & 12526.5(8)   &  $1,2^+$  \\     
9292.6(12)  & 9284.22(14)  & $2^+$     &     11317(3) & 11314.4(15)   & $(3)^+$       & 12576(3)        & 12577(3)     &  $2^+$   \\      
            & 9299.77(24)  &           &     11387.7(14) & 11389.8(11)  & $1^-$       & \textit{12670}* & 12669.9(2)  & $2$  \\           
\textit{9460}*  & 9457.81(4)   & $(3)^+$   &                 & 11394(4)     &          &  12738(3)       & 12737.1(9)  & $2^+$    \\      
    9520(3)  & 9516.28(4)   & $(4^+)$   &     11453.2(21) & 11452.51(13)     & $2^+$   &  12819(4)       & 12815.9(6)  & $1^+,2^+$  \\    
            & 9527.8(21)   & $(6^+)$   &                  & 11457(3)         & $0^+$   &  12854(3)       & 12850.3(8)  &  \\              
9837.2(25)  & 9828.11(11)  & $1^+$     &           11520.3(23) & 11518.2(6)   & $2^+$  &  12924(4)       & 12919.7(7)  & $2^+, 3, 4^+$  \\
    9977(4)  & 9967.19(22)  & $1^+$    &           11688.7(14) & 11695.6(6)   & $4^+$  & 12965(4) & 12961.9(8)  & $1^+, 2^+$     \\           
    10021(3) & 10027.97(9)  & $5^-$    &                  11823(3)    & 11827(4)     &       \\    
10055(3) & 10058.54(16) & $(1,2)^+$    &                                 & 11862(5)     &       \\        
10163.2(19) & 10161(3)     & $0^+$     &                   11857(3)           & 11864.9(13)  & $1^-$ \\    
10328.1(18) & 10333.29(13) &           &                  11935(3)    & 11931.2(6)   &      \\ 
\textit{10358}* & 10360.51(13) & $2^+$ &     11989.3(14)        & 11987.72(10) & $2^+$      \\                                                      
    10572.7(21) & 10576.02(7)  & $4^+$ &     12014(3)           & 12015.2(8)   & $3^-$      \\
    \bottomrule
    \bottomrule
  \end{tabular}
\end{table}
\end{lscapenum}
\restoregeometry

\subsection{The Energies Reported by Hale \textit{et al}.}
\label{sec:hale_discussion}

Our energies are in fair agreement with those previously reported. However, in the astrophysical region of interest there is significant disagreement between the present values and those of Ref.~\cite{hale_2004}. Of particular concern is the state corresponding to the $138$-keV resonance, whose mean values falls $\approx 9$ keV below what is reported in Ref.~\cite{hale_2004}. Furthermore, the previous measurement was also performed at TUNL using the SPS. 

Studying the information reported in Ref.~\cite{hale_2004} it was noticed that energies were only reported for a region of $\approx 400$ keV. Outside of this region the identity of twelve states were assumed for the calibration. Of these twelve states, the most interior, i.e, the last states before the interpolated region, were states identified as $11330$ keV and $12184$ keV. Comparing the spectrum from this work and that shown in Fig.~3 of Ref.~\cite{hale_2004}, the state labeled $11330$ keV in their spectrum corresponded to the state that our calibration identified as $11317(3)$ keV. Ref.~\cite{firestone_2007} lists two states around this energy range, one with $E_x = 11314.4(15) $ keV and the other $E_x = 11330.2(10)$ keV. Neither of these states has a definite spin parity in the current evaluation, but the compilation of Ref.~\cite{endt_eval_1990}, which was used as a reference in the previous study, identified the lower energy state as $(3,4)^+$  and the higher as $(2^+ \text{-} 4^+)$. These assignments seem to be in tension with the $(p, p^{\prime})$ angular distribution of Ref.~\cite{zwieglinski_1978}, which assigns the lower lying state $\ell=3$ giving $J^{\pi} = 3^{-}$. However, Ref.~\cite{Warburton_1981} reports $\log ft = 5.19(14)$ for $^{24}$Al$(\beta^+)$ ( ground state $J^{\pi} = 4^+$), which based on the empirical rules derived in Ref.~\cite{Raman_1973} requires an allowed decay giving $(3, 4, 5)^+$ for this state. In light of these discrepancies, it is hard to reach a firm conclusion about the identity of the state populated in this work and Ref.~\cite{hale_2004}.

One method to resolve the disagreement is to recalibrate our data using the calibration states of the previous study. This cannot be considered a one-to-one comparison because of the Bayesian method used to calibrate the focal plane, but it should show the impact of misidentifying the state around $11320$ keV. To be specific, I consider three sets of energies:
\begin{enumerate}
    \item The results of this work from Section \ref{sec:energy_cal_na} (Set $\#1$).
    \item The peak centroids of this work energy calibrated using the calibration states of Hale \textit{et al.} (Set $\#2$).
    \item The energies reported in Ref.~\cite{hale_2004} (Set $\#3$).
\end{enumerate}
The results shown in Table \ref{tab:hale_comp_energies} report these three sets of energies. As can be seen, using the same calibration for our data (Set $\#2$) produces consistent results with the previous study (Set $\#3$).

The above discussion presents the evidence that led to the decision to exclude excitation energies of Ref.~\cite{hale_2004} from the recommended energies of the current work. There is a reasonable cause to do this at the current time, but further experiments are needed to firmly resolve this issue. 

\begin{table}
\centering
\setlength{\tabcolsep}{8pt}
\caption{ \label{tab:hale_comp_energies} Comparison of the $^{24}$Mg excitation energies measured in this work (Set $\# 1$), the excitation energies derived from our data if the calibration of Hale \textit{et al.} is used (Set $\# 2$), and finally the energies Hale \textit{et al.} reported in Ref.~\cite{hale_2004} (Set $\# 3$). These results indicate that the state close to $11320$ keV was previously misidentified, and, as a result, led to systematically higher excitation energies.}
\begin{threeparttable}
\begin{tabular}{lllllllll}
\toprule
\toprule
Set $\# 1$     & Set $\# 2$ &   Set $\# 3$ \\ \hline
$11688.7(14$) & $11695(3)$     & $11698.6(13)$ \\
$11823(3)$    & $11828(3)$     & $11831.7(18)$ \\
$11857(3)$    & $11860.1(19)$  & $11862.7(12)$ \\
$11935(3)$    & $11937.5(17)$  & $11936.5(12)$ \\
              &                & $11965.3(12)^{\dagger}$ \\
$11989.3(14)$ & $11991.2(17)$  & $11992.9(12)$ \\
$12014(3)$    & $12016.2(16)$  & $12019.0(12)$ \\
$12050(3)$    & $12051.4(17)$  & $12051.8(12)$ \\
\bottomrule
\bottomrule
\end{tabular}
\begin{tablenotes}
\item[$\dagger$] Ref.~\cite{hale_2004} reports this state, which appears as an unresolved peak in their spectrum. The current study does not find a corresponding peak in the same region.
\end{tablenotes}
\end{threeparttable}
\end{table}

\subsection{Suggested Energies for Astrophysically Relevant States }

The recommended energies based on the measurements of this work, the compilation of Section \ref{sec:energy_level_update}, and the removal of Ref.~\cite{hale_2004} from consideration are presented below. Note that states not measured in this work are listed for completeness. All values come from a weighted average, except for the $11695$-keV state. This state has extreme tension between the two most precise measurements, which are those of this work and those of Ref.~\cite{schmalbrock_1983}. To reflect this disagreement in the uncertainty, the \textit{expected value method} of Ref.~\cite{BIRCH_2014} was used.

\begin{table}[]
\centering
\setlength{\tabcolsep}{8pt}
\caption{ \label{tab:recommened_energies} The recommended resonance energies for astrophysically relevant states. These energies were derived from the energies of this work and the compiled energies of Section \ref{sec:energy_level_update}, and using the $Q$-value calculated from Ref.~\cite{Wang_2017}.}
\begin{tabular}{ll}
\toprule
\toprule
$E_x$ (keV)      & $E_r$ (keV)        \\ \hline
$11389.6(12)$  & $-303.09(12)$ \\
$11452.9(4) $  & $-239.8(4)$   \\
$11521.1(14)$  & $-171.6(14)$  \\
$11695(5)   $  & $2(5)$        \\
$11729.8(16)$  & $37.1(16)$    \\
$11825(3)   $  & $132(3)$      \\
$11860.8(14)$  & $168.1(14)$   \\
$11933.06(19)$ & $240.37(19)$  \\
$11966.7(5) $  & $274.2(8)$    \\
$11988.45(6)$  & $295.76(6)$   \\
$12016.8(5) $  & $324.1(5)$    \\
$12051.3(4) $  & $358.6(4)$    \\
$12119.5(9) $  & $426.8(9)$    \\
$12183.3(1) $  & $490.6(1)$    \\
$12341.0(4) $  & $648.3(4)$    \\
$12405.3(3) $  & $712.6(3)$    \\
$12528.5(6) $  & $835.8(6)$    \\
$12576(3)   $  & $883(3) $     \\
$12669.9(4)  $ & $977.2(4)$    \\
$12738.8(7)  $ & $1046.1(7)$   \\
$12817.77(19)$ & $1125.08(19)$ \\
$12852.2(5)  $ & $1159.5(5) $  \\
$12921.6(4)  $ & $1228.9(4) $  \\
$12963.9(5)  $ & $ 1271.2(5) $ \\
\bottomrule
\bottomrule
\end{tabular}
\end{table}

\section{Background Subtraction for the Region of Interest}
\label{sec:background_subtraction}

Energy calibration revealed that the $11823$-keV and $11857$-keV states were obscured at multiple angles by the peak corresponding to the $7276$ keV state in $^{15}$O. This state arises from $^{14}$N contamination in the target, and its overlap with the $11857$ keV state can be seen in Fig.~\ref{fig:calibrated_na_energies}. Due to the unknown spin and parity of the $11823$ keV state, this contaminant peak presents a major challenge because it overlaps most strongly at the lower angles, which, in turn, have the most impact on the inferred $\ell$ assignment. In order to accurately subtract the contribution from the contaminant peak, it was decided to analyze these three states simultaneously using a Bayesian model. Because the spin and parity of the $11857$ keV state ($1^{-}$) and the $7276$ keV state (${7/2}^+$) of $^{15}$O are known, simultaneous analysis of this region can guarantee that the extracted yields are consistent with all known information.

Since the three states are not clearly resolved, it is necessary to constrain the shape, area, and location of the contaminant peak. The data at $\theta_{lab} = 13^{\circ}$ and $15^{\circ}$ are the only spectra which show a clear signal from the nitrogen contamination. Thus, it was decided to use the average of the peak parameters at these two angles to model the contribution of the nitrogen peak at the other angles. The centroid location can, in principle, be predicted from the energy calibration. However, the internal calibration for the sodium states does not accurately account for energy loss in the target. It was necessary to systematically shift the predicted peak locations by $\approx 4$ channels in order to match the observations at $13^{\circ}$ and $15^{\circ}$. Additionally, at several angles there were sufficient statistics in the background runs on the $^{\textnormal{nat}}$C target to check this correction. The $4$ channel offset was found to give excellent agreement with these observations. This agreement also indicates that the nitrogen is primarily located in the carbon backing.  

The predicted area of the $^{15}$O state for the angles, at which it was not directly observed, was calculated using DWBA. Absolutely determining this quantity is infeasible without knowing the amount of nitrogen in the target and the spectroscopic factor of the state. However, the ratio of the DWBA cross sections gives a scaling factor that can convert the observed yields at $13^{\circ}$ and $15^{\circ}$ to predicted yields at the angles of interest. Furthermore, a ratio reduces the dependence on the chosen optical potential for the calculations. Thus, the predicted yield at an angle $\theta_i$ is given by:
\begin{equation}
    \label{eq:predicted_yield_N}
    \frac{dY}{d \Omega}(\theta_i) = \bigg[ \frac{d \sigma}{d \Omega}\underset{\textnormal{DWBA}}{(\theta_i)} / \frac{d \sigma}{d \Omega}\underset{\textnormal{DWBA}}{(\theta_{13, 15})} \bigg] \frac{dY}{d \Omega}(\theta_{13, 15}),
\end{equation}
where $\theta_{13,15}$ denotes either $13^{\circ}$ or $15^{\circ}$. The results of Ref.~\cite{bertone} indicate that the $7276$ keV state of $^{15}$O is characteristic of an $\ell = 2$ transfer. A zero-range DWBA calculation was carried out for a $1d_{7/2}$ state using the Becchetti and Greenlees $^{3}$He global optical potential, Ref.~\cite{b_g_3he}, and the Daehnick $d$ optical potential \cite{daehnick_global}. Another calculation was carried out for the potentials listed in Ref.~\cite{bertone}. As expected these two results were in excellent agreement. The deviation was $ < 1 \%$ for angles $< 20^{\circ}$ . The global calculations were adopted, and the yields from both $13^{\circ}$ and $15^{\circ}$ were scaled for angles $5^{\circ} \text{-} 11^{\circ}$. The average between these two predictions was adopted. These yields were then transformed back into the lab frame, and Eq.~\ref{eq:number_of_reactions} and Eq.~\ref{eq:number_of_beam} were used to calculate the expected number of counts in the focal plane spectrum.     

\begin{figure}
    \centering
    \includegraphics[width=.75\textwidth]{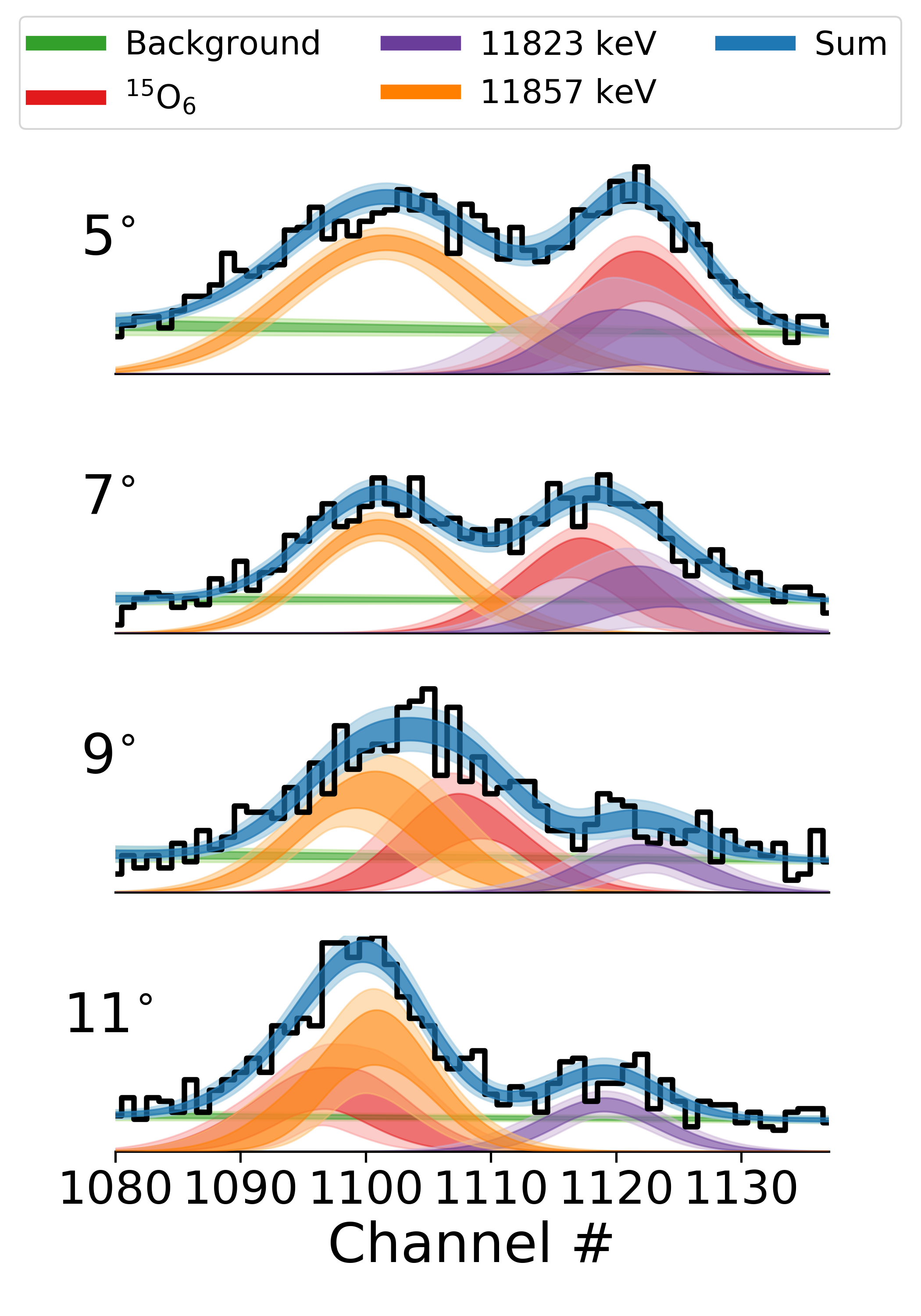}
    \caption{Results of the Bayesian peak fitting. The bin numbers have been shifted to approximately align the $11823$ and $11857$ keV peaks. This procedure shows the clear kinematic shift of the $^{15}$O$_6$ peak, and the corresponding change in the shape of the doublet. Dark bands are $68 \%$ credibility intervals, and the light bands are $95 \%$ intervals. }
    \label{fig:background_subtraction_roi}
\end{figure}

\begin{figure}
    \centering
    \includegraphics[width=.8\textwidth]{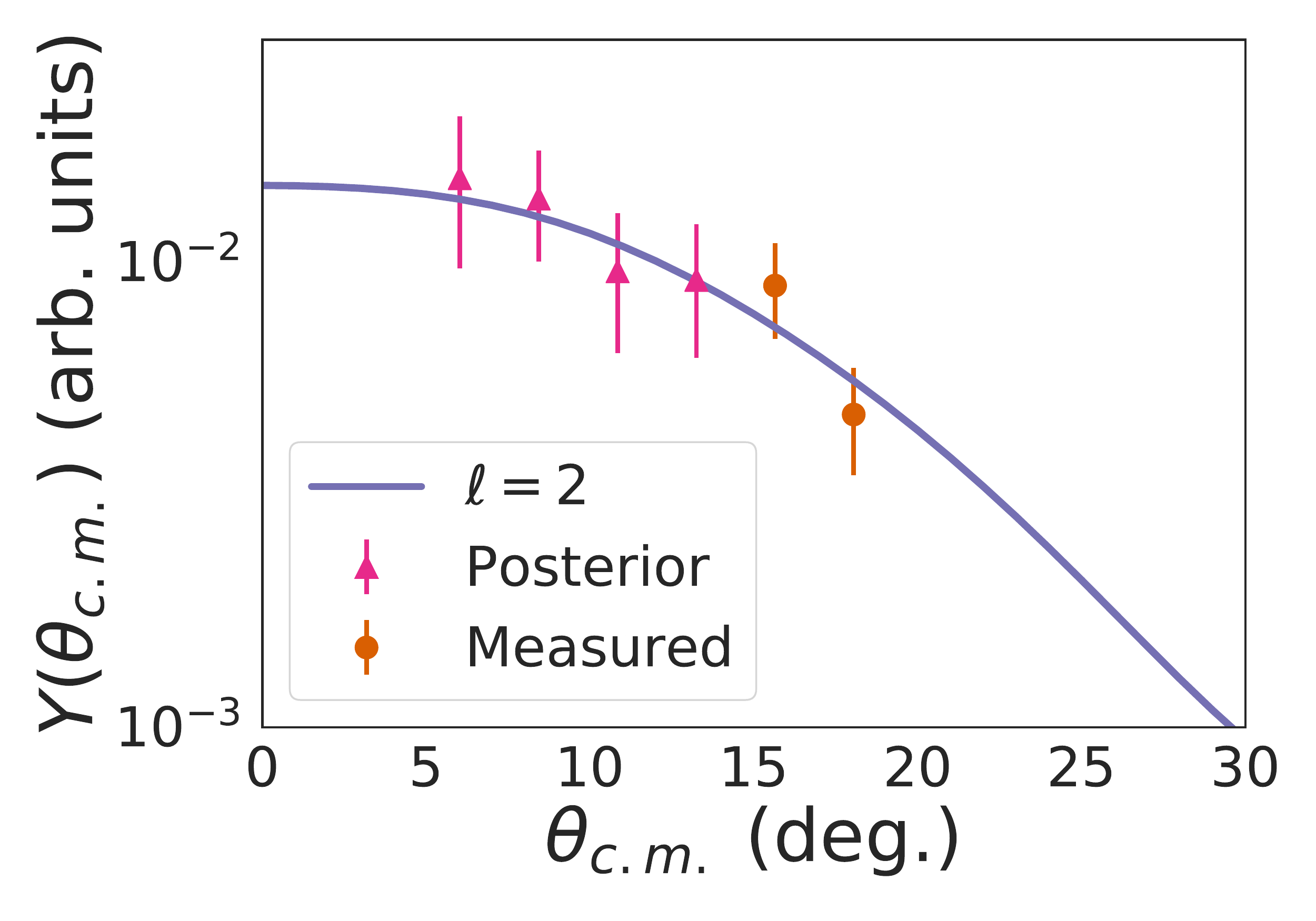}
    \caption{DWBA prediction for the $7276$ keV state in $^{15}$O  using the global potentials described above. The results of the posterior values for the yields are the pink triangles, while the measured yields are shown in orange. This shows that the background subtraction is consistent with the known information about this state.}
    \label{fig:dwba_14N}
\end{figure}

A Bayesian model was constructed to account for all of the above considerations. The priors for the parameters of the contaminant peak are given by the predictions of the energy calibration, the observation of the FWHM at $13^{\circ}$ and $15^{\circ}$, and the DWBA calculations for the areas. The DWBA areas were assigned an uncertainty of $30 \%$. The priors for the states of interest were constructed based on the rough location of the peaks in the doublet, the widths of nearby states, and flat priors for the area. A linear background was also considered, and its priors were broad normal distributions. The likelihood function was selected to be a Poisson distribution for each bin:
\begin{equation}
    \label{eq:poisson_like}
    \textnormal{Poisson}(\mathbf{C}, \mathbf{D}) = \prod_i^N \frac{D_i^{C_i}}{C_i!}e^{-D_i},
\end{equation}
where $C_i$ is the number of counts in bin $i$, $D_i$ is the total number of predicted counts in bin $i$, and $N$ is the total number of bins. The model becomes:
\begin{align}
  \label{eq:background_bayesian_model}
    & \textnormal{Contaminant Priors:} \nonumber \\
    & c^{\prime} \sim \mathcal{N}(c_{cal}, \sigma^2_{cal}) \nonumber \\
    & w^{\prime} \sim \mathcal{N}(3.878, \{0.794\}^2) \nonumber \\
    & A^{\prime} \sim \mathcal{N}(A_{\textnormal{DWBA}}, \{ 0.30A_{\textnormal{DWBA}}  \}^2) \nonumber \\
    & \textnormal{Other Priors:} \nonumber \\
    & c_{j} \sim \mathcal{N}(c^{obs}_{j}, 5.0^2_{j}) \nonumber \\
    & w_j \sim \mathcal{N}(w_{obs}, 1) \nonumber \\
    & A_j \sim \textnormal{Uniform}(0, 10^4) \\
    & a \sim \mathcal{N}(1, 100^2) \nonumber \\
    & b \sim \mathcal{N}(100, 100^2) \nonumber \\
    & \textnormal{Function:} \nonumber \\
    & B_{i} = a x_{i} + b \nonumber \\
    & f(x_i ; A, w, c) = \frac{A}{\sqrt{2 \pi w^2}} e^{\frac{(x_i-c)^2}{2 w^2}} \nonumber \\
    & S_i = f(x_i; A^{\prime}, w^{\prime}, c^{\prime}) + \sum_j f(x_i; A_j, w_j, c_j)  \nonumber \\
    & D_{i} = S_i + B_i  \nonumber \\
    & \textnormal{Likelihood:} \nonumber \\
    & \textnormal{Poisson}(\mathbf{C}, \mathbf{D}), \nonumber
\end{align}
where the primed variables correspond to the contaminant peak, the bin number is denoted $i$, the $x$ position of the bin is $x_i$, $cal$ are the predictions of the energy calibration, \say{DWBA} are the areas derived from the $\textnormal{DWBA}$ scaling, $obs$ refers to the observed widths of nearby peaks and the rough position of the doublet peaks, $j$ is the index that runs over the two peaks, $a$ is the slope of the linear background, $b$ is its intercept, $B_i$ is the total number of background counts in bin $i$, and $S_i$ is the total number of signal counts in a bin $i$ (contaminant peak plus $^{24}$Mg peaks). This model is applied to each angle individually, and the credibility intervals for resulting peaks are shown in Fig.~\ref{fig:background_subtraction_roi}. As a check, the posterior distributions for the number of counts in the contamination peak were used to recalculate the expected yields, and compare to the DWBA cross section. This calculation is shown in Fig.~\ref{fig:dwba_14N}, and it shows that the posterior values are consistent with the theoretical angular distribution that generated their priors. Once the background subtraction yielded the parameters for the peaks, the centroids were plugged into the energy calibrations and the yields were calculated. At $5^{\circ}$, the $11823$-keV state had a posterior area that was consistent with zero. This angle was treated as an upper limit, and was excluded from the weighted average for the energy and subsequent DWBA calculations.  

{}

\section{Yield Determination}

Extraction of $C^2S$ for a state requires that the absolute scale of the differential cross section is known. Uncertainties associated with the target thickness and beam integration in the current experiment make the determination of an absolute scale subject to uncontrolled systematic effects. Beam and target effects can be removed by normalizing the data from the focal plane to the $^{3}$He elastic scattering measured by the monitor detector positioned at $45^{\circ}$. An absolute scale can be established by inferring an overall normalization from the comparison of the relative elastic scattering angular distribution collected in the focal plane to the optical model predictions. Similar approaches can be found in Refs.~\cite{hale_2001, hale_2004, vernotte_optical}     

To clarify the point of the relative measurement, lets assume that two detectors measure the reaction products of a single reaction on a single target. In this case, detector $1$ is sensitive to the differential cross section $\frac{d \sigma}{d \Omega}( \theta_1 )$ at laboratory angle $\theta_1$, while detector $2$ measures $\frac{d \sigma}{d \Omega}( \theta_2 )$ at laboratory angle $\theta_2$. The ratio of these two measurements will, therefore, be insensitive to shared properties of the cross section at both angles. These include target stoichiometry, target thickness, and charge collection. Thus, if a relative measurement is made, the ratio will be free from the influence of these effects at the cost of no longer being directly comparable to theoretical calculations of the cross section. Considering the above, it can be seen that all that needs to be calculated in the case of a relative measurement is the differential reaction yield for both detectors:
\begin{equation}
    \label{eq:reaction_yield}
    \frac{dY}{d \Omega} = \frac{N_R}{N_B \Omega},
\end{equation}
where $N_R$ is the number of reactions, $N_B$ is the number of beam particles, and $\Omega$ is the detector solid angle. Returning to the $^{23}$Na$(^3 \textnormal{He}, d)$ data, $N_R$ is proportional to the area of a peak on the focal plane or the monitor detector. This area needs to be corrected for the livetime of the DAQ:
\begin{equation}
    \label{eq:number_of_reactions}
    N_R = \frac{A}{t_{live}}.        
\end{equation}
$N_B$ is derived from the BCI reading, which is a scalar value that represents the number of BCI pulses per coulomb. The BCI full-scale for this experiment was set to  $10^{-10}$ C$/$pulse. Combining these terms:
\begin{equation}
    \label{eq:number_of_beam}
    N_B = \frac{BCI \times full\text{-}scale}{eq},
\end{equation}
where $q$ is the charge state of the beam and $e$ is the elementary charge. Finally, the geometric solid angles of the detectors are $\Omega_{FP} = 1.00(4)$ msr and $\Omega_{Si} = 4.23(4)$ msr for the focal plane and monitor detector, respectively.

\subsection{Monitor Detector Analysis}

\begin{figure}
    \centering
    \includegraphics[width=\textwidth]{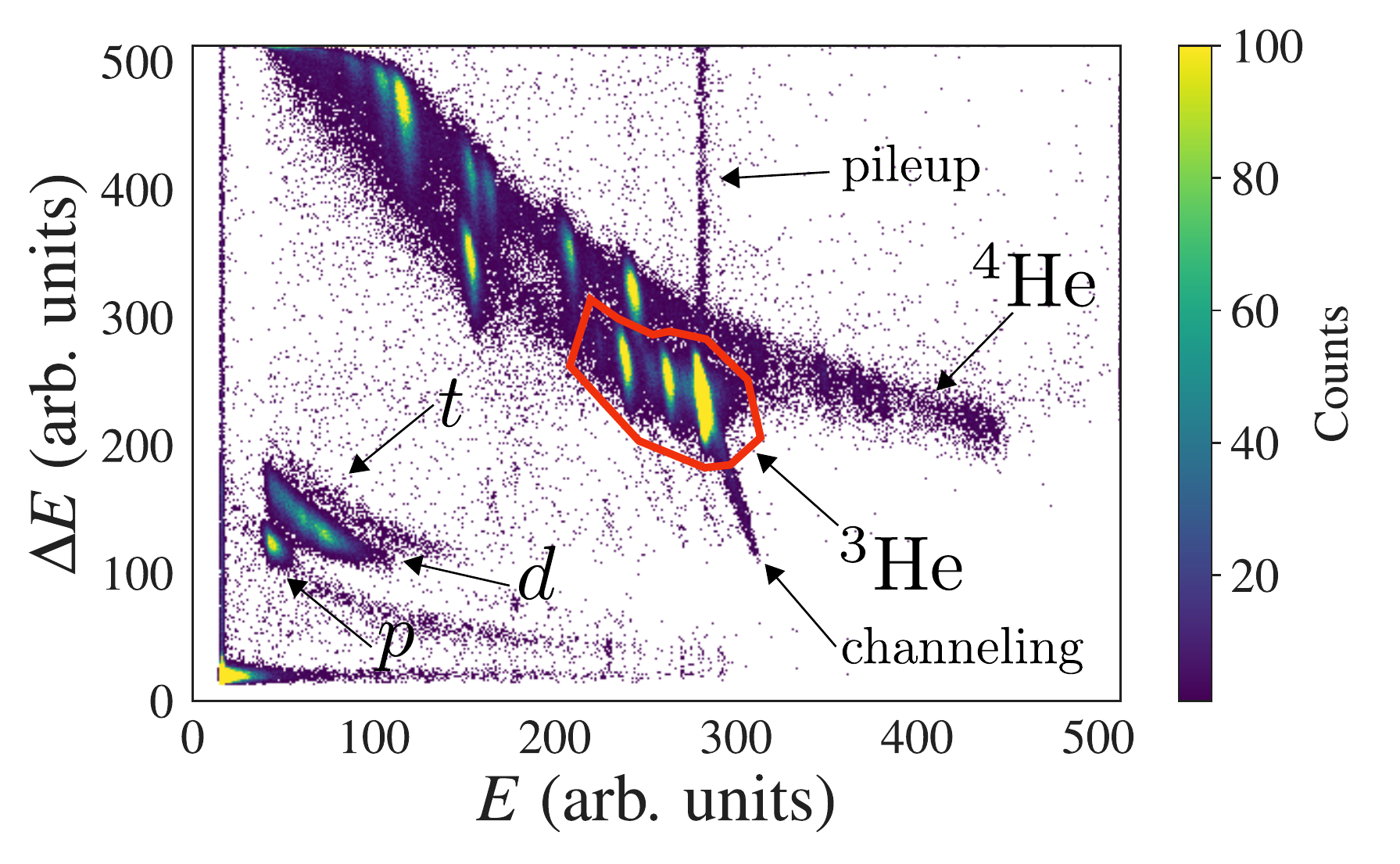}
    \caption{2D $E/ \Delta E$ for the monitor detector. This spectrum was taken at $45^{\circ}$. The orange contour is drawn around the elastic scattering peaks. It can be seen that the intensity of the Br peaks caused significant pile up in the $\Delta E$ detector in addition to channeling.}
    \label{fig:e_de_si}
\end{figure}

Extraction of the areas for peaks in the focal plane was already discussed in Section \ref{sec:peak_fitting_fp}. However, the relative method also requires detailed analysis of the monitor detector spectrum, with the goal being to extract the counts from $^{3}$He elastically scattering off of $^{23}$Na. Fig.~\ref{fig:e_de_si} shows the spectrum obtained with the monitor detector positioned at $45^{\circ}$, as it was throughout the experiment. Although the $^3$He overlaps mildly with the $^4$He band, it was still possible to gate effectively on the elastic scattering states. 

\begin{figure}
    \centering
    \includegraphics[width=\textwidth]{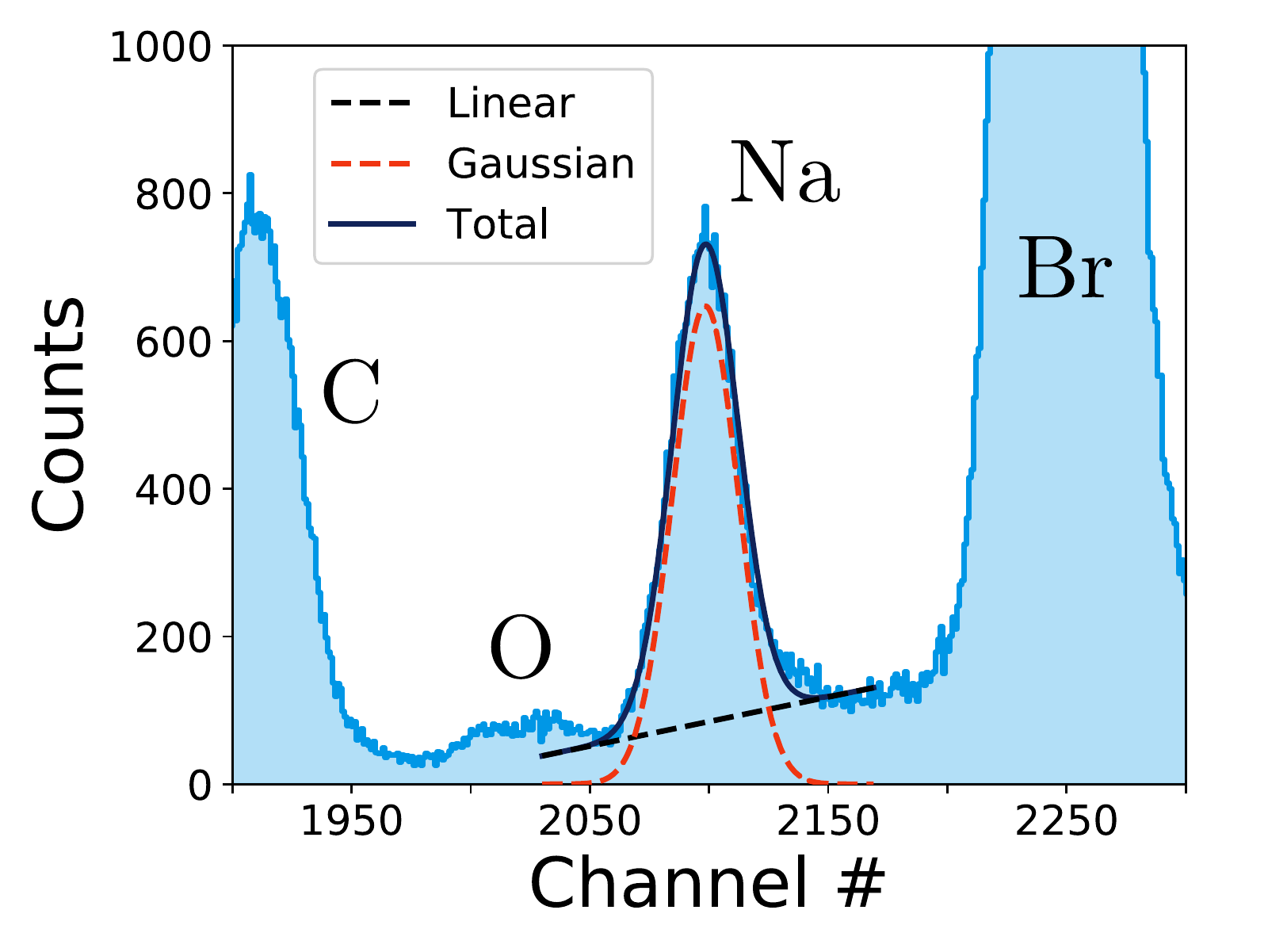}
    \caption{Monitor detector spectrum for a single run. An example of the Gaussian fit used to deduce the area of the sodium peak is shown.}
    \label{fig:si_spectrum}
\end{figure}

Peak fitting was performed on the gated $E$ spectrum. An energy calibration was not performed owing to the simplicity of the region of interest. In particular, the states that can be easily identified and used for calibration are the elastic scattering peaks, which, in turn, are the only peaks of interest. An energy calibration would also allow the creation of a $\Delta E + E$ spectrum that could be used to improve resolution, but the resolution of the gated $E$ spectrum was sufficient to make this step unnecessary. The monitor detector yields were analyzed for each individual DAQ run file. These run files were started and stopped approximately every hour during data taking. Examining the data in this granular fashion was done in order to look for systematic target degradation, which, if present, would need to be accounted for in the analysis of the focal plane spectrum as well. Gaussian fits with linear backgrounds were used to extract the area of the carbon and sodium peaks. An example fit for one of the run files is shown in Fig.~\ref{fig:si_spectrum}. The extracted counts for the sodium peak were highly sensitive to the selected fitting region. This sensitivity was caused by the tail of the bromine peak. Changes in the fit regions were capable of producing changes of $\approx 10 \%$ in the extracted area. To reduce variations between spectra, the same region was used in the fit each time. 

\begin{figure}
    \centering
    \includegraphics[width=\textwidth]{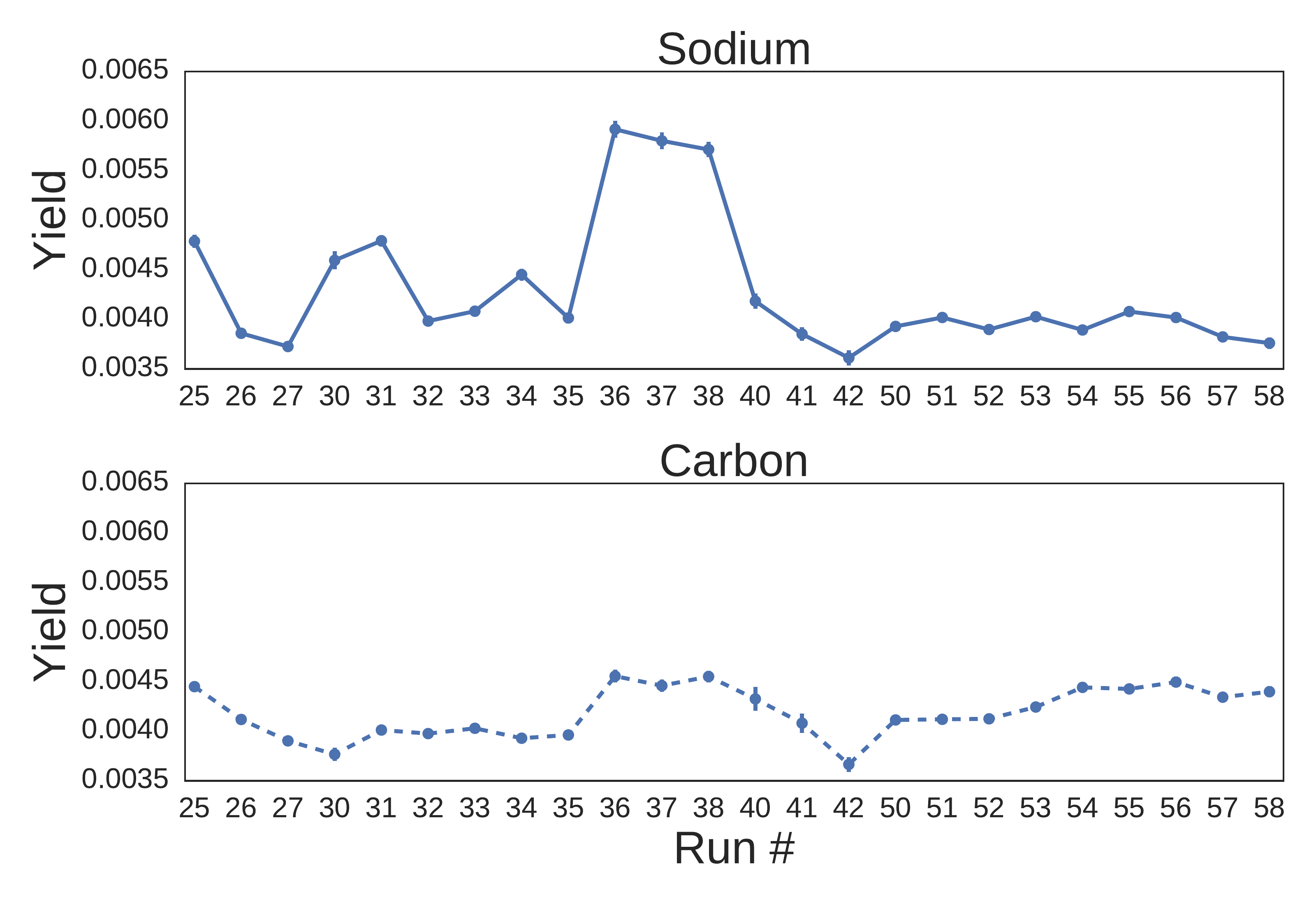}
    \caption{Yield fluctuations for the carbon and sodium peaks. All of these runs used the same NaBr target. These runs constitute the data for angles $9^{\circ} \text{-} 15^{\circ}$.}
    \label{fig:yield fluctuations}
\end{figure}

It can be seen in Fig.~\ref{fig:yield fluctuations} that the sodium peak experienced far more variance than the carbon peak. The statistical uncertainties for all of these points amount to less than $2 \%$ because of the forward angle of the detector. Several causes for this behavior were explored such as changing beam angle, DAQ issues, and target effects. The only satisfactory explanation comes from target effects. This conclusion is based on the relatively low variation in the carbon yield, the majority of the strength of the carbon yield comes from the target backing, and the gradual reduction in variation of the sodium yield. A plausible explanation in light of these facts is that the evaporation of the NaBr targets produced a highly nonuniform layer of material. This nonuniformity increased the magnitude of the fluctuations that were seen in the carbon backing. Once the target had been exposed to beam for some period of time, the nonuniform layer was sputtered off, and the fluctuations in the yields decreased. 

Accounting for the fluctuations in the monitor detector yields was done by doing a simple average of the observations at each angle. These averaged yields represent systematic effects, but it is clear our measurements scatter significantly from angle to angle. Thus, the run to run variation was treated as an additional statistical uncertainty when used to normalize the focal plane data. This approach gives significant additional contributions to the total uncertainties for $5^{\circ}$ and $9^{\circ}$, where the scatter amounts to $24 \%$ and $14 \%$, respectively. The average yields and their associated uncertainties are shown in Table \ref{tab:silicon_table}. Close inspection of these data support the hypothesis that the targets initially had a nonuniform layer, with the highest percentage uncertainties corresponding to the first angle measured with each target (note that $3^{\circ}$ was measured after $17^{\circ}$).    

\begin{table}
\centering
  \setlength\tabcolsep{6pt}
  \def\arraystretch{1.2}
\caption{\label{tab:silicon_table} Average yield values for the monitor detector at each SPS angle. The monitor detector was fixed at an angle of $\theta_{mon} = 45^{\circ}$.}
\begin{tabular}{lcl}
\toprule
\toprule
$\theta_{SPS}$ & NaBr Target & Average Yield (msr$^{-1}$)             \\ \hline 
$3^{\circ}$              & $\#3$           & $2.84(12) \times 10^{-9}$ \\
$5^{\circ}$              & $\#1$           & $1.9(4)\times 10^{-9}$    \\
$7^{\circ}$              & $\#1$           & $2.18(12)\times 10^{-9}$  \\
$9^{\circ}$              & $\#2$           & $2.6(4)\times 10^{-9}$    \\
$11^{\circ}$             & $\#2$           & $2.75(22)\times 10^{-9}$  \\
$13^{\circ}$             & $\#2$           & $3.70(6)\times 10^{-9}$   \\
$15^{\circ}$             & $\#2$           & $2.54(5) \times 10^{-9}$  \\
$17^{\circ}$             & $\#3$           & $2.79(18) \times 10^{-9}$ \\
$19^{\circ}$             & $\#3$           & $2.57(15) \times 10^{-9}$ \\
$21^{\circ}$             & $\#3$           & $2.40(7) \times 10^{-9}$  \\
\bottomrule
\bottomrule
\end{tabular}
\end{table}

\subsection{Final Form of the Data}
\label{sec:scale_of_na_data}

With the extraction of the monitor detector yields, it is now possible to express all of the angular distributions in terms of the ratio:
\begin{equation}
    \label{eq:yield_ratio}
    \frac{dY}{d \Omega}(\theta_i) = \bigg[ \frac{d Y}{d \Omega}_{FP}{(\theta_i)} / \frac{d Y}{d \Omega}_{mon} \bigg],
\end{equation}
for the yield of a peak measured by the spectrograph at center of mass angle $i$. The dependence of these quantities on the solid angle means they must be converted to the center of mass frame before the ratio is taken. Each ratio comes with an uncertainty given by:
\begin{equation}
    \label{eq:yield_ratio_unc}
    \sigma_{Y} = \sqrt{\sigma_{FP}^2 + \sigma_{mon}^2},
\end{equation}
where $\sigma_{FP}^2$ is the statistical uncertainty of the yield measured on the focal plane, while $\sigma_{mon}^2$ is the statistical uncertainty in the monitor detector coming from both the counting uncertainty and the averaging of the runs.   

\section{Bayesian DWBA Analysis}

All of the measured angular distributions are expressed in terms of a ratio to the elastic scattering measured in the monitor detector at $\theta_{mon} = 45^{\circ}$. In order to extract spectroscopic factors from these data, an absolute normalization must be obtained for these data. Following the Bayesian method of Section \ref{sec:bay_dwba}, this is done by estimating the posterior distribution for $\eta$, which normalizes the measured elastic scattering data for the entrance channel to the predictions of the chosen optical potential. Since this analysis simultaneously extracts $\eta$ as well as $C^2S$, the uncertainty in the normalization will naturally be reflected in the uncertainty of $C^2S$.     

As will be shown, the present elastic scattering data set differs substantially from the $^{70}$Zn$(d, d)$. It is therefore necessary to modify the Bayesian model to account for these differences. 

\subsection{Global Potential Selection}

The first attempts to fit the elastic scattering data 
used the optical model from the lab report of Beccehetti and Greenless \cite{b_g_3he}. The imaginary depth of this potential for a beam of $E_{^{3}\textnormal{He}}=21.249$ MeV on $^{23}$Na is $36$ MeV. This is nearly twice as deep as the values reported in the more recent works of Trost \textit{et al.} \cite{TROST_1980}, Pang \textit{et al.} \cite{pang_global} and Liang \textit{et al.} \cite{Liang_2009}. Although these works prefer a surface potential, the work of Vernotte \textit{et al.} \cite{vernotte_optical} is parameterized by a volume depth, and also favours depths around $20$ MeV. The overly deep well depth is a major issue for the Bayesian analysis method because the priors are centered around the global values (see Section \ref{sec:model}). It was observed that the data preferred a lower depth, thereby causing a bi-modal posterior, with one mode centered around the global depth and the other resulting from the influence of the data.  Based on these results, a decision was made to use the Liang potential (Ref.~\cite{Liang_2009}) because of its applicability in the present mass and energy range. It should be noted that the potential as presented in Ref.~\cite{Liang_2009} has an imaginary spin-orbit part, but there is little evidence from their data set to support its inclusion in the optical model. Therefore, the imaginary portion of the spin-orbit has been excluded from this analysis.  Finally, these results suggest against the future use of the Beccehetti and Greenless $^3$He and $t$ optical potential.  

All of the potentials used in the following analysis are listed in Table~\ref{tab:opt_parms_na}. The deuteron potential is the non-relativistic \say{L} potential from Ref.~\cite{daehnick_global}. Since the region of interest is $11$-$12$ MeV, the outgoing deuterons will have an energy of $E_d \approx E_{^3\textnormal{He}} + Q_{(^3\textnormal{He}, d)} - E_x$. The potential was thus calculated for deuteron scattering at $15.5$ MeV. The bound state spin-orbit term was set to roughly satisfy $\lambda = 25$ with $\lambda \approx 180 V_{so} / V$ for values of $V$ in the above energy range.

\begin{table*}[ht]
\centering
\begin{threeparttable}[e]
\caption{\label{tab:opt_parms_na}Optical potential parameters used in this work before inference.}
\setlength{\tabcolsep}{4pt} 
\begin{tabular}{ccccccccccccccccc}
\toprule[1.0pt]\addlinespace[0.3mm] Interaction  & $V$ & $r_{0}$ & $a_{0}$ & $W$ & $W_{s}$ & $r_{i}$ & $a_{i}$ & $r_{c}$ & $V_{so}$ & $r_{so}$ & $a_{so}$ \\
                                                 & (MeV) & (fm) & (fm) & (MeV) & (MeV) & (fm) & (fm) & (fm) & (MeV) & (fm) & (fm)\\ \hline\hline\addlinespace[0.6mm]

\hspace{0.15cm} $^{3}$He $+ ^{23}$Na \tnote{a} & $117.31$ & $1.18$ & $0.67$ & & $19.87$ &$1.20$ & $0.65$ & $1.29$ & $2.08$ & $0.74$ & $0.78$ \\
$d$ $+$ $^{24}$Mg\tnote{b} & $88.1$ & $1.17$ & $0.74$ & $0.30$  & $12.30$ & $1.32$  &  $0.73$ & $1.30$ & $6.88$ & $1.07$ & $0.66$ & \\
\hspace{0.1cm}$p$ $+$ $^{23}$Na & \tnote{c} & $1.25$ & $0.65$ & & & & & $1.25$ & $6.24$ & $1.25$ & $0.65$  \\[0.2ex]
\bottomrule[1.0pt]
\end{tabular}
\begin{tablenotes}
\item[a] Global potential of Ref. \cite{Liang_2009}.
\item[b] Global potential of Ref. \cite{daehnick_global}.
\item[c] Adjusted to reproduce binding energy of the final state.
\end{tablenotes}
\end{threeparttable}
\end{table*}

\subsection{Elastic Scattering}

As stated previously, elastic scattering yields were measured for $15^{\circ} \text{-} 55^{\circ}$ in $5^{\circ}$ steps and finally at $59^{\circ}$, for a total of $10$ angles. The yields at each angle were normalized to those measured by the monitor detector. A further normalization to the Rutherford cross section was applied to the elastic scattering data to ease the comparison to the optical model calculations.   

Low angle elastic scattering cross sections in normal kinematics can be collected to almost arbitrary statistical precision, with the present data having statistical uncertainty
of approximately $2 \text{-} 7 \%$. It is likely that in this case the residuals between these data and the optical model predictions are dominated by theoretical and experimental systematic uncertainties. The Bayesian model can thus be modified to consider an additional unobserved uncertainty in the elastic channel:
\begin{equation}
  \label{eq:elastic_unc}
  \sigma_{elastic, i}^{\prime 2} = \sigma_{elastic, i}^2 + \bigg(f_{elastic} \frac{d \sigma}{d \Omega}_{optical, i} \bigg)^2,
\end{equation}
where the experimentally measured uncertainties, $\sigma_{elastic, i}$, at angle $i$ have been added in quadrature with an additional uncertainty coming from the predicted optical model cross section. This prescription is precisely the same procedure that is used for the additional transfer cross section uncertainty from Eq.~\ref{eq:unc}. However, the elastic data are the only meaningful constraints for the optical model parameters. With only $10$ data points, a weakly informative prior on $f_{elastic}$ would remove the predictive power of these data. Thus, an informative prior must be chosen. For this work, I chose the form:
\begin{equation}
  \label{eq:elastic_f}
  f_{elastic} \sim \textnormal{HalfNorm}(0.10^2).
\end{equation}
This quantifies the expectation that the data will have residuals with the theoretical prediction of about $10 \%$. The above prior was found to provide the best compromise between the experimental uncertainties, which led to unphysical optical model parameters, and solutions above $f = 50 \%$ where the data become non-predictive.  

\begin{figure}
    \centering
    \includegraphics[width=\textwidth]{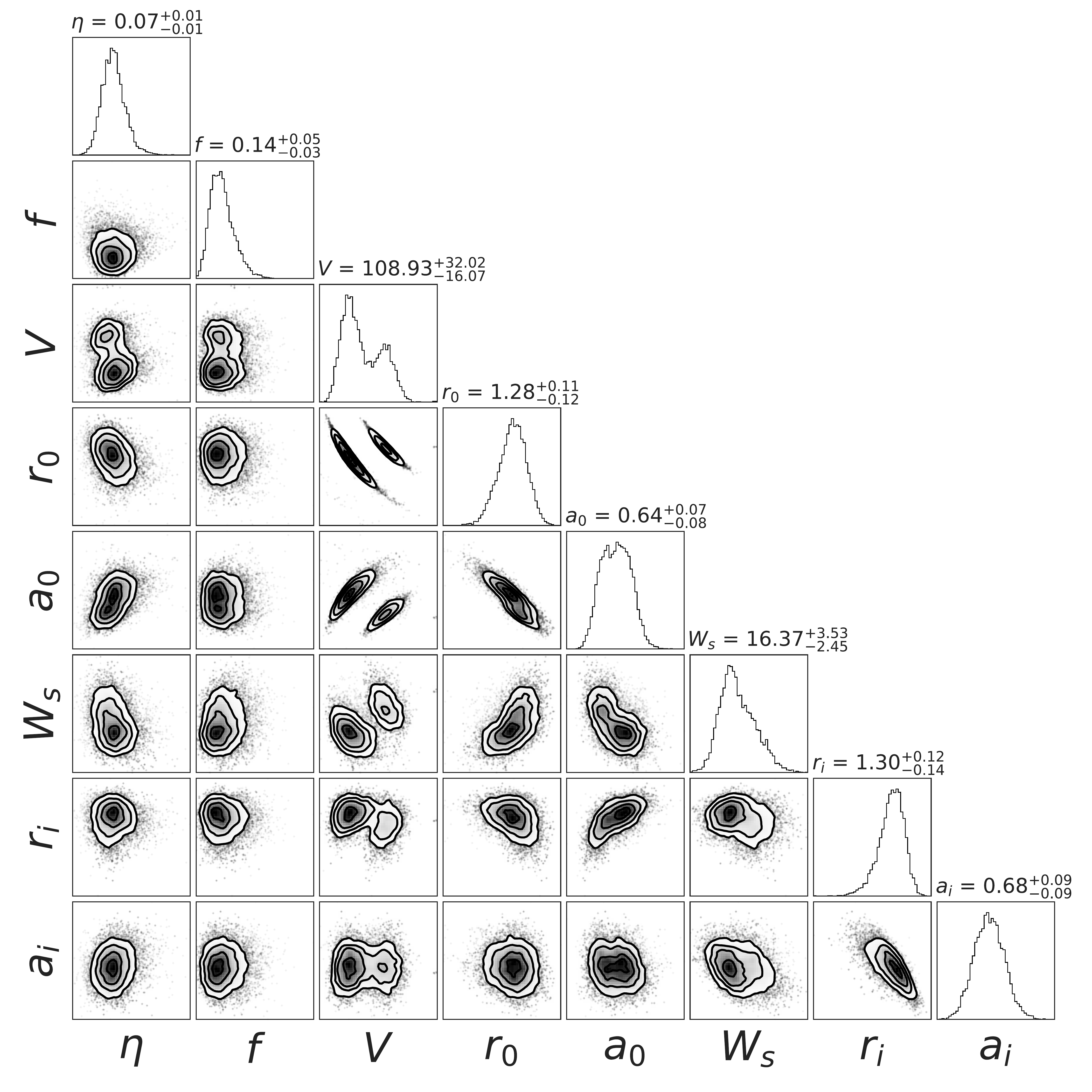}
    \caption{Pair correlation plot of the posterior samples for the nested sampling run. The discrete ambiguity is prominent in the $^{3}$He $+ ^{23}$Na data.}
    \label{fig:corner_discrete}
\end{figure}

Once the above parameter was included, the data could be reliably fit. However, it became clear that the discrete ambiguity was a serious issue in this analysis. Recall that, in Section \ref{sec:bay_dwba} the biasing of the entrance channel potential priors towards their expected physical values was sufficient to remove any other mode from the posterior. For the present data, the potential priors did little to alleviate this issue. The nested sampling algorithm in \texttt{dynesty} was used to explore both of these modes, so that methods to meaningfully distinguish them could be investigated. Nested sampling, as opposed to MCMC, can explore multi-modal distributions with ease \cite{speagle2019dynesty}, but as previously mentioned is not suited towards precise posterior estimation. A run was carried out with $1000$ live points, and required over $5 \times 10^6$ likelihood calls. The pair correlation plot of these samples is shown in Fig.~\ref{fig:corner_discrete}, and the impacts of the discrete ambiguity can be seen. 

\begin{figure}
    \centering
    \includegraphics[width=.8\textwidth]{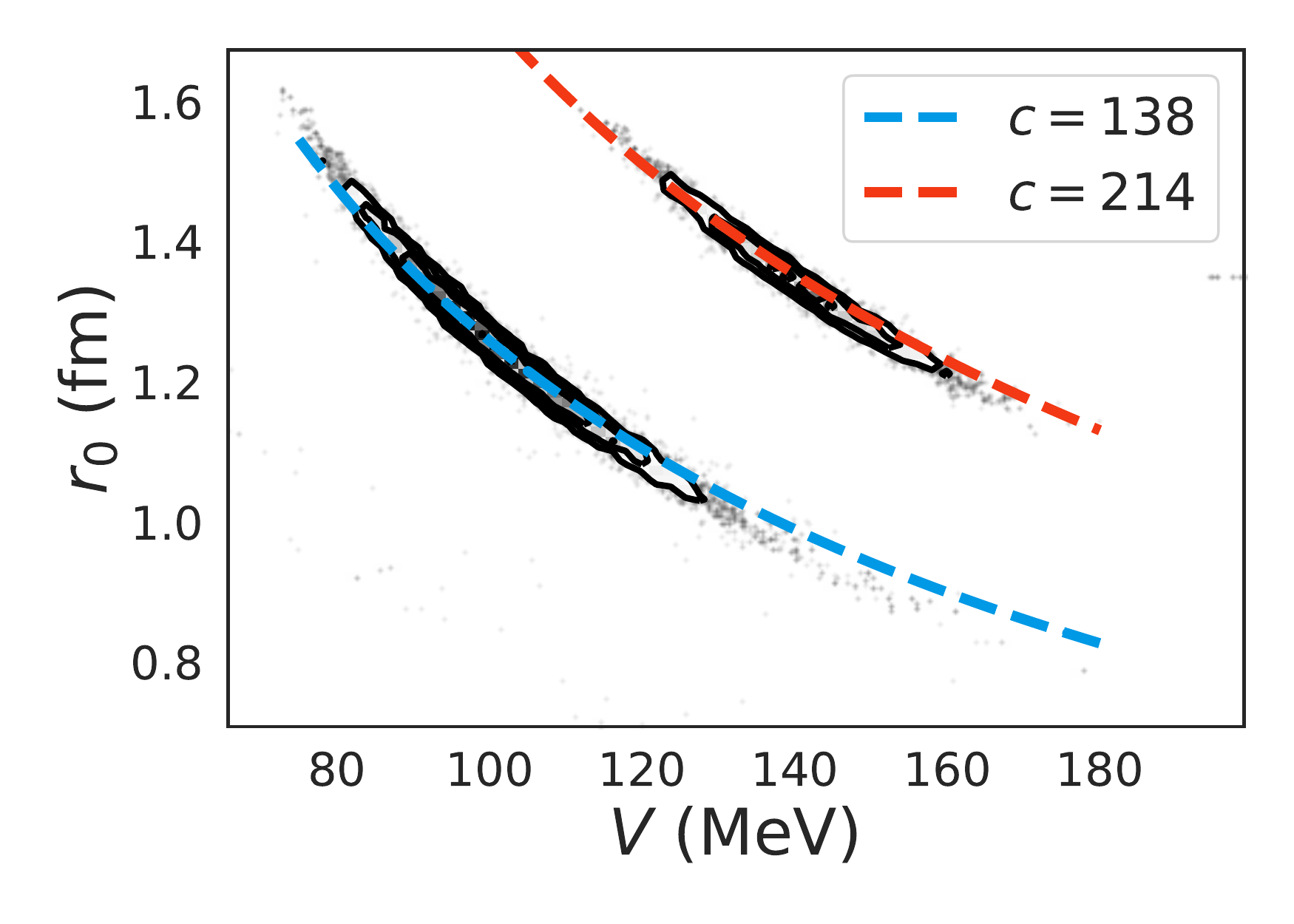}
    \caption{The discrete ambiguity as seen in the $V$ versus $r_0$ correlations between the samples of the nested sampling calculation. The colored lines show the description of the correlation based on the analytic form $Vr_0^n = c$. The value of $c$ provides a way to distinguish these modes.  }
    \label{fig:vr_constant_samples}
\end{figure}

\begin{figure}
    \centering
    \includegraphics[width=.8\textwidth]{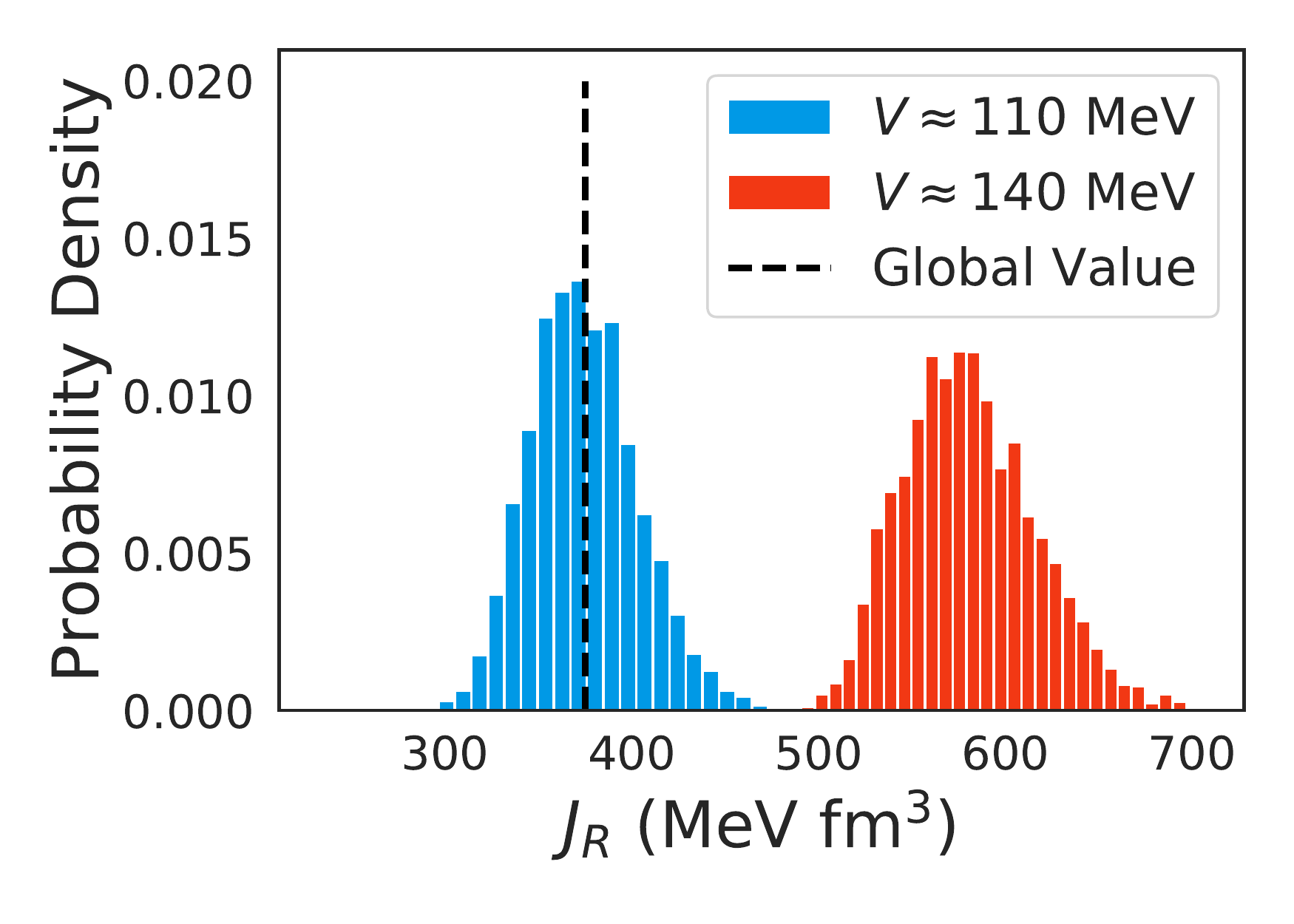}
    \caption{Values from the volume integral of the real potential as calculated using Eq.~\ref{eq:j_analytical} and the samples from the nested sampling calculation. The discrete ambiguity causes two well separated peaks to appear.}
    \label{fig:j_r_hist}
\end{figure}

Two different options were explored to differentiate the modes. The first was a simple selection of the modes based on the continuous ambiguity,  $Vr_0^n = c$. Fig.~\ref{fig:corner_discrete} shows that the correlation between $V$ and $r_0$ can cleanly resolve the two modes, while the other correlations have significant overlap between them. In this approach, the constant, $c$, is calculated for each mode, with the exponent, $n=1.14$, taken from Ref.~\cite{vernotte_optical}. It was found that the correlation in the samples was well described by this relation as shown in Fig.~\ref{fig:vr_constant_samples}. The second method that was investigated utilized the volume integral from Eq.~\ref{eq:j_int}. Ref.~\cite{varner} gives an analytical form of this integral:
\begin{equation}
 \label{eq:j_analytical}
   J_R = \frac{1}{A_P A_T} \frac{4 \pi}{3} R_0^3 V \bigg[ 1 + \bigg( \frac{\pi a_0}{R_0}\bigg)^2 \bigg],
\end{equation}
where $R_0 = r_0 A_T^{1/3}$. Calculating $J_R$ for the samples in each mode resulted in two clearly resolved peaks, as shown in Fig.~\ref{fig:j_r_hist}. 

It was decided to use the first method, and exclude the other modes via a uniform distribution based on the relationship between $V$ and $r_0$. This calculation had the advantages of being relatively simple and only involving two variables. The method based on $J_R$ has the advantage that the global values well predict the location of the peak, but its dependence on $a_0$ makes its possible effect on the posterior less clear. Integrating the $Vr_0^n$ relation into the Bayesian method requires a probability distribution be specified. A uniform distribution that covered $\pm 30 \%$ around $c$ of the physical mode was chosen. I have intentionally avoided the word prior because this condition clearly does not represent a belief about the parameter $c$ before inference. Rather, this is a \textit{constraint} enforced on the posterior to limit the inference to the physical mode \cite{Wu_2019}. It should be emphasized that the posterior distributions of all the parameters will be conditioned on $c$, i.e., $P(\theta|D, c)$. The constraint is written:
\begin{equation}
    \label{eq:c_constraint}
    c \sim \textnormal{Uniform}(c_0(1-0.30), c_0(1+0.30)),
\end{equation}
where $c_0$ is the value that is roughly centered around the lower mode. In this case $c_0 = 132.9$. As long as the distribution in Eq.~\ref{eq:c_constraint} covers all of the physical mode and excludes the unphysical ones, the value of $c_0$ and the width of the distribution should be understood to be arbitrary.

\subsection{Transfer Considerations}

A majority of the states of astrophysical interest lie above the proton threshold, and are therefore unbound. Calculation of the overlap functions, as discussed in Section \ref{sec:overlaps}, is done by using a single particle potential, with its Woods-Saxon depth adjusted to reproduce the binding energy. For unbound states, an analogous procedure would be to adjust the well depth to produce a resonance centered around $E_r$. FRESCO does not currently support a search routine to vary $V$ to create a resonance condition, meaning that $V$ would have to be varied by hand until a phase shift of $\pi/2$ is observed. Such a calculation is obviously time consuming. An alternative is the weak binding approximation. This approach assumes that the wave function of resonance scattering resembles the wave function of a loosely bound particle, typically with a binding energy on the order of $E_{bind} = 1$ keV. Studies have shown that this approximation performs well for states within $\approx 500$ keV of the particle threshold, and reproduce the unbound calculations to within $1 \%$ \cite{Kankainen_2016, KAHL_2019}. There are indications that the validity of this approximation depends on the $\ell$ value. The reasoning is that states with higher $\ell$ values more closely resemble bound states, due to the influence of the centrifugal barrier, and therefore are better described by the approximation \cite{Poxon_Pearson_2020}. For this work, DWBA calculations for states above the proton threshold were carried out with the weak binding approximation.            

Further complications arise from the non-zero ground state of $^{23}$Na ($J^{\pi} = 3/2^+$). In this case, angular distributions can be characterized by a mixture of $\ell$ transitions. Although in principle every allowed $\ell$ transition can contribute, practically speaking, it is difficult to unambiguously determine all but the lowest two $\ell$ contributions because of the rapidly decreasing cross section with increasing $\ell$ \cite{hodgson1971}.
Ignoring the light particle spectroscopic factor, the relationship between the experimentally measured differential cross section and the DWBA prediction becomes: 
\begin{equation}
  \label{eq:mixed_l}
  \frac{d \sigma}{d \Omega}_{exp} = C^2S \bigg[ \alpha \frac{d \sigma}{d \Omega}_{\textnormal{DWBA}, \ell_1} + (1-\alpha) \frac{d \sigma}{d \Omega}_{\textnormal{DWBA}, \ell_2}   \bigg],
\end{equation}
where $\alpha$ is defined such that $C^2S_{\ell_1} = C^2S \alpha$ 
and $C^2S_{\ell_2} = C^2S (1 - \alpha)$. Note that the values for $\ell$ must still obey parity conservation, meaning the most probable combinations for $(^3$He$, d)$ are $\ell = 0 + 2$ and $\ell = 1 + 3$. Incorporating multiple $\ell$ transfers into the Bayesian framework requires assigning a prior to $\alpha$. The above definitions make it clear that $\alpha = [0, 1]$; therefore, an obvious choice is:
\begin{equation}
    \label{eq:alpha_prior}
    \alpha \sim \textnormal{Uniform}(0, 1).
\end{equation}

\subsection{Bayesian Model for $^{23}$Na$(^{3}\textnormal{He}, d)^{24}$Mg}
\label{sec:spec_factors}

Before explicitly defining the Bayesian model for the DWBA analysis, the points made above should be reiterated for clarity. 

\begin{enumerate}
    \item The measured elastic scattering uncertainties have been added in quadrature with an inferred theoretical uncertainty.
    \item The $^3$He optical model used has a severe discrete ambiguity. A constraint based on the continuous ambiguity has been added to the model to select the physical mode.
    \item Due to the non-zero ground state of $^{23}$Na, the transfer cross section can have contributions from multiple $\ell$ values.
    \item Cross sections decrease rapidly with increasing $\ell$, which means the relative contributions can only be reliably determined for two distinct $\ell$ values.
    \item The relative contributions are weighted according to a parameter $\alpha$ that is uniformly distributed from $0$ to $1$.
\end{enumerate}

Folding these additional parameters and considerations in the Bayesian model of Section \ref{sec:bay_dwba} gives:
\begin{align}
  \label{eq:dwba_model_na}
 & \textnormal{Parameters:} \nonumber \\
 & n = 1.14 \nonumber \\
 & c_0 = 132.9 \nonumber \\
 & \textnormal{Priors:} \nonumber \\
 & \boldsymbol{\mathcal{U}}_{\textnormal{Entrance}} \sim \mathcal{N}(\mu_{\textnormal{central}, k}, \{0.20 \, \mu_{\textnormal{central}, k}\}^2) \nonumber \\
 & \boldsymbol{\mathcal{U}}_{\textnormal{Exit}} \sim \mathcal{N}(\mu_{\textnormal{global}, k}, \{0.10 \, \mu_{\textnormal{global}, k}\}^2) \nonumber \\
 & f \sim \textnormal{HalfNorm}(1) \nonumber \\
 & f_{elastic} \sim \textnormal{HalfNorm}(0.10^2) \nonumber \\
 & \delta D_0^2 \sim \mathcal{N}(1.0, 0.15^2) \nonumber \\
 & C^2S \sim \textnormal{HalfNorm}(1.0^2) \nonumber \\
 & g \sim \textnormal{Uniform}(-10, 10) \nonumber \\
 & \textnormal{Functions:} \\
 & \eta = 10^{g}  \nonumber \\
 & c = \mathcal{U}_{\textnormal{Entrance}, (k=0)} \big(\mathcal{U}_{\textnormal{Entrance}, (k=1)}\big)^n \nonumber \\
 & \frac{d \sigma}{d \Omega}^{\prime}_{\textnormal{Optical}, j} = \eta \times \frac{d \sigma}{d \Omega}_{\textnormal{Optical}, j} \nonumber \\
 & \frac{d \sigma}{d \Omega}^{\prime}_{\textnormal{DWBA}, i} = \eta \times \delta D_0^2 \times C^2S \times \frac{d \sigma}{d \Omega}_{\textnormal{DWBA}, i} \nonumber \\
 & \sigma_i^{\prime 2} = \sigma_{\textnormal{Transfer}, i}^2 +  \bigg(f\frac{d \sigma}{d \Omega}^{\prime}_{\textnormal{DWBA}, i}\bigg)^2 \nonumber \\
 & \sigma_{elastic, i}^{\prime 2} = \sigma_{elastic, i}^2 + \bigg(f_{elastic} \frac{d \sigma}{d \Omega}_{optical, i} \bigg)^2 \nonumber \\
 & \textnormal{Likelihoods:} \nonumber \\
 & \frac{d \sigma}{d \Omega}_{\textnormal{Transfer}, i} \sim \mathcal{N}\bigg(\frac{d \sigma}{d \Omega}^{\prime}_{\textnormal{DWBA}, i}, \sigma_i^{\prime \, 2} \bigg) ,  \nonumber \\
 & \frac{d \sigma}{d \Omega}_{\textnormal{Elastic}, j} \sim \mathcal{N}\bigg(\frac{d \sigma}{d \Omega}^{\prime}_{\textnormal{Optical}, j}, \sigma_{elastic, i}^{\prime 2} \bigg) ,  \nonumber \\
 & \textnormal{Constraint:} \nonumber \\
 & c \sim \textnormal{Uniform}(c_0(1-0.30), c_0(1+0.30)), \nonumber
\end{align}
where the index $k$ runs over the optical model potential parameters, $i$ and $j$ denote the elastic scattering and transfer cross section angles, respectively, and $\mathcal{U}_{\textnormal{Entrance}, (k=0,1)}$ are the real potential depth and radius for the entrance channel. Note that the prior for $C^2S$ has also changed from Eq.~\ref{eq:model}. Since the astrophysical states of interest are highly excited states in $^{24}$Mg, the majority of the strength for the shell they occupy has been exhausted. The expectation is that $C^2 << 1$, and the width of the prior can safely be reduced from $n_{nucleons}$ to $1.0$. In the case of a mixed $\ell$ transfer, the model has the additional terms:
\begin{align}
    \label{eq:model_mixed_l}
 & \textnormal{Prior:} \nonumber \\
 &  \alpha \sim \textnormal{Uniform}(0, 1) \nonumber \\
 & \textnormal{Function:} \\
 & \frac{d \sigma}{d \Omega}^{\prime}_{\textnormal{DWBA}, i} = \eta \times \delta D_0^2 \times C^2S \times \bigg[ \alpha \frac{d \sigma}{d \Omega}_{\textnormal{DWBA}, \ell_1} + (1-\alpha) \frac{d \sigma}{d \Omega}_{\textnormal{DWBA}, \ell_2}   \bigg] \nonumber,
\end{align}
where the definition for $\frac{d \sigma}{d \Omega}^{\prime}_{\textnormal{DWBA}, i}$ is understood to replace all other occurrences of that variable in Eq.~\ref{eq:dwba_model_na}. Note that the individual cross sections, $\frac{d \sigma}{d \Omega}_{\textnormal{DWBA}, \ell_1}$ and $\frac{d \sigma}{d \Omega}_{\textnormal{DWBA}, \ell_2}$, are calculated using the same optical potential.  

\subsection{Results}

The above Bayesian model was applied to the eleven states observed in the astrophysical region of interest. For each state, \texttt{emcee} was run with $400$ walkers taking $8000$ steps, giving a total of $3.2 \times 10^6$ samples. Of these samples, the first $6000$ steps were discarded as burn in, and the last $2000$ steps were thinned by $50$ for $16000$ final samples. These $16000$ samples were used to estimate the posterior distributions for $C^2S$, and to construct the differential cross sections shown in Fig.~\ref{fig:mcmc_cs_na}. An example of the simultaneous fit obtained for the elastic scattering data is shown in Fig.~\ref{fig:elastic_fit_na}. All of the data have been plotted as a function of their relative value (Section \ref{sec:scale_of_na_data}). The normalization $\eta$ was found to be $\eta = 0.075^{+0.007}_{-0.006}$, which shows that the absolute scale of the data, despite the influence of the optical model parameters, can be established with a $9 \%$ uncertainty.

The values obtained for $(2J_f+1)C^2S$ in this work are listed in Table \ref{tab: na_c2s_table}. There is general agreement between our values and those of Ref.~\cite{hale_2004}, which provides further evidence that the absolute scale of the data is well established. However, for the three $2^+$ states that show a mixture of $\ell = 0 + 2$, the current values are consistently lower. In these cases, the Bayesian method demonstrates that considerable uncertainty is introduced when a mixed $\ell$ transfer is present. The origin of this effect merits a deeper discussion, which I will now present. 

The posterior distributions for $(2J_f+1)C^2S$ from states with unique $\ell$ transfers were found to be well described by log-normal distributions. Estimations of these distributions can be made by deriving the log-normal parameters $\mu$ and $\sigma$ from the samples. An example of the agreement of the posterior samples with a log-normal distribution is shown in Fig.~\ref{fig:11390_log_normal}. The $med.$ and $f.u.$ quantities calculated from these parameters are listed in Table \ref{tab: na_c2s_table}. It can be seen that states that have a unique $\ell$ transfer show factor uncertainties of $f.u. \approx 1.30$, or, rather, a $30 \%$ uncertainty. On the other hand, states that show a mixed $\ell$ transition vary from $f.u. = 1.40 \text{-} 2.00$. However, it can be seen in Fig.~\ref{fig:11453_log_normal}, that while the individual $\ell$ components have a high factor uncertainty or deviate strongly from a log-normal distribution, their sum shares the same properties as the states with a single $\ell$ transfer. In other words, the overall spectroscopic factor still has a $30 \%$ uncertainty. This situation is analogous to the one encountered in Section \ref{sec:mc_rates}. The mean grows linearly with each component of the sum while the uncertainty roughly grows as the square root. This fact requires, without appealing to the current Bayesian methods, that the individual $\ell$ components must have a greater percentage uncertainty than their sum. Since previous studies, like those of Ref.~\cite{hale_2004}, assume a constant uncertainty with the extraction of spectroscopic factors, each $\ell$ component is assigned the same percentage uncertainty. The above discussion highlights that this assumption cannot be true, regardless of the statistical method. The influence of optical model parameters limits the precision of the total normalization of the cross section; thereby, giving an upper limit on the precision that can be expected from the components. These results indicate that applying a standard $\chi^2$ fit to a mixed $\ell$ transfer might not accurately extract the individual spectroscopic factors if optical model uncertainties are ignored.            

I will now discuss the results and the relevant previously reported information for each of these states.

\begin{figure}
    \centering
    \includegraphics[width=.8\textwidth]{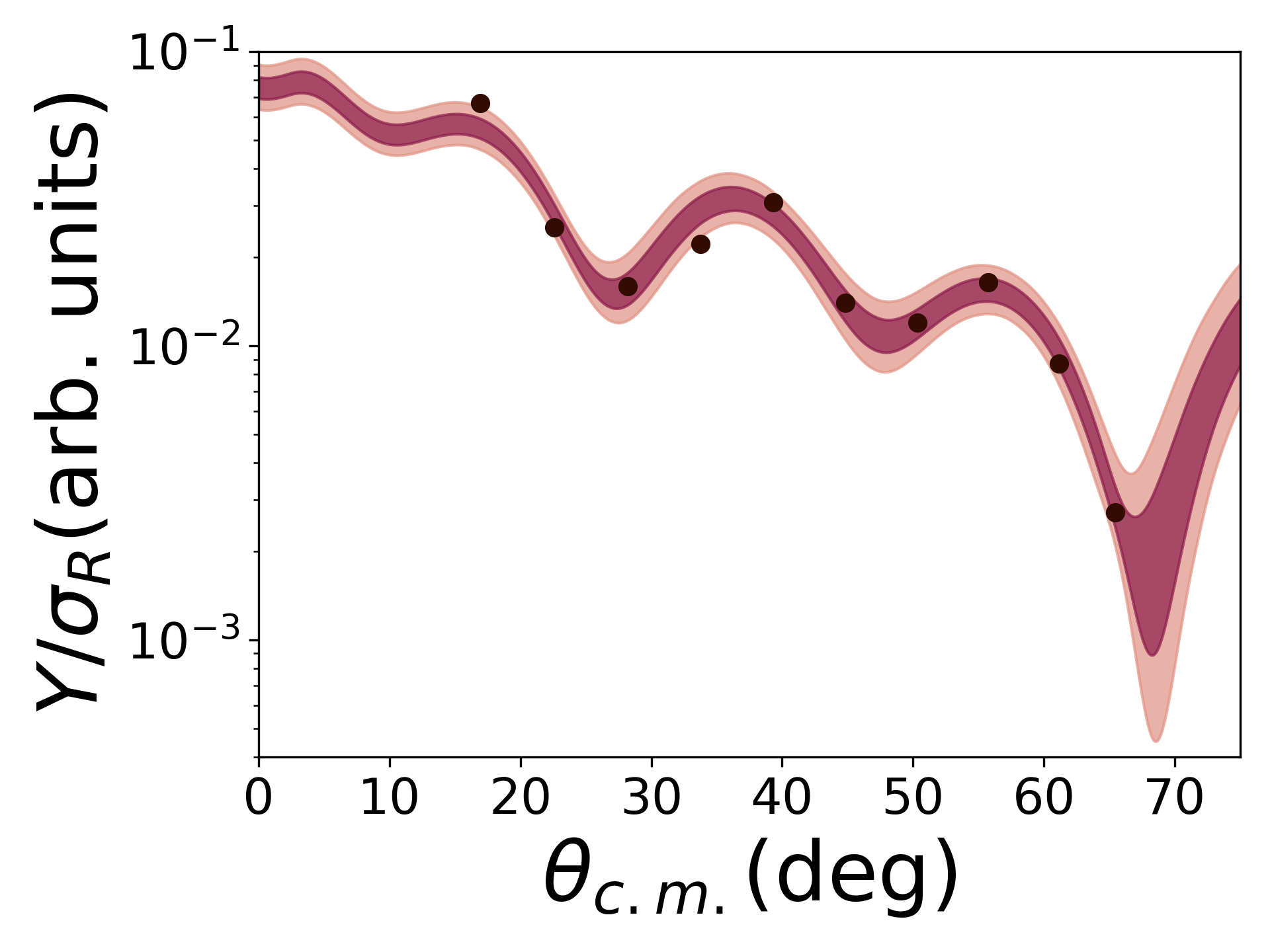}
    \caption{The credibility intervals obtained for the elastic scattering fit. The dark and light purple bands show the $68 \%$ and $95 \%$ credibility intervals, respectively. The measured error bars are smaller than the points, while the adjusted uncertainty of Eq.~\ref{eq:elastic_unc} that is inferred from the data is not shown.}
    \label{fig:elastic_fit_na}
\end{figure}

\afterpage{
\clearpage
\null
\hspace{0pt}
\vfill
\captionof{figure}{The DWBA calculations for the states of $^{24}$Mg. The $68 \%$ and $95 \%$ credibility intervals are shown in purple and light purple, respectively. Only data point up to the first minimum were considered, and they are shown in dark brown. For the $11825$-keV state, the $68 \%$ bands are shown for the $\ell=0\text{-}3$
transfers.}
\label{fig:mcmc_cs_na}
\vfill
\newpage
\clearpage
\begin{figure}
  \ContinuedFloat\centering
  \captionsetup[subfigure]{labelformat=empty}
    \vspace{-1\baselineskip}
    \begin{subfigure}[t]{0.45\textwidth}
        \includegraphics[width=\textwidth]{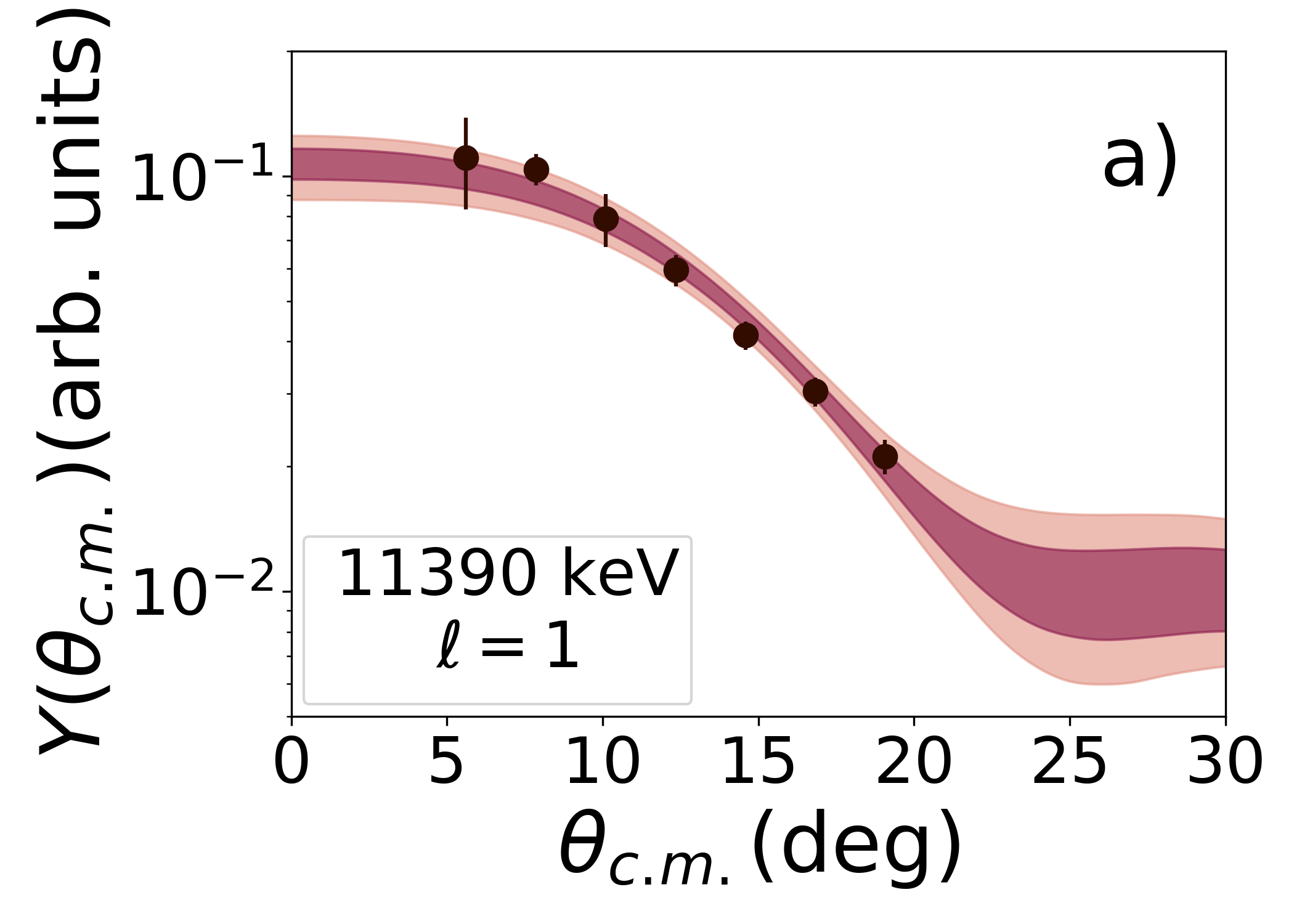}
        \caption{\label{fig:11390_fit}}
      \end{subfigure}
          \vspace{-1\baselineskip}
    \begin{subfigure}[t]{0.45\textwidth}
      \includegraphics[width=\textwidth]{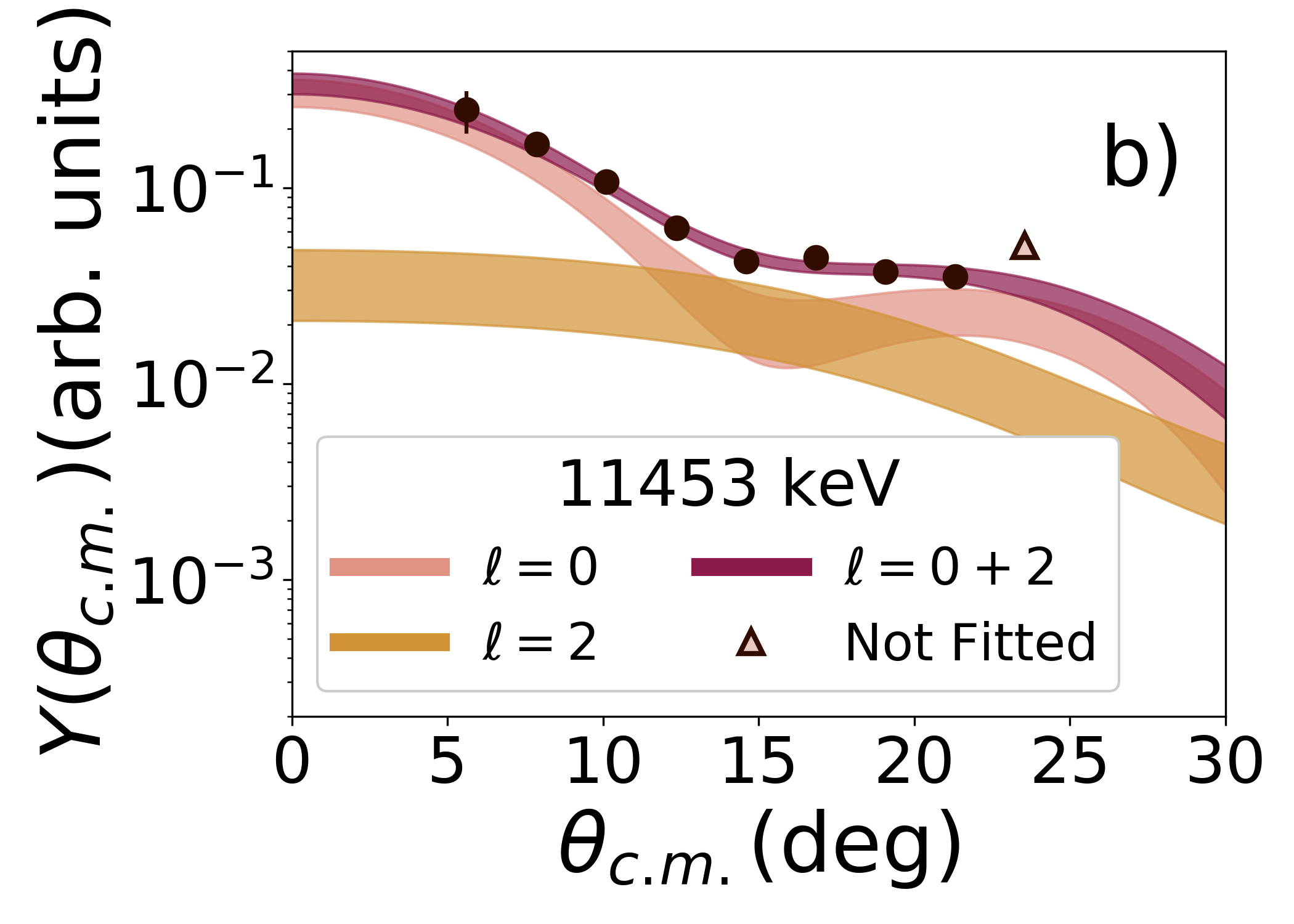}
          \vspace{-1\baselineskip}
      \caption{\label{fig:11453_fit}}
    \end{subfigure}
    \vspace{-1\baselineskip}
    \begin{subfigure}[t]{0.45\textwidth}
      \includegraphics[width=\textwidth]{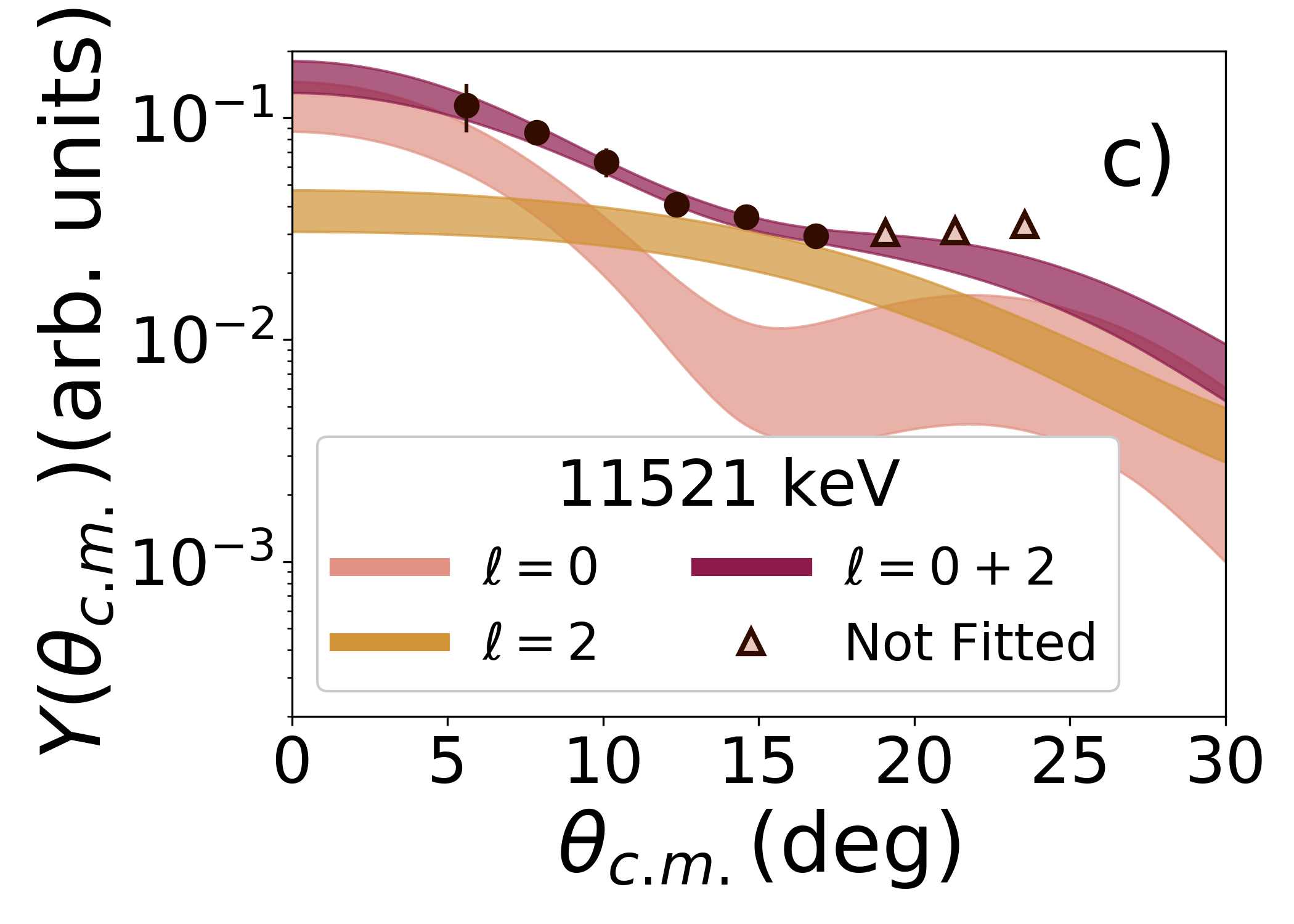}
          \vspace{-1\baselineskip}
      \caption{\label{fig:11521_fit}}
    \end{subfigure}
    \begin{subfigure}[t]{0.45\textwidth}
      \includegraphics[width=\textwidth]{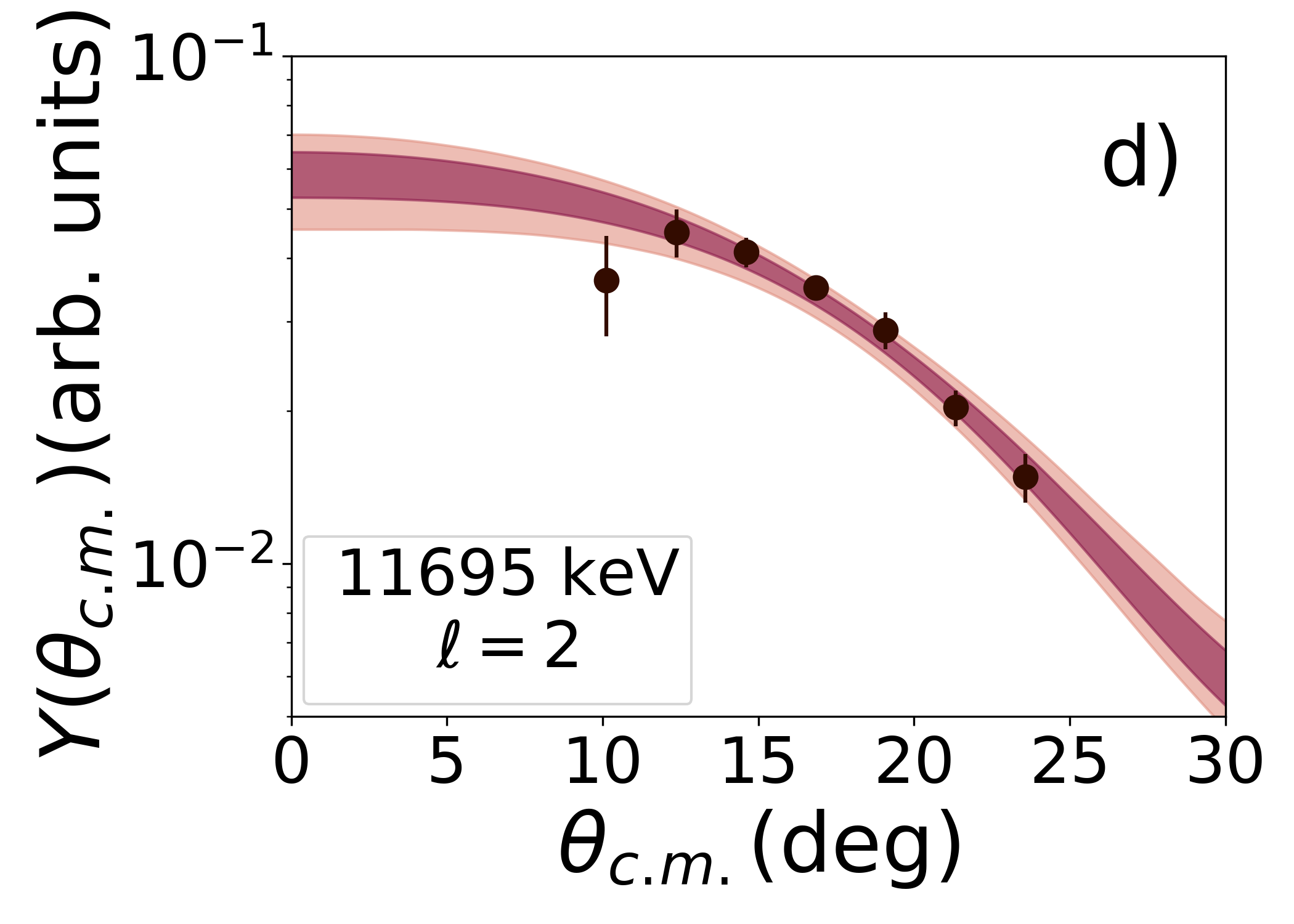}
          \vspace{-1\baselineskip}
      \caption{\label{fig:11695_fit}}
    \end{subfigure}
    \vspace{-1\baselineskip}
    \begin{subfigure}[t]{0.45\textwidth}
      \includegraphics[width=\textwidth]{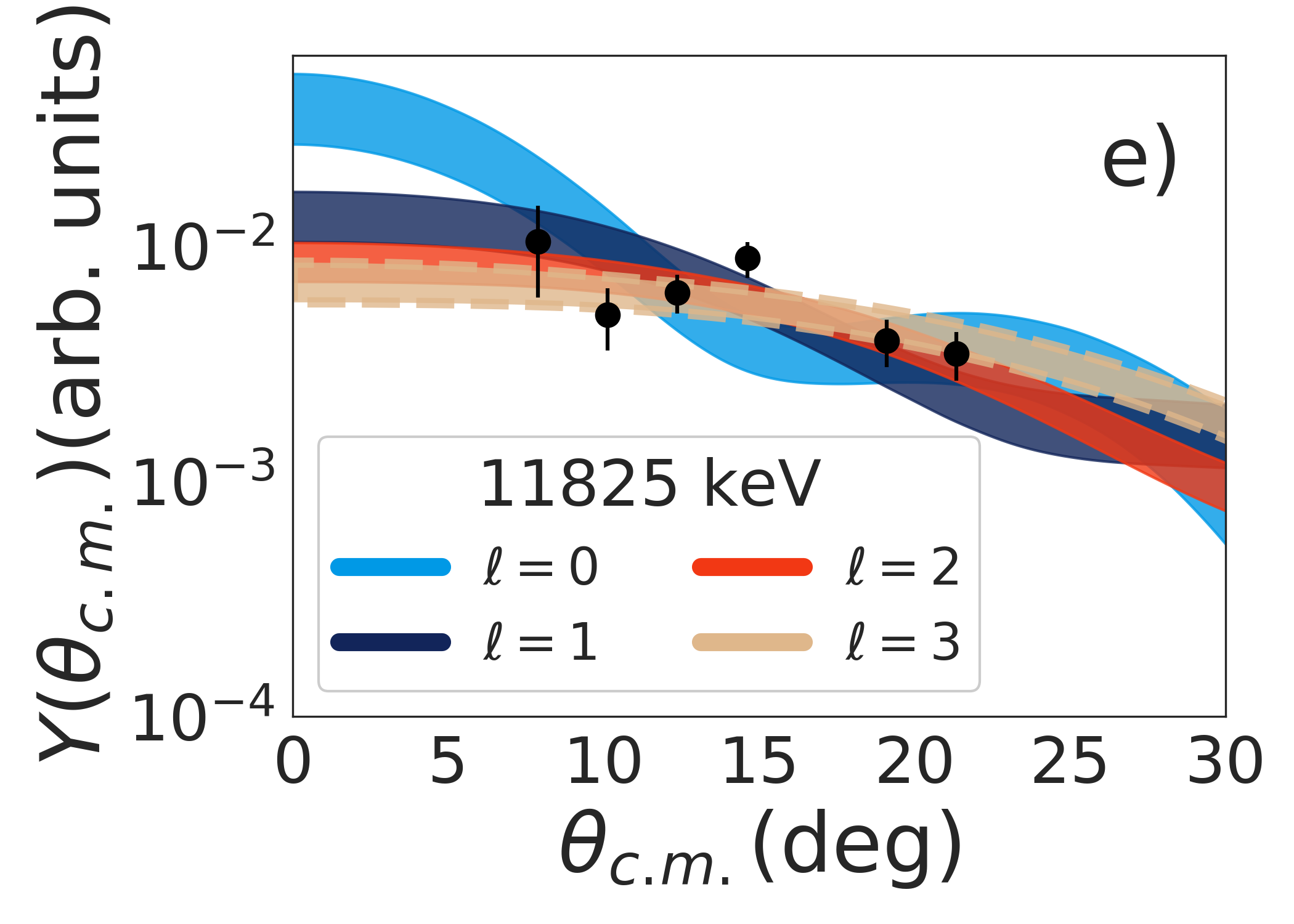}
          \vspace{-1\baselineskip}
      \caption{\label{fig:11824_fit}}
    \end{subfigure}
    \begin{subfigure}[t]{0.45\textwidth}
      \includegraphics[width=\textwidth]{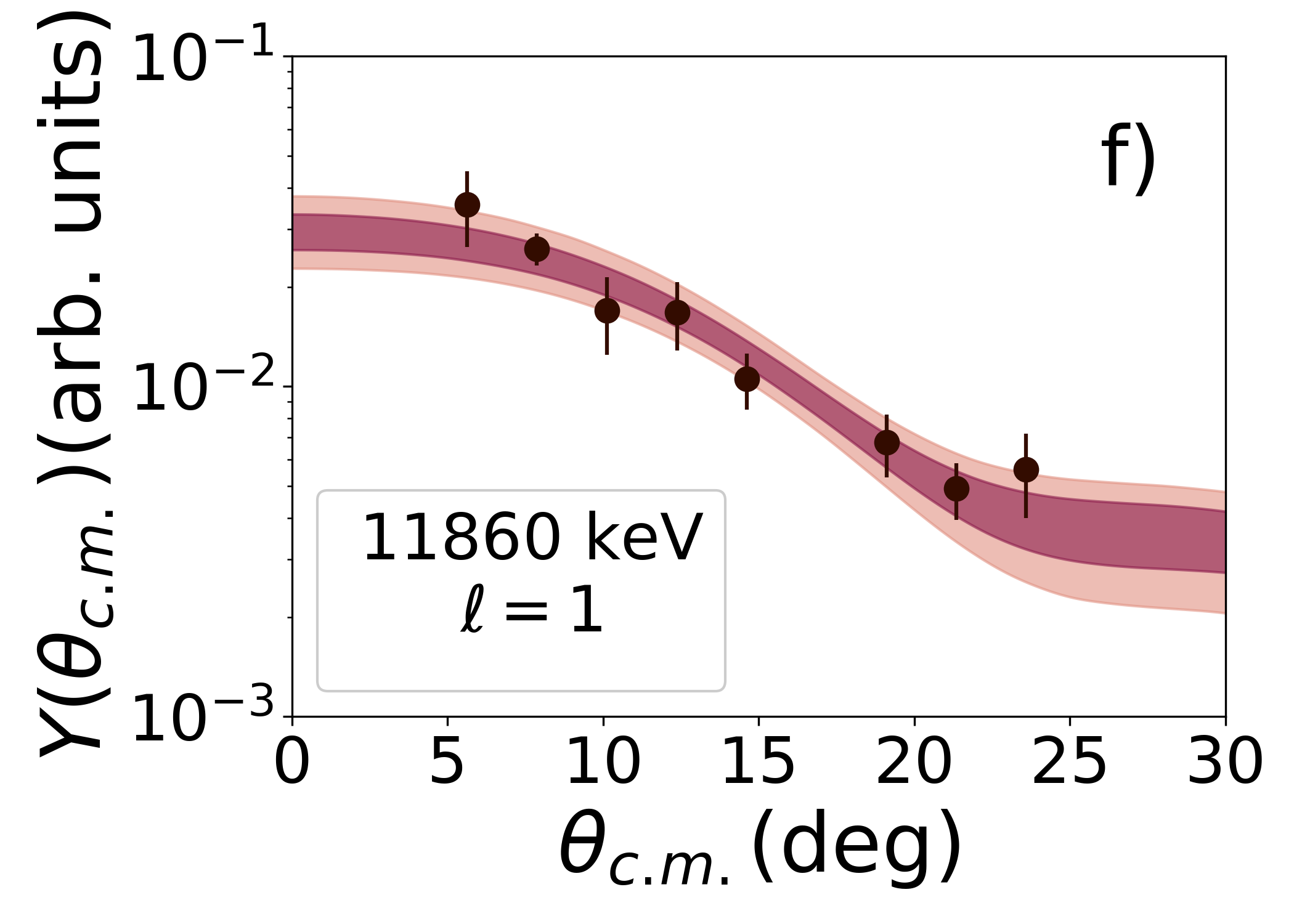}
          \vspace{-1\baselineskip}
      \caption{\label{fig:11860_fit}}
    \end{subfigure}
\end{figure}
\begin{figure}\ContinuedFloat
  \centering
  \captionsetup[subfigure]{labelformat=empty}
    \begin{subfigure}[t]{0.45\textwidth}
      \includegraphics[width=\textwidth]{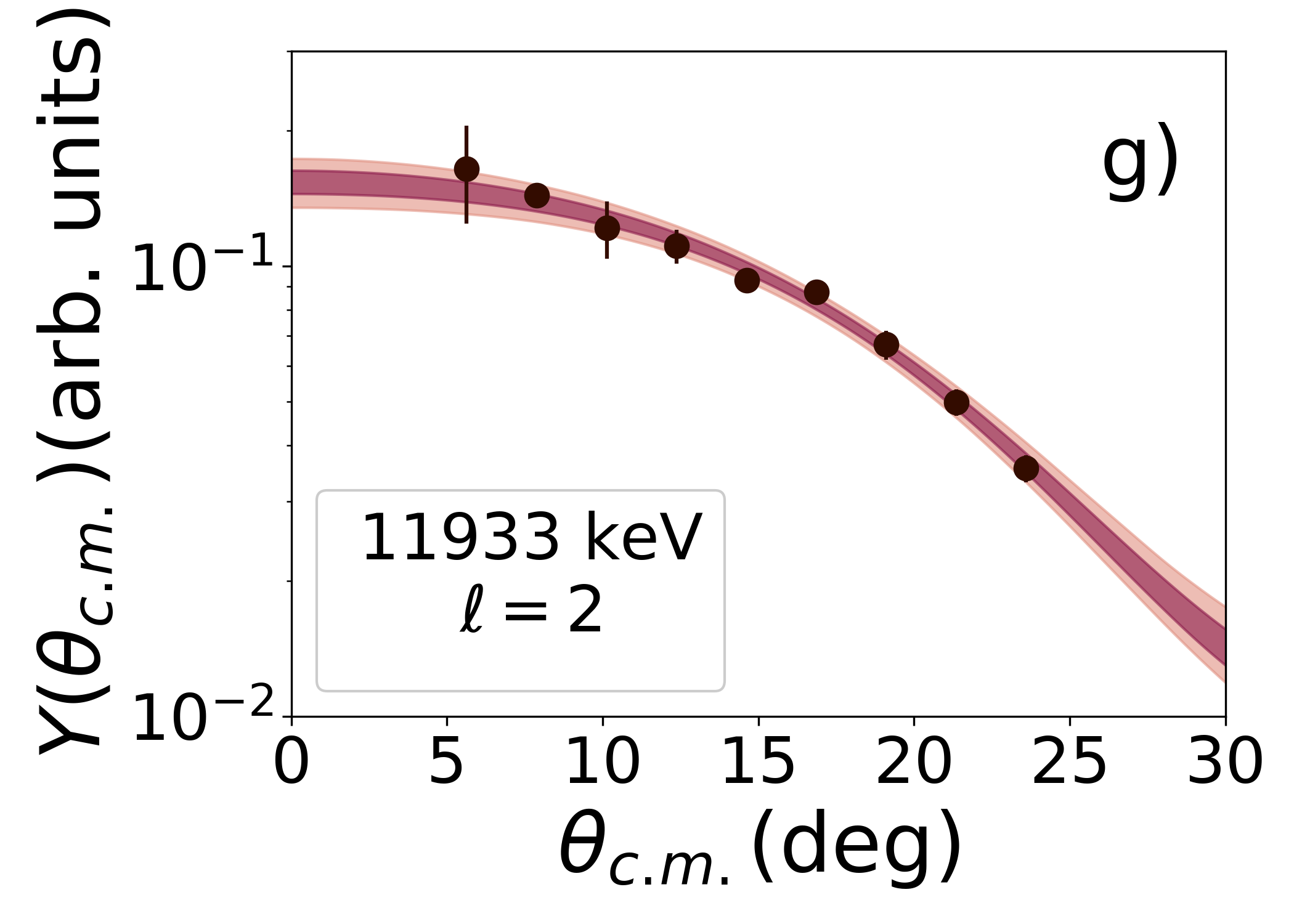}
          \vspace{-1\baselineskip}
      \caption{\label{fig:11933_fit}}
    \end{subfigure}
    \begin{subfigure}[t]{0.45\textwidth}
      \includegraphics[width=\textwidth]{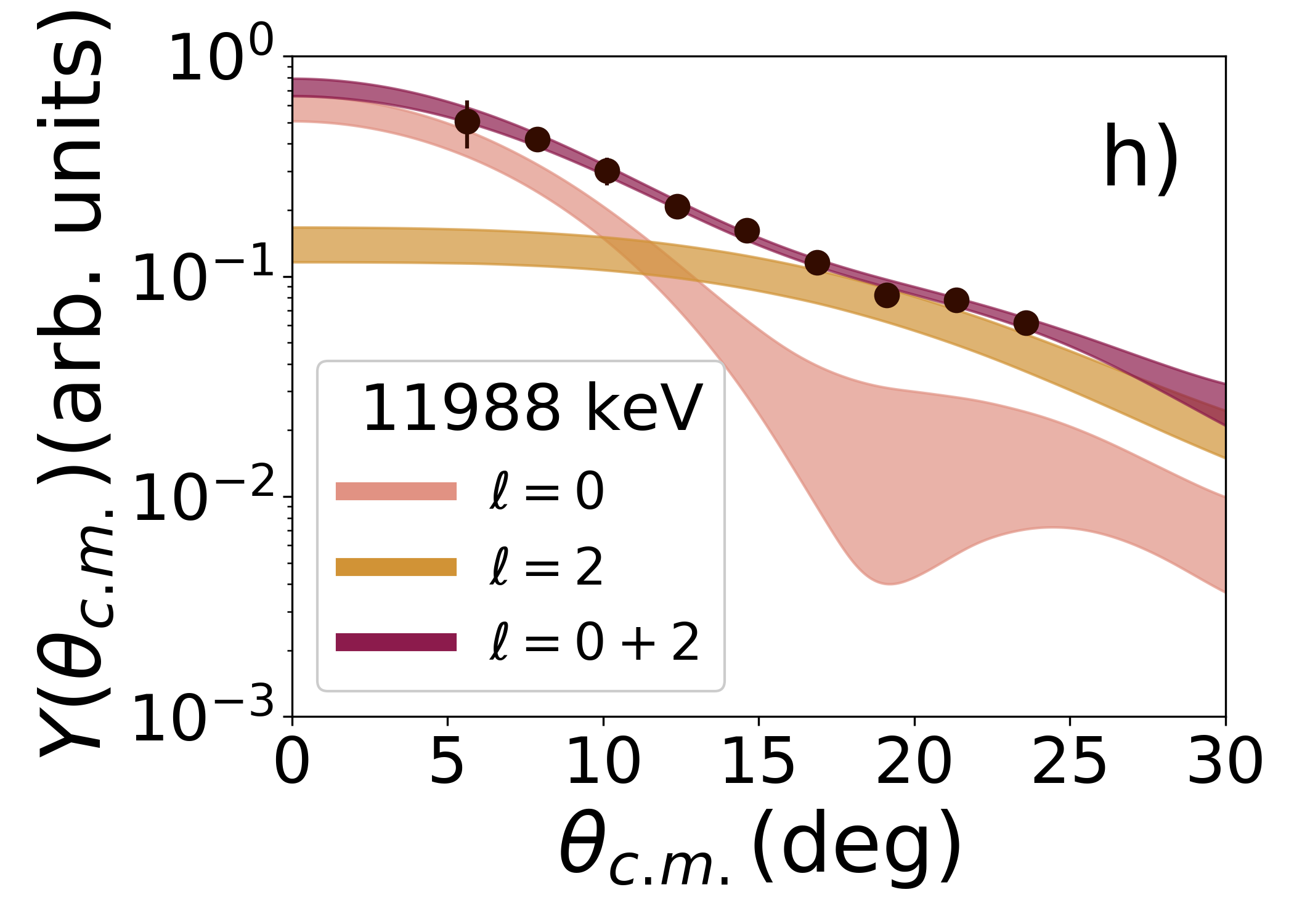}
          \vspace{-1\baselineskip}
      \caption{\label{fig:11988_fit}}
    \end{subfigure}
        \begin{subfigure}[t]{0.45\textwidth}
      \includegraphics[width=\textwidth]{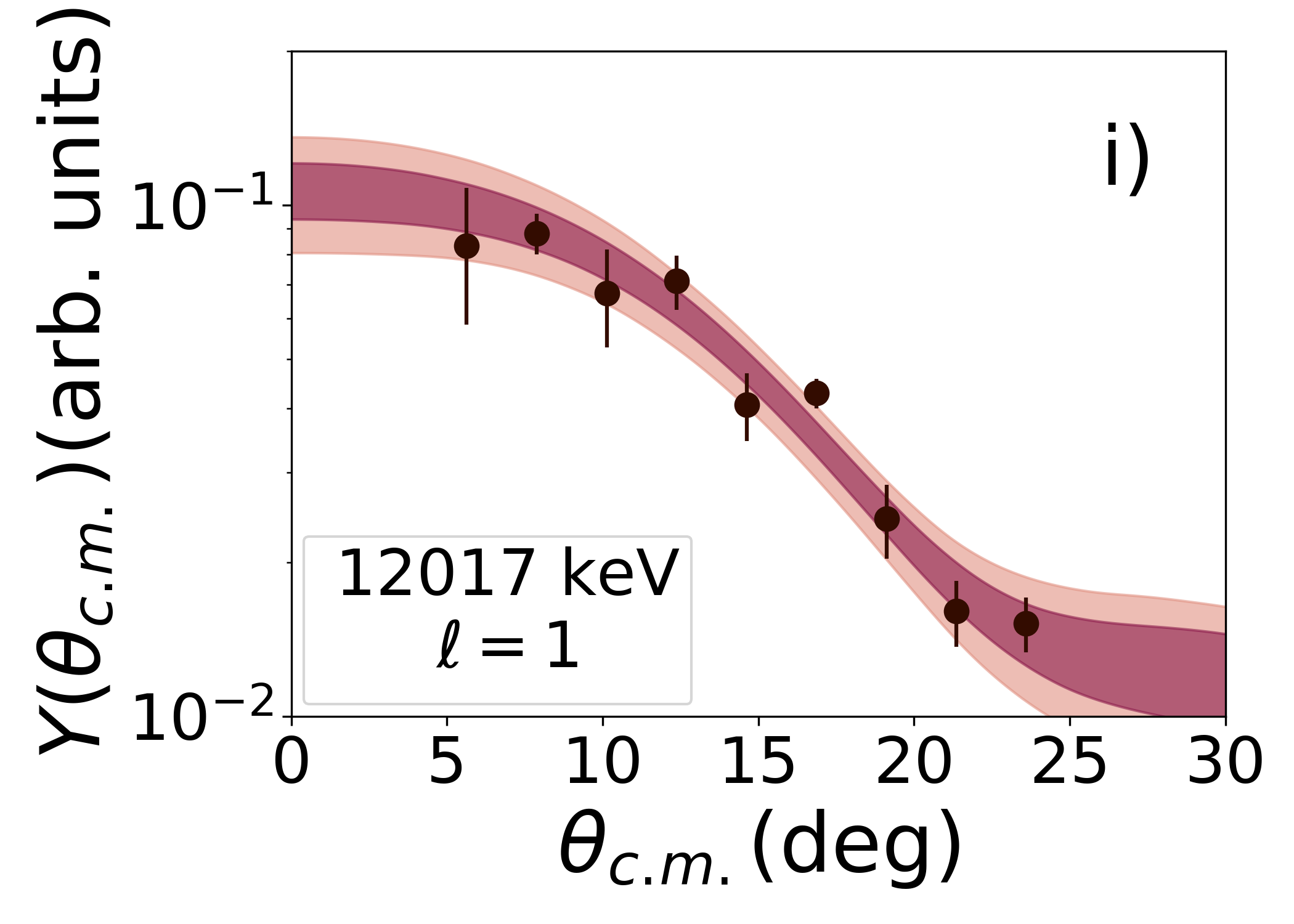}
          \vspace{-1\baselineskip}
      \caption{\label{fig:12017_fit}}
    \end{subfigure}
            \begin{subfigure}[t]{0.45\textwidth}
      \includegraphics[width=\textwidth]{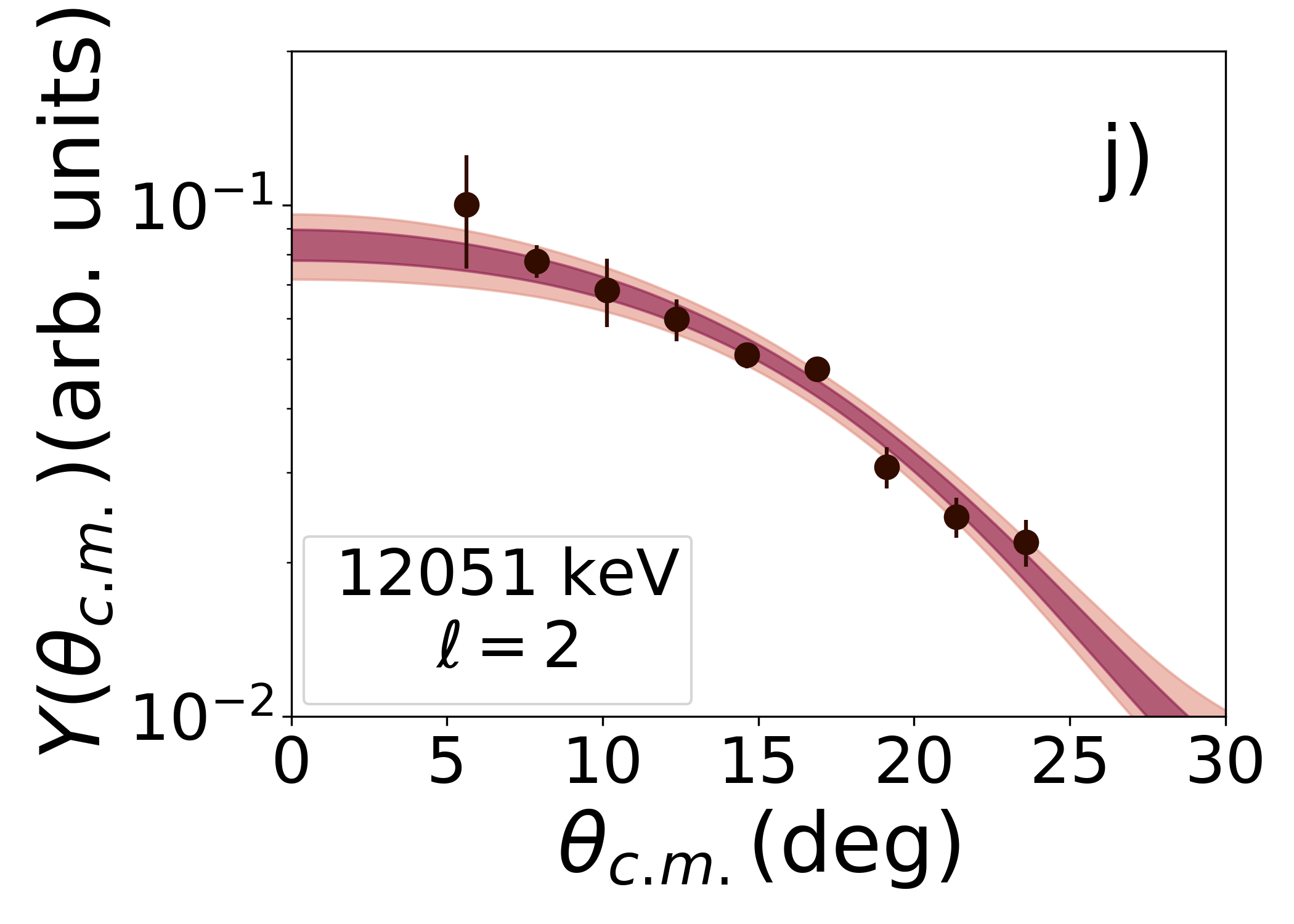}
          \vspace{-1\baselineskip}
      \caption{\label{fig:12051_fit}}
    \end{subfigure}
            \begin{subfigure}[t]{0.45\textwidth}
      \includegraphics[width=\textwidth]{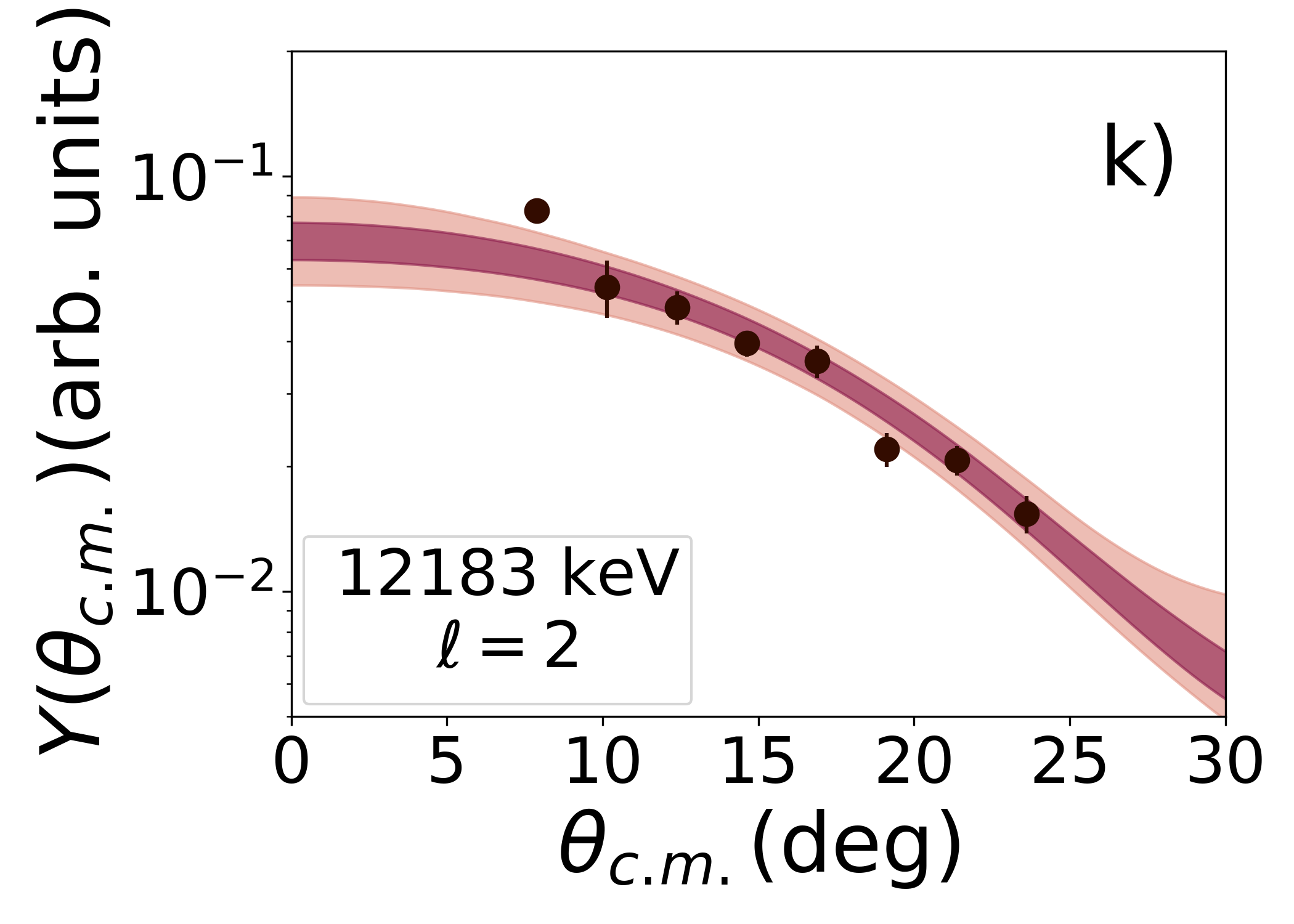}
          \vspace{-1\baselineskip}
      \caption{\label{fig:12183_fit}}
    \end{subfigure}
    \vspace{-1.8\baselineskip}
\end{figure}
\clearpage
}
\newpage

\begin{figure}
    \centering
    \includegraphics[width=.8\textwidth]{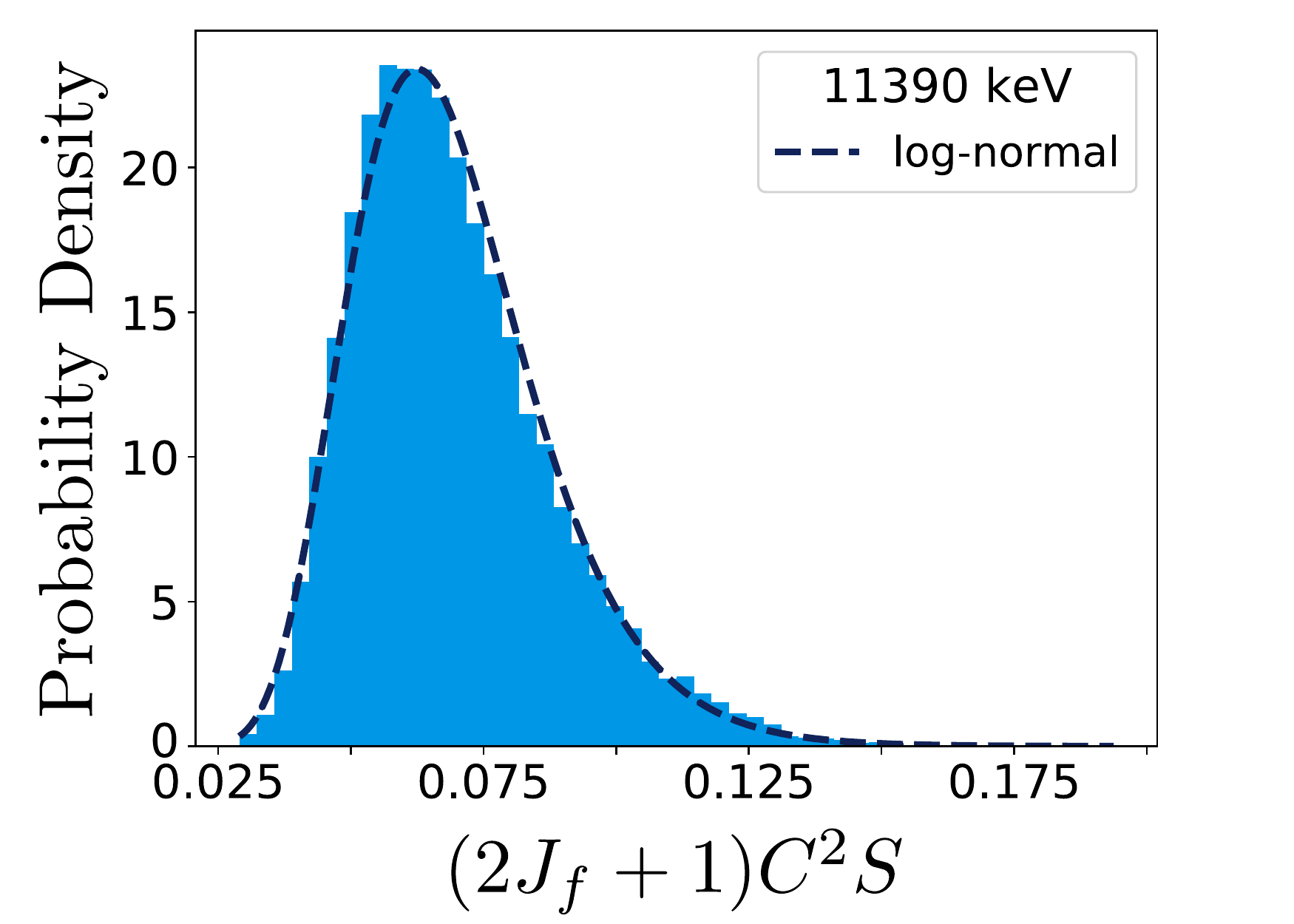}
    \caption{The posterior samples of $(2J_f+1)C^2S$ for the $11390$-keV state. The dark blue dashed line shows the corresponding log-normal distribution.}
    \label{fig:11390_log_normal}
\end{figure}

\begin{figure}
    \centering
    \includegraphics[width=\textwidth]{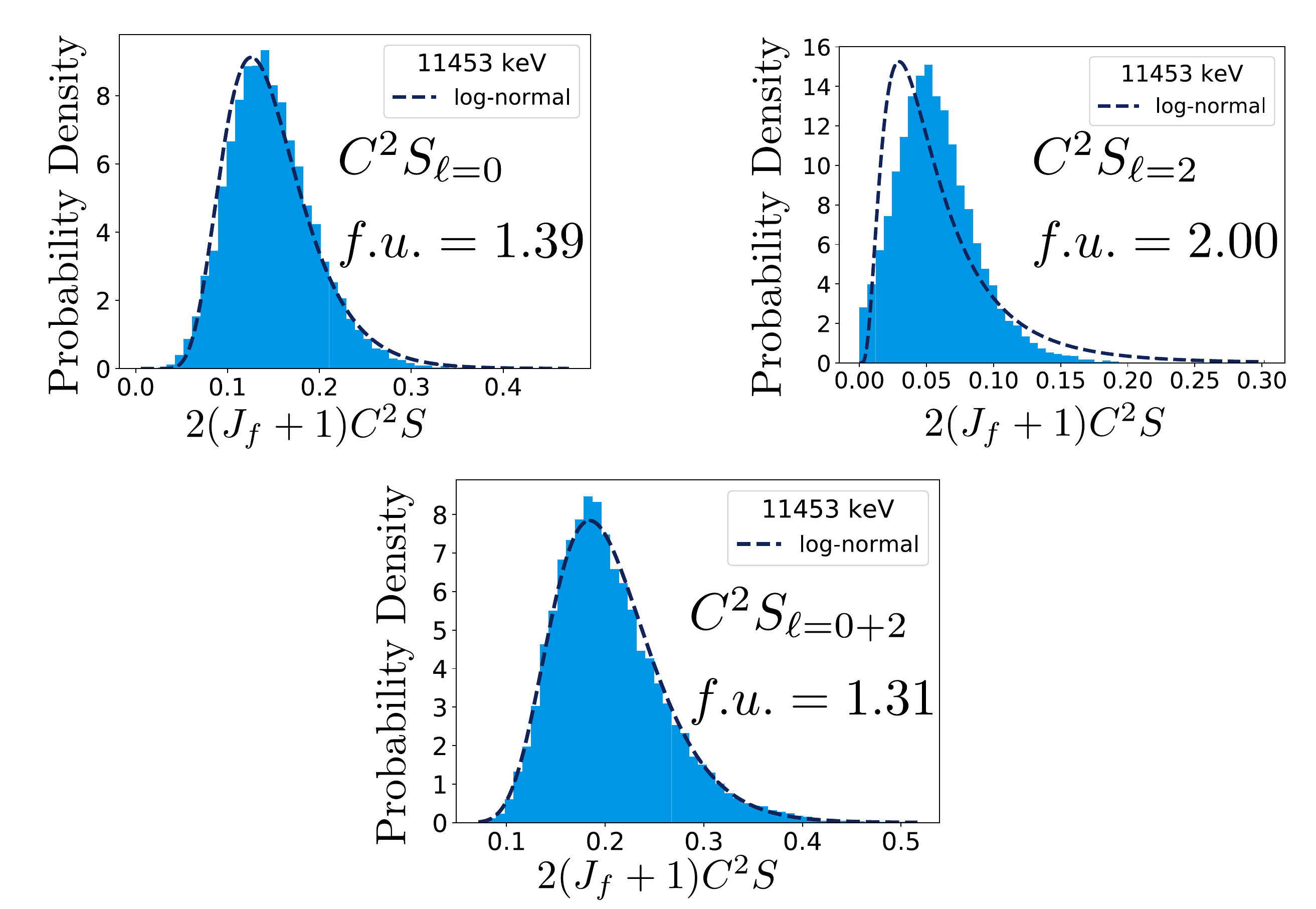}
    \caption{Plots of the three posterier distributions associated with the $11453$-keV spectroscopic factor. The top two plots show the individual $\ell$ $(2J_f+1)C^2S$ samples and their corresponding log-normal distributions. The bottom shows the sum of the two components and its log-normal distribution. It can be seen that the sum has the same factor uncertainty as transfers described by a single $\ell$ value.}
    \label{fig:11453_log_normal}
\end{figure}

\begin{table}
\centering
  \setlength{\tabcolsep}{5pt}
  \caption{ \label{tab: na_c2s_table} The values of $(2J_f+1)C^2S$ that were derived in this work compared to those of Ref.~\cite{hale_2004}. All values for this work give the $68 \%$ credibility interval from the posterior estimation. Additionally, the parameters of the corresponding log-normal distribution are listed. All spin parity information, except that of the $11825$-keV state, is taken from Ref.~\cite{firestone_2007}, and are updated based on the current observations.}
\begin{threeparttable}
  \begin{tabular}{lllllll}
    \toprule
    \toprule
    $E_x$ (keV) & $J^{\pi}$ & $\ell$ & $(2J_f+1)C^2S$ & $med.$   & $f.u.$& Ref.~\cite{hale_2004}  \\ \hline \vspace{-2mm}
  \\\vspace{2mm}
$11390$     & $1^-$    & $1$  &$0.066^{+0.021}_{-0.015}$ & $0.067$ & $1.30$ & $0.06$ \\ \vspace{2mm}
$11453$     & $2^+$    & $0+2$ & $0.14^{+0.05}_{-0.04}$ + $0.05^{+0.03}_{-0.02}$ &   $0.14$ + $0.048$      & $1.39$ + $2.00$ & $0.24$ + $0.16^{\dagger}$  \\ \vspace{2mm}
$11521$     & $2^+$    & $0+2$ & $0.05^{+0.03}_{-0.02}$ + $0.057^{+0.024}_{-0.018}$ &  $0.055$ + $0.056$ & $1.61$ + $1.51$ & $0.10^{\ddagger}$ \\ \vspace{2mm}
$11695$     & $4^+$    & $2$ & $0.085^{+0.025}_{-0.018}$ & $0.086$  & $1.29$ & $0.11$ \\ \vspace{2mm} 
{$11825$}         &    & $0$ & $0.023^{+0.012}_{-0.007}$ & $0.024$  & $1.52$ & $0.039$ \\ \vspace{2mm}
        &    & $1$ & $0.010^{+0.004}_{-0.003}$ & $0.010$  & $1.40$ & $0.009$ \\ \vspace{2mm}
        &    & $2$ & $0.014^{+0.005}_{-0.003}$ & $0.014$  & $1.36$ & $0.015$ \\ \vspace{2mm}
        &    & $3$ & $0.025^{+0.009}_{-0.006}$ & $0.025$  & $1.36$ & $0.024$ \\ \vspace{2mm}
$11860$     & $1^-$    & $1$ & $0.022^{+0.007}_{-0.005}$ & $0.022$  & $1.32$ & $0.026$ \\ \vspace{2mm}
$11933$     & $(2 \text{-} 4)^+$ & $2$ & $0.23^{+0.07}_{-0.05}$ & $0.24$ & $1.30$ & $0.25$      \\ \vspace{2mm}
$11988$     & $2^+$ & $0+2$ & $0.26^{+0.10}_{-0.07}$ + $0.24^{+0.10}_{-0.07}$  & $0.26$ + $0.24$ & $1.40$ + $1.45$ & $0.42$ + $0.33$      \\ \vspace{2mm}
$12017$     & $3^-$ & $1$ & $0.20^{+0.06}_{-0.04}$ & $0.20$ & $1.30$ & $0.13$      \\ \vspace{2mm}
$12051$     & $4^+$ & $2$ & $0.13^{+0.04}_{-0.03}$ & $0.14$ & $1.30$ & $0.13$      \\ \vspace{2mm}
$12183$     & $(1,2^+)$ & $2$ & $0.12^{+0.04}_{-0.03}$ & $0.12$ & $1.34$ & $0.13$      \\
    \bottomrule
    \bottomrule
\end{tabular}
\begin{tablenotes}
\item[$\dagger$] Ref.~\cite{hale_2004} assumed a doublet. The $(2J_f+1)C^2S$ values were taken from these two states.
\item[$\ddagger$] Ref.~\cite{hale_2004} assumed a doublet, with a portion of the strength assigned to a negative parity state.
\end{tablenotes}
\end{threeparttable}
\end{table}

\subsubsection{The $11391$-keV State; $-303$-keV Resonance}

This state has been reported in several studies, and is known to have a spin parity of $J^{\pi}=1^-$. Our measurements
confirm an $\ell=1$ nature to the angular distribution, making it a candidate for a subthreshold $p$-wave resonance. A higher lying state with unknown spin-parity has been reported in Ref.~\cite{vermeer_1988} at $E_x = 11394(4)$ keV. The current evaluation states that the
$^{25}\textnormal{Mg}(^{3}\textnormal{He}, ^{4}\!\textnormal{He})^{24}\textnormal{Mg}$ measurement of Ref.~\cite{El_Bedewi_1975} also observes this
higher state at $11397(10)$ keV, but their angular distribution gives an $\ell=1$ character, indicating it would be compatible with the lower $J^{\pi}=1^-$ state. Ref.~\cite{hale_2004} finds a similar peak in their spectrum, but considered it a doublet because of the ambiguous shape of the angular distribution, which was caused primarily by the behaviour of the data above $20^{\circ}$. Considering the excellent agreement between our data and an $\ell=1$ transfer, only the state at $11389.6(12)$ keV with $J^{\pi}=1^-$ was considered to be populated. The present calculation assumes a $2p_{3/2}$ transfer and is shown in Fig.~\ref{fig:11390_fit}.

\subsubsection{The $11453$-keV State; $-240$-keV Resonance}

Two states lie in the region around $11.45$ MeV, with the lower assigned $J^{\pi}=2^+$ and the upper $J^{\pi}=0^+$. The only study that reports a definitive observation of the $0^+$, $11460(5)$ keV state is the $(\alpha,\alpha_0)$ of Ref.~\cite{goldberg_54}. The current study and that of Ref.~\cite{hale_2004} indicate that there is a state around $E_x = 11452$ keV that shows a mixed $\ell = 0 + 2$ angular distribution. Since the ground state of $^{23}$Na is non-zero, this angular distribution can be the result of a single $2^+$ state, and the $\ell=2$ component cannot be unambiguously identified with the higher lying $0^+$ state. The $(p,p^{\prime})$ measurement of Ref.~\cite{zwieglinski_1978} notes a state at $11452(7)$ keV with $\ell=2$. The excellent agreement between our excitation energy and the gamma ray measurement of Ref.~\cite{endt_1990} leads us to assume the full strength of the observed peak comes from the $2^+$ state. The calculation shown in Fig.~\ref{fig:11453_fit} assumes transfers with quantum numbers $2s_{1/2}$ and $1d_{5/2}$.

\subsubsection{The $11521$-keV State; $-172$-keV Resonance}

Another sub-threshold $2^+$ state lies at $11521.1(14)$ keV. A state with unknown spin-parity was observed at $11528(4)$ keV in Ref.~\cite{vermeer_1988}, but has not been seen on other studies. Based on the measured $\Gamma_{\gamma}/\Gamma \approx 1$, and assuming the efficacy of the methods in Ref.~\cite{vermeer_1988}, there is a high likelihood this unknown state has an unnatural parity. The present angular distribution, Fig.~\ref{fig:11521_fit}, is indicative of a mixed $\ell = 0 + 2$ assignment. Thus, the observation is associated with the $2^+$ state at $11521.1(14)$ keV, and transfers were calculated using $2s_{1/2}$ and $1d_{5/2}$.      

\subsubsection{The $11695$-keV State; $2$-keV Resonance}

For our measurement this state was partially obscured by a contaminant peak from the ground state of $^{17}$F coming from $^{16}$O$(^{3} \textnormal{He}, d)^{17}$F for $\theta_{Lab} < 9^{\circ}$. Previous measurements have established a firm $4^+$ assignment, and our angular distribution is consistent with an $\ell =2$ transfer. The fit for a $1d_{5/2}$ transfer is shown in Fig.~\ref{fig:11695_fit}.  

\subsubsection{The $11825$-keV State; $132$-keV Resonance}
\begin{figure}
  \centering
  \includegraphics[width=.6\textwidth]{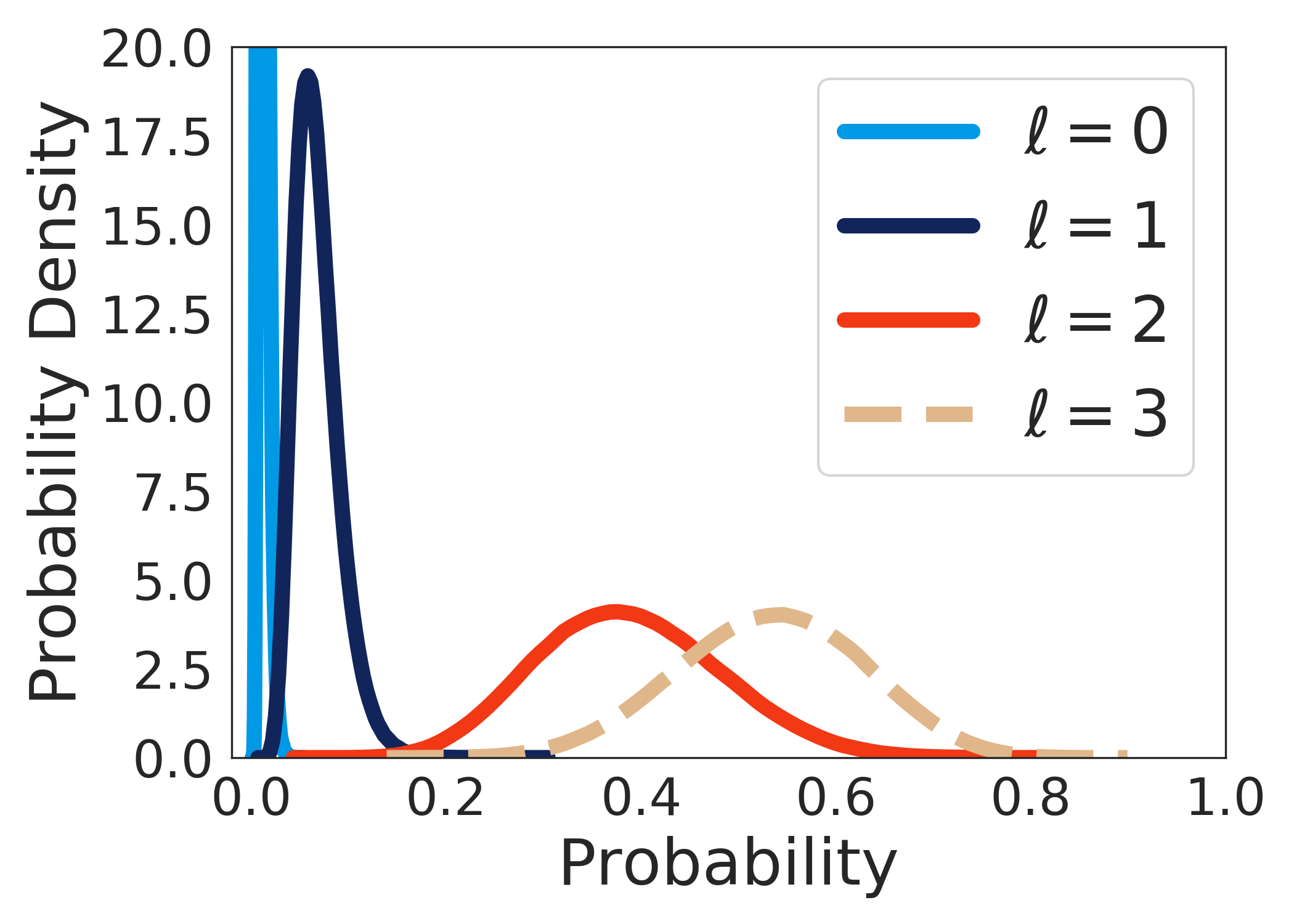}
  \caption{The distributions from the nested sampling algorithm for the most likely $\ell$ values for the $11825$-keV state.}
  \label{fig:l_comp_probs_na}
\end{figure}

As discussed in Section \ref{sec:background_subtraction}, this state is obscured at several angles by the fifth excited state of $^{15}$O. The previous constraints on its spin parity come from the comparison of the extracted spectroscopic factors for each $\ell$ value in Ref.~\cite{hale_2004} and the upper limits established in Ref.~\cite{Rowland_2004} and subsequently Ref.~\cite{Cesaratto_2013}.
This DWBA analysis finds an angular distribution consistent with Ref.~\cite{hale_2004}, which it should be noted experienced similar problems with the nitrogen contamination, but with the Bayesian model comparison methods presented in Section \ref{sec:bay_dwba},  constraints can be set based purely on the angular distribution. All of the considered $\ell$ transfer are shown in Fig.~\ref{fig:11824_fit}, and were calculated assuming $2s_{1/2}$, $2p_{3/2}$, $1d_{5/2}$, and $1f_{7/2}$ transfers, respectively. The results of the nested sampling calculations, which give the relative probabilities of each transfer, are presented in Table~\ref{tab:probs} and shown in Fig.~\ref{fig:l_comp_probs_na}. One key difference between this calculation and that of the $3.70$-MeV state in Section \ref{sec:bay_dwba}, is that the adopted values were taken to be the mean instead of the median. Since the statistical errors of the nested sampling are normally distributed  in $\ln Z$, the resulting probabilities are distributed log-normally. The choice of the mean instead of the median then amounts to selecting the arithmetic mean instead of the geometric mean, which ensures $\sum_{\ell} P(\ell) = 1$.

\begin{table*}[]
  \centering
  \setlength{\tabcolsep}{12pt}
    \caption{\label{tab:probs} Results of the model comparison calculations for the $11825$ keV state. For each $\ell$ value, I list the $\log{Z}$ value calculated with nested sampling, the median Bayes factor when compared to the most likely transfer $\ell=3$, and the mean probability of each transfer.}

  \begin{tabular}{llll}
    \toprule
    \toprule
    $\ell$     & $\log{Z}_{\ell}$              & $B_{3 \ell}$         & $P(\ell)$         \\ \hline \vspace{-2mm}
    \\
           $0$    &      $44.226(294)$           & $47.79$            & $1 \%$           \\ 
           $1$    &      $45.990(289)$           & $8.20$             & $7 \%$ \\ 
           $2$    &      $47.762(323)$           & $1.39$             & $39 \%$ \\
           $3$    &      $48.093(293)$           & $1.00$             & $53 \%$ \\ 
    \bottomrule
    \bottomrule
\end{tabular}
\end{table*}

\subsubsection{The $11860$-keV State; $168$-keV Resonance}

There are two states within a few keV of one another reported to be in this region. One is known to have $ J^{\pi}=1^-$, and has been populated in nearly all of the experiments listed in Table~\ref{tab:energy_comp}. The other state is reported to decay to the $6^+$, $8114$-keV state, with a $\gamma$-ray angular distribution that favors an assignment of $8^+$ \cite{branford_1972}. The later polarization measurements of Ref.~\cite{wender_1978} support the assignment of $8^+$. For our experiment, the tentative $8^+$ state is likely to have a negligible contribution to the observed peak, and the angular distribution in Fig.~\ref{fig:11860_fit} is consistent with a pure $\ell=1$ transfer. The calculation assumed $2p_{5/2}$.

\subsubsection{The $11933$-keV State; $240$-keV Resonance} 

The $11933$-keV State does not have a suggested spin assignment in the current ENSDF evaluation \cite{firestone_2007}. The compilation of Ref.~\cite{endt_eval_1990} lists a tentative $(2-4)^+$. The 
$\ell=2$ angular distribution from the $(^4 \textnormal{He}, ^3\textnormal{He})$ measurement of Ref.~\cite{El_Bedewi_1975} suggests $(0\text{-}4)^+$. The $0^+$ and $1^+$ assignments are ruled out by the $\gamma$-decay of this state to the $J^{\pi} = 2^+$ $1368$-keV and $J^{\pi} = 4^+$ $4122$-keV states observed in Ref.~\cite{Berkes_1964}. The results of this work indicate an $\ell=2$ transfer. Schmalbrock \textit{et. al} suggest that this state could be the analogue to a $T=1$ state with spin $3^+$ in $^{24}$Na \cite{schmalbrock_1983}. Based on these observations, and the satisfactory ability to describe the angular distribution with $\ell=2$, a $1d_{5/2}$ transfer was calculated, and is shown in Fig.~\ref{fig:11933_fit}.    

\subsubsection{The $11988$-keV State; $295$-keV Resonance}

As can be seen in Table \ref{tab:energy_comp}, the $11988$-keV State has been observed in multiple experiments, including the high precision $\gamma$-ray measurement of Ref.~\cite{endt_1990}. A spin parity of $2^{+}$ has been assigned based on the inelastic measurement of Ref.~\cite{zwieglinski_1978}. The current fit is shown in Fig.~\ref{fig:11988_fit} and assumes a mixed $\ell = 0+2$ transition with $2s_{1/2}$ and $1d_{5/2}$.

\subsubsection{The $12017$-keV State; $324$-keV Resonance}

The $12017$-keV state is known to have $J^{\pi}=3^-$, which was established from the angular distributions of Ref.~\cite{Kuperus_1963, Fisher_1963} and confirmed by the inelastic scattering of Ref.~\cite{zwieglinski_1978}. Our angular distribution is consistent with an $\ell=1$ transfer, which rules out the $j = 1/2$ possibility, giving a unique single particle state with $2p_{3/2}$. The fit is shown in Fig.~\ref{fig:12017_fit}.

\subsubsection{The $12051$-keV State; $359$-keV Resonance}

The angular distributions of Ref.~\cite{Fisher_1963} established $J^{\pi}=4^+$ for the $12051$-keV state, which was later confirmed by the inelastic scattering of Ref.~\cite{zwieglinski_1978}. The angular distribution of the present work is well described by a transfer of $1d_{5/2}$, which is shown in Fig.~\ref{fig:12051_fit}.   

\subsubsection{The $12183$-keV State; $491$-keV Resonance}

Ref.~\cite{MEYER_1972} observed that the $12183$-keV state $\gamma$-decays to $0^+$, $2^+$, and $1^{+}$ states, which permits values of $(1,2^{+})$. The angular distribution of Ref.~\cite{hale_2004} permits either $\ell = 0$ or $\ell = 0+2$ transfers, which requires the parity of this state be positive. The current work finds an angular distribution consistent with a pure $\ell=2$ transfer. The calculation of the $1d_{5/2}$ transfer is shown in Fig.~\ref{fig:12183_fit}.


\section{Proton Partial Widths}

The spectroscopic factors extracted in Section \ref{sec:spec_factors} are only an intermediate step in the calculation of the $^{23}$Na$(p, \gamma)$ reaction rate. As discussed in Section \ref{sec:calc_partial_widths}, $C^2S$ can be thought of as a scale factor that converts single-particle quantities to physical ones. From the proton spectroscopic factors of this work, proton partial widths can be calculated using Eq.~\ref{eq:proton_partial_width}. Additionally, in the case of a mixed $\ell$ transfer, the total proton width is calculated using:
\begin{equation}
    \label{eq:mixed_l_proton_width}
    \Gamma_p = \sum_{\ell} \Gamma_{p, \ell}.
\end{equation}
However, in the current case the $\ell=2$ single particle widths, $\Gamma_{sp}$, are typically two orders of magnitude lower than the $\ell=0$ ones, making them negligible in the calculations presented below.

\subsection{Bound State Uncertainties}

Before presenting the results of this work, it is important to discuss the uncertainties that could impact the determination of $\Gamma_p$. One of the largest is the bound state parameters used to define the overlap function. Since the overlap function is extremely sensitive to the choice of Woods-Saxon radius and diffuseness parameters, the extracted spectroscopic factor can vary considerably. This dependence has been discussed extensively in the literature, for a review, see Ref.~\cite{2014_Tribble}. Section \ref{sec:spec_factor_dis_cu} confirmed this strong dependence in a Bayesian framework. If the uncertainties of $C^2S$ are independent from those of $\Gamma_{sp}$, then single-particle transfer reaction experiments that determine spectroscopic factors will be unable to determine $\Gamma_p$ with the precision needed for astrophysics applications.

Ref.~\cite{bertone} noted an important consideration for the calculation of $\Gamma_p$ from $C^2S$ and $\Gamma_{sp}$. If these quantities are calculated using the \textit{same} bound state potential parameters, the variation in $C^2S$ is anticorrelated with that of $\Gamma_{sp}$. Thus, the product of these two quantities, i.e., $\Gamma_{p}$, has a reduced dependence on the chosen bound state potentials. Using the same bound state parameters for both parameters Refs.~\cite{hale_2001, hale_2004} found variations in $\Gamma_p$ of $\approx 5 \%$. With the Bayesian methods of this study, it is interesting to investigate whether this anticorrelation still holds in the presence of optical model uncertainties and using the bound state MCMC samples that are inherently correlated with all the model parameters.

\begin{figure}
    \centering
    \includegraphics[width=.8\textwidth]{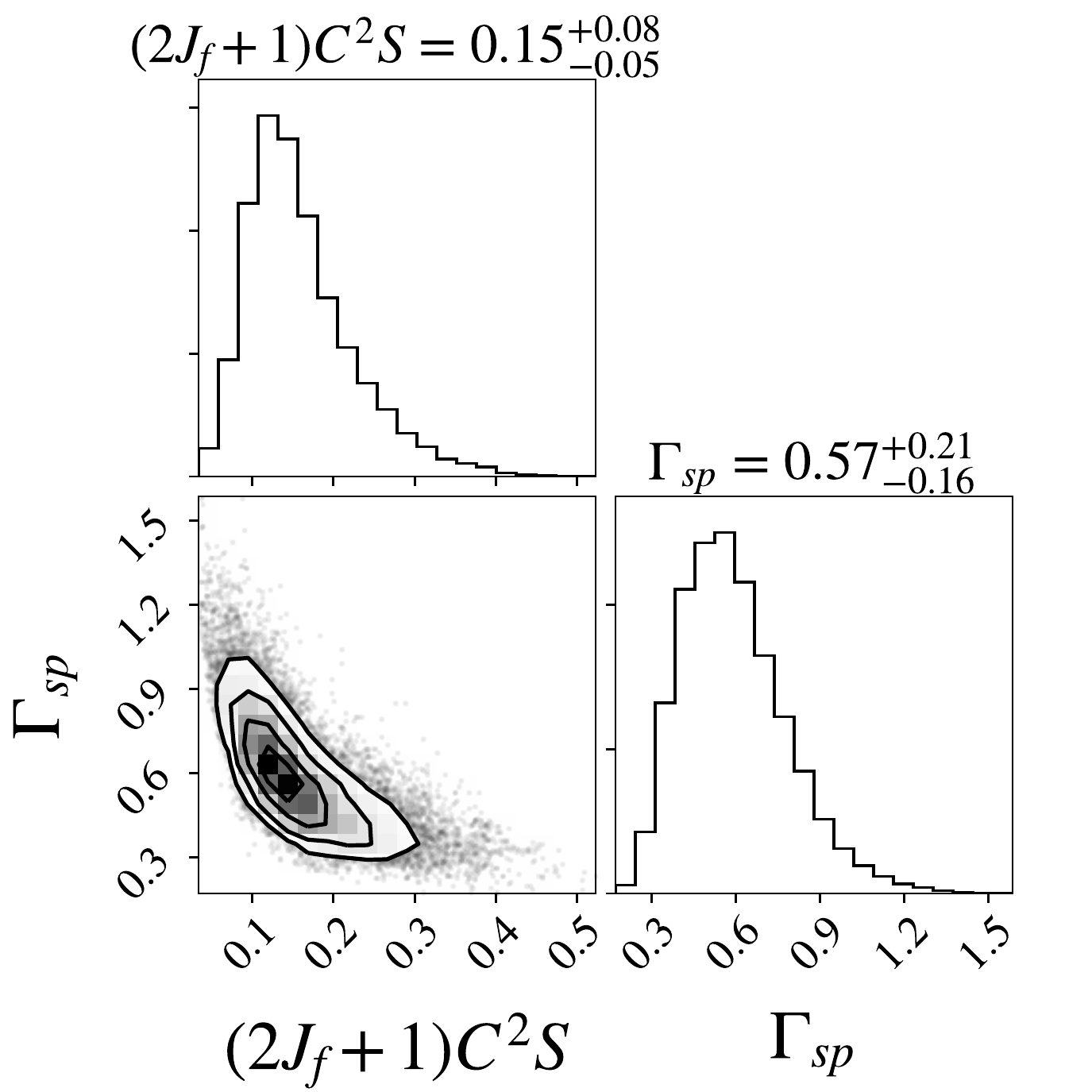}
    \caption{Pair correlation plot for the MCMC posterior samples of $\Gamma_{sp}$ and $(2J_f+1)C^2S$. A strong anticorrelation exists when the same bound state parameters are used to calculate both quantities.}
    \label{fig:corner_g_sp_sf}
\end{figure}

Modifications were made to the code \texttt{BIND} so that it could be run on a set of tens of thousands of bound state samples to produce a set of $\Gamma_{sp}$ samples. Due to the numerical instability of the algorithm for low energy resonances, the potential impact of the weak binding approximation, and the difficulties for mixed $\ell$ transitions, the state selected for this calculation need to have a $ 500 \gtrapprox E_r \gtrapprox 100$ keV, $\ell \geq 2$, and a known spin parity. The only such state is at $E_x = 12051$ keV ($E_r = 358$ keV). A new MCMC calculation was carried out using the same model as Eq.~\ref{eq:dwba_model_na} with the additional parameters for the bound state $r_0$ and $a_0$. These were given priors:
\begin{align}
    \label{eq:bound_state_priors}
    & r_0 \sim \mathcal{N}(1.25, 0.125^2) \\
    & a_0 \sim \mathcal{N}(0.65, 0.065^2). \nonumber 
\end{align}
The sampler was again run with $400$ walkers taking $8000$ steps. The final $2000$ steps were thinned by $50$ giving $16000$ posterior samples. These samples were then plugged into \texttt{BIND} to produce the $16000$ samples of $\Gamma_{sp}$. Since these samples all come directly from the MCMC calculation they naturally account for the variations in the optical model parameters as well as $C^2S$. First it is worth establishing the bound state parameters influence on the uncertainty of $C^2S$. The log-normal distribution well described this distribution and had a factor uncertainty of $f.u.=1.50$ increased from $f.u.=1.30$ in the case of fixed bound state parameters. The pair correlation plot for $(2J_f+1)C^2S$ versus $\Gamma_{sp}$ is shown in Fig.~\ref{fig:corner_g_sp_sf}. The resulting distribution gives $(2J_f+1)\Gamma_{p} = 0.083^{+0.025}_{-0.018}$ eV, while the value calculated using fixed bound state parameters gives $(2J_f+1)\Gamma_{p} = 0.082^{+0.025}_{-0.018}$ eV. The cancellation between the variation in $\Gamma_{sp}$ and $C^2S$ is nearly exact in this case, with the resulting uncertainty being $30 \%$ in both calculations. This relation still requires further study using Bayesian methods, particularly the influence of the bound state quantum numbers $n$ and $j$ which cannot be determined from the transfer data, but for the present work the potential influence of the bound state parameters on $\Gamma_p$ is considered negligible.     

\subsection{Subthreshold Resonances}

Three of the observed states lie close enough to the proton threshold to be astrophysically relevant. As mentioned in Section \ref{sec:calc_partial_widths}, $P_{\ell}$ and therefore $\Gamma_{sp}$ cannot be calculated for subthreshold states. Instead these resonances will be integrated by \texttt{RatesMC} using $\theta^2 = C^2S \theta_{sp}^2$. $\theta_{sp}^2$ can be calculated using the fits provided in either Ref.~\cite{ILIADIS_1997} or Ref.~\cite{BARKER_1998}. I have adopted the fit of Ref.~\cite{ILIADIS_1997}. It should be noted that this fit was derived using the bound state parameters $r_0 = 1.26$ fm and $a_0 = 0.69$ fm which differ from those used in this work. The impact of this difference was investigated by using higher lying states where values of $\theta_{sp}^2$ could also be calculated using \texttt{BIND}. The maximum observed deviation was $10 \%$, which is in decent agreement with the expected accuracy of the fit as mentioned in Ref.~\cite{ILIADIS_1997}. The values of $\theta^2$ for this work are shown in Table \ref{tab:subthresh_resonance_table}. Besides the $68 \%$ credibility intervals, the mean and standard deviation are also reported for the samples. These values correspond to Eq.~\ref{eq:lognormal_mean} and the square root of Eq.~\ref{eq:lognormal_variance}, respectively, and are the required inputs for \texttt{RatesMC}, which derives the log-normal parameters based on the mean and standard deviation.           

\begin{table*}
\centering
  \setlength{\tabcolsep}{10pt}
  \caption{\label{tab:subthresh_resonance_table} Reduced width calculations for the observed subthreshold resonances. All $\theta_{sp}^2$ values were calculated using the fit of Ref.~\cite{ILIADIS_1997} and should be considered to have a $10 \%$ systematic uncertainty. The $68 \%$ credibility intervals of the samples are presented in the fifth column, while the last column gives their mean and standard deviation, which are the inputs required by \texttt{RatesMC}.}
  \begin{tabular}{llllll}
    \toprule
    \toprule
    $E_x$(keV) & $E_r$(keV)  & $J^{\pi}$   & $\theta_{sp}^2$ & $(2J_f + 1) \theta^2 $ & $E[x]$   \\ \hline
    \\ [-1.5ex]
    
$11389.6(12)  $      &    $-303.1(12)$     &  $1^-$      &  $0.738$      &     $0.049^{+0.016}_{-0.011}$ & $0.051(14)$  \\ [0.8ex] 
$11452.9(4)   $      &   $-239.8(4)$       &  $2^+$      &  $0.654$      &     $0.09^{+0.03}_{-0.03}$ & $0.10(3)$           \\ [0.8ex]
$11521.1(14)  $      &   $-171.6(14)$      &  $2^+$ &  $0.639$  &    $0.035^{+0.018}_{-0.013}$ & $0.038(18)$             \\ [0.8ex]
    \bottomrule
    \bottomrule
  \end{tabular}
\end{table*}

\subsection{Resonances Above Threshold}

Eight resonances were observed above the proton threshold and below $500$ keV. Except for $E_r = 2$, all of the $\Gamma_{sp}$ values were calculated using \texttt{BIND}. \texttt{BIND} calculations were carried out with the Woods-Saxon potential parameters $r_0 = 1.25$ fm, $a_0 = 0.65$ fm, $r_c = 1.25$ fm, $V_{so} = 6.24$, and channel radius of $1.25$ fm. The low resonance energy of $E_r = 2$ presented numerical challenges for \texttt{BIND}, so it was calculated using the fit of Ref.~\cite{ILIADIS_1997}. The shift in energy from $5$ keV from Ref.~\cite{hale_2004} to $2$ keV in this work is also an excellent example of the extreme energy dependence of the partial widths. If this resonance has an energy of $5$ keV, the single particle width is on the order of $\Gamma_{sp} \approx 10^{-59}$ eV, while the updated energy gives $\Gamma_{sp} \approx 10^{-97}$ eV.   

\begin{table*}
\centering
  \setlength{\tabcolsep}{8pt}
  \caption{ \label{tab:gamma_p_table} Proton partial widths derived from this work. The values of $\Gamma_{sp}$ from \texttt{BIND} are listed for reference. $(2J_f+1)\Gamma_p$ values are given in terms of their $68 \%$ credibility intervals. Finally the mean and standard deviation are listed for use in \texttt{RatesMC} calculations.}
\begin{threeparttable}
  \begin{tabular}{llllll}
    \toprule
    \toprule
    $E_x$(keV) & $E_r$(keV)  & $J^{\pi}$   & $\Gamma_{sp}$(eV) & $(2J_f + 1) \Gamma_p $(eV) & $E[x]$  \\ \hline
    \\ [-1.5ex]
$11695(5)     $      &    $2(5)$           & $4^+$  &   $2.737 \times 10^{-97}$ $^{\dagger}$     &  $2.3^{+.7}_{-0.5} \times 10^{-98}$              & $2.4(6) \times 10^{-98}$ \\ [0.8ex]
    $11825(3)     $      &    $132(3)$         &   $\ell=0$     &   $9.785 \times 10^{-04}$     &   $2.3^{+1.2}_{-0.7} \times 10^{-5}$ & $2.6(12) \times 10^{-5}$             \\ [0.8ex]
          &             &   $\ell=1$     &   $2.072 \times 10^{-04}$     &   $2.0^{+0.8}_{-0.6} \times 10^{-6}$ & $2.2(8) \times 10^{-6}$             \\ [0.8ex]
          &             &   $\ell=2$     &   $4.425 \times 10^{-06}$     &   $6.0^{+2.1}_{-1.5} \times 10^{-8}$ & $6.3(20) \times 10^{-8}$             \\ [0.8ex]
          &             &   $\ell=3$     &   $5.492 \times 10^{-08}$     &   $1.4^{+0.5}_{-0.3} \times 10^{-9}$ & $1.4(4) \times 10^{-9}$             \\ [0.8ex]
$11860.8(14)  $      &    $168.1(14)$      & $1^-$  &   $5.894 \times 10^{-3}$ &  $1.3^{+0.4}_{-0.3} \times 10^{-4}$ & $1.4(4) \times 10^{-4}$                \\ [0.8ex]
$11933.06(19) $      &    $240.37(19)$     & $(2 \text{-} 4)^+$       &   $1.034 \times 10^{-2}$     &   $2.4^{+0.7}_{-0.5} \times 10^{-3}$ & $2.5(7) \times 10^{-3}$ \\ [0.8ex]
$11988.45(6)  $      &   $295.76(6)$       & $2^+$  &  $15.39$    &  $4.0^{+1.5}_{-1.1}$ & $4.2(15)$                      \\ [0.8ex]
$12016.8(5)   $      &    $324.1(5)$       & $3^-$       &  $8.550$      &  $1.7^{+0.5}_{-0.4}$ & $1.8(5)$              \\ [0.8ex]
$12051.3(4)   $      &     $358.6(4)$      & $4^+$       &   $6.141 \times 10^{-1}$     &  $8.2^{+2.5}_{-1.8} \times 10^{-2}$ & $8.6(24) \times 10^{-2}$               \\ [0.8ex]
$12183.3(1)   $      &    $490.6(1)$       &  $(1,2)^+$      &  $9.318$      &    $1.1^{+0.4}_{-0.3}$ & $1.2(4)$           \\ [0.8ex]

    \bottomrule
    \bottomrule
  \end{tabular}
\begin{tablenotes}
\item[$\dagger$] Calculated using $\theta_{sp}^2$ from the fit of Ref.~\cite{ILIADIS_1997} to avoid the numerical instability of \texttt{BIND} at $2$ keV. An additional $10 \%$ systematic uncertainty should be considered. 
\end{tablenotes}
  
\end{threeparttable}
\end{table*}

\subsection{Discussion}

The literature for $\omega \gamma$ and $\Gamma_p$ values is extensive. Ref.~\cite{hale_2004} compiled and corrected the measurements for stopping powers and target stoichiometry. Using the compiled values and the recent measurement of Ref.~\cite{BOELTZIG_2019}, a comparison can be made from the current work and previous measurements. The unknown spin and parity of many of these states makes direct comparison to $\omega \gamma$ subject to large uncertainties, so as an alternative $(2J_f + 1) \Gamma_p$ has been deduced from $\omega \gamma$ where possible. This of course requires knowledge of $\Gamma_{\gamma}/\Gamma$, which is only known for a select few of the observed resonances. I will now detail the information used for these resonances.

\subsubsection{$132$-keV Resonance}

The $132$-keV Resonance was measured directly for the first time at LUNA and is reported in Ref.~\cite{BOELTZIG_2019}. The value from that work is $\omega \gamma =  1.46^{+0.58}_{-0.53} \times 10^{-9}$ eV. Using $\Gamma_{\gamma}/\Gamma = 0.95(4)$ from Ref.~\cite{vermeer_1988} implies $(2J_f+1)\Gamma_p = 1.23^{+0.49}_{-0.45} \times 10^{-8}$ eV. The upper limit reported in Ref.~\cite{Cesaratto_2013} can also be used for comparison and yields $(2J_f+1)\Gamma_p \leq 4.35 \times 10^{-8}$ eV. The closest value from this work is the $\ell = 2$ transfer which gives $(2J_f+1)\Gamma_p = 6.3^{+2.1}_{-1.5} \times 10^{-8}$ eV. The disagreement between our value and that of LUNA is stark, and a significant amount of tension exists with the upper limit of Ref.~\cite{Cesaratto_2013}. 

\subsubsection{$168$-keV Resonance}

The $168$-keV Resonance has a proton width of $(2J_f+1)\Gamma_p = 1.83(39) \times 10^{-4}$ eV as discussed in Ref.~\cite{hale_2004}. This value is in good agreement with the current work $(2J_f+1)\Gamma_p = 1.3^{+0.4}_{-0.3} \times 10^{-4}$ eV.

\subsubsection{$240$-keV Resonance}

Using the resonance strength measured in Ref.~\cite{BOELTZIG_2019} of $\omega \gamma = 4.82(82) \times 10^{-4}$ eV and $\Gamma_{\gamma}/\Gamma > 0.7$ from Ref.~\cite{vermeer_1988}, $(2J_f+1)\Gamma_p$ has a lower limit of $3.86(66) \times 10^{-3}$ eV, which is in mild tension with the transfer value of $2.5(7) \times 10^{-3}$ eV.

\subsubsection{$295$-keV Resonance}

Ref.~\cite{BOELTZIG_2019} measured $\omega \gamma = 1.08(19) \times 10^{-4}$ eV, while Ref.~\cite{vermeer_1988} gives $\Gamma_{\gamma}/\Gamma = 0.70(9)$. In this case, $(2J_f+1)\Gamma_p = 1.2(2) $ eV. The current value is in significant disagreement with $(2J_f+1)\Gamma_p = 4.0^{+1.5}_{-1.1}$ eV.

\subsection{$490$-keV Resonance}

The $490$-keV Resonance is considered a standard resonance for the $^{23}$Na$(p, \gamma)$ reaction, and has a value of $9.13(125) \times 10^{-2}$ eV \cite{PAINE_1979}. Unfortunately, $\Gamma_{\gamma}/\Gamma$ is not known. However, an upper limit for $\omega \gamma_{(p, \alpha)}$ has been set at $\leq 0.011$ eV \cite{hale_2004}. The ratio of the two resonances strengths can set an upper limit for $\Gamma_{\alpha}/\Gamma_{\gamma}$:
\begin{equation}
    \label{eq:resonance_strength_ratio}
    \frac{\omega \gamma_{(p, \alpha)}}{\omega \gamma_{(p, \gamma)}} = \frac{\Gamma_{\alpha}}{\Gamma_{\gamma}}.
\end{equation}
Plugging in the values gives $\Gamma_{\alpha}/\Gamma_{\gamma} \leq 0.12 $. Assuming $\Gamma_p << \Gamma_{\gamma}$, $\Gamma_{\gamma}/\Gamma \geq 0.89$. The current value for $(2J_f+1)\Gamma_p = 1.1^{+0.4}_{-0.3}$ eV which can be compared to the upper limit of the standard resonance of $(2J_f+1)\Gamma_p = 0.821(112)$ eV. If we assume the $\alpha$ channel is completely negligible, $(2J_f+1)\Gamma_p = 0.730(100)$ eV. The standard resonance value appears to be consistent with the current work.

\subsubsection{Final Remarks}

The above comparisons make it clear that the agreement between the current experiment and previous measurements is inconsistent. Of particular concern are the $132$-keV and $295$-keV resonances, in which the disagreement is at a high level of significance. However, the measurement of Ref.~\cite{BOELTZIG_2019} at LUNA used the $295$-keV resonance as a reference during the data collection on the $132$-keV resonance, which could explain some correlation between those resonance strengths when compared to this work.  Furthermore, the updated resonance energy of $132$ keV compared to the previously assumed $138$ keV could move the beam off of the plateau of the target yield curve, but the magnitude of this effect is difficult to estimate. However, the measurement of Ref.~\cite{Cesaratto_2013} has an upper limit that is consistent with the LUNA value and is in tension with the current work. The upper limit also assumed the $138$-keV resonance energy, but appeared to use a much thicker target ($\approx 30$ keV) than the LUNA measurement ($\approx 15$ keV) making it less sensitive to the resonance energy shift. All of this discussion presupposes that the proton state has $\ell = 2$ and that our observed angular distribution arises completely from a direct reaction mechanism. If the spin is one of the other possible values, the current results will differ by over an order of magnitude, which could indicate the observed yields have significant contributions from a compound reaction mechanism.

\section{The $^{23}$Na$(p, \gamma)$ and $^{23}$Na$(p, \alpha)$  Reaction Rates}

It was briefly mentioned in the previous section that there exists a formidable amount of data relevant to the $^{23}$Na$(p, \gamma)$ reaction rate. The compiled values of Ref.~\cite{hale_2004} make up the majority of the current STARLIB rate. A detailed reanalysis of this rate is likely needed, but is well beyond the scope of the current work. For now two issues of relevance to globular cluster nucleosynthesis will be investigated:

\begin{enumerate}
    \item The impact of our recommended energies.  
    \item The potential impacts of the observed discrepancy in the proton partial widths for the $132$-keV resonance.  
\end{enumerate}

\subsection{Energy Update}

\begin{figure}
    \centering
    \includegraphics[width=.45\textwidth]{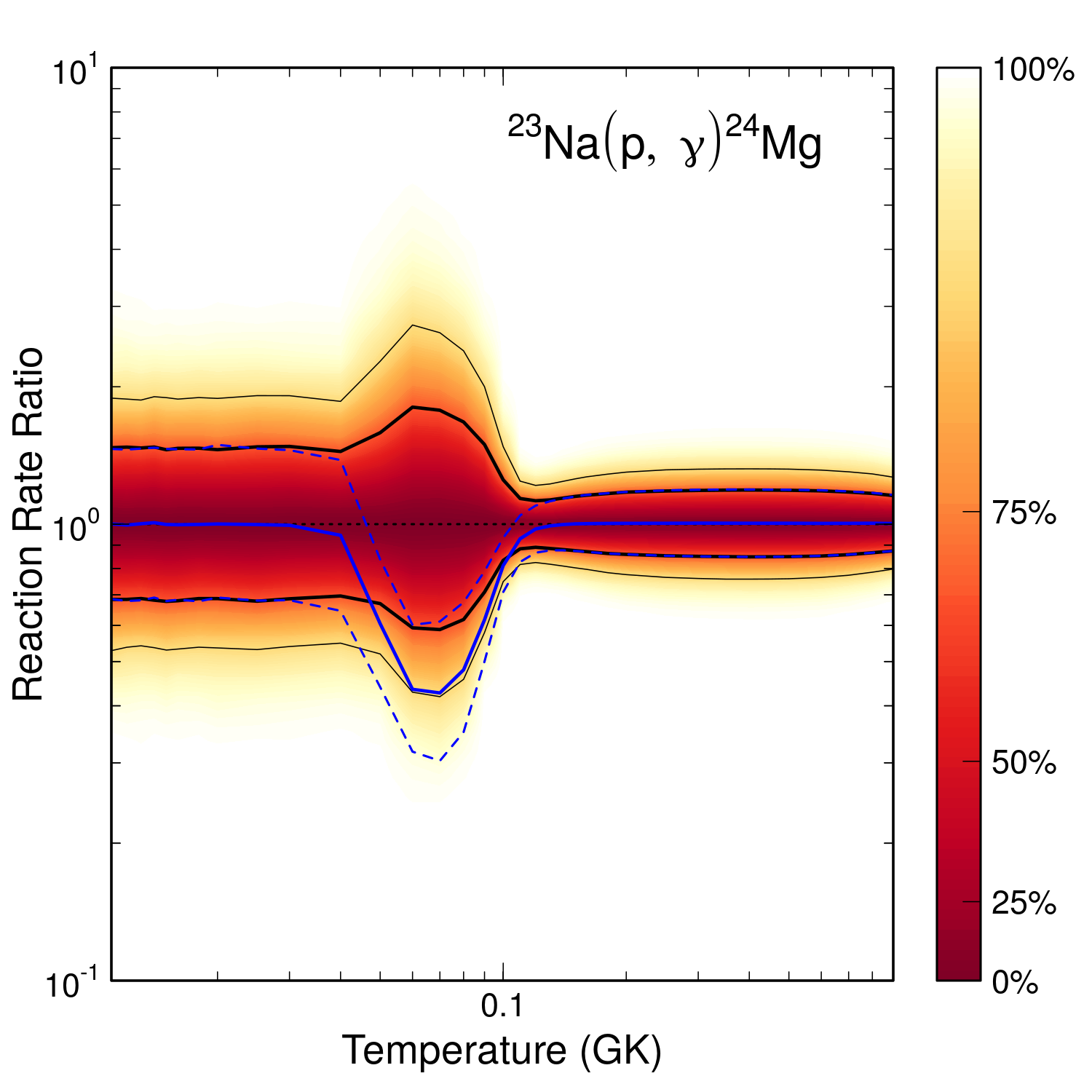}\includegraphics[width=.45\textwidth]{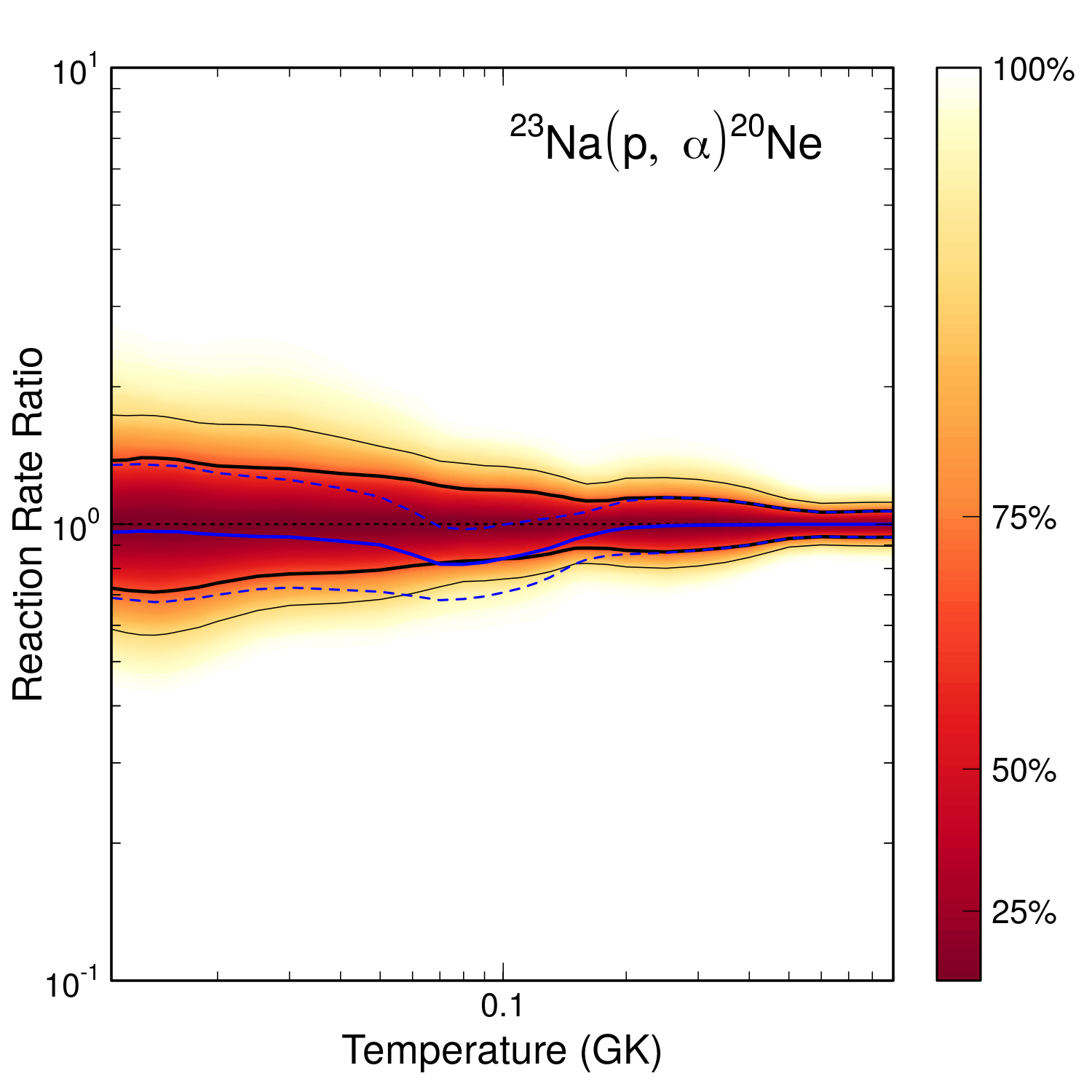}
    \caption{(left) The $(p, \gamma)$ reaction rate ratio. The rate has been normalized to its median, and the contours show the relative uncertainty as a function of temperature. The solid blue line is the recommend rate of Ref.~\cite{BOELTZIG_2019}, and the dashed blue lines show its $68 \%$ coverage. (right) The reaction rate ratio plot for the $(p, \alpha)$ rate. Again the blue line shows the previous rate.}
    \label{fig:compare_p_g_energy_update}
\end{figure}

The resonance energies presented in Table \ref{tab:recommened_energies} were substituted into the \texttt{RatesMC} input files provided by STARLIB for the $(p, \gamma)$ and $(p, \alpha)$ rates. Particle partial widths were scaled as needed to reflect the new energies. The resulting rates, normalized to their median are shown in Fig.~\ref{fig:compare_p_g_energy_update}. The blue lines show the $68 \%$ coverage of the rates as determined by LUNA \cite{BOELTZIG_2019}. The influence of the new energy for the $132$-keV resonance on the $(p, \gamma)$ rate can be clearly seen. Recall that the resonance energy enters the rate exponentially, and in this case the $5$-keV shift in energy is responsible for the rate increasing by a factor of $2.5$ for temperatures of $70 \text{-} 80$ MK. The impact of the new energies on the $(p, \alpha)$ rate are more modest. While there is a factor of $1.25$ increase, the new rate is well within the uncertainty of the previous rate.    

\begin{figure}
    \centering
    \includegraphics[width=\textwidth]{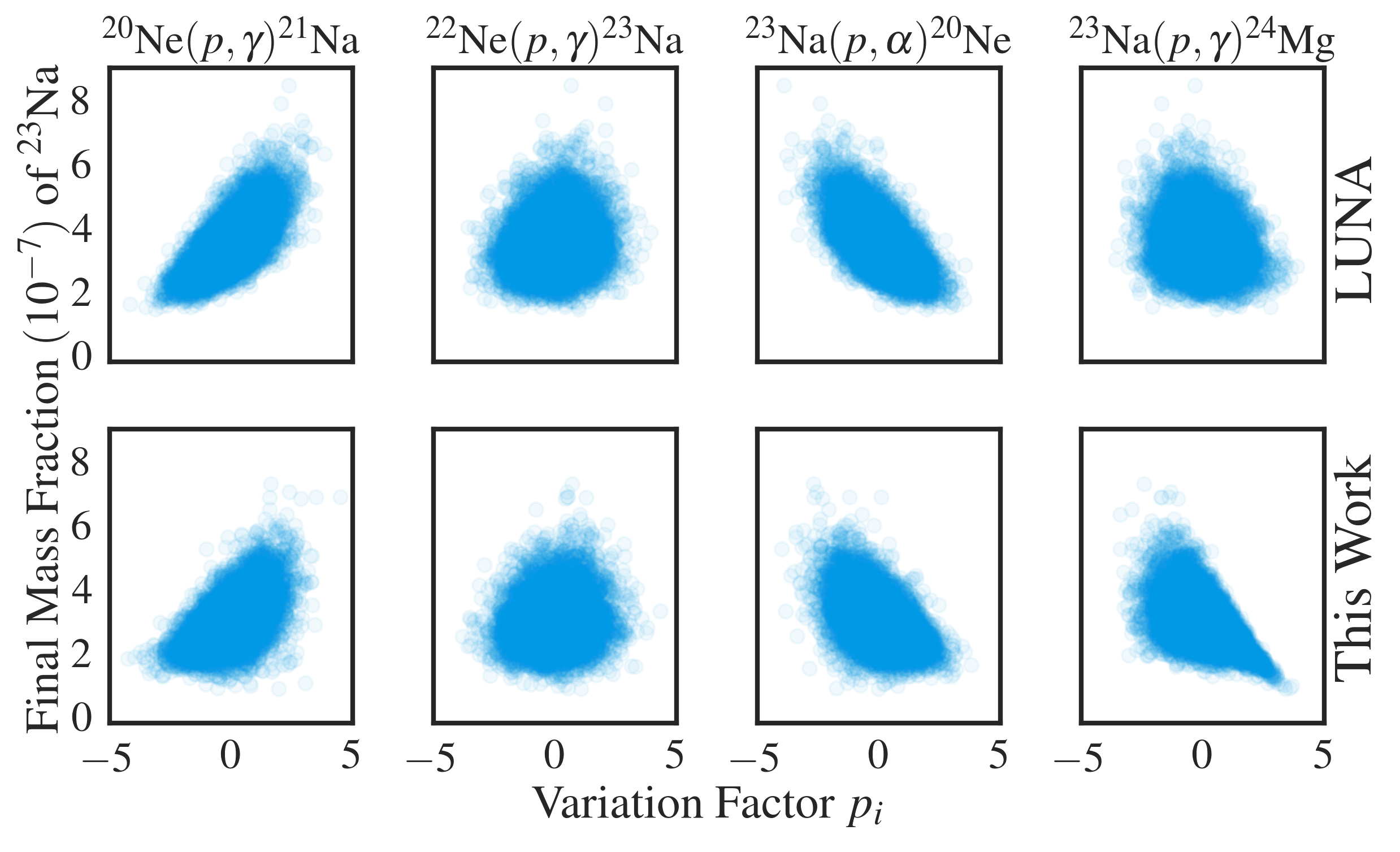}
    \caption{Pair correlation plots for the rate variation factor and the final mass fraction of sodium. The top row shows the four most influential reactions when using the rates of Ref.~\cite{BOELTZIG_2019}, while the bottom row uses the updated energies of this work. The new energy of the $132$-keV resonance in $^{23}$Na$(p, \gamma)$ is shown to dramatically increase the dependence of the final sodium abundance on the $^{23}$Na$(p, \gamma)$ rate.}
    \label{fig:na_dot_plot}
\end{figure}

\begin{figure}
    \centering
    \includegraphics[width=.6\textwidth]{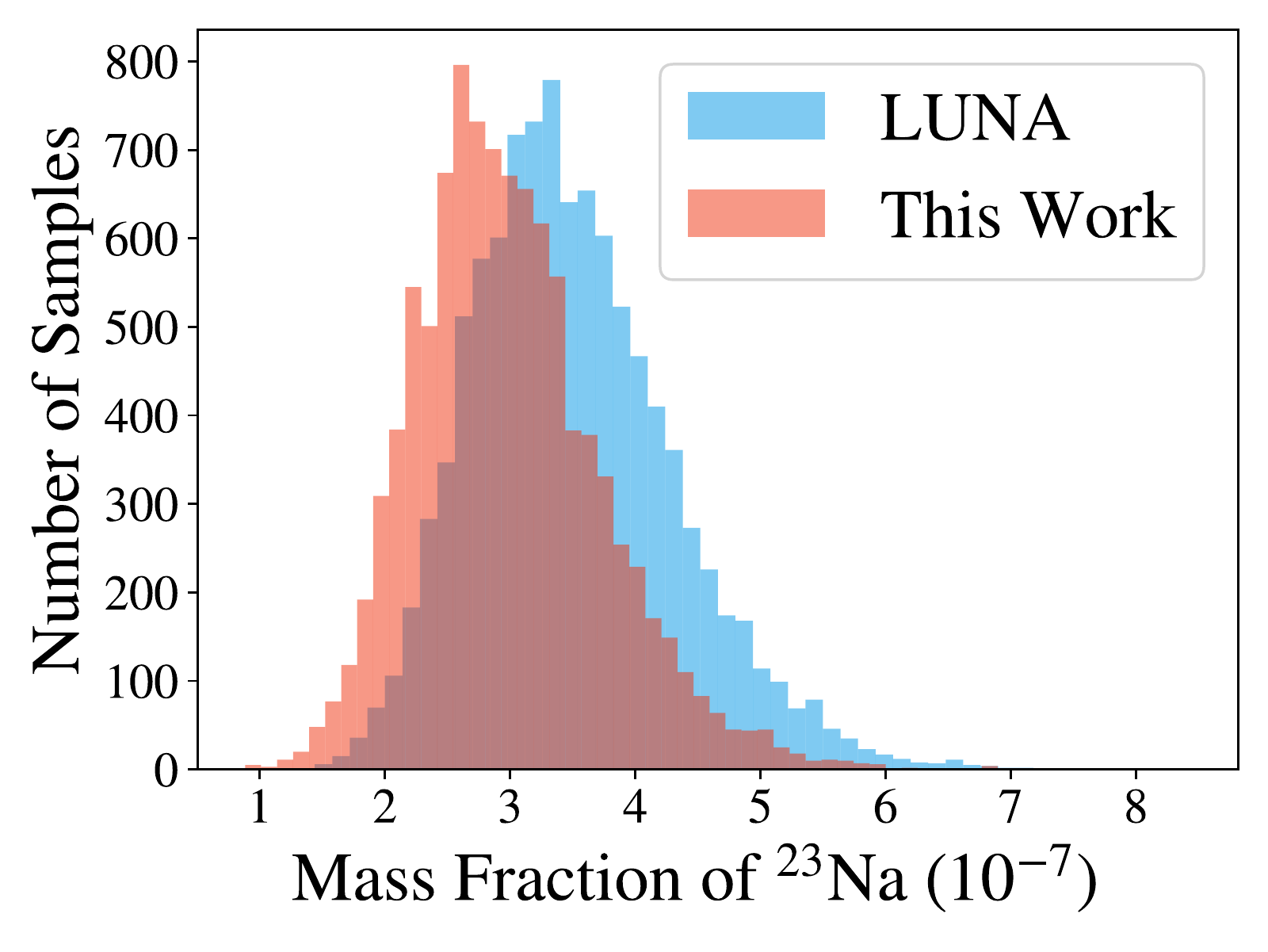}
    \caption{Histrograms of the samples from the Monte Carlo reaction network runs for both rates of Ref.~\cite{BOELTZIG_2019} and this work. While the spread of the two histograms is similar, the updated rate causes sodium to be destroyed more quickly.}
    \label{fig:na_abundance_comp}
\end{figure}

In order to assess the impact of the increase of the $(p, \gamma)$ rate, a Monte Carlo network calculation was carried out by integrating the updated rates into STARLIB. Remaining relatively agnostic towards the potential source of Na enrichment in globular clusters, a simple single zone calculation was carried out with $T = 75$ MK, $\rho = 10$ g/cm$^3$, and an initial composition taken from Ref.~\cite{iliadis_2016}. The initial mass fraction of H was $X_{\textnormal{H}} = 0.75 $, and the network was run until this mass fraction fell to $10 \%$ of its initial value. Two Monte Carlo runs were performed for the LUNA rates and the rates of this work, respectively. $10^4$ iterations were taken for each set of rates. The Spearman rank-order coefficient was used to identify correlations between the rate variation factor, $p_i$, and the final $^{23}$Na abundance. The four most influential reaction rates are shown in Fig.~\ref{fig:na_dot_plot}, notice the dramatic increase in correlation for the updated $^{23}$Na$(p, \gamma)$ rate, which is accompanied by a mild weakening of the correlation from $^{23}$Na$(p, \alpha)$ and $^{20}$Ne$(p, \gamma)$. Figure \ref{fig:na_abundance_comp} shows the distribution of the final mass fractions from the $10^4$ network runs. The peak of the histogram for the updated rates is shifted downward, with a comparable spread in $^{23}$Na to the samples using the LUNA rates. Both of these figures show that the increase in the $(p, \gamma)$ rate found in this work leads to an overall reduction in the final abundance of sodium. It is worth emphasizing that the majority of this change is just from the $5$-keV shift in the $132$-keV resonance.           

\subsection{Transfer Measurement as the Only Constraint on the $132$-keV Resonance}

\begin{figure}
    \centering
    \includegraphics[width=.6\textwidth]{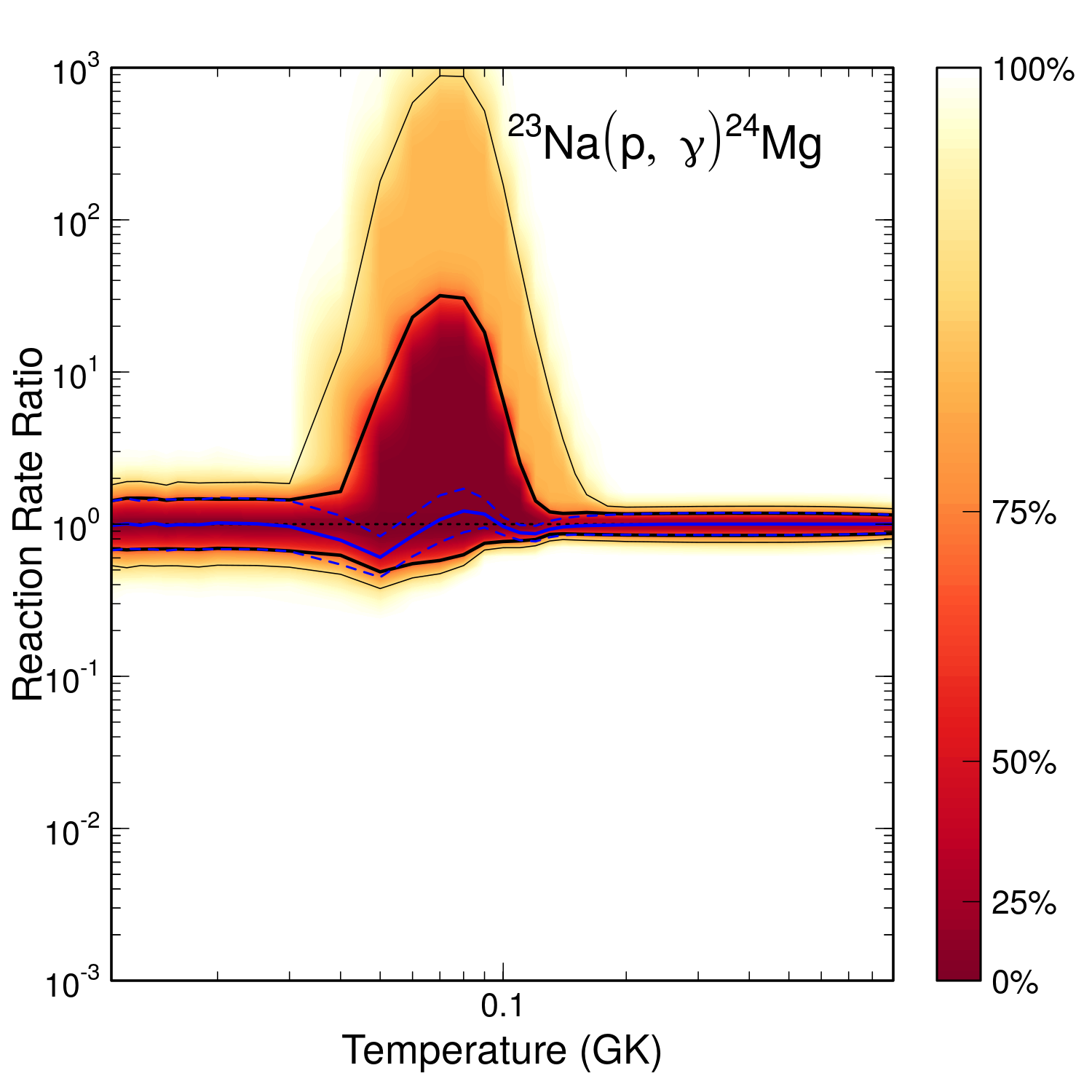}
    \caption{The $(p, \gamma)$ reaction rate ratio using just the constraints of the current work. The rate has been normalized to its median, and the contours show the relative uncertainty as a function of temperature. The solid blue line is the recommend rate of Ref.~\cite{BOELTZIG_2019}, and the dashed blue lines show its $68 \%$ coverage. The highly asymmetric distribution around $70$ MK is caused by the improbable but allowed $\ell=0,1$ transfers.}
    \label{fig:compare_p_g_l_update}
\end{figure}

The large discrepancy between the proton partial widths derived in this work compared to those inferred from Ref.~\cite{BOELTZIG_2019} is concerning, but ultimately the dependence of the current proton partial widths on the assumptions of DWBA does not provide a strong argument to discount the findings of the direct studies. However, it is instructive to further examine the results of this work in order to demonstrate how assigning probabilities to $\ell$ values impacts the reaction rate. If it was the case that the only available constraints were the proton partial widths derived in this work, then the probability of each transfer derived from the Bayesian model comparison calculation would have to be accounted for in the Monte Carlo reaction rate calculation. Fortunately, \texttt{RatesMC} can sample these probabilities as described in Ref.~\cite{2014_Mohr}. The result of plugging the mean probabilities given in Table \ref{tab:probs} into the rate calculation is shown in Fig.~\ref{fig:compare_p_g_l_update}. This figure shows the orders of magnitude uncertainty introduced by the distinct partial widths. The $\approx 1\%$ chance of an $\ell = 0$ transfer gives a factor of $1000$ potential increase in the reaction rate in the tails of the distribution. Note that the high asymmetry is caused by only considering $\ell =0 \text{-} 3$ transfers. This calculation demonstrates the vital role of direct measurements in nuclear astrophysics. The uncertainties inherent to transfer reactions can compound rapidly if a clear $\ell$ value cannot be determined for a single influential resonance.         

\section{Conclusions}

 Utilizing the high resolution capabilities of the TUNL SPS, astrophysically import excited states in $^{24}$Mg were observed. Careful calibration and compilation of previous results give a significantly lower resonance energy for the $132$-keV resonance. This resonance has the single largest contribution to the $(p, \gamma)$ reaction rate at temperatures important for globular cluster nucleosynthesis. Angular distributions were analyzed using the Bayesian DWBA methods of Chapter \ref{chap:bay_dwba}, and spectroscopic factors were extracted. This chapter has presented an analysis that is the first of its kind, where Bayesian methods were used to accurately determine uncertainties at every step of the transfer reaction analysis. The astrophysical impact of these uncertainties was briefly investigated. While there is still a need for an updated rate evaluation, the results of this experiment indicate that significant uncertainties are still present in both the $^{23}$Na$(p, \gamma)$ and $^{23}$Na$(p, \alpha)$ reaction rates, and as a result our knowledge of the Na-O anticorrelation in globular clusters is still limited by the nuclear physics.

%% file: Chapter-7/Chapter-7.tex
\chapter{Summary and Conclusions}
\label{chap:potassium}

The Na-O and K-Mg elemental abundance anomalies in globular clusters remain open questions. The astrophysical site that produces these signatures is still unknown, and as a result indicates that our current theories of stellar evolution are incomplete. Chapter 1 provided a brief introduction to these phenomena, focusing on the history of the astronomical observations. The continued improvement in observational techniques has revealed that these anomalies are a result of globular clusters containing multiple stellar population, with the oldest population of stars enriching approximately $30 \%$ of the newest generation by some unknown mechanism. While the burning site responsible for the Na-O and K-Mg are most likely different, our current understanding of globular cluster nucleosynthesis is hampered by our imprecise knowledge of both the sodium and potassium destroying reactions $^{23}$Na$(p, \gamma)$ and $^{39}$K$(p, \gamma)$, respectively. 

Chapter 2 outlined how stellar burning processes are linked to nuclear properties that can be measured in the laboratory. At the low temperatures thought to characterize the pollution sites of globular clusters, thermonuclear reaction rates are dominated by resonant reactions. The coulomb barrier makes direct study of these reactions at the very least difficult, and frequently, as is the case of the lowest lying resonances, impossible. By connecting resonance parameters to nuclear structure, it becomes possible to constrain these rates using transfer reactions. In particular, transfer reaction can constrain the energy, spin parity, and particle partial widths of astrophysically relevant resonances.

Thermonuclear reactions rates are dependent on experimentally measured quantities, and Chapter 3 reviewed the propagation of experimental uncertainties through reaction rate calculations using Monte Carlo techniques. By collecting a network of reaction rates and their associated uncertainties, the impact of individual rates can be assessed. This procedure was demonstrated with the results of a reevaluation of the $^{39}$K$(p, \gamma)$ rate. By including all known experimental data, this rate was shown to have uncertainties that are too large to make precise predictions required to study the K-Mg abundance anomaly. Future experimental studies are required to better determine this rate.        

Chapter 4 provided an overview of the TUNL tandem lab. The author's efforts in recommissioning the Split-pole spectrograph and the accompanying focal plane detector were discussed. This work provided the opportunity to study transfer reaction with excellent energy resolution, a necessity for nuclear astrophysics. Chapter 5 detailed two Bayesian techniques that accurately quantify the uncertainties from spectrograph experiments. The energy calibration method accounts for deviations in the data from the fit function, which has long been an issue for spectrograph measurements. These additional uncertainties are critical when deducing excitation energies that will be used to calculate resonance energies for astrophysically important states. Bayesian DWBA makes it possible to extract spectroscopic factors and assign $\ell$ values taking into account the uncertainties arising from the phenomenological optical potentials used by the theory.

Finally, all of the above techniques were combined to indirectly study the $^{23}$Na$(p, \gamma)$ reaction using $^{23}$Na$(^3\textnormal{He}, d)$. The single most import resonance in $^{23}$Na$(p, \gamma)$ for globular cluster nucleosynthesis has previously been determined to lie at $138$ keV. The excellent energy resolution of the Split-pole spectrograph allowed the state corresponding to this resonance to be observed at multiple angles. A thorough energy calibration was carried out using updated values and the Bayesian method of Chapter 5. These considerations resulted in a new suggested resonance energy of $E_r = 132(3)$ keV, which is over $5$ keV lower than previously suggested. Bayesian DWBA was carried out for 13 states close to the proton threshold. Spectroscopic factors were generally found to be in agreement with Ref.~\cite{hale_2004}. Probabilities were assigned to each allowed $\ell$ transfer for the $132(3)$-keV resonance. Proton partial widths were derived, and while decent agreement was found for some states, the $132$-keV resonance was found to be, at best, a factor of three different from those Ref.~\cite{BOELTZIG_2019}. The reason for this discrepancy is unclear at this time. The reaction rate for both $^{23}$Na$(p, \gamma)$ and $^{23}$Na$(p, \alpha)$ were updated for the energies of this work. Modest changes in the $^{23}$Na$(p, \alpha)$ rate were observed, while a factor of $2.5$ increase in the median rate of the $^{23}$Na$(p, \gamma)$ reaction was determined. This increase is due almost solely to the new recommended energy of the $132$-keV resonance. Finally, it was shown that the constraints from the transfer reaction measurement are insufficient to precisely determine the rate because of the ambiguous $\ell$ assignment for the $132$-keV resonance. These large uncertainties underscore the need for accurate direct measurements, while also indicate a need for a better understanding of the uncertainties in indirect studies.

Several directions exist for future work. The Bayesian DWBA methods of Chapter \ref{chap:bay_dwba} are still in their infancy. The assumptions on the prior distributions at this time are motivated primarily to account for global studies while making the problem computationally tractable. However, deriving these global values using a Bayesian method would allow future transfer measurements to more accurately account for parameter uncertainties. The discrepancy between the direct measurements and the partial widths extracted in Chapter \ref{chap:sodium} is of particular concern. Systematic studies that investigate the reliability of estimating partial widths from transfer reactions are needed. The disagreement between the excitation energies of this work and the study of Ref.~\cite{hale_2004} requires further study using independent methods. An attractive option is a particle-$\gamma$ coincidence study to provide precise energies with different systematic effects than those of the spectrograph measurements. If the resonance energy deduced in this work is found to be reliable, then further direct studies are likely needed to verify the results of measurements carried out which assumed the previous, higher resonance energy. Finally, the $^{39}$K$(^{3}\textnormal{He}, d)$ reaction has been successfully carried out at TUNL, and analysis of this experiment will provide useful constraints on the rate as discussed in Chapter 3.         

%% file: Appendix-A/Appendix-A.tex
\chapter{Comments on the Measurement of Vermeer}
\label{chap:rant_on_energies}

The measurement of Ref.~\cite{vermeer_1988} has potential issues that make it difficult to properly assess its conclusions. Though it was ultimately included in our energy compilation, the following observations should be noted. 

The study of Ref.~\cite{vermeer_1988} performed a measurement of $^{12}$C$(^{16}$O$, \alpha)^{24}$Mg at several different beam energies using a silicon surface barrier detector to detect the heavy recoils and a MDM-2 spectrograph to detect the $\alpha$ particles. The energies reported in that study come from the focal plane detector in the MDM-2 spectrograph. However, the states used to calibrate this detector are not mentioned. The only mention of the process of energy calibration is:
\begin{quote}
The centroids of clearly-resolved peaks were used for energy
calibration. Typically about 10 or 15 peaks were used to derive a fourth order polynomial for excitation energy as a function of channel number. The excitation energies used for calibration
purposes are those of ref. 13.   
\end{quote}
Their \say{ref 13} is the compilation of Endt (Ref.~\cite{ENDT_1978}). We are left with the impression that a large number of calibration states were used, with different states being selected for each beam energy. What is unclear is if these states were included in the reported energies.  As was mentioned in Section \ref{sec:energy_level_update}, calibration states should never be considered independent measurements. In the absence of information on which states were used for calibration, it was assumed that the calibration states were properly excluded from the reported values, and as a consequence it was included in the compilation of the present study. However, until the calibration states used in the study are clearly determined, the reported values should be viewed with some skepticism.